\documentclass[11pt,a4paper,twoside]{report}

\usepackage[ngerman,english]{babel}

\usepackage{dcolumn}                                                 

\usepackage{amsfonts}

\usepackage{amsmath}                                            

\usepackage{amssymb}

\usepackage{amsthm}

\usepackage{amsbsy}

\usepackage{bbm}                                                                    

\usepackage{color}

\usepackage{slashed}

\usepackage{colonequals}

\usepackage{multirow, booktabs}

\usepackage{graphicx}

\usepackage{afterpage}

\usepackage{exscale}

\usepackage{subfigure} 

\usepackage{axodraw4j}                                                     

\usepackage{pstricks}                                                   

\usepackage{cancel}

\usepackage{a4wide}

\usepackage{array}
\newcolumntype{x}[1]{>{\raggedleft\hspace{0pt}}p{#1}}

\usepackage[square,comma,numbers,sort&compress]{natbib}

\usepackage[small,labelfont=bf,margin=20pt,tableposition=top]{caption}
\usepackage{fancyhdr}
\usepackage[ps2pdf=true,colorlinks,pdfpagelabels]{hyperref}
\usepackage[figure,table]{hypcap}                                  
\hypersetup{
   bookmarksnumbered,
   pdfstartview={FitV},
   pdfpagemode={UseOutlines},
   pdfauthor={Kai Schmitz},
   pdftitle={The B-L Phase Transition -- Implications for Cosmology and Neutrinos},
   pdfsubject={Doctoral Thesis},
   citecolor={blue},
   linkcolor={blue},
   urlcolor={blue},
   filecolor={blue}}

\newcommand{\BmL}{$B$$-$$L$\ }

\allowdisplaybreaks[1]  

\begin{document}


\title{
\vspace*{-4.55cm}
{\normalsize\normalfont DESY-THESIS-2012-039
\hfill\mbox{}\\Juli 2012
\hfill\mbox{}\\}
\vspace{2.5cm}
\begin{huge}\textbf{The \BmL Phase Transition}\end{huge}\\[4mm]
\begin{LARGE}\textbf{Implications for Cosmology and Neutrinos}\end{LARGE} \vspace{10mm}}

\author{
\begin{LARGE}\textbf{Dissertation}\end{LARGE}\\[8mm]
\begin{LARGE}\textbf{zur Erlangung des Doktorgrades}\end{LARGE}\\[8mm]
\begin{LARGE}\textbf{des Departments Physik}\end{LARGE}\\[8mm]
\begin{LARGE}\textbf{der Universität Hamburg}\end{LARGE}\\[16mm]
\begin{Large}\textbf{vorgelegt von}\end{Large}\\[6mm]
\begin{Large}\textbf{Kai Schmitz}\end{Large}\\[6mm]
\begin{Large}\textbf{aus Berlin}\end{Large}\\[16mm]
\begin{Large}\textbf{Hamburg}\end{Large}\\[6mm]
\begin{Large}\textbf{2012}\end{Large}\vspace*{-1cm}}

\date{}

\pdfbookmark[0]{Titlepage}{title}

\maketitle


\newpage
\thispagestyle{empty}
\section*{}
\vfill

\begin{tabbing}
\hspace{8.5cm} \=\kill
Gutachter der Dissertation:
\> Prof.\ Dr.\ Wilfried~Buchm\"uller \\[1mm]
\> Prof.\ Dr.\ Mark~Hindmarsh \\[1mm]
\> Prof.\ Dr.\ G\"unter~Sigl \\[3mm]
Gutachter der Disputation:
\> Prof.\ Dr.\ Wilfried~Buchm\"uller \\[1mm]
\> Prof.\ Dr.\ Jan~Louis \\[3mm]
Datum der Disputation:
\> 9.\ Juli 2011 \\[3mm]
Vorsitzender des Prüfungsausschusses:
\> Prof.\ Dr.\ Georg~Steinbr\"uck\\[3mm]
Vorsitzender des Promotionsausschusses:
\> Prof.\ Dr.\ Peter~Hauschildt\\[3mm]
Dekan der Fakultät für Mathematik, \> \\
Informatik und Naturwissenschaften:
\> Prof.\ Dr.\ Heinrich~Graener
\end{tabbing}

\newpage


\renewcommand\abstractname{Abstract}

\noindent\begin{minipage}[t]{\linewidth}
\phantomsection
\addcontentsline{toc}{chapter}{Abstract}

\begin{abstract}


We investigate the possibility that the hot thermal phase of the early universe
is ignited in consequence of the \BmL phase transition, which represents the cosmological
realization of the spontaneous breaking of the Abelian gauge symmetry associated with
$B$$-$$L$, the difference between baryon number $B$ and lepton number $L$.
%
%
Prior to the \BmL phase transition, the universe experiences a stage of hybrid inflation.
Towards the end of inflation, the false vacuum of unbroken \BmL symmetry decays, which entails
tachyonic preheating as well as the production of cosmic strings.
Observational constraints on this scenario require the \BmL phase transition
to take place at the scale of grand unification.
The dynamics of the \BmL breaking Higgs field and the \BmL gauge degrees
of freedom, in combination with thermal processes, generate an abundance
of heavy (s)neutrinos.
These (s)neutrinos decay into radiation, thereby reheating the universe, generating the
baryon asymmetry of the universe and setting the stage for the thermal production of gravitinos.
The \BmL phase transition along with the (s)neutrino-driven reheating process
hence represents an intriguing and testable mechanism to generate the initial conditions
of the hot early universe.
%
%
We study the \BmL phase transition in the full supersymmetric Abelian Higgs model,
for which we derive and discuss the Lagrangian in arbitrary and unitary gauge.
As for the subsequent reheating process, we formulate the complete set of
Boltzmann equations, the solutions of which enable us to give a detailed and time-resolved
description of the evolution of all particle abundances during reheating.
Assuming the gravitino to be the lightest superparticle (LSP), the requirement of consistency
between hybrid inflation, leptogenesis and gravitino dark matter implies relations between
neutrino parameters and superparticle masses, in particular a lower bound on the gravitino
mass of $10\,\textrm{GeV}$.
As an alternative to gravitino dark matter, we consider the case of very heavy gravitinos,
which are motivated by hints for the Higgs boson at the LHC.
We find that the nonthermal production of pure wino or higgsino LSPs,
i.e.\ weakly interacting massive particles (WIMPs), in heavy gravitino
decays can account for the observed amount of dark matter,
while simultaneously fulfilling the constraints imposed by
primordial nucleosynthesis and leptogenesis, within a range of LSP,
gravitino and neutrino masses.
%
%
Besides its cosmological implications, the spontaneous breaking of \BmL
also naturally explains the small observed neutrino masses via the seesaw mechanism.
Upon the seesaw model we impose a flavour structure of the Froggatt-Nielson type which,
together with the known neutrino data, allows us to strongly constrain yet
undetermined neutrino observables.


\end{abstract}

\addtocounter{page}{2}
\cleardoublepage
\end{minipage}
\newpage\mbox{}\thispagestyle{empty}\newpage


\noindent\begin{minipage}[t]{\linewidth}
{\selectlanguage{ngerman}
\phantomsection
\addcontentsline{toc}{chapter}{Zusammenfassung}

\begin{abstract}


Wir untersuchen die M\"oglichkeit, dass die hei\ss e thermische Phase
des fr\"uhen Universums in Folge des \BmL Phasen\"ubergangs, welcher die kosmologische
Umsetzung der spontanen Brechung der mit $B$$-$$L$, der Differenz von
Baryonenzahl~$B$ und Leptonenzahl~$L$, verkn\"upften Abelschen Eichsymmetrie
darstellt, enz\"undet wird.
%
%
Vor dem \BmL Phasen\"ubergang durchlebt das Universum einen Abschnitt der Hybridinflation,
gegen deren Ende das falsche Vakuum ungebrochener \BmL Symmetrie zerf\"allt,
was tachyonisches Vorheizen sowie die Produktion kosmischer Strings nach sich zieht.
Aus Beobachtungen gewonnene Einschr\"ankungen dieses Szenarios erfordern es, dass der
\BmL Phasen\"ubergang bei der Skala der Gro\ss en Vereinheitlichung stattfindet.
Die Dynamik des \BmL brechenden Higgsfeldes und der \BmL Eichfreiheitsgrade, zusammen
mit thermischen Prozessen, generiert ein Vorkommen an schweren (S)neutrinos.
Diese (S)neutrinos zerfallen in Strahlung, wodurch sie das Universum aufheizen,
die Baryonenasymmetrie des Universums erzeugen und der thermischen Produktion von Gravitinos
den Weg ebnen.
Der \BmL Phasen\"ubergang stellt folglich mitsamt dem (S)neutrino-getriebenen Aufheizprozess
einen \"uberzeugenden und testbaren Mechanismus zur Erzeugung der Anfangsbedingungen des
hei\ss en fr\"uhen Universums dar.
%
%
Wir studieren den \BmL Phasen\"ubergang im vollst\"andigen supersymmetrischen
Abelschen Higgsmodel, f\"ur welches wir die Lagrangedichte in beliebiger und unit\"arer
Eichung herleiten und diskutieren.
In Hinblick auf den anschlie\ss enden Aufheizprozess formulieren wir den kompletten
Satz an Boltzmanngleichungen, deren L\"osungen uns zu einer detaillierten
und zeitaufgel\"osten Beschreibung aller Teilchenh\"aufigkeiten 
verhelfen.
Angenommen, das Gravitino ist das leichteste Superteilchen (LSP), so impliziert die
Forderung nach Konsistenz zwischen Hybridinflation, Leptogenese und Gravitino-Dunkler-Materie
Beziehungen zwischen Neutrinoparametern und Superteilchenmassen, insbesondere eine untere
Schranke an die Gravitinomasse von $10\,\textrm{GeV}$.
Als Alternative zu Gravitino-Dunkler-Materie betrachten wir den Fall
sehr schwerer Gravitinos, die durch Hinweise auf das Higgs-Boson am LHC motiviert sind.
Wir stellen fest, dass die nichtthermische Produktion reiner Wino- oder Higgsino-LSPs,
d.h.\ schwach wechselwirkender massereicher Teilchen (WIMPs), in den Zerf\"allen schwerer
Gravitinos f\"ur die beobachtete Menge an Dunkler Materie innerhalb einer Bandbreite von
LSP-, Gravitino- und Neutrinomassen aufkommen und zugleich den von primordialer
Nukleosynthese und Leptogenese auferlegten Einschr\"ankungen gen\"ugen kann.
%
%
Abgesehen von ihren kosmologischen Auswirkungen, erkl\"art die spontane \BmL Brechung
auch in nat\"urlicher Weise die kleinen beobachteten Neutrinomassen verm\"oge des
Seesaw-Mechanismus.
Wir erlegen dem Seesaw-Model eine Flavour-Struktur vom Froggatt-Nielsen-Typ auf,
welche es uns zusammen mit den bekannten Neutrinodaten erlaubt, bislang unbestimmte
Neutrinoobservablen stark einzuschr\"anken.


\end{abstract}

}
\addtocounter{page}{4}
\cleardoublepage
\end{minipage}
\newpage\mbox{}\thispagestyle{empty}\newpage
\selectlanguage{english}


\cleardoublepage

\phantomsection
\addcontentsline{toc}{chapter}{Contents}
\tableofcontents

\clearpage

\phantomsection
\addcontentsline{toc}{chapter}{List of Figures}
\listoffigures


\clearpage
\pagenumbering{arabic}

\chapter{Introduction}


The cosmic microwave background (CMB) radiation \cite{Komatsu:2010fb}
and the primordial abundances of the light nuclei \cite{Nakamura:2010zzi}
provide direct evidence for a hot thermal phase in the early universe.
While the CMB represents a full-sky picture of the hot early universe close to
its minimal temperature, primordial nucleosynthesis allows us to
probe the history of the universe up to the first tenth of a second after the
big bang.
Going further back in time, beyond the generation of the light elements, the theoretical
extrapolation becomes increasingly uncertain.
Up to temperatures slightly above the electroweak scale, we are still able to make an
educated guess about the evolution of the universe based on the established and well-tested
physics of the standard model of particle physics.
At temperatures around the scale of quantum chromodynamics (QCD), we thus
expect the occurrence of a phase transition, in the course of which
quarks and gluons become confined into hadrons.
Similarly, one presumes a phase transition around the electroweak scale, which causes
the Higgs boson, the electroweak gauge bosons as well as all fermions expect
for neutrinos to acquire a mass via the Higgs mechanism.
If indeed realized in the early universe, the electroweak phase transition would
correspond to the cosmological implementation of electroweak symmetry breaking.
Meanwhile, in anticipation of new insights from observations and experiments as to
the physics beyond the standard model, we are at present merely able to speculate about
the nature of the processes taking place at even higher energy scales.
While the conclusive identification of a successor to the standard model
is still pending, we know for sure that some processes must occur in the very
early universe, which cannot be accounted for by the known laws of physics.


A clear indication for physics beyond the standard model is the present composition
of the universe \cite{Nakamura:2010zzi}.
First of all, it is astonishing that all matter in the universe which can be more or less
well described by standard model physics seems to be almost exclusively made out of baryons
and hardly out of antibaryons.
This cosmic asymmetry between matter and antimatter calls for a nonequilibrium process
in the hot early universe, in which a primordial baryon asymmetry is generated before
baryons and antibaryons decouple from the hot plasma.
Furthermore, as ordinary matter contributes with only $5\,\%$ to the energy budget of
the universe, we are led to the conclusion that $95\,\%$ of the energy of the universe
reside in unknown forms of matter and energy,\textit{ viz.}\ dark matter and dark energy.
Dark matter, which encompasses $27\,\%$ of the total energy of the universe,
is able to clump under the influence of the gravitational force and thus plays
a crucial role in the theory of structure formation.
Studies of the cosmic density perturbations, seen for instance in the
CMB \cite{Komatsu:2010fb} or the distribution of galaxies in the neighbourhood
of the Milky Way \cite{Percival:2009xn}, imply that dark matter has to be present
in the early universe long before the end of the hot thermal phase.
Barring a modification of general relativity, the remaining
$73\,\%$ of the energy of the universe have to be attributed to some form of dark energy,
which is commonly identified as the energy of the vacuum and as such explained in terms
of a cosmological constant $\Lambda$.
Further evidence for new physics derives from the properties of the CMB.
The minute temperature anisotropies of the CMB exhibit correlations on scales
exceeding the sound horizon at the time of photon decoupling and thus point to
a mechanism in the very early universe capable of generating primordial metric
fluctuations with super-horizon correlations.
Finally, on the particle physics side, the flavour oscillations
among the three standard model neutrino species \cite{Fukuda:2000np,Ahmad:2001an}
represent the clearest evidence for physics beyond the standard model.
These oscillations indicate that neutrinos have tiny, but nonzero masses,
although the standard model stipulates them to be massless.


The most popular solution to the problem of the primordial density perturbations
as well to other puzzles related to the initial conditions of big bang cosmology
is inflation \cite{Guth:1980zm,Linde:1981mu,Albrecht:1982wi}---a stage of accelerated expansion
in the very early universe driven by the energy of the vacuum.
During inflation, the quantum fluctuations of a scalar field, the so-called inflaton field,
are stretched to super-horizon scales, whereupon they freeze, remaining basically unchanged in
shape until the onset of structure formation.
The dynamics of the inflaton field correspond to those of an ensemble of inflaton
particles in a coherent quantum state at zero temperature.
Assuming inflation to be the source of the primordial perturbations, one arrives at the
question as to the origin of the entropy of the hot plasma filling the universe during
the thermal phase.
In summary, we conclude that contemporary particle cosmology faces the task to
explain the origin of the hot early universe as well as its initial conditions,
i.e.\ the entropy of the thermal bath, the primordial baryon asymmetry and the abundance
of dark matter.


In this thesis, we put forward the idea that the emergence of the hot thermal
universe might be closely related to the decay of a false vacuum of unbroken \BmL symmetry,
where \BmL denotes the difference between baryon number $B$ and lepton number $L$.
In such a scenario, the energy of the false vacuum drives a stage of
hybrid inflation \cite{Linde:1991km,Linde:1993cn},
ending in a phase transition, in the course of
which the Abelian gauge symmetry $U(1)_{B-L}$ becomes spontaneously broken.
Guided by the expectation that phase transitions might be in fact common phenomena
in the early universe, we hence propose that also the very origin of the hot thermal phase
has to be attributed to a phase transition, \textit{viz.} the \BmL phase transition as we
shall refer to it from now on.
To be very clear about this point, we stress that the \BmL phase transition
represents the cosmological realization of spontaneous \BmL breaking,
similarly as the electroweak phase transition represents the cosmological
realization of electroweak symmetry breaking.


Hybrid inflation ending in the spontaneous breaking of a local symmetry
is an attractive scenario of inflation, as it establishes a connection
between cosmology and particle physics.
The symmetry breaking at the end of inflation may, in particular, be identified
as an intermediate stage in the breaking of the gauge group of some
theory of grand unification (GUT) down to the gauge group of the standard model.
In this sense, the \BmL phase transition may be easily embedded into a grander scheme
based on a GUT theory featuring \BmL as an additional gauge
symmetry.
A prime example in this context are GUT theories with gauge group $SO(10)$
\cite{Buchmuller:1991ce}.
We also note that incorporating \BmL into the gauge group of the theory is an
almost trivial extension of the standard model.
As it turns out, the global $U(1)_{B-L}$ is already an anomaly-free
symmetry of the standard model Lagrangian \cite{'tHooft:1976up,'tHooft:1976fv}.
Upon the introduction of three generations of right-handed neutrinos,
it is then readily promoted to a local symmetry \cite{Babu:1989tq,Babu:1989ex}.
Unfortunately, the statistical properties of the CMB temperature anisotropies rule out
the simplest nonsupersymmetric version of hybrid inflation \cite{Komatsu:2010fb}.
For this reason we will consider supersymmetric $F$-term hybrid inflation
\cite{Copeland:1994vg,Dvali:1994ms} in this thesis.
Apart from its usual inner-theoretical and aesthetic virtues, including
supersymmetry into our analysis also has an important phenomenological advantage.
Invoking a discrete symmetry such as matter \cite{Dimopoulos:1981zb} or
$R$ parity \cite{Farrar:1978xj} renders the lightest superparticle (LSP)
stable, turning it into an excellent particle candidate
for dark matter \cite{Pagels:1981ke,Goldberg:1983nd,Ellis:1983ew}.
Moreover, supersymmetry implies that each right-handed neutrino
pairs up with a complex scalar to form a chiral multiplet.
In the course of the \BmL phase transition, these neutrino multiplets acquire
Majorana mass terms, such that after symmetry breaking the physical neutrino states
consist of three heavy Majorana neutrinos $N_i$ and three heavy complex sneutrinos
$\tilde{N}_i$.
The \BmL phase transition hence also sets the stage for the seesaw mechanism
\cite{Minkowski:1977sc,Yanagida:1979as,Glashow:1979nm,GellMann:1980vs,Mohapatra:1979ia},
which elegantly explains the tiny masses of the standard model neutrinos.


The decay of the false vacuum at the end of hybrid inflation is accompanied
by tachyonic preheating \cite{Felder:2000hj,Felder:2001kt}
and the production of topological defects in the form of cosmic strings
\cite{Vilenkin:1984ib,Hindmarsh:1994re,Hindmarsh:2011qj}.
Successful hybrid inflation in combination with the
nonobservation of cosmic strings requires that the \BmL phase transition
indeed has to take place at the GUT scale \cite{Battye:2006pk,Nakayama:2010xf}.
Tachyonic preheating denotes the rapid transfer of the false vacuum energy
into a gas of nonrelativistic \BmL Higgs bosons, entailing the nonadiabatic production
of all particles coupled to the Higgs field \cite{GarciaBellido:2001cb}.
After the \BmL phase transition the energy density of the universe is dominated by
the abundance of Higgs bosons, which slowly decay into heavy neutrinos and sneutrinos.
In combination with tachyonic preheating, the dynamics of the \BmL gauge degrees
of freedom (DOFs) as well as thermal processes, the decay of the Higgs boson and its
superpartners produces an abundance of heavy (s)neutrinos.
These (s)neutrinos subsequently decay into radiation, thereby generating the entropy
of the hot thermal phase, i.e.\ reheating the universe.


Hence, an important consequence of the \BmL phase transition is that the reheating
process is driven by the decay of the heavy (s)neutrinos.
This in turn automatically yields baryogenesis via a mixture of nonthermal
and thermal leptogenesis \cite{Fukugita:1986hr}.
Our work is thus closely related to previous studies on thermal leptogenesis
\cite{Plumacher:1997ru,Buchmuller:2004nz} as well as on nonthermal leptogenesis
via inflaton decay \cite{Lazarides:1991wu,Asaka:1999yd,Asaka:1999jb,HahnWoernle:2008pq}.
Furthermore, the fact that the reheating process is (s)neutrino-driven results
in the temperature scale of reheating, i.e.\ the reheating temperature, being
determined by the (s)neutrino lifetime and therefore directly related to
(s)neutrino parameters.
Of course, the final baryon asymmetry is also determined by (s)neutrino parameters and
so we arrive at the remarkable conclusion that the initial conditions of the hot early
universe cannot be freely chosen, but are fully controlled by the parameters of a Lagrangian,
which could in principle be measured by particle physics experiments and astrophysical
observations.
The \BmL phase transition is hence not only a particularly simple
mechanism for the generation of the initial conditions of the hot early universe,
it is also testable in present-day and future experiments.


Assuming supersymmetry to be a local symmetry, the particle spectrum also features
the gravitino---the spin-$3/2$ superpartner of the spin-$2$ graviton, which acts as
the gauge field of local supersymmetry transformations.
Due to the high reheating temperatures reached after the \BmL phase transition,
thermal gravitino production during the reheating process is unavoidable
\cite{Bolz:2000fu,Pradler:2006qh}.
Depending on the superparticle mass spectrum, this may lead to various
cosmological problems.
As for a light stable gravitino, inelastic scatterings in the thermal bath
may produce the gravitino so efficiently that it overcloses the universe.
Meanwhile, the late-time decay of an unstable gravitino may alter the abundances
of the light elements and thus spoil the successful theory of big bang nucleosynthesis (BBN)
\cite{Weinberg:1982zq,Ellis:1984er,Kawasaki:2004yh,Kawasaki:2004qu,Jedamzik:2006xz}.
To avoid these problems, we will consider two particular superparticle mass spectra
in this thesis.
In the first case, we will assume the gravitino to be the LSP
with a mass of $\mathcal{O}(10..100)\,\textrm{GeV}$,
as it typically arises in scenarios of gravity- or gaugino-mediated supersymmetry breaking.
Gravitino dark matter can then be thermally produced at a reheating temperature
compatible with leptogenesis \cite{Bolz:1998ek}.
In the second case, we will take the gravitino to be the heaviest superparticle
with a mass of $\mathcal{O}(10..1000)\,\textrm{TeV}$.
Such large gravitino masses are realized in anomaly mediation,
which is a promising scenario of supersymmetry breaking, given the recent hints
by the LHC experiments ATLAS and CMS that the Higgs boson may
have a mass of about $125\,\textrm{GeV}$ \cite{ATLAS:2012ae,Chatrchyan:2012tx}.
A gravitino heavier than roughly $10\,\textrm{TeV}$ can be consistent with primordial
nucleosynthesis and leptogenesis \cite{Weinberg:1982zq,Gherghetta:1999sw,Ibe:2004tg},
thus allowing us to circumvent all cosmological gravitino problems.
In our second scenario, the nonthermal production of pure wino or higgsino LSPs, i.e.\ weakly
interacting massive particles (WIMPs), in the decay of heavy, thermally produced gravitinos
accounts for the relic density of dark matter.


In this thesis, we study the \BmL phase transition in the full supersymmetric
Abelian Higgs model, for which derive the complete Lagrangian in arbitrary and
unitary gauge.
From this Lagrangian, we cannot only infer the decay rates of all particles under study,
but also read off how the corresponding mass eigenvalues evolve with time in the course of
spontaneous symmetry breaking.
These time-dependent masses are an important input to the calculation of the
particle abundances produced during tachyonic preheating.
In order to describe the reheating process subsequent to the \BmL phase transition,
we derive the Boltzmann equations for all particle species of interest.
To facilitate our calculations, we treat the various contributions to the respective
heavy (s)neutrino abundances separately, i.e.\ we formulate a separate Boltzmann equation
for each contribution.
Thanks to this novel technical procedure, we are able to solve a subset of
Boltzmann equations analytically.
Solving the remaining equations numerically, we obtain a detailed and time-resolved
picture of the evolution of all particle abundances during reheating.
An interesting result of our analysis is that the competition between cosmic
expansion and entropy production leads to an intermediate period of constant reheating
temperature, during which the baryon asymmetry as well as the thermal gravitino abundance
are produced.
The final results for these two quantities as well as the reheating
temperature turn out to be rather insensitive to the influence of the extra
superparticles not contained in the supersymmetric standard model.
Likewise, the decay of the \BmL gauge DOFs shortly after preheating hardly affects the
final outcomes of our calculations.
Based on these observations, we conclude that the investigated scenario of reheating
is quite robust against uncertainties in the underlying theoretical framework.


Successful hybrid inflation and leptogenesis constrain the viable range of
neutrino mass parameters.
Combining these constraints with the requirement that dark matter be made out of
gravitinos, we find relations between neutrino parameters and superparticle masses,
in particular a lower bound on the gravitino mass of $10\,\textrm{GeV}$.
Similarly, we infer relations between the masses of the dark matter particle,
the gravitino and the standard model neutrinos in the case of WIMP dark matter.
Requiring consistency between hybrid inflation, leptogenesis, dark matter
and BBN, we derive upper and lower bounds on the LSP
mass as well as lower bounds on the gravitino mass, all of which depend on
the lightest neutrino mass.
For instance, given that the lightest neutrino has a mass of
$0.05\,\textrm{eV}$, a higgsino LSP would have to be lighter than $900\,\textrm{GeV}$,
while the gravitino would need to have a mass of at least $10\,\textrm{TeV}$.


Our quantitative analysis of the reheating process by means of Boltzmann equations
is based on a flavour model \cite{Buchmuller:1998zf} of the Froggatt-Nielsen type
\cite{Froggatt:1978nt}.
Generally speaking, Froggatt-Nielsen flavour models are able to reconcile the large
quark and charged-lepton mass hierarchies and the small quark mixing angles with
the observed small neutrino mass hierarchies and the large neutrino mixing angles
in a natural way.
In this thesis, we point out that the Froggatt-Nielsen flavour structure, which we employ for
our analysis, together with the known neutrino data, strongly constrains yet undetermined
parameters of the neutrino sector.
Treating unknown $\mathcal{O}(1)$ parameters as random variables, we obtain surprisingly
sharp predictions for the smallest mixing angle,
$\sin^2\left(2\theta_{13}\right) = 0.07_{-0.05}^{+0.11}$, the smallest neutrino mass,
$m_1 = 2.2_{-1.4}^{+1.7} \times 10^{-3}\,\textrm{eV}$, and one Majorana phase,
$\alpha_{21} / \pi = 1.0_{-0.2}^{+0.2}$.


This thesis is organized as follows.
In Ch.~\ref{ch:cosmology}, we briefly review the basics of early universe cosmology,
which we require as background material for our further discussion.
We outline how the present composition of the universe calls for new physics beyond
the standard model, discuss the main observational evidence for the hot thermal phase
in the early universe, i.e.\ the CMB and BBN, and shortly touch on the other phase transitions,
which we expect to take place in the early universe, i.e.\ the QCD and the electroweak phase
transition.
Finally, we review the electroweak sphaleron process, which is a crucial
ingredient to leptogenesis.
The reader acquainted with these rudiments of particle cosmology is invited to skip
our introductory chapter and directly proceed with Ch.~\ref{ch:framework}.


In Ch.~\ref{ch:framework}, we develop a theoretical framework for a consistent cosmology,
which addresses most of the problematic issues alluded to in Ch.~\ref{ch:cosmology}.
First, we motivate supersymmetric $F$-term hybrid inflation as an attractive
inflationary scenario and compile several useful formulae, which we need for our later
analysis of the production of cosmic strings.
Then we turn to the seesaw mechanism and the right-handed neutrinos.
We introduce the superpotential for all quark and lepton superfields
and subsequently use it to derive the mass and mixing matrices in the lepton sector.
Next, we motivate leptogenesis as the most promising scenario of baryogenesis
and elaborate on the two superparticle mass spectra, which we consider in this thesis.
In the latter part, we particularly emphasize how the spectra under study circumvent
the cosmological gravitino problems.
Finally, we assemble all pieces of the puzzle and outline how the \BmL phase transition
at the end of inflation gives rise to a consistent cosmology.
We summarize all mechanisms for the production of particles during preheating and reheating
and illustrate how the fact that the reheating process is driven by
the decay of heavy (s)neutrinos directly implies relations between neutrino
and superparticle masses.
In conclusion, we present our Froggatt-Nielsen flavour structure and parametrize
our entire model in terms of flavour charges.


In Ch.~\ref{ch:neutrinos}, we employ Monte-Carlo techniques to study the
dependence of yet undetermined neutrino observables on the unknown
$\mathcal{O}(1)$ factors contained in the Froggatt-Nielsen model.
After a few technical remarks on our procedure, we list the surprisingly precise
predictions for the various parameters in the neutrino sector and demonstrate
that we are partly even able to reproduce them analytically.


In Ch.~\ref{ch:model}, we lay the theoretical foundation for our study of the
\BmL phase transition and the subsequent reheating process.
To be able to describe the dynamics of all physical particle species after
spontaneous symmetry breaking, we require the Lagrangian of the supersymmetric
Abelian Higgs model in unitary gauge.
In a first step, we therefore derive the Lagrangian of a general supersymmetric
Abelian gauge theory in arbitrary gauge.
Then we evaluate this Lagrangian in unitary gauge for our specific field content,
which readily provides us with all time-dependent mass eigenvalues and decay rates
that we require for our further analysis.


In Ch.~\ref{ch:phasetransition}, we discuss the nonperturbative dynamics during
the decay of the false vacuum of unbroken $B$$-$$L$.
First, we estimate the abundance of cosmic strings
produced during the \BmL phase transition and restrict the parameters of hybrid
inflation based on the requirement of successful inflation and the fact that no
observational indications of effects related to cosmic strings have been found so far.
In the second section of Ch.~\ref{ch:phasetransition}, we introduce the quench
approximation for the waterfall transition at the end of hybrid inflation and
generalize the common waterfall conditions \cite{Linde:1993cn}, which only apply
to the original, nonsupersymmetric variant of hybrid inflation, to the supersymmetric case.
Furthermore, we compute the particle abundances generated during preheating.


In Ch.~\ref{ch:reheating}, we study the reheating process subsequent
to the \BmL phase transition by means of Boltzmann equations.
For a series of particle species, we first formulate template Boltzmann equations
serving as proxies for their actual Boltzmann equations.
After solving these template equations analytically and in full generality,
we then apply our findings to our actual scenario.
Moreover, we develop techniques to describe the evolution of the gravitational
background analytically and to track the evolution of the temperature of the hot plasma
by means of its own Boltzmann equation.
In the next section, assuming the gravitino to be the LSP,
we present the solutions of the Boltzmann equations for
a representative choice of parameter values.
Apart from a comprehensive discussion of the evolution of all particle abundances,
we motivate a particular definition of the reheating temperature and
check the robustness of the reheating process against small changes in the
theoretical setup.
In the third section of Ch.~\ref{ch:reheating}, we finally carry out a scan
of the parameter space, from which we infer relations between neutrino and
superparticle masses.
To some extent, we are again able to reproduce our results analytically.
For all important quantities we provide useful fit formulae.


In Ch.~\ref{ch:wimp}, we consider the production of WIMP dark matter in
the decay of heavy, thermally produced gravitinos.
After a short comment on the competition between our nonthermal WIMP production
mechanism and thermal WIMP freeze-out, we present constraints on the neutralino,
gravitino and neutrino masses and sketch the prospects for the experimental confirmation
of our scenario.


In Ch.~\ref{ch:conclusions}, we conclude and summarize our results.
Furthermore, we give an outlook as to the possible directions into which
the analysis presented in this thesis could be extended.
The three appendices contain important supplementary material.
In App.~\ref{ch:conventions}, we summarize the formalism of Boltzmann equations
and discuss the properties of particle species in kinetic and thermal equilibrium.
In App.~\ref{ch:scatterings}, we provide the proof for an important relation,
which is needed in the derivation of the Boltzmann equation for the lepton asymmetry
and which is related to the $CP$ violation in $2$-to-$2$ scattering processes with
heavy (s)neutrinos in the intermediate state.
In App.~\ref{ch:gravitinos}, we derive an analytical expression for the abundance
of thermally produced gravitinos and illustrate how our quantitative discussion
in Ch.~\ref{ch:reheating} is easily generalized to gluino masses other than the
one we employ in our analysis.


The discussion in Chs.~\ref{ch:neutrinos}, \ref{ch:reheating} and \ref{ch:wimp}
is based on two projects in collaboration with
Wilfried Buchm\"uller and Gilles Vertongen as well as on three projects in
collaboration with Wilfried Buchm\"uller and Valerie Domcke, the results of which
were respectively first published in Refs.~\cite{Buchmuller:2010yy,Buchmuller:2011mw}
and Refs.~\cite{Buchmuller:2011tm, Buchmuller:2012wn,Buchmuller:2012bt}.



\chapter{Early Universe Cosmology}
\label{ch:cosmology}


The main intention of this thesis is to motivate and investigate the \BmL phase transition
as the possible origin for the thermal phase of the hot early universe.
Before we are ready to do so, we have to acquaint ourselves with the observational
evidence for this phase and understand which physical processes have or may have
taken place in it.
For this reason we shall provide a brief review of early universe cosmology
in this chapter, thereby compiling the background material for the further discussion.
We will first discuss the present composition of the universe (cf.\ Sec.~\ref{sec:cmpstuni})
and then some of the main events in the thermal history of the universe in reverse chronological
order (cf.\ Sec.~\ref{sec:htthmlphs}).
We would like to emphasize that in this introductory chapter we will crudely restrict ourselves
to aspects which are relevant for our purposes.
More balanced and comprehensive presentations of the topic are for instance provided in
standard textbooks~\cite{Kolb:1990vq,Dodelson:2003ft,Mukhanov:2005sc} or
dedicated review articles~\cite{Trodden:2004st,Rubakov:2005tx,Nakamura:2010zzi}.


\newpage

\section{Composition of the Universe}
\label{sec:cmpstuni}


Over the last years the observational progress has marked the advent
of the era of precision cosmology.
The combined data exhibits an impressive consistency and is in very good agreement with
the currently accepted concordance model of big bang cosmology, the Lambda-Cold Dark Matter
($\Lambda$CDM) model.
Major evidence for this standard scenario of big bang cosmology derives
from several cosmological observations, the most eminent being perhaps
(i) the observed primordial abundances of the light elements, matching very well the
theoretical prediction from BBN~\cite{Nakamura:2010zzi},
(ii) the angular power spectrum of the temperature anisotropies in the CMB
as measured by the Wilkinson Microwave Anisotropy Probe (WMAP)
satellite~\cite{Komatsu:2010fb},
(iii) the imprint of baryonic acoustic oscillations (BAOs) in the local distribution of matter
as seen in galaxy surveys~\cite{Percival:2009xn},
(iv) direct measurements of the cosmic expansion rate, i.e.\ the Hubble parameter $H_0$, by the
Hubble Space Telescope~\cite{Riess:2009pu}, and
(v) distance measurements based on type Ia supernovae (SNe)~\cite{Hicken:2009dk,Kessler:2009ys}.


All observed cosmological phenomena are consistent with the assumption that our universe
is spatially flat~\cite{Komatsu:2010fb,Kowalski:2008ez}.
Indeed, combining the data on CMB anisotropies, BAOs and $H_0$ shows that presently,
at $95\,\%\,\textrm{CL}$, the total energy density of the universe $\rho_{\textrm{tot}}$
does not deviate by more than $1\,\%$ from the critical energy density $\rho_c$
that is required for exact spatial flatness.
In the following we shall hence neglect the possibility of a small
spatial curvature and assume that $\rho_{\textrm{tot}} = \rho_c$,~which
is equivalent to saying that all density parameters $\Omega_i$ sum to unity,
\begin{align}
\Omega_{\textrm{tot}} = \sum_i \Omega_i = \sum_i \frac{\rho_i}{\rho_c} =
\frac{\rho_{\textrm{tot}}}{\rho_c} = 1 \,.
\label{eq:omegatot}
\end{align}
This sum receives contributions from three different forms of
energy or matter: radiation, matter and dark energy.
In the present epoch the energy in radiation from beyond our galaxy is dominated by the
photons of the CMB.
Relic neutrinos which are presumed to be present in the current universe as a remnant of
the hot early universe either belong to radiation or matter, depending on their absolute masses.
The matter component splits into a small baryonic and a large dark nonbaryonic fraction.
We shall now discuss in turn how photons, neutrinos, baryonic matter, dark matter and dark
energy respectively contribute to $\Omega_{\textrm{tot}}$.


\subsection{CMB Photons}


In the early 1990s the Cosmic Background Explorer (COBE) satellite experiment
was the first precision measurement to confirm two key features of the CMB.
Since COBE we know that the CMB has an almost perfect Planckian spectrum
\cite{Mather:1993ij,Mather:1998gm} and that it is highly isotropic,
with its temperature fluctuating across the sky only at the level of $10^{-5}$ \cite{Smoot:1992td}.
Together, these findings provide strong evidence for a hot thermal phase in
the early universe preceded by an inflationary era (cf.\ Sec.~\ref{subsec:CMB}).
The mean CMB temperature is $T_\gamma^0 = 2.7255(6)\,\textrm{K}$ \cite{Fixsen:2009ug}.
Given the thermal black-body distribution of the CMB photons, this temperature directly implies
the following entropy, number and energy densities 
\begin{align}
s_{\gamma}^0 \simeq 1500 \,\textrm{cm}^{-3} \,,\quad
n_{\gamma}^0 \simeq 410 \,\textrm{cm}^{-3}  \,,\quad
\rho_{\gamma}^0 \simeq 260 \,\textrm{meV}\,\textrm{cm}^{-3} \,.
\label{eq:snrhogamm0}
\end{align}
The present value of the critical energy density is determined by the current
expansion rate.
With the aid of the dimensionless Hubble parameter $h$, which is defined through the relation
$H_0 = 100 \,h \,\textrm{km}/\textrm{s}/\textrm{Mpc}$, we are able to write
$\rho_c^0$ as
\begin{align}
\rho_c^0 = \frac{3 M_P^2}{8\pi} H_0^2 \simeq 10.54 \, h^2 \, \textrm{keV}\textrm{cm}^{-3}\,,
\label{eq:rhoc0}
\end{align}
with $M_P \simeq 1.22 \times 10^{19}\,\textrm{GeV}$ denoting the Planck mass.
The $\Lambda$CDM fit to the combined CMB, BAO and $H_0$ data gives
$h \simeq 0.704$ \cite{Komatsu:2010fb}, such that
$\rho_{\textrm{c}}^0 \simeq 5200 \,\textrm{eV}\,\textrm{cm}^{-3}$,
which results in a photon density parameter
\begin{align}
\Omega_{\gamma}^0 \simeq 5 \times 10^{-5} \,.
\end{align}
Barring some unknown form of dark radiation \cite{Archidiacono:2011gq},
the only other significant contribution to the present-day entropy density
in radiation comes from neutrinos.\footnote{Note that in the recent cosmic past,
shortly after the onset of star formation, the entropy contained in black holes has come
to dominate over the entropy in radiation \cite{Egan:2009yy}.\smallskip}
We thus conclude that photons are responsible for a large fraction of the
radiation entropy in the current universe,
but contribute only to a negligible extent to the total energy density.


\subsection{Relic Neutrinos}
\label{subsec:neutrinos}


In the hot early universe neutrinos are produced and kept in thermal equilibrium
via weak interactions.
Around a temperature $T \sim 1 \,\textrm{MeV}$ the rate of these interactions drops
below the Hubble rate, causing the neutrinos to decouple from the thermal bath
and evolve independently of all other species afterwards.
The presence of a relic abundance of primordial neutrinos in the current universe
is hence a fundamental prediction of the hot big bang scenario.
It is doubtful whether this cosmic neutrino background (CNB) will ever be directly observed,
as the low-energetic CNB neutrinos interact only extremely weakly \cite{Ringwald:2009bg}.
By contrast, a series of physical processes in the early universe such as BBN, the evolution
of the CMB temperature anisotropies or the formation of matter structures
on large scales are fortunately sensitive to the influence of primordial neutrinos, which
provides us with compelling indirect evidence for their
existence \cite{Hannestad:2006zg,Smith:2011es}.


The observed oscillations between the three neutrino flavours
\cite{Fukuda:2000np,Ahmad:2001an} indicate that neutrinos have small
masses\footnote{In the following discussion we shall restrict ourselves to the relic
abundance of primordial neutrinos. If neutrinos are Dirac fermions, the abundance
of antineutrinos should at each time be approximately the same as the abundance of neutrinos.}.
This has a direct impact on their evolution after decoupling.
If neutrinos were massless, their temperature $T_\nu$ would decrease for the most
part in parallel to the photon temperature $T_\gamma$ as the universe continues to expand.
Only at photon temperatures around the electron mass $m_e \simeq 511\,\textrm{keV}$,
$T_{\gamma}$ and $T_{\nu}$ would behave slightly differently.
Around $T_{\gamma} \sim m_e$, the thermal production of electrons and positrons
begins to cease.
$e^+e^-$ annihilations into photons then deposit the entire energy formerly contained
in electrons and positrons in the photon component, which slows down the decline of
$T_\gamma$ for a short time, but not the decline of $T_\nu$.
For massless neutrinos entropy conservation would imply
$T_\nu^0 = \left(4/11\right)^{1/3} T_{\gamma}^0 \simeq 1.9\,\textrm{K}$
and neutrinos would presently have a density $\Omega_\nu^0 \simeq 3 \times 10^{-5}$.
The energy density of massive neutrinos, however, experiences a slower
redshift due to the cosmic expansion than the energy density of massless
neutrinos.
While the energy of a massless neutrino goes to zero
as the universe expands, the energy $E_{\nu_i}$ of a neutrino mass eigenstate
with mass $m_{\nu_i}\neq0$ asymptotically approaches $m_{\nu_i}$.
Once the energy of a massive neutrino is dominated by its mass rather
its momentum, it becomes nonrelativistic.
For sufficiently large neutrino masses, the energy contained in nonrelativistic
neutrinos thus outweighs by far the energy of neutrinos that are still relativistic,
such that the present neutrino density is well described by
\begin{align}
\Omega_\nu^0 h^2 \simeq \frac{m_{\nu,\textrm{tot}}}{94 \,\textrm{eV}} \,,\quad
m_{\nu,\textrm{tot}} = \sum_{\nu_i} m_{\nu_i} \,,
\label{eq:Omeganuh2}
\end{align}
where the sum runs over all mass eigenstates that have turned nonrelativistic at some
value of $T_\gamma$ below $1\,\textrm{MeV}$, i.e., given the measured
mass squared differences, over at least two out of three states.
The lower bound on the sum of neutrino masses implied by the mass squared differences is
roughly $0.05\,\textrm{eV}$, so that $\Omega_\nu^0 \gtrsim 1 \times 10^{-3}$.
On the other hand, several cosmological observations constrain $m_{\nu,\textrm{tot}}$ from above.
Massive free-streaming neutrinos damp the growth of matter fluctuations and could thus
leave an imprint in large-scale structure (LSS) observables \cite{Hu:1997mj,Lesgourgues:2006nd}.
So far, no effects from neutrino masses have yet been observed.
Instead, combining data from galaxy surveys, WMAP, BAO, $H_0$ and type Ia SNe,
one is able to put an upper limit of
$0.28\,\textrm{eV}$ on $m_{\nu,\textrm{tot}}$ \cite{Thomas:2009ae}, which
corresponds to $\Omega_\nu^0 \lesssim 6 \times 10^{-3}$.

After leaving thermal equilibrium, most neutrinos never again interact
with other particles.
The entropy and total number of neutrinos hence remain practically unchanged after
decoupling, which is why we speak of the neutrinos as being \textit{frozen out}.
At the time neutrinos decouple, they are relativistic.
Their entropy and number densities thus subsequently always evolve
as the corresponding densities of massless neutrinos would do, independently of the
fact that neutrinos are actually massive, turning nonrelativistic at lower temperatures.
Because of this peculiar thermal history, neutrinos represent a prime example for
what is often referred to as \textit{hot relics}.
With the aid of the would-be temperature of massless neutrinos,
$T_\nu^0 \simeq 1.9\,\textrm{K}$,
we then obtain $s_\nu^0 \simeq 1400 \,\textrm{cm}^{-3}$ and
$n_\nu^0 \simeq 340 \,\textrm{cm}^{-3}$.

In conclusion, we find that also neutrinos contribute only to a negligibly small
extent to the total energy density of the universe,
\begin{align}
1 \times 10^{-3} \lesssim \Omega_\nu^0 \lesssim 6 \times 10^{-3} \,,
\label{eq:omgeanu0bounds}
\end{align}
which follows from Eq.~\eqref{eq:Omeganuh2} and the bounds on the total neutrino mass,
\begin{align}
0.05\,\textrm{eV} \lesssim m_{\nu,\textrm{tot}} \lesssim 0.28 \,\textrm{eV} \,.
\label{eq:mnutotrange}
\end{align}
In return, their entropy density is almost as large as the one of the CMB photons.
The present radiation entropy density $s_R^0$, comprising the
photon entropy density and the entropy densities of all hot relics,
i.e.\ neutrinos in the standard hot big bang scenario, then turns out to be
\begin{align}
s_R^0 = s_\gamma^0 + s_\nu^0 \simeq 2900\,\textrm{cm}^{-3}\,.
\end{align}
Note that, by definition,  $s_R^0$ can also be written as
the entropy of a thermal bath with an effective number of degrees of
freedom $g_{*,s}^0$ at temperature $T_{\gamma}^0$,
\begin{align}
s_R^0 = \frac{2\pi^2}{45} g_{*,s}^0 \left(T_\gamma^0\right)^3 \,,\quad
g_{*,s}^0 = 2 + \frac{7}{8} \cdot 3 \cdot 2 \cdot \frac{4}{11} = \frac{43}{11} \,.
\label{eq:sRgstars0}
\end{align}
The entropy associated with this density directly corresponds to the entropy inherent in
the thermal bath during the hot phase of the early universe.
A conclusive explanation for its origin is still lacking and it is a major task of modern
particle cosmology to explore possible sources for this primordial entropy.
A key motivation of this thesis is to demonstrate that the spontaneous breaking of \BmL
at the end of inflation represents a viable scenario for its generation.


\subsection{Baryonic Matter}
\label{subsec:baryncmttr}


All forms of matter in the universe that can be more or less well described by standard
particle physics, such as gas clouds, stars, planets, black holes, etc., are baryonic, i.e.\
made out of ordinary atoms, whose nuclei are composed of protons and
neutrons.\footnote{In order to ensure that the universe as a whole is electrically charge neutral,
there has to be present one electron for each proton in the universe.
As a single proton is, however, roughly $1800$ times heavier than an electron, the contribution
from electrons to the total energy presently stored in matter is negligibly small,
which is why we will not consider it any further.}
The present abundance of these baryons, or more precisely nucleons, is conveniently
parametrized in terms of the baryon-to-photon ratio $\eta_b$,
\begin{align}
\Omega_b^0 h^2 = \frac{m_N}{\rho_c^0 / h^2} \,n_\gamma^0 \, \eta_b^0 \simeq
\frac{1}{273} \left(\frac{\eta_b^0}{10^{-10}}\right) \,,\quad
\eta_b^0 = \frac{n_b^0}{n_\gamma^0} \,.
\end{align}
where $m_N \simeq 940 \,\textrm{MeV}$ is the mass of a single nucleon,
$n_b^0$ denotes the present number density of baryons, and where we have used the value
for $n_\gamma^0$ stated in Eq.~\eqref{eq:snrhogamm0}.
In the standard BBN scenario with three generations of relativistic neutrinos,
the primordial abundances of the light nuclei are solely controlled by the
baryon-to-photon ratio (cf.\ Sec.~\ref{subsec:BBN}).
The measurement of these abundances hence provides us with an observational handle on $\eta_b^0$.
Matching the observed abundances with the theoretical BBN prediction, one
finds at $95\,\%\,\textrm{CL}$ \cite{Nakamura:2010zzi}
\begin{align}
\textrm{BBN:} \qquad 5.1 \times 10^{-10} \leq \eta_b^0 \leq 6.5 \times 10^{-10} \,,\quad
0.019 \leq \Omega_b^0 h^2 \leq 0.024 \,.
\label{eq:BBNetab0}
\end{align}
One of the key predictions of standard cosmology is that between BBN and
the decoupling of the CMB the number of baryons as well as the photon
entropy are conserved such that the baryon-to-photon ratio remains unchanged
between these two processes.
This prediction can be observationally tested as the CMB power spectrum is fortunately
very sensitive to the physical baryon density $\rho_b \propto \Omega_b h^2$
(cf.\ Sec.~\ref{subsec:CMB}).
Fitting the $\Lambda$CDM model to the CMB data yields \cite{Komatsu:2010fb}
\begin{align}
\textrm{CMB:} \qquad \eta_b^0 \simeq \left(6.18 \pm 0.14\right) \times 10^{-10} \,,\quad
\Omega_b^0 h^2 = 0.02260 \pm 0.00053 \,,
\label{eq:omegab0CMB}
\end{align}
which is consistent with the BBN result in Eq.~\eqref{eq:BBNetab0} and hence serves
as yet another endorsement of the standard picture.
The agreement between the two determinations of $\eta_b^0$ is particular
remarkable in so far as they probe completely different physical processes
occurring in two widely separated epochs.
Due to its high precision, we will from now on, after some additional rounding,
use the CMB value as our estimate for the present baryon-to-photon ratio,
$\eta_b^{\textrm{obs}} = 6.2 \times 10^{-10}$, which corresponds to
a baryon density parameter $\Omega_b^0 \simeq 4.6 \times 10^{-2}$.


Depending on the perspective, we are led to the conclusion that the present abundance
of baryons in the universe is either exceptionally low or high.
First of all, it is surprising that BBN and the CMB concordantly imply that only
a fraction of roughly $5\,\%$ of the total energy of the universe resides in baryons.
In view of the fact that our universe appears to be spatially flat, one might rather
expect a baryon density parameter $\Omega_b^0 \simeq 1$.
The low abundance of baryons is hence an indication for the presence of other
nonbaryonic forms of matter or energy, viz. dark matter and dark energy,
that account for $95\,\%$ of the energy budget of the universe.
On the other hand $\Omega_b^0$ is remarkably large compared to the theoretical
expectation.\footnote{It
is also large compared to the observed abundance of luminous matter.
The density parameter of stars is smaller than $\Omega_b^0$ by one order of magnitude,
$\Omega_{\textrm{stars}} \simeq 2.7 \times 10^{-3}$ \cite{Fukugita:2004ee}.
Most baryons are thus optically dark, probably contained in some diffuse
intergalactic medium~\cite{Cen:1998hc}.}
In the early universe the baryon-to-photon ratio freezes out when the baryons
decouple from the thermal bath at temperatures of $\mathcal{O}(10..100)\,\textrm{MeV}$.
Assuming that the universe is locally baryon-antibaryon symmetric down to
temperatures of this magnitude, the annihilation of baron-antibaryon pairs shortly before
decoupling would dramatically reduce the abundances of both baryons and antibaryons.
In consequence of this \textit{annihilation catastrophe} the present baryon-to-photon would be
nine orders of magnitude smaller than the observed value, $\eta_b^0 \simeq 5 \times 10^{-19}$
\cite{Kolb:1990vq,Buchmuller:2005eh}.
The most reasonable way out of the annihilation catastrophe is the possibility that
the universe possesses a baryon-antibaryon asymmetry at temperatures of
$\mathcal{O}(100)\,\textrm{MeV}$.
The excess of baryons over antibaryons at the time of annihilation would then explain
the large observed baryon abundance.


Further evidence for a primordial baryon asymmetry comes from the fact that
the observable universe seems to contain almost exclusively matter and
almost no antimatter.\footnote{Antiparticles of cosmic origin such as antiprotons
and positrons are seen in cosmic rays.
Their fluxes are, however, consistent with the assumption that they are merely
secondaries produced in energetic collisions of cosmic rays with the interstellar
medium rather than primordial relics.\smallskip}
If there were to exist large areas of antimatter in the universe, annihilation processes
along the boundaries between the matter and antimatter domains would result in characteristic
gamma ray signals.
As no such signals have yet been observed, the local abundance of antimatter can
be tightly constrained on a multitude of length scales, ranging from our solar
system, to galaxies and clusters of galaxies.
X- and gamma-ray observations of the Bullet Cluster, a system of two colliding galaxy clusters,
put for instance an upper bound of $3 \times 10^{-6}$ on the local antimatter fraction, thus
ruling out serious amounts of antimatter on scales of $\mathcal{O}(20)\,\textrm{Mpc}$,
which are the largest scales directly probed so far~\cite{Steigman:2008ap}.
Furthermore, assuming that matter and antimatter are present in equal shares
on cosmological scales, one can show that the matter domain we inhabit
virtually has to cover the entire visible universe \cite{Cohen:1997ac}.


The absence of antimatter in our universe thus allows for a different interpretation
of the baryon-to-photon ratio $\eta_b^0$.
As the ratio of photons to antibaryons is practically zero, $\eta_b^0$ can also be
regarded as a measure for the baryon asymmetry of the universe (BAU),
\begin{align}
\eta_b^0  = \frac{n_b^0}{n_\gamma^0} \rightarrow \frac{n_b^0 - n_{\bar{b}}^0}{n_\gamma^0} \,.
\end{align}
To emphasize this different interpretation of the baryon-to-photon ratio
we will write $\eta_B^0$ instead of $\eta_b^0$ in the following, where the subscript $B$
is supposed to refer to the total baryon number of the universe.
Again, standard cosmology lacks an explanation for the origin of this primordial asymmetry.
A second key motivation for this thesis is hence to identify a natural mechanism
for the dynamical generation of the BAU that can be consistently embedded into an overall
picture of the early universe.
As we will demonstrate, leptogenesis after nonthermal neutrino production in
the decay of \BmL Higgs bosons represents a viable and particularly attractive option.


\subsection{Dark Matter}
\label{subsec:DM}


A plethora of astrophysical and cosmological observations indicates that next to ordinary matter
some form of dark matter (DM), i.e.\ nonluminous and nonabsorbing matter which reveals
its existence only through its gravitational influence on visible matter, is ubiquitously
present in the universe.\footnote{For recent reviews on dark matter, cf.\ for instance
Refs.~\cite{Bertone:2004pz,Einasto:2009zd,Hooper:2009zm,Bertone:2010zz}.
Another ansatz to account for the various observed, but unexplained gravitational effects
is to modify the theory of general relativity.
While modifications of gravity (cf.\ in particular Refs.~\cite{Milgrom:1983ca,Bekenstein:2004ne})
are often able to explain isolated phenomena, they usually struggle to give a
consistent description of all observed phenomena, which is why we will not consider
them any further in this thesis.\smallskip}
Direct evidence for dark matter derives from all observable length scales.
The rotation curves of spiral galaxies as well as
the velocity dispersions of stars in elliptical galaxies
probe the abundance of dark matter on the scale of individual
galaxies.\footnote{Seminal works in this field have been the observations
by Vera Rubin and Kent Ford, who measured the rotation curve of the
Andromeda Nebula in 1970~\cite{Rubin:1970zz},
as well as by Sandra Faber and Robert Jackson, who studied stellar velocities
in elliptical galaxies in 1976~\cite{Faber:1976sn}.\smallskip}
This applies in particular to our own galaxy, whose rotation curve in combination
with other data allows to determine the fraction of dark matter in the neighborhood
of our solar system quite precisely \cite{Catena:2009mf}.
On the scale of clusters of galaxies, peculiar galaxy velocities in viralized galaxy
clusters, X-ray observations of the hot intracluster gas and gravitational lensing
effects on background galaxies point to large amounts of dark
matter.\footnote{The first astronomer to stumble upon the problem of the
\textit{missing mass} in galaxy clusters was Fritz Zwicky.
In 1933, observations of the Coma Cluster led him to conclude that the galaxies
in the cluster should actually fly apart, if there were not
large amounts of invisible matter present in it, holding them together~\cite{1933AcHPh...6..110Z}.
Zwicky is hence usually credited as the discoverer of dark matter.\smallskip}
Especially compelling evidence for dark matter comes from detailed studies
of the Bullet Cluster, whose dynamics can only be understood if it is assumed
to be predominantly composed of very weakly self-interacting dark matter~\cite{Clowe:2006eq}.
Finally, on cosmological scales the presence of dark matter is implied by the theory of
structure formation.
If the presently observed LSS of matter in the universe was to be traced back
only to the density fluctuations of ordinary baryonic matter at the time of photon decoupling,
the temperature anisotropies in the CMB would have to be at the level of $10^{-3}$.
However, the fact that they are actually two orders of magnitude smaller indicates
that baryonic density perturbations can, in fact,
not be the source of the required primordial wells of the gravitational potential.
Instead these potential wells have to be attributed to some form of nonbaryonic dark matter that,
unimpeded by photon pressure, is able to start clumping way before decoupling.
Furthermore, numerical simulations of structure formation show that most dark matter has to
be cold at the onset of structure formation, i.e.\ has to turn nonrelativistic long before
the energy in matter begins to dominate over the energy in radiation.\footnote{As light
neutrinos turn nonrelativistic only at very late times in the cosmological evolution, they
represent, in fact, a form of hot dark matter in the current universe.}

By now the overwhelming observational evidence has firmly established the notion that
nonbaryonic cold dark matter (CDM) is the prevailing form of matter in the universe.
It is thus one of the key ingredients to the $\Lambda$CDM model.
Strong support for the CDM picture is again provided by the CMB
power spectrum, which is next to the baryon density $\rho_b$ also
sensitive to the total matter density $\rho_m \propto \Omega_m h^2$
(cf.\ Sec.~\ref{subsec:CMB}).
Assuming dark matter to be cold and nonbaryonic, the combined CMB, BAO and $H_0$ data
allow for a precise determination of $\Omega_m h^2$~\cite{Komatsu:2010fb},
\begin{align}
\Omega_m^0 h^2 = 0.1349 \pm 0.0036 \,,
\label{eq:omegam0CMB}
\end{align}
which is roughly six times larger than the present baryon density $\Omega_b^0 h^2$ as
inferred from the primordial abundances of the light elements or the CMB power spectrum.
With the aid of Eqs.~\eqref{eq:omegab0CMB} and \eqref{eq:omegam0CMB},
the present density parameter of dark matter then turns out to
be\footnote{Later on we shall use a rounded version of the value
in Eq.~\eqref{eq:omegaDM0CMB}, namely $\Omega_{\textrm{DM}}^{\textrm{obs}}h^2 = 0.11$.\smallskip}
\begin{align}
\Omega_{\textrm{DM}}^0 h^2 = \Omega_m^0 h^2 - \Omega_b^0 h^2 = 0.1123 \pm 0.0036 \,,\quad
\Omega_{\textrm{DM}}^0 \simeq 0.227 \,.
\label{eq:omegaDM0CMB}
\end{align}


We thus know quite certainly that dark matter accounts for roughly $23 \,\%$ of the
energy budget of the universe.
The nature and the origin of dark matter have, however, remained mysterious puzzles so far.
At the present stage we are merely able to constrain to some extent its properties.
First of all, the mismatch between determinations of $\Omega_b^0 h^2$ and $\Omega_m^0 h^2$,
i.e.\ the present abundances of baryons in particular and of matter in general,
as well as arguments based on the theory of structure
formation indicate that dark matter has to be cold and nonbaryonic for the most
part.\footnote{Certain scenarios of warm dark matter or mixed dark matter which is composed of
a mixture of cold, warm and or hot components, are also admissible~\cite{Viel:2005qj,Jedamzik:2005sx}.
Likewise, also small amounts of baryonic matter in the form of massive compact halo objects
(MACHOs)~\cite{Paczynski:1985jf,Tisserand:2006zx} and or cold molecular gas
clouds~\cite{DePaolis:1994sb} may well contribute to the dark matter in galaxy halos.}
As it is \textit{dark}, the particles constituting dark matter are usually assumed to
be electrically neutral.
Similarly, if these particles carried colour charge, they would strongly interact with
baryons, thus altering, for instance, the predictions of BBN and the appearance of the CMB.
Hence the dark matter particles are assumed to be colour-neutral.
Finally, they have to be perfectly stable or at least sufficiently long-lived in order to
explain the presence and influence of dark matter on cosmological time scales up to the current
epoch.
Interestingly, no known particle fulfills all these requirement
and thus the existence of dark matter is one of the strongest indications for
physics beyond the standard model.
Particle cosmology now faces the task to identify which hypothetical new elementary
particles could serve as dark matter particles, embed dark matter into a consistent
picture of the cosmological evolution, and explain in particular how its present
abundance is generated (cf.\ Eq.~\eqref{eq:omegaDM0CMB}).
Therefore, the third key motivation of this thesis is to demonstrate that several
well-motivated dark matter scenarios can actually be easily realized, if reheating
after inflation is triggered by the \BmL phase transition.
For the most part, we will consider a scenario in which thermally produced gravitinos
account for dark matter.
In Ch.~\ref{ch:wimp}, we will then turn to a setup in which either higgsinos
or winos represent the constituents of dark matter.


\subsection{Dark Energy}
\label{subsec:DE}


A crucial result of our discussion so far is that dark matter, baryonic matter,
neutrinos and photons together account for only roughly $27\,\%$ of the energy
budget of the universe.
The remaining $73\,\%$ have to be attributed to some form of dark energy
that, as opposed to dark matter, does not cluster under the influence of gravity.
At the present stage we almost do not know anything about the nature and the origin
of dark energy, whereby dark energy represents one of the greatest mysteries of modern physics.
At least some light on the properties of dark energy is shed by the fact that
the expansion of our universe is currently accelerating.\footnote{The accelerated
expansion of our universe became evident for the first time in measurements of the
distance-redshift relation of high-redshift type Ia SNe in
1998~\cite{Riess:1998cb,Perlmutter:1998np}.\smallskip}
As matter and radiation on their own always lead to either a decelerating expansion
or an accelerated contraction, the dark energy has to be responsible for the
observed acceleration.
Assuming that dark energy can be described as a perfect fluid, just as all other forms of
matter and energy in the universe, the requirement that it be the source of the
accelerated expansion constrains its equation of state,
$\omega = p_{\textrm{DE}} / \rho_{\textrm{DE}} < -1/3$, where $p_{\textrm{DE}}$ and
$\rho_{\textrm{DE}}$ denote the pressure and the energy density of dark energy, respectively.
In other words: the accelerated expansion indicates that dark energy has a negative pressure.


There are several attempts to explain the presence of dark energy.
Many approaches assume, for instance, that dark energy corresponds to the energy of a
scalar field moving in some specific potential.
Depending on whether this field has a canonical kinetic term or not, dark energy is then
often referred to as quintessence~\cite{Ratra:1987rm,Wetterich:1987fm} or
$k$ essence~\cite{ArmendarizPicon:2000ah}.
An alternative possibility is that dark energy is entirely illusory, being in fact
an artifact of an incorrect treatment of gravity.
In this view, general relativity has to be modified in such a way
that the accelerated expansion can be accounted for without any recourse to dark
energy~\cite{Jain:2007yk,Clifton:2011jh}.
The simplest solution, however, is provided by Einstein's cosmological constant $\Lambda$.
Including a $\Lambda$ term in the field equations of general relativity
corresponds to adding a constant vacuum energy density $\rho_\Lambda  = \Lambda /\kappa$
with $\kappa = 8\pi/M_P^2$ and equation of state $\omega = -1$ to the energy budget
of the universe.
Although this ansatz is the least sophisticated one, it is consistent with all observations
and thus, along the lines of Occam's razor, the explanation of choice for dark energy in the
$\Lambda$CDM model.\footnote{Naively one might expect the energy density of the vacuum to be
related to the Planck scale, $\rho_\Lambda \sim M_P^4$.
Interpreting dark energy as the energy of the vacuum, one then has to explain
why $\rho_\Lambda \simeq 0.73 \rho_c^0 \sim 10^{-123} M_P^4$.
For a classic discussion of this so far unsolved problem cf.\ Ref.~\cite{Weinberg:1988cp}.}
Our earlier results for the density parameters of all other forms of matter and energy
in the $\Lambda$CDM model then allow us to calculate the density parameter of
dark energy~\cite{Komatsu:2010fb},
\begin{align}
\Omega_\Lambda^0 = \Omega_{\textrm{tot}}^0 - \Omega_{\textrm{DM}}^0 -
\Omega_b^0 - \Omega_\nu^0 - \Omega_\gamma^0
 = 0.728^{+0.015}_{-0.016} \,.
\end{align}
Finally, we remark that fitting the CMB, BAO and the SNe data from Ref.~\cite{Hicken:2009dk}
to a relaxed version of the $\Lambda$CDM model, in which $\Omega_{\textrm{tot}}$ and
$\omega$ are allowed to differ from $1$ and $-1$, respectively, yields a dark
matter equation of state $\omega = - 0.999^{+0.057}_{-0.056}$~\cite{Komatsu:2010fb},
which is in excellent agreement with the assumption of a cosmological constant.
For the moment being, as long as there is no commonly accepted explanation of
dark energy in sight, we thus settle for a rather pragmatic approach and adopt
the notion of a cosmological constant in this thesis, keeping
in mind that it should be regarded as a placeholder for a future theory of
dark energy that is still to come.


\subsection{Stages in the Expansion History}
\label{subsec:stages}


The identification of the key items in the cosmic energy inventory as well as
the determination of their respective contributions $\Omega_i^0$ to the total
energy density mark milestones of modern cosmology.
Together with the current expansion rate $H_0$, the density parameters $\Omega_i^0$
fully determine the present state of the universe on all scales on which the
cosmological principle holds.
On top of that, they also allow to trace the evolution of the universe
back in time up to temperatures of $\mathcal{O}(1)\,\textrm{MeV}$, i.e.\
until weak interactions begin to bring about interchanges between the
abundances of the different species.
Below the threshold for $e^+e^-$ pair production, $T \ll m_e$,
the energy densities of photons, matter and dark energy can,
for instance, be written as functions of the cosmological redshift $z$ in the following way,
\begin{align}
T \ll m_e \,:\qquad
\rho_i(z) = \rho_c^0 \,\Omega_i^0 \left(1 + z \right)^{3(1+\omega_i)}
\,,\quad i = \gamma,m,\Lambda\,,
\label{eq:rhoiz}
\end{align}
with $\omega_i$ denoting the coefficient in the equation of state of species $i$.
We respectively have $\omega_\gamma = 1/3$, $\omega_m = 0$ and $\omega_\Lambda = -1$.
The energy density of a nonrelativistic neutrino species with typical
momentum $p_{\nu_i}$  and mass $m_{\nu_i}$ evolves
similarly to the matter energy density $\rho_m$,
\begin{align}
p_{\nu_i}(z) \lesssim m_{\nu_i}\,:\qquad \omega_{\nu_i} \approx 0 \,,\quad
\rho_{\nu_i} (z) \approx
\frac{\rho_c^0}{h^2}\frac{m_{\nu_i}}{94\,\textrm{eV}}\left(1+z\right)^3 \,.
\label{eq:rhonuinonrel}
\end{align}
Once the typical neutrino momenta $p_{\nu_i}$ begin to exceed $m_{\nu_i}$, the respective
neutrino species becomes relativistic,%
\footnote{Given the allowed range of the total neutrino mass
(cf.\ Eq.~\eqref{eq:mnutotrange}), matching the two expressions for $\rho_{\nu_i}$
in Eqs.~\eqref{eq:rhonuinonrel} and \eqref{eq:rhonuirel} and solving for $z$
shows that the heaviest neutrino, which eventually contributes most
to $\Omega_\nu^0$, turns nonrelativistic at a redshift of $\mathcal{O}(10..100)$.}
so that its energy density henceforth runs in parallel to $\rho_\gamma$,
\begin{align}
m_{\nu_i} \lesssim p_{\nu_i}(z) \ll m_e \,:\qquad
\omega_{\nu_i} \approx 1/3 \,,\quad
\rho_{\nu_i} (z) \approx \frac{7}{8}\left(\frac{4}{11}\right)^{4/3}\rho_\gamma(z) \,.
\label{eq:rhonuirel}
\end{align}
The density of the total radiation energy is given as usual,
$\rho_R(z) = g_{*,\rho}(z)/g_\gamma \,\rho_\gamma(z)$,
with $g_{*,\rho}$ counting the effective number of relativistic degrees of freedom.

In the present epoch dark energy dominates the total energy of the universe,
$\Omega_\Lambda^0 \gtrsim \Omega_m^0 \gg \Omega_\nu^0 \simeq \Omega_\gamma^0$.
However, as the energy densities of radiation, matter and dark energy scale differently
with redshift $z$, this changes as we go back in time.
First, at $z = z_\Lambda$ the energy contained in matter catches up with dark energy,
$\rho_m\left(z_\Lambda\right) = \rho_\Lambda\left(z_\Lambda\right)$.
Then, at $z = z_{\textrm{eq}}$ radiation takes eventually over as the dominant form of energy in
the universe, $\rho_R\left(z_{\textrm{eq}}\right) = \rho_m\left(z_{\textrm{eq}}\right)$.
The above scaling relations for the energy densities $\rho_i$ imply
\begin{align}
z_\Lambda = \left(\frac{\Omega_\Lambda^0}{\Omega_m^0}\right)^{1/3} - 1 \simeq 0.39 \,,\quad
z_{\textrm{eq}} = \frac{g_{*,\rho}^0}{g_{*,\rho}^{eq}}\frac{\Omega_m^0}{\Omega_\gamma^0} - 1
\simeq 3200 \,,
\end{align}
where we have used that $g_{*,\rho}^0 = 2$ and
$g_{*,\rho}^{eq} = 2 +  7/8 \cdot 3 \cdot 2 \cdot (4/11)^{4/3} \simeq 3.36$.
These two redshifts correspond to the following photon temperatures,
\begin{align}
T_\Lambda = T_\gamma\left(z_\Lambda\right)
\simeq 3.8\,\textrm{K} \simeq 0.33 \,\textrm{meV} \,,\quad
T_{eq} = T_\gamma\left(z_{\textrm{eq}}\right) \simeq 8800\,\textrm{K} \simeq 0.76 \,\textrm{eV} \,,
\label{eq:TLambdaTeq}
\end{align}
as well as to the following values of the cosmic time $t$,
\begin{align}
t_\Lambda = t\left(z_\Lambda\right) \simeq 9.6 \,\textrm{Gyr}\,,\quad
t_{eq} = t\left(z_{\textrm{eq}}\right) \simeq 56\,\textrm{kyr}\,,
\label{eq:tLambdateq}
\end{align}
which are to be compared to the age of the universe,
$t_0 =13.75 \pm 0.13\,\textrm{Gyr}$~\cite{Komatsu:2010fb}.

In summary, we conclude that the universe experiences at least three
dynamically different stages in its expansion history.
(i) In the very recent cosmic past, $z < z_\Lambda$, the energy of the universe
is dominated by the vacuum contribution, which, due to its negative pressure, causes
the expansion to accelerate.
(ii) Between $z = z_\Lambda$ and $z = z_{\textrm{eq}}$ most energy is contained in pressureless matter.
Note that it is in this epoch that matter structures are able to form in the
universe.\footnote{Curiously enough, the matter-dominated era lasts sufficiently long
to allow for the formation of such complex structures as galaxies, solar systems and human
beings, which, from the perspective of mankind, appears to be a
fortunate \textit{cosmic coincidence}.
The question of why dark energy becomes relevant exactly at the present time, i.e.\
why presently $\Omega_\Lambda \sim \Omega_m$ rather than $\Omega_\Lambda \ll \Omega_m$
or $\Omega_\Lambda \gg \Omega_m$, is one of the greatest puzzles of modern cosmology.
Cf.\ e.g.\ Ref.~\cite{Zlatev:1998tr}.}
(iii) For $z > z_{\textrm{eq}}$ radiation is the most abundant form of energy in the universe.
When speaking of the \textit{hot thermal phase of the early universe} or the
\textit{hot early universe}, we actually refer to this phase of radiation domination.
During the radiation-dominated era the universe is filled by a hot plasma in thermal equilibrium
that becomes increasingly hotter and denser as one goes further back in time.
In the approximation of a constant number of relativistic degrees of freedom
$g_{*,\rho}$, the temperature $T \equiv T_\gamma$ of the thermal bath scales inversely
proportional to $t^{1/2}$,
\begin{align}
T(t) \approx \left(\frac{90M_P^2}{32\pi^3 g_{*,\rho}\,t^2}\right)^{1/4} \simeq
0.86\,\textrm{MeV} \left(\frac{43/4}{g_{*,\rho}}\right)^{1/4}
\left(\frac{1\,\textrm{s}}{t}\right)^{1/2} \,,
\label{eq:Ttrelation}
\end{align}
where we have normalized $g_{*,\rho}$ to its value at the time of neutrino decoupling.
As the temperature continues to rise, more and more particle species reach
thermal equilibrium with the bath, causing $g_{*,\rho}$ to increase.
Turning this picture around, we may equivalently say that in the hot early universe
various species decouple one after another from the thermal bath in consequence of the
declining temperature.
These departures from thermal equilibrium shape the present state of the universe.
Up to now we have already discussed the decoupling of neutrinos
at $T \sim 1\,\textrm{MeV}$ and the decoupling of baryons at ${T \sim 10..100\,\textrm{MeV}}$.
As we will see later on, similar nonequilibrium processes at even higher
temperatures may be responsible for the relic density of dark matter
and the baryon asymmetry of the universe.
In fact, the very aim of this thesis is to describe a possible origin for
the hot thermal phase of the early universe, namely the spontaneous breaking of
\BmL at the end of inflation, that naturally entails the simultaneous generation of entropy,
baryon asymmetry and dark matter.


\section{The Hot Thermal Phase}
\label{sec:htthmlphs}


\begin{figure}[t]
\begin{center}
\includegraphics[width=0.95\textwidth]{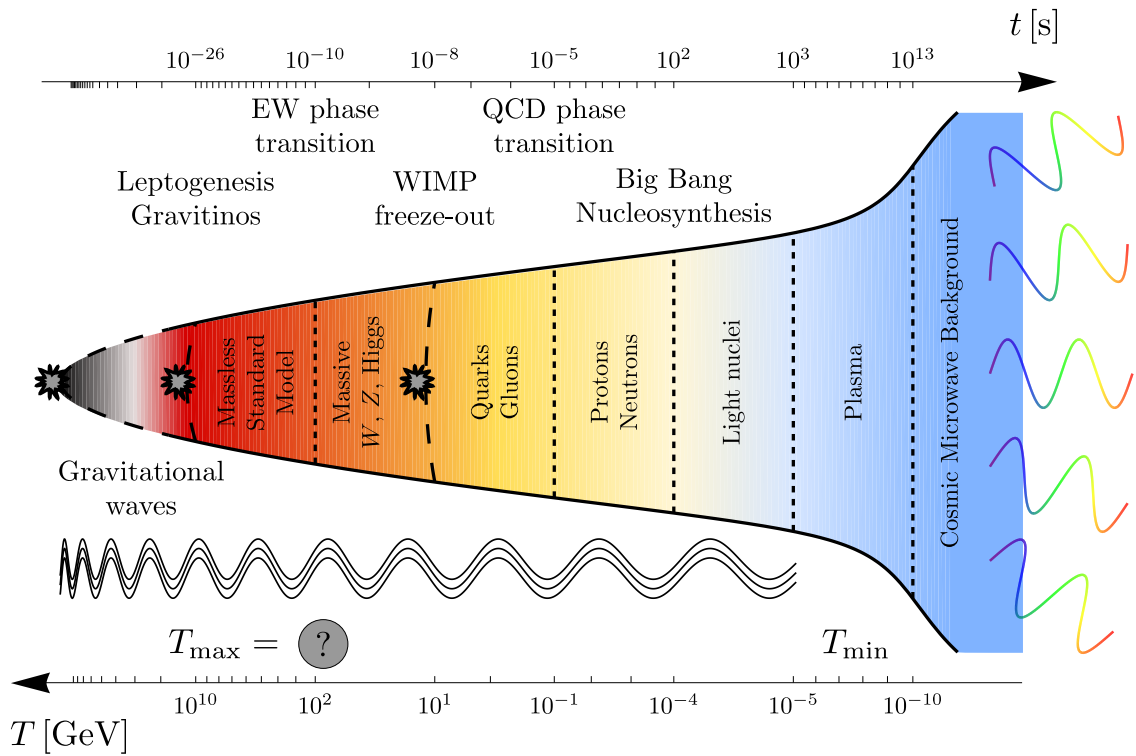}\medskip
\caption[Timeline of the hot thermal phase of the early universe]{Timeline of
the hot thermal phase of the early universe illustrating (i) the relation between
the temperature of the thermal bath $T$ and the cosmic time $t$ (cf.\ Eq.~\eqref{eq:Ttrelation}),
(ii) the chronology of several important, partly hypothetical nonequilibrium processes,
(iii) a representative selection of those forms of matter or energy that are
respectively involved in these processes, and (iv) several possibilities for the reheating
temperature after inflation (cf.\ Sec.~\ref{sec:byndsm}).}
\label{fig:history}
\end{center}
\end{figure}

The hot early universe represents the stage for a great variety of
physical processes taking place over an enormous range of energy scales
(cf.\ Fig.~\ref{fig:history} for an overview of the main events in its
thermal history).
As a final preparation before turning to our own scenario,
we shall now discuss in more detail the decoupling of the CMB,
primordial nucleosynthesis, the QCD and the electroweak phase transition
as well as electroweak sphalerons.


\subsection{The Cosmic Microwave Background}
\label{subsec:CMB}


Towards the end of the radiation-dominated phase, at temperatures of
$\mathcal{O}(1)\,\textrm{eV}$, protons, i.e.\ hydrogen nuclei, are kept
in thermal equilibrium via the steady interplay of radiative recombination
and photoionization processes.
However, as the plasma cools in the course of the expansion, photoionization becomes
less efficient, the hydrogen nuclei begin to bind free electrons into neutral
atoms and the ionization fraction of hydrogen freezes out at a vanishingly small value.
This process is usually referred to as hydrogen recombination.\footnote{Prior to
hydrogen recombination, at $T \sim 0.5\,\textrm{eV}$, helium decouples in a
similar way.
As hydrogen is still fully ionized at this time, the universe remains opaque after
helium recombination.\smallskip}
Due to the high abundance of thermal photons in the plasma it takes place at a temperature
significantly below the binding energy of hydrogen, $B_\textrm{H} = 13.6\,\textrm{eV}$.
In fact, the temperature has to drop to $T_{\textrm{rec}} \simeq 0.30\,\textrm{eV}$
until the fractional ionization reaches a value of $10\,\%$.
As the abundance of free electrons continues to decrease even further, the rate $\Gamma_\gamma$
of Thomson scatterings between thermal photons and plasma electrons falls below
the Hubble rate $H$.
At $T_{\textrm{dec}} \simeq 0.26\,\textrm{eV}$ the mean free photon path
equals the Hubble radius $H^{-1}$, or equivalently $\Gamma_\gamma = H$,
and most photons scatter for the last time.
This moment of \textit{last scattering} marks the time when the
photons decouple and the universe becomes transparent to radiation.
After decoupling the photons freely propagate until they eventually reach us
in the form of CMB radiation.
In this sense the CMB represents a full-sky picture of the early universe at a temperature
of $T_{\textrm{dec}} \simeq 0.26\,\textrm{eV}$, i.e.\ at a redshift
$z_{\textrm{dec}} \simeq 1100$ and a cosmic time $t_{\textrm{dec}} \simeq 360\,\textrm{kyr}$.

To be precise, the decoupling of the CMB actually occurs during
the matter-dominated era (cf.\ $T_{eq}$ in Eq.~\eqref{eq:TLambdaTeq}).
But as its origin is inextricably linked with the thermal history of the universe,
it represents nonetheless one of the main physical phenomena associated with the
hot big bang~\cite{Dicke:1965zz}.
In particular, the fact that the CMB has an almost perfect Planckian spectrum
may be regarded as key evidence for an early stage
during which the universe was filled by a hot plasma in thermal equilibrium.
Alternative attempts to explain the origin of the CMB, such as the idea put forward
by the proponents of the steady state theory proposing that the CMB may in fact be
starlight thermalized by dust grains, typically end up with a superposition of
blackbody spectra corresponding to different temperatures.

The CMB not only provides striking evidence for the hot thermal phase,
as we have seen in Sec.~\ref{sec:cmpstuni}, it also allows to precisely
determine a multitude of cosmological parameters that enter into the theoretical description
of the early universe.\footnote{For reviews on the physics of the CMB
and its potential to constrain cosmological models, cf.\ for instance
Refs.~\cite{Hu:2001bc,Samtleben:2007zz}.}
The primary CMB observable encoding cosmological information is the variation
of the CMB temperature across the sky, which is conveniently characterized by
the angular power spectrum $C_\ell$ of the relative temperature
fluctuations,
%
\begin{align}
\frac{\delta T}{T_0}\left(\mathbf{n}\right) =
\sum_{\ell = 2}^\infty \sum_{m=-\ell}^{+\ell} a_{\ell m} Y_{\ell m}\left(\mathbf{n}\right)
\,,\quad
\left<a_{\ell m}^* a_{\ell'm'}\right> = C_\ell \, \delta_{\ell\ell'} \delta_{mm'} \,.
\end{align}
Except for the dipole anisotropy, which is interpreted as being due to the
motion of the earth relative to the absolute CMB rest frame, the CMB temperature
anisotropies directly correspond to the density perturbations inherent in the
baryon-photon fluid at the time of last scattering.
Several physical processes leave their imprint in the observed power spectrum.
(i) The tight coupling between photons and baryons leads to higher temperatures
in regions of high baryon density.
(ii) Photons that have to climb out of potential wells after decoupling are
gravitationally redshifted.
This translates into a shift of the observed
with respect to the intrinsic temperature fluctuation,
which is usually referred to as the Sachs-Wolfe effect~\cite{Sachs:1967er}.
Similarly, decaying gravitational potentials traversed by the CMB photons
on their way from the surface of last scattering to the observer induce small
boots in the observed CMB temperature.
This is known as the integrated Sachs-Wolfe effect.
(iii) The non-zero velocity of the plasma at decoupling results in a Doppler
shift in the frequency of the CMB photons.
(iv) Perturbations in the gravitational potential, induced by the
growing density fluctuations of dark matter, as well as photon pressure
drive acoustic oscillations in the photon-baryon fluid, which gives rise to a
series of acoustic peaks in the CMB power spectrum.\footnote{Perturbations
in the photon-baryon fluid can only evolve causally as long as they extend over
scales smaller than the sound horizon.
This explains the position of the first acoustic peak in the CMB power spectrum.
It is located at an angular scale of roughly $1^\circ$ or equivalently
at $\ell \sim 200$, which corresponds to the angular diameter of the sound
horizon at last scattering.}
These four effects, but in particular the acoustic peaks,
are very sensitive to the parameters of the underlying cosmology.
Barring a few degeneracies, the CMB power spectrum encodes
information about at least ten basic cosmological parameters.

First of all, four parameters characterize the power spectra of
primordial density fluctuations as well as primordial gravitational waves.
These primordial scalar and tensor perturbations, as they are also referred
to, eventually evolve into the CMB temperature fluctuations.
The parameters characterizing their power spectra, $P_s$ and $P_t$, hence determine
the initial conditions for the evolution of the CMB anisotropies.
Usually, $P_s$ and $P_t$ are taken to be power-laws,
\begin{align}
P_s(k) = A_s \left(\frac{k}{k_*}\right)^{n_s - 1} \,,\quad
P_t(k) = A_t \left(\frac{k}{k_*}\right)^{n_t} \,,\quad
r = \frac{A_t}{A_s} \,,
\end{align}
where $k$ is the comoving momentum scale and $k_*$ stands for an arbitrary
reference scale.
Technically, $P_s$ denotes the power spectrum of the curvature perturbation~$\mathcal{R}$,
which measures the spatial curvature of a comoving slicing of spacetime.
$P_t$ represents in fact the sum of two power spectra, $P_+$ and $P_\times$,
which respectively account for the two physical polarization modes $h_+$
and $h_\times$ of the general traceless and transverse spatial metric perturbation.
Note that due to rotational invariance $P_+ = P_\times = P_t / 2$.
The great virtue of the three perturbations $\mathcal{R}$, $h_+$ and $h_\times$
is that they are time-independent at early times, i.e.\ as long as they extend over
scales larger than the Hubble radius $H^{-1}$.
So far, the CMB data has revealed no sign of tensor modes.
Thus, only the curvature perturbation amplitude $A_s$ as well as the scalar spectral
index $n_s$ have been measured up to now.
Neglecting potential tensor contributions and using a reference scale
$k_* = 0.002 \,\textrm{Mpc}^{-1}$, the combined WMAP, BAO and $H_0$ data
yields \cite{Komatsu:2010fb},
\begin{align}
A_s = \left(2.441^{+0.088}_{-0.092}\right) \times 10^{-9} \,,\quad
n_s = 0.963 \pm 0.012 \,,\quad
\label{eq:AsnsWMAP}
\end{align}
For comparison, the COBE data implies an amplitude $A_s \simeq 2.28 \times 10^{-9}$
at roughly the same scale $k_*$.
This result is usually referred to as the COBE normalization of the scalar power
spectrum~\cite{Bunn:1996py}.
WMAP, BAO and the SNe data from Ref.~\cite{Hicken:2009dk} together yield a tight upper bound
on the tensor-to-scalar ratio, $r < 0.20$ at $95\,\%\,\textrm{CL}$.
A measurement of the tensor spectral index $n_t$ is beyond the scope of any experiment
in the near future.
In single-field slow-roll models of inflation
(cf.\ Sec.~\ref{sec:byndsm}) $n_t$ does not represent an independent parameter
in any case.
It is rather directly related to the tensor-to-scalar ratio via the
\textit{consistency relation}, $n_t = -r/8$, which reduces the number of
free parameters fixing the initial conditions of the CMB anisotropies to three.
The background cosmology setting the stage for the evolution of the CMB anisotropies
is described by at least five parameters:
the expansion rate $H_0$, the energy densities of matter and baryons,
or equivalently $\Omega_m^0 h^2$ and $\Omega_b^0 h^2$, the density parameter
of dark energy $\Omega_{\textrm{DE}}^0$, and the coefficient $\omega$ in the
equation of state for dark energy.
In Sec.~\ref{sec:cmpstuni}, we discussed in detail the numerical values
of these parameters according to the CMB data in combination with other cosmological
observations.
Finally, one astrophysical parameter influences the CMB power spectrum:
the integrated optical depth $\tau$, which characterizes the amount of CMB photons
that undergo Thomson scattering owing to the reionization of the universe in the recent
cosmic past.
$\tau$ completes the set of standard parameters usually included in analyses of
the CMB power spectrum.
Beyond this set further parameters, such as the density of massive neutrinos
$\Omega_\nu^0 h^2$ or the running of the scalar spectral index $d \,n_s/d \ln k$,
may be taken into account as well.

While the CMB stands out as one of the main pillars of the picture of the big bang,
it also shows very plainly some of the severe problems big bang cosmology is facing
with regard to its initial conditions.
First of all, the observation that presently $\Omega_{\textrm{tot}}$ does not
deviate by more than $1\,\%$ from unity gives rise to the \textit{flatness problem}.
In a decelerating universe the deviation from exact flatness
always grows as some power of the cosmic
time.\footnote{Given a scale factor $a \propto t^p$,
${\Omega_{\textrm{tot}} - 1}$ scales like $\dot{a}^{-2} \propto t^{2(1-p)}$.
During the phases of radiation and matter domination we respectively have
$p = 1/2$ and $p = 2/3$.}
The total density parameter $\Omega_{\textrm{tot}}$ of a universe
exhibiting a small, but non-zero curvature in the present epoch
must hence approach unity to arbitrary precision as one goes back in time.
In other words, the initial value of $\Omega_{\textrm{tot}}$ must be unnaturally
fine-tuned.
Second, at the time of last scattering the past or particle horizon,
i.e.\ the distance scale characterizing the radial extent of causally connected domains,
is of $\mathcal{O}(100)\,\textrm{Mpc}$ corresponding to an angular diameter of
$\mathcal{O}(1^\circ)$ in the sky.
By contrast, the CMB is highly isotropic across the entire sky, which is to say
that at the time of decoupling the photon temperature is almost perfectly homogeneous
over a huge number of causally disconnected regions.
Again, this high degree of homogeneity can only be achieved by an unnatural
fine-tuning of the initial conditions, a puzzle which is known as the \textit{horizon problem}.
Furthermore, the minute deviations from an exactly isotropic temperature, that
we do observe in the CMB, finally lead to the third and perhaps most severe problem.
The mechanism responsible for the high degree of homogeneity over a multitude
of causally disconnected regions also has to explain why the temperature fluctuations
around the homogeneous background are precisely at the level of $10^{-5}$ and, in particular,
why they are correlated over scales exceeding the causal horizon at decoupling.
This problem may be translated into the following two fundamental questions:
(i) what is the origin of the primordial scalar and tensor perturbations and
(ii) which statistical properties do they have?
As we will see in Sec.~\ref{sec:byndsm}, all these three problems concerning the initial
conditions of the hot big bang can be successfully solved in inflationary cosmology.


\subsection{Primordial Nucleosynthesis}
\label{subsec:BBN}


Primordial or big bang nucleosynthesis (BBN), i.e.\ the generation of the
light elements during the first $20\,\textrm{min}$ of the radiation-dominated era,
represents the earliest testable nonequilibrium process in the history
of the universe which can be accounted for by well-understood standard model physics
only (cf.\ Fig.~\ref{fig:history}).\footnote{For reviews on BBN, cf.\ for
instance Refs.~\cite{Olive:1999ij,Iocco:2008va}.}
At present it hence provides the deepest reliable probe of the early universe.
The overall agreement of the observed primordial abundances of the light elements
with the predictions of BBN serves as a strong corroboration of hot big bang cosmology,
underpinning our picture of the early universe to a similar extent as the anisotropies
in the CMB.


Before the onset of BBN, at temperatures $T \gg 1\,\textrm{MeV}$ or
correspondingly at times $t \ll 1\,\textrm{s}$, the weak interactions
$n\,\nu_e \leftrightarrows p\,e^-$, $n\,e^+ \leftrightarrows p\,\bar{\nu}_e$,
and $n \leftrightarrows p\,e^-\,\bar{\nu}_e$ keep the neutron-to-proton
ratio $n/p$ in thermal equilibrium, $n/p = e^{-Q/T}$ with
$Q = m_n - m_p = 1.293 \,\textrm{MeV}$ denoting the neutron-proton mass difference.
Around a temperature of $1\,\textrm{MeV}$ the rate of neutron-proton interconversion
processes $\Gamma_{np}$ eventually drops below the Hubble rate $H$ and the neutron-to-proton
ratio freezes out at $n/p \simeq 1/6$.
Subsequent to freeze-out, $n/p$ still continues to decrease due to neutrons undergoing
$\beta^-$ decay, $n\rightarrow p\,e^-\,\bar{\nu}_e$.
At the time the neutrons decouple from the thermal bath, the temperature has already
fallen below the binding energy of deuterium,
$T \sim 1\,\textrm{MeV} < \Delta_{\textrm{D}} \simeq 2.23\,\textrm{MeV}$.
The synthesis of deuterium, however, does not yet commence because of the
large abundance of highly energetic photons that immediately dissociate each
newly formed deuterium nucleus.
This delay in the production of the light elements is referred to as the
\textit{deuterium bottleneck}.
It is overcome once the number of photons per baryon above the deuterium
photodissociation threshold has decreased below unity, which happens at
a temperature $T\sim 0.1\,\textrm{MeV}$ or roughly at the end of the
first three minutes.
The breaking of the deuterium bottleneck marks the onset of BBN.
At last deuterium can be efficiently produced and further processed into
heavier elements such as helium-3, helium-4 and lithium-7.


Independently of the nuclear reaction rates, virtually all free neutrons
end up bound in helium-4, which is the most stable one among the light elements.
At $T \sim 0.1\,\textrm{MeV}$ the neutron-to-proton ratio has decreased to
$n/p \simeq 1/7$ and the primordial mass fraction of helium-4 can be estimated as
\begin{align}
Y_{\textrm{p}} = \frac{4 \,n_{{}^4\textrm{He}}}{n_b} \approx
\frac{4\left(n_n/2\right)}{n_p + n_n} = \frac{2\,n/p}{1 + n/p} \simeq 25 \,\%\,,
\label{eq:Yphelium4}
\end{align}
which corresponds to a ratio by number of helium-4 to hydrogen of
${}^4\textrm{He}/\textrm{H} \simeq 8\,\%$.
Deuterium, helium-3 and lithium-7 are produced in much smaller numbers.
At the end of BBN around $t \sim 20\,\textrm{min}$, when the temperature has dropped
to $T \sim 0.03 \,\textrm{MeV}$ and most nuclear reactions have become inefficient,
$\textrm{D}/\textrm{H}$ and ${}^3\textrm{He}/\textrm{H}$ are of $\mathcal{O}\left(10^{-5}\right)$,
while ${}^7\textrm{Li}/\textrm{H}$ is of $\mathcal{O}\left(10^{-10}\right)$.
The complicated network of nuclear reactions that lead to these primordial
abundances is described by a coupled system of kinetic
equations that needs to be solved numerically~\cite{Wagoner:1966pv,Cyburt:2001pp}.
Besides the temperature $T$ or equivalently the cosmic time $t$,
the Hubble rate and the nuclear reaction rates that enter into these equations
are functions of only one cosmological parameter: the number density of
baryons $n_b$ during BBN.
As $n_b$ is directly related to the present value of the
BAU, $n_b = n_\gamma \,g_{*,s}/g_{*,s}^0 \,\eta_B^0$,
this explains why the observed primordial abundances of the light elements
give us a handle on $\eta_B^0$ (cf.\ Sec.~\ref{subsec:baryncmttr}).


The abundance of primordial deuterium is inferred from spectra of high-redshift
quasar absorption systems, while primordial helium-4 is observed
in low-metallicity regions of ionized hydrogen.
The spectra of old metal-poor, i.e.\ population {\footnotesize II} stars in the spheroid
of our galaxy allow to determine the primordial abundance of lithium-7.
All in all, the theoretical BBN predictions match the observed abundances of deuterium,
helium-4 and lithium-7 quite well within the $\eta_B^0$ range stated in
Eq.~\eqref{eq:BBNetab0}.\footnote{Data on helium-3 solely derives from the solar system and
high-metallicity regions of ionized hydrogen in our galaxy, which makes it difficult to infer
its primordial abundance.
On top of that, the theory of stellar helium-3 synthesis is in conflict with
observations.
For these two reasons, helium-3 is usually not used as a cosmological probe.}
An obvious curiosity, however, is that the lithium-7 abundance points to a value of
$\eta_B^0$ that is smaller by at least $4.2 \,\sigma$ than the value jointly
favoured by the abundances of deuterium and helium-4.
This discrepancy is known as the \textit{lithium problem}~\cite{Cyburt:2008kw}
and potentially indicates effects of new physics.


Leaving aside the lithium problem, we conclude that BBN is able to correctly predict
the primordial abundances of the light elements over a range of nine orders magnitude.
This success is a milestone of big bang cosmology, encouraging us to believe
that the laws of physics which we are able to test in laboratory experiments
also apply to the very first moments of the universe.
We are thus confident that modern particle physics allows us to speculate
about the history of the universe at still earlier times, $t \ll 1\,\textrm{s}$,
although as of now we have no means of observationally accessing them.
Furthermore, the success of BBN provides us with a powerful tool to constrain
deviations from the standard cosmology.


The helium-4 abundance, for instance, is very sensitive to the value
of $g_{*,\rho}$ and thus the presence of additional relativistic species
during BBN~\cite{Steigman:1977kc}.
Increasing $g_{*,\rho}$ above its standard value entails a faster Hubble expansion,
which results in the neutrons decoupling at earlier times.
The neutron-to-proton ratio then freezes out a correspondingly higher temperature,
leading to a larger abundance of primordial helium-4 (cf.\ Eq.~\eqref{eq:Yphelium4}).
Deviations from the standard value of $g_{*,\rho}$ are usually parametrized in
terms of an effective number of neutrino species
$N_{\textrm{eff}} = N_{\textrm{eff}}^{\textrm{st}} + \Delta N_{\textrm{eff}}$.
Before $e^+e^-$ annihilation, $N_{\textrm{eff}}^{\textrm{st}}$ is given as
$N_{\textrm{eff}}^{\textrm{st}} =3.046$~\cite{Mangano:2005cc}
and $g_{*,\rho}$ is related to $N_{\textrm{eff}}$
through $g_{*,\rho} = 2 + 7/8 \cdot \left(4 + N_{\textrm{eff}}\cdot2\right)$.
In turns out that the primordial helium-4 mass fraction scales with $\Delta N_{\textrm{eff}}$
as $\Delta Y_{\textrm{p}} \simeq 0.013 \Delta N_{\textrm{eff}}$~\cite{Bernstein:1988ad},
which allows to place limits on $N_{\textrm{eff}}$ by means of the measured
abundance of primordial helium-4.
In combination with the seven-year WMAP data, one finds $N_{\textrm{eff}} < 4.2$ at
$95\,\%\,\textrm{CL}$~\cite{Mangano:2011ar}.


Likewise, the late-time decay of a massive nonrelativistic particle which is not
included in the standard BBN scenario may as well alter the primordial
abundances of the light elements.
Similarly to additional relativistic species, the presence of such a particle
modifies the expansion rate prior to its decay.
On top of that, if the new particle dominates the energy density of the universe
at the time of its decay, a significant amount of entropy is produced while its decay
products thermalize.
This changes the time-temperature relationship and results in a diluted baryon-to-photon ratio.
Based on these effects, one can derive an upper bound on the lifetime
of the decaying particle or equivalently a lower bound on the
temperature of the thermal bath at the time the entropy production is
completed~\cite{Kawasaki:1999na}.
If the process of entropy production shortly before BBN is identified
with the reheating of the universe after inflation, this lower bound on the
temperature corresponds to the lowest possible value of the reheating temperature
$T_{\textrm{RH}}^{\textrm{min}}$ (cf.\ Sec.~\ref{subsec:infltn}).
Combining the observed primordial abundances of deuterium and helium-4
with CMB and LSS data, one obtains $T_{\textrm{RH}}^{\textrm{min}} \simeq 4 \,\textrm{MeV}$
at $95\,\%\,\textrm{CL}$~\cite{Hannestad:2004px}.


Independently of whether a long-lived massive particle dominates the energy density
of the universe or not, it may after all spoil the success of standard BBN through
the cascade processes induced by its decay.
Charged particles or photons emitted in radiative decays of the long-lived particle
entail electromagnetic showers~\cite{Cyburt:2002uv}.
Sufficiently energetic photons produced in these showers are then able to
photodisintegrate previously formed light nuclei.
Moreover, given appropriate couplings and on condition that they are kinematically
allowed, decays into colour-charged particles trigger
hadronic cascade processes~\cite{Kawasaki:2004yh}.
These involve energetic pions, kaons, neutrons, protons as well as the
corresponding antiparticles, all of which are able to react with the light nuclei in various ways.
The hadrons emitted in the decays of the long-lived particle induce, for instance,
extraordinary interconversion processes between the background nucleons.
This leads to an enhancement of the neutron-to-proton ratio after neutron decoupling
and thus to a larger abundance of helium-4.
At the same time, the energetic hadrons are also able to dissociate background helium-4 nuclei
and to produce the other light elements nonthermally.
If the decaying particle is electrically charged, it can
form bound states with background nuclei, which again changes the nuclear reaction rates.
Especially, the production of lithium-6 may be catalyzed in this way~\cite{Pospelov:2006sc}.


In order to determine the net effect of a long-lived massive particle on the primordial abundances
of the light elements, it is necessary to compute the distributions of the various decay
products of the decaying particle as functions of time.
These spectra then allow to calculate the rates of the photo- and hadrodissociation,
neutron-proton interconversion, and nonstandard production processes induced by the decay
of the unstable particle.
Requiring the impact of the decaying particle to remain small, such that the consistency between
the theoretical BBN predictions and the astrophysical observations is maintained, one can
derive constraints on the mass, lifetime and abundance of the unstable particle prior to its
decay~\cite{Kawasaki:2004qu,Jedamzik:2006xz}.
In Ch.~\ref{ch:wimp} we will in particular consider bounds on the properties
of a very heavy gravitino decaying shortly before BBN~\cite{Kawasaki:2008qe}.


\subsection{Phase and Topological Transitions}
\label{subsec:transitions}


The synthesis of the light elements marks the earliest process in the hot early
universe that is firmly established on the basis of observations.\footnote{Recall that
BBN enables us to trace the evolution of the hot thermal phase up to temperatures as high as
$T_{\textrm{RH}}^{\textrm{min}} \simeq 4\,\textrm{MeV}$
or equivalently cosmic times as early as $t \simeq 0.05\,\textrm{s}$ (cf.\ Sec.~\ref{subsec:BBN}).
\smallskip}
The exact nature of all nonequilibrium processes occurring prior to BBN, such as the
generation of the BAU or the primordial metric perturbations, are currently still subject to
speculations.
On the other hand, the standard model of particle physics describes
the interactions of elementary particles with great precision all the way up to the
TeV scale.
Based on standard model physics one is thus able to make an educated guess about the
history of the universe up to $T \sim 1\,\textrm{TeV}$ or equivalently
$t\sim 10^{-13}\,\textrm{s}$.
In the following we shall in particular elaborate on the phase and topological transitions
which presumably take place in the very early universe.


\subsubsection{QCD Phase Transition}


At temperatures well above the scale of quantum chromodynamics
(QCD),\footnote{The QCD scale $\Lambda_{\textrm{QCD}}$ corresponds to the energy scale
at which, according to its renormalization group running in perturbative QCD,
the strong coupling constant $g_s$ formally diverges.}
$T \gg \Lambda_{\textrm{QCD}} \simeq 220\,\textrm{MeV}$, most quarks, antiquarks and
gluons interact only very weakly with each other.
Instead of being bound in baryons or mesons, they freely propagate through the thermal bath
as independent degrees of freedom, forming what is referred to as a quark-gluon plasma.
However, as the temperature decreases, the strong force becomes increasingly stronger,
until at a temperature $T \sim 100\,\textrm{MeV}$ all colour-charged particles get
confined in hadrons, i.e.\ pions for the most part.
This transition from the quark-gluon plasma to hadronic matter is known as
the QCD or quark-gluon phase transition (cf.\ Fig.~\ref{fig:history}).
Its order parameter, $\xi_{\textrm{QCD}}$, keeping track of the progress of the QCD
transition as it unfolds, is given by the vacuum expectation value (VEV) of the quark condensate
operator, $\xi_{\textrm{QCD}} = \big<q_L q_R + \bar{q}_L \bar{q}_R\big>$.
While $\xi_{\textrm{QCD}}$ initially vanishes, it is of
$\mathcal{O}\big(\Lambda_{\textrm{QCD}}^3\big)$
at the end of the QCD phase transition.
As the quark condensate operator transforms nontrivially under chiral transformations,
we conclude that the QCD phase transition entails the spontaneous breaking of
the global chiral symmetry in the quark sector.
We also note that, according to numerical lattice calculations, the QCD phase
transition is most likely a smooth crossover rather than a first or second
order phase transition.
One thus expects that it does not leave any observationally detectable
imprint in the cosmic evolution.


\subsubsection{Electroweak Phase Transition}


The QCD phase transition is believed to be preceded by the electroweak (EW)
phase transition, occurring close to the Fermi or electroweak scale
$v_{\textrm{EW}} \simeq 174\,\textrm{GeV}$ (cf.\ Fig.~\ref{fig:history}).
At temperatures $T \gg v_{\textrm{EW}}$ all standard model particles are massless
and the universe is said to be in the symmetric phase.
The order parameter of the electroweak phase transition, $\xi_{\textrm{EW}}$,
is identified with the VEV of the Higgs product operator $H^\dagger H$,
with $H$ denoting the standard model Higgs doublet,
$\xi_{\textrm{EW}} = \big<H^\dagger H\big>$.
By definition, $\xi_{\textrm{EW}}$ vanishes in the symmetric phase.
Once the temperature drops below a critical value $T_{\textrm{EW}}$,
the Higgs boson $h \in H$, the electroweak gauge bosons and all fermions
except for neutrinos acquire masses through the Higgs mechanism.
This is reflected in the order parameter $\xi_{\textrm{EW}}$ obtaining a
nonzero value that approaches $v_{\textrm{EW}}^2$ in the zero-temperature
limit.
Both the explicit value of $T_{\textrm{EW}}$ as well as the order of the electroweak phase
transition depend on the Higgs boson mass $m_h$.
The LHC experiments ATLAS and CMS recently presented hints that the Higgs boson may
have a relatively large mass,
$ m_h \simeq 125\,\textrm{GeV}$~\cite{ATLAS:2012ae,Chatrchyan:2012tx}.
Based on this value for $m_h$, one finds
$T_{\textrm{EW}} \simeq 170\,\textrm{GeV}$~\cite{Mukhanov:2005sc}.
Furthermore, given $ m_h \simeq 125\,\textrm{GeV}$, the electroweak phase transition
turns out to be a smooth crossover without any dramatic cosmological consequences.


After the phase transition the universe is in the Higgs
phase and the electroweak symmetry is said to be spontaneously broken to the
electromagnetic symmetry,
\begin{align}
SU(2)_W \times U(1)_Y \rightarrow U(1)_{\textrm{EM}}\,.
\end{align}
This terminology is, however, not quite correct as $H^\dagger H$ transforms as
a singlet under all gauge transformations, so that the electroweak symmetry remains
intact even after the electroweak phase transition.
What happens instead is a rearrangement of the physical degrees of freedom,
proceeding in such a way that after the phase transition the electroweak symmetry
is realized in a nonlinear fashion.
It would hence be more appropriate to speak of the electroweak symmetry as being
\textit{hidden} subsequent to the phase transition.
However, as it is more common to refer to it as being \textit{broken},
we will adopt this terminology in the following.\footnote{Likewise, when referring
to some Higgs product operator $s^\dagger s$ acquiring a VEV $v$,
we will also sometimes write $v = \left<s\right>$, although we actually mean
$v = \left<s^\dagger s\right>^{1/2}$.\smallskip}
We note that this discussion applies in particular also to the \BmL
phase transition, during which the \BmL gauge symmetry actually becomes hidden
rather than broken.


\subsubsection{Electroweak Instanton and Sphaleron Transitions}


As the temperature approaches the electroweak scale, also nonperturbative
processes which simultaneously violate baryon number $B$ and lepton number $L$
gain in importance.
Their emergence is a direct consequence of the fact that the electroweak
dynamics are governed by a chiral and non-Abelian gauge theory.
First of all, we note that both global $U(1)_B$ and $U(1)_L$ transformations represent
accidental symmetries of the standard model Lagrangian.
Hence, both $B$ and $L$ are conserved in the standard model
at the classical level.
Due to the chiral nature of the electroweak interactions, they are, however,
violated at the quantum level through the triangle anomaly,
which results in the divergences of the baryon and lepton number currents, $J_B^\mu$ and
$J_L^\mu$, being nonzero~\cite{'tHooft:1976up,'tHooft:1976fv},
\begin{align}
\partial_\mu J_B^\mu = \partial_\mu J_L^\mu =
\frac{N_f}{32\pi^2} \epsilon^{\mu\nu\sigma\tau}
\left(-g_W^2 \textrm{Tr}\,W_{\mu\nu} W_{\sigma\tau}
+ g_Y^2 B_{\mu\nu} B_{\sigma\tau}\right) \,.
\label{eq:anomaly}
\end{align}
Here, $N_f$ counts the number of fermion families, $\epsilon^{\mu\nu\sigma\tau}$ represents
the Levi-Civita symbol in four dimensions, $W_{\mu\nu}^a$ and $B_{\mu\nu}$
are the field strength tensors of the weak and hypercharge gauge fields, and
$g_W$ and $g_Y$ denote the corresponding gauge couplings.
The second ingredient to the nonconservation of $B$ and $L$ is
the complicated structure of the vacuum of the $SU(2)_W$ gauge theory.
As for any non-Abelian gauge theory, the $SU(2)_W$ vacuum
manifests itself in infinitely many, homotopically distinct,\footnote{Gauge
configurations belonging to different homotopy classes are transformed into
each other via \textit{large} gauge transformations.} pure gauge configurations, each
of which is characterized by a specific integer topological charge or Chern-Simons
number $N_{\textrm{CS}}$.
An important observation is that distinct realizations of the $SU(2)_W$ vacuum differing
by $\Delta N_{\textrm{CS}} = 1$ are connected to each other via a non-contractible
loop in field configuration space~\cite{Manton:1983nd}.
The field configuration of highest energy along this path is known as the
sphaleron~\cite{Klinkhamer:1984di}.
Corresponding to a saddle-point of the energy functional of the gauge-Higgs system,
the sphaleron represents a classical, spatially localized and static, but unstable
solution of the electroweak field equations.
Its energy $E_{\textrm{sph}}$ determines the height of the potential barrier by
which two adjacent realizations of the $SU(2)_W$ vacuum are separated,
\begin{align}
E_{\textrm{sph}}(T) \simeq \frac{8\pi}{g_W} \sqrt{2}v_{\textrm{EW}}(T) \,,\quad
v_{\textrm{EW}}(T) = \xi_{\textrm{EW}}^{1/2}(T) \,.
\label{eq:EsphvEWT}
\end{align}


Now combining the nontrivial topology of the $SU(2)_W$ vacuum with
the fact that the currents $J_B^\mu$ and $J_L^\mu$ have nonzero divergences
(cf.\ Eq.~\eqref{eq:anomaly}), one can show that both $B$ and $L$ are violated
in topological vacuum transitions,
\begin{align}
\Delta B = \Delta L = N_f \,\Delta N_{\textrm{CS}} \,.
\end{align}
In the standard model, in which we have $N_f = 3$, the smallest jump in $B$ and $L$
is hence $\Delta B = \Delta L = \pm 3$.
The difference between $B$ and $L$ is, by contrast, always conserved in topological
transitions.
This is also evident from the vanishing divergence of the \BmL current,
$\partial_\mu J_{\textrm{\BmL}}^\mu = \partial_\mu J_B^\mu - \partial_\mu J_L^\mu = 0$
(cf.\ Eq.~\eqref{eq:anomaly}).


Topological transitions between different realizations of the
$SU(2)_W$ vacuum come in two different varieties.
One possibility is tunneling \textit{through} the potential barrier via $SU(2)_W$ instantons.
The instanton rate is, however, proportional to
$\exp\left(-16\pi^2/g_W^2\right) \sim 10^{-170}$ and thus severely suppressed.
This is to say that in the standard model $B$- and $L$-violating processes are
completely negligible at low temperature.
On the other hand, in the hot plasma filling the universe during the
radiation-dominated era, thermal fluctuations can lead to sphaleron transitions
\textit{over} the potential barrier~\cite{Kuzmin:1985mm}.
In the Higgs phase, the sphaleron rate is proportional to
$\exp\left(- E_{\textrm{sph}}/T\right)$~\cite{Arnold:1987mh}
and hence becomes unsuppressed as soon as $T \gtrsim E_{\textrm{sph}}$.
Although the barrier is large at zero temperature,
$E_{\textrm{sph}} \simeq 10\,\textrm{TeV}$, it rapidly melts away as
the temperature approaches the critical value $T_{\textrm{EW}}$ from below
(cf.\ Eq.~\eqref{eq:EsphvEWT}).
That is why sphalerons already reach thermal equilibrium at a temperature
slightly below the critical temperature, $T_{\textrm{sph}}^{\textrm{min}} \sim
T_{\textrm{EW}}-10\,\textrm{GeV}$~\cite{KlapdorKleingrothaus:1999bd},
rather than at temperatures as high as $10\,\textrm{TeV}$.
Conversely, we may say that at $T_{\textrm{sph}}^{\textrm{min}}$ the sphaleron
processes freeze-out, so that for $T \ll T_{\textrm{sph}}^{\textrm{min}}$ both
$B$ and $L$ are conserved.
This means in particular that at the latest around $T = T_{\textrm{sph}}^{\textrm{min}}$
the baryon asymmetry is fixed to its present value.


During the restoration of the electroweak symmetry the potential barrier vanishes completely.
Hence, the actual sphaleron configuration in the sense of a saddle-point of the energy
functional no longer exists in the symmetric phase.
Instead, at $T > T_{\textrm{EW}}$, topological transitions occur due to thermal fluctuations
in the electroweak gauge fields.
In the following we shall, however, refer to these transitions as sphaleron processes
nonetheless.
In the symmetric phase, sphaleron transitions occur at rate per
unit volume $\Gamma_{\textrm{sph}} / V \propto \alpha_W^5 T^4$~\cite{Arnold:1999ux}
where $\alpha_W = g_W^2 / \left(4\pi\right)$.
This result can be used to show that sphalerons are in thermal equilibrium
up to a temperature $T_{\textrm{sph}}^{\textrm{max}} \sim 10^{12}\,\textrm{GeV}$.
At higher temperatures the sphaleron rate is again outweighed by the expansion rate.


Above the electroweak scale all standard model gauge and Yukawa interactions as well
as the electroweak sphaleron and QCD instanton processes are in thermal equilibrium.
This implies relations between the chemical potentials of all fermions and Higgs particles,
which, together with the requirement that the total hypercharge of the thermal bath be zero,
can be used to derive the sphaleron-driven equilibrium values of $B$ and
$L$~\cite{Khlebnikov:1988sr},
\begin{align}
\textrm{$B = C_{\textrm{sph}} (B$$-$$L)$} \,,\quad
\textrm{$L = \left(C_{\textrm{sph}} - 1\right) (B$$-$$L)$} \,,\quad
C_{\textrm{sph}} = \frac{8 N_f + 4 N_H}{22 N_f + 13 N_H} \,,
\label{eq:BLeqCsph}
\end{align}
with $N_H$ denoting the number of Higgs doublets.
The standard model (SM) only contains one Higgs doublet $H$, while in its minimal
supersymmetric extension, the minimal supersymmetric standard model (MSSM),
two Higgs doublets, $H_u$ and $H_d$, are required in order to ensure anomaly freedom,
\begin{align}
\textrm{SM:} \quad N_H = 1\,,\: C_{\textrm{sph}} = \frac{28}{79} \qquad
\textrm{MSSM:} \quad N_H = 2\,,\: C_{\textrm{sph}} = \frac{8}{23} \,.
\label{eq:CsphSMMSSM}
\end{align}
From Eq.~\eqref{eq:BLeqCsph} we conclude that if $\textrm{\BmL}=0$,
sphaleron processes always completely wash out any baryon asymmetry,
which is generated in some nonequilibrium process at
$T \gg T_{\textrm{sph}}^{\textrm{min}}$.
By contrast, as \BmL is conserved in topological transitions, any primordial
\BmL asymmetry is guaranteed to survive until sphaleron freeze-out.
From this perspective, the baryon asymmetry, which we presently observe in the universe,
points to a nonequilibrium process above the electroweak scale that is responsible
for the generation of a primordial \BmL asymmetry.
As we will see in Sec.~\ref{subsec:baryogenesis}, leptogenesis is a prime candidate for
such a process.
Let us denote the time when the $B$$-$$L$-violating process, i.e.\ leptogenesis
in our case, terminates by $t_f$.
The present value of the baryon asymmetry or baryon-to-photon ratio $\eta_B$
is then related to the primordial \BmL in the following way,
\begin{align}
\eta_B^0 = \left.\frac{n_B}{n_\gamma}\right|_{t_0} =
C_{\textrm{sph}} \left.\frac{n_{\textrm{\BmL}}}{n_\gamma}\right|_{t_0} =
C_{\textrm{sph}} \frac{g_{*,s}^0}{g_{*,s}}
\left.\frac{n_{\textrm{\BmL}}}{n_{\gamma}}\right|_{t_f} \,.
\label{eq:etaB0}
\end{align}



\chapter{Framework for a Consistent Cosmology}
\label{ch:framework}


As we have seen in the previous chapter, a series of astrophysical and cosmological
observations yields direct evidence for a hot thermal stage in the early history of
our universe.
Owing to the successful theory of BBN, our picture of this phase is rather well
established up to times as early as $t \sim 0.1\,\textrm{s}$.
Beyond that, standard model physics allows us to make an educated guess
about the further evolution of the universe up to temperatures of $\mathcal{O}(1)\,\textrm{TeV}$,
which are reached at $t \sim 10^{-13} \,\textrm{s}$.
However, we have also seen that, despite the impressive achievements in recent years,
modern cosmology still faces a multitude of serious problems, all of which
point to new physics beyond the standard model.


In this chapter, we shall thus contrive a consistent cosmological scenario, based on the
idea that the hot early universe is ignited by the spontaneous breaking of $B$$-$$L$,
which provides us with answers to several questions of early universe cosmology in one go.
First, we will motivate various extensions of the standard model,
each of which has respectively been put forward in order to address one of these
questions individually (cf.\ Sec.~\ref{sec:byndsm}).
Then we will assemble all the pieces of the puzzle and outline how the \BmL
phase transition at the end of inflation gives rise to a consistent cosmology
(cf.\ Sec.~\ref{sec:orgnunvrs}).
Finally, we will introduce a phenomenological flavour model based on an
Abelian Froggatt-Nielsen flavour symmetry, which will enable us to study this
cosmology in quantitative terms.


\newpage


\section{Beyond the Standard Model}
\label{sec:byndsm}


The theoretical framework, within which we will develop a consistent cosmology,
shall feature the following phenomena beyond the standard model:
(i) supersymmetric $F$-term hybrid inflation as the key to resolving the flatness and horizon
problems as well as to explaining the origin of the primordial metric fluctuations
(cf.\ Sec.~\ref{subsec:CMB}),
(ii) the type I seesaw mechanism in order to account for the small masses of the standard model
neutrinos (cf.\ Sec.~\ref{subsec:neutrinos}),
(iii) leptogenesis as the process generating the primordial \BmL asymmetry in the early
universe (cf.\ Secs.~\ref{subsec:baryncmttr} and \ref{subsec:transitions}), and
(iv) gravitinos or other weakly interacting massive particles (WIMPs) as particle
candidates for dark matter (cf.\ Sec.~\ref{subsec:DM}).\footnote{Meanwhile,
we will ignore the question as to the nature of dark energy, the
coincidence problem (cf.\ Secs.~\ref{subsec:DE} and \ref{subsec:stages})
as well as the lithium problem in BBN (cf.\ Sec~\ref{subsec:BBN}).\smallskip}
Let us now discuss each of these phenomena in turn.


\subsection{Inflation}
\label{subsec:infltn}


\subsubsection{Basics of Inflation}


Inflation denotes a stage of accelerated cosmic expansion taking place
in the very early universe during which gravity acts as a repulsive
force.\footnote{The first inflationary model
(\textit{old inflation}) was proposed by Alan Guth in 1980~\cite{Guth:1980zm}.
Subsequently, it was further developed (\textit{new inflation})
by Andrei Linde~\cite{Linde:1981mu} as well as Andreas Albrecht and Paul Steinhardt~\cite{Albrecht:1982wi}.
For reviews on inflation,
cf.\ for instance Refs.~\cite{Lyth:1998xn,Linde:2005ht}.\smallskip}
It provides viable solutions to the flatness and horizon problems, if during inflation
physical scales are stretched by at least a factor of $\mathcal{O}\left(10^{29}\right)$.
To see that, note that in an accelerating universe the total density parameter
$\Omega_{\textrm{tot}}$ always asymptotically approaches unity.
Hence, $\Omega_{\textrm{tot}} = 1$ is a future attractor of any inflationary universe,
irrespectively of its concrete initial conditions.
After inflation the deviation from exact spatial flatness is, in particular, such small
that even at present $\Omega_{\textrm{tot}}$ is still very close to unity, although
the decelerated expansion during radiation and matter domination actually causes
$\left|\Omega_{\textrm{tot}} - 1\right|$ to grow again.
Moreover, the inflationary paradigm implies that, as a result of the immense cosmic expansion,
the entire observable universe in fact originates from a single homogeneous, causally connected
patch.\footnote{The observable universe may as well emerge from an \textit{inhomogeneous},
causally connected domain.
Inflation would then simply have to last longer, so that initial inhomogeneities
are stretched to physical scales which presently still exceed the size of the observable universe.}
Before the onset of inflation, the entire universe is hence in thermal contact after all,
which resolves the horizon problem.


The requirement that the scale factor has to increase by at least a factor of
$\mathcal{O}\left(10^{29}\right)$ during inflation can be translated into the condition
that at the beginning of inflation the deviation from the vacuum equation of state
must not exceed $1\,\%$.
In field theory such an equation of state, $\omega \simeq -1$, can be easily realized
by a homogeneous real scalar field, the \textit{inflaton} field $\varphi$, which slowly
rolls down its potential.
The inflaton dynamics are governed by the Klein-Gordon equation in an expanding
Friedmann-Lema\^itre background,
\begin{align}
\ddot{\varphi} + 3 H \dot{\varphi} + V'\left(\varphi\right) = 0 \,,\quad
V'\left(\varphi\right) = \frac{d}{d\varphi} V\left(\varphi\right)\,.
\label{eq:KleinGordon}
\end{align}
For a sufficiently flat scalar potential $V$, the two slow-roll conditions are satisfied:
(i) the energy of the scalar field is dominated by its potential energy,
$\frac{1}{2}\dot{\varphi}^2 \ll V$, and
(ii) the acceleration of the scalar field is negligibly small compared to the friction
as well as to the gradient term in its equation of motion,
$\big|\ddot{\varphi}\big| \ll 3 H \big|\dot{\varphi}\big|, \big|V'\big|$.
Eq.~\eqref{eq:KleinGordon} then reduces to
\begin{align}
3 H \dot{\varphi} + V'\left(\varphi\right) \approx 0 \,,
\label{eq:KleinGordonSR}
\end{align}
and the equation of state approximately corresponds to the one of the vacuum.
In the slow-roll approximation, the scale factor $a$ grows exponentially fast,
\begin{align}
a(t) \approx a_e \exp\left[H_I\left(t_e-t\right)\right] \approx e^{N_e} \,,
\end{align}
where $H_I$ denotes the value of the Hubble parameter after neglecting
all contributions to the total energy density $\rho_{\textrm{tot}}$ except
for the vacuum energy density $\rho_0$ and $N_e$ stands for the number of
\textit{e-folds} by which the universe expands between a given time $t$ and the
end of inflation at time $t_e$,
\begin{align}
H(t) \approx H_I = \left(\frac{8\pi}{3 M_P^2} \rho_0\right)^{1/2} \,,\quad
N_e = \int_{t}^{t_e} dt' H\left(t'\right) \approx H_I \left(t_e - t\right) \,.
\label{eq:HINe}
\end{align}


A key argument for attributing inflation to a slowly-rolling scalar field is that
it comes with a built-in mechanism to generate the primordial metric
fluctuations imprinted in the CMB.
Quantum fluctuations of the scalar field arising on scales below the Hubble horizon,
$H^{-1} \approx H_I^{-1} = \textrm{const.}$, are stretched in the course of inflation
to ever larger physical scales.
As they cross the horizon, the fluctuations \textit{freeze-in}, which means that their
amplitudes remain preserved as soon as they become sensitive to curvature effects.
Once inflation ends, the primordial fluctuations re-enter the horizon, now in
the form of the classical scalar metric perturbations which lay the foundation of the
CMB anisotropies.
Similarly, tensor metric perturbations stretched to super-horizon scales during
inflation give rise to the primordial gravitational waves which also affect the
physics of the CMB.


Slow-roll inflation generally predicts nearly scale-invariant and almost perfectly
Gaussian power spectra for both primordial scalar and tensor perturbations.
In any model of slow-roll inflation, the inflationary observables $A_s$, $n_s$, $r$, and $n_t$
can be conveniently calculated from the slow-roll parameters  $\epsilon_V$ and $\eta_V$,
\begin{align}
\epsilon_V = \frac{1}{2\kappa} \left(\frac{V'}{V}\right)^2 \,,\quad
\eta_V = \frac{1}{\kappa} \frac{V''}{V} \,,\quad
\kappa = \frac{8\pi}{M_P^2} \,.
\end{align}
The two slow-roll conditions stated above are equivalent to the requirement
that $\epsilon_V$ and $\eta_V$ be very small, $\epsilon_V,\left|\eta_V\right| \ll 1$.
If that requirement is fulfilled, one finds
\begin{align}
P_s(k) \approx \left.\frac{\kappa^2V}{24\pi^2\epsilon_V}\right|_{aH = k} \,,\quad
P_t(k) \approx \left.\frac{2\kappa^2V}{3\pi^2}\right|_{aH = k} \,,
\label{eq:PsPt}
\end{align}
where for a given scale $k$ the right-hand sides of both equations are to be evaluated
at the respective time when the scale exits the Hubble horizon during inflation.
From Eq.~\eqref{eq:PsPt} one can easily deduce:
\begin{align}
A_s = P_s(k_*) \,,\quad
n_s = 1+ 2 \eta_V - 6 \epsilon_V \,,\quad
r = 16 \epsilon_V \,,\quad
n_t = -2 \epsilon_V \,,
\label{eq:Asnsrnt}
\end{align}
where $\epsilon_V$ and $\eta_V$ are understood to be evaluated at $k = k_*$.


At the end of inflation the universe is flat, homogeneous and isotropic.
Its energy density only receives contributions from the vacuum as well as the
homogeneous inflaton field $\varphi$, which may be regarded as a classical condensate
of inflaton particles.
The energy densities of other particles, any pre-existent baryon asymmetry,
the abundance of topological defects as well as any primordial curvature are
completely diluted during inflation.\footnote{The rigorous formulation of
this statement goes by the name of the \textit{no-hair} theorem~\cite{Wald:1983ky}.\smallskip}
Subsequent to inflation, the energy contained in the vacuum and the inflaton field is
converted into the energy of a hot thermal plasma.
This process, connecting the inflationary stage with the radiation-dominated era, is known
as the \textit{reheating} of the universe and a large fraction of this thesis is devoted
to a detailed study of its dynamics.
The characteristic temperature scale of reheating is referred to as the reheating
temperature $T_{\textrm{RH}}$.
It is a measure for the highest temperature ever reached in the hot early universe
and thus represents one of the most important parameters of early universe cosmology.
In the context of our cosmological framework, it is in particular closely related to
the origin of the BAU and the nature of dark matter.
Note that, in Fig.~\ref{fig:history}, we have indicated several possibilities for the actual
value of the reheating temperature.


\subsubsection{Supersymmetric \boldmath{$F$}-Term Hybrid Inflation}


Among the multitude of inflationary models present in the literature, we shall consider
supersymmetric $F$-term hybrid inflation in thesis.
In general, models of hybrid inflation feature next to the inflaton field $\varphi$
at least one second scalar field, the \textit{waterfall} field $\sigma$, which is stabilized during
inflation due to its interactions with the inflaton field and whose potential energy
dominates the total energy budget.\footnote{Note that hybrid inflation differs in this
respect from ordinary single-field inflation, during which $\rho_{\textrm{tot}}$
receives its largest contribution from the potential energy of the inflaton field itself.}
However, once $\varphi$ drops below a critical value $\varphi_c$, the waterfall field $\sigma$
becomes destabilized and inflation ends in a phase transition, during which
$\varphi$ and $\sigma$ evolve from the unstable \textit{false} vacuum to the stable
\textit{true} vacuum.
Depending on the transformation behaviour of the field $\sigma$, this phase transition at
the end of inflation may be accompanied by the spontaneous breakdown of some global or local
symmetry of the Lagrangian.
Hybrid inflation hence offers a natural setting for spontaneous symmetry
breaking (SSB) in the course of the cosmic evolution.
In this sense, it allows to establish a connection between cosmology and particle physics
and thus represents a particularly attractive scenario of inflation.
The symmetry breaking at the end of hybrid inflation may, in particular,
be identified as an intermediate stage in the breaking of the gauge group $G_{\textrm{GUT}}$
of some grand unified theory (GUT) down to the standard model gauge group $G_{\textrm{SM}}$.
The scalar sector, containing among other fields the two scalars $\varphi$
and $\sigma$, would then be determined by the particle content of the respective GUT theory
and inflation would turn out to be a mere implication of particle physics.
As we will discuss in more detail below, we shall, of course, presume that it is
the difference between baryon and lepton number \BmL which is broken
at the end of inflation.


The first and simplest model of hybrid inflation was proposed by Andrei Linde in
the early 1990s~\cite{Linde:1991km,Linde:1993cn}.
Rather than embedding SSB into a cosmological context, Linde's original motivation,
however, was to construct a model in which inflation ends differently as compared to
the standard scenarios featuring a first-order phase transition or a slow-roll motion
gradually becoming faster and faster.
Indeed, under quite generic conditions, the field $\sigma$ evolves very rapidly,
though continuously, towards the true vacuum during the phase transition
(cf.\ Sec.~\ref{sec:preheating}).
Hybrid inflation hence typically ends very abruptly, which is why the phase transition
marking its end is usually referred to as a \textit{waterfall} transition.
Unfortunately, Linde's model predicts a scalar spectral index $n_s \geq 1$,
with the deviation from an exactly flat spectrum, $n_s = 1$, generically being very
small.
As observations point to $n_s = 0.963 \pm 0.012$ (cf.\ Eq.~\eqref{eq:AsnsWMAP}),
it is therefore disfavoured at the level of at least $3\,\sigma$.
This conflict with the observational data can, however, be resolved in
supersymmetric versions of hybrid inflation.


Let us consider supersymmetric $F$-term hybrid inflation, which is implemented by
the following superpotential~\cite{Copeland:1994vg,Dvali:1994ms},
\begin{align}
W_{B-L} = \frac{\sqrt{\lambda}}{2}\Phi \left(v_{B-L}^2 - 2 S_1 S_2\right) \,.
\label{eq:WBL}
\end{align}
Here, $\lambda$ is a dimensionless coupling constant, $v_{B-L}$ represents a mass scale,
and $\Phi$, $S_1$ and $S_2$ are chiral superfields.
$W_{B-L}$ is the simplest renormalizable superpotential allowing for the
spontaneous breaking of a $U(1)$ symmetry of the Lagrangian.
We identify this $U(1)$ with the global $U(1)_{B-L}$, which is preserved in
standard model interactions even at the quantum level
(cf.\ Sec.~\ref{subsec:transitions}),\footnote{In Sec.~\ref{subsec:seesaw}, we will
see that $U(1)_{B-L}$ can, in fact, also be promoted to a local symmetry.} and assign
the following \BmL charges to the three chiral superfields:
$q_\Phi = 0$ as well as $q_S \equiv q_{S_2} = - q_{S_1} = 2$.
Hence, $\Phi$ represents a \BmL singlet, while the two fields
$S_1$ and $S_2$ transform as conjugates of each other.
The radial component $\varphi$ of the complex scalar
$\phi = \varphi/\sqrt{2}e^{i\theta}$ contained in $\Phi$ plays the role
of the inflaton.
The \BmL Higgs boson or waterfall field $\sigma$ corresponds to one of the
four real scalar degrees of freedom contained in 
the superfields $S_1$ and $S_2$ (cf.\ Sec.~\ref{sec:beforeSSB}).
In the following, we shall collectively refer to all component fields of
$\Phi$, $S_1$ and $S_2$ as the symmetry breaking sector.
Given the above \BmL charge assignments, we remark that $W_{B-L}$
is \textit{natural} in the strong sense that it is the most general renormalizable
superpotential which is compatible with the $U(1)_{B-L}$ as well as with
$R$~symmetry.
As the superpotential itself has to carry $R$ charge $R\left(W_{B-L}\right) = 2$,
we know that $R\left(\Phi\right) = 2$ and $R\left(S_1 S_2\right) = 0$, which means
that $R\left(S_1\right) = R\left(S_2\right) = 0$ in the simplest case.
These $R$ charge assignment forbid all terms involving the inflaton field $\Phi$
except for the linear term in Eq.~\eqref{eq:WBL} and hence enforce the absence of
undesirable inflaton couplings.
Meanwhile, the \BmL charges ensure that $S_1$ and $S_2$ can only appear in the
superpotential in the form of their product $S_1 S_2$.


During inflation the scalar mass eigenstates contained in $S_1$ and $S_2$ have masses
$m_{s_\pm}^2 = \lambda/2\left(\varphi^2 \mp v_{B-L}^2\right)$
(cf.\ Sec.~\ref{sec:beforeSSB}) and are thus stabilized at the origin for large
inflaton field values.
The inflaton-Higgs system is hence in the false vacuum state,
$\left<\Phi\right> \neq 0$ and $\left<S_1\right> = \left<S_2\right> = 0$,
and the scalar potential energy is dominated by the vacuum energy
density $\rho_0$,
\begin{align}
S_{1,2}\rightarrow 0 \,:\quad
V \rightarrow \left|W_{,\Phi}\right|^2 = \frac{1}{4}\lambda v_{B-L}^4 = \rho_0 \,,\quad
W_{,\Phi} = \frac{\partial W}{\partial\phi}
\label{eq:Fterm}
\end{align}
As $\rho_0$ originates from the $F$-term contribution $\left|W_{,\Phi}\right|^2$ to
the scalar potential, the model defined through Eq.~\eqref{eq:WBL}
is known as $F$-term hybrid inflation.
Correspondingly, models in which $\rho_0$ stems from a $D$-term
contribution to the scalar potential are referred to as $D$-term hybrid
inflation~\cite{Binetruy:1996xj,Halyo:1996pp}.
Note that, while \BmL is conserved during inflation, supersymmetry is spontaneously
broken by $\rho_0$.
This situation is reversed at the end of inflation.
Once the inflaton field $\varphi$ drops below the critical value $\varphi_c = v_{B-L}$,
the mass squared of the waterfall field turns negative, i.e.\ the waterfall field
becomes tachyonically unstable, which triggers the transition to the true vacuum.
In the true ground state, supersymmetry is no longer broken, but \BmL
is broken spontaneously,
$\left<\Phi\right> = 0$ and $\left<S_1\right> = \left<S_2\right> = v_{B-L}/\sqrt{2}$.
This phase transition entailing the breaking of \BmL gives this thesis its
name.
In Sec.~\ref{sec:beforeSSB}, we will compute the full Lagrangian governing the
\BmL phase transition; in Ch.~\ref{ch:phasetransition}, we will then elaborate
in more detail on the various nonperturbative processes associated with it.


An appealing feature of supersymmetric hybrid inflation is that it comes with
an intrinsically flat inflaton potential.
Along the inflationary trajectory, $S_{1,2} = 0$, the ordinary tree-level
scalar potential is constant (cf.\ Eq.~\eqref{eq:Fterm}).
Furthermore, assuming a canonical K\"ahler potential, one can show that, as $W_{B-L}$
is linear in the inflaton field $\Phi$, the supergravity (SUGRA) corrections to the inflaton
mass $m_\varphi$ exactly cancel at tree-level~\cite{Copeland:1994vg}.
This provides a solution to the notorious \textit{eta problem}, with which
inflationary models based on supergravity usually have to struggle.
As supergravity typically induces inflaton masses of order the Hubble rate,
$m_\varphi^2 \sim H_I^2 = 8\pi\rho_0/\left(3M_P^2\right)$, the contribution from
the inflaton mass to the slow-roll parameter $\eta_V$ can become dangerously large,
$\eta_V = m_\varphi^2 M_P^2/\left(8\pi V\right) + ... \sim 1$.
Given the superpotential $W_{B-L}$, together with a minimal K\"ahler potential, the
eta problem is now partly resolved.\footnote{Of course, other scalar fields present
in the early universe and having values of $\mathcal{O}(M_P)$, such as moduli fields
from string theory, could still spoil inflation through their contributions to $\eta_V$.}
The actual slope of the inflaton potential, causing $\varphi$ to roll towards
its critical value $\varphi_c$, is induced at the one-loop level by the
Coleman-Weinberg potential~\cite{Coleman:1973jx},
\begin{align}
V_{\textrm{CW}} = \frac{1}{64\pi^2} \textrm{STr}
\left[M^4 \ln\left(\frac{M^2}{\Lambda^2}\right)\right] \,,\quad
\textrm{STr}\,M^2 = \sum_s (-1)^{2s}\left(2s+1\right) \textrm{Tr}\,M_s^2\,,
\label{eq:VCWdef}
\end{align}
where $\Lambda$ is a renormalization scale and $M^2$ denotes the total mass
matrix squared of our theory, which receives contributions from particles
of all possible spins $s$, $M^2 = \bigoplus_s M_s^2$.
Thanks to the spontaneous breaking of supersymmetry by the vacuum energy density
$\rho_0$, the supertrace in Eq.~\eqref{eq:VCWdef} ends up being
nonzero.
The mass splitting among the mass eigenstates contained in $S_1$ and $S_2$
results in
\begin{align}
V_{\textrm{CM}} = \frac{\lambda \rho_0}{16\pi^2}\left[
\ln\left(\frac{\lambda\varphi^2}{2\Lambda^2}\right) +
\frac{1}{2}\sum_{n=\pm1}\left(1 + n \,x^2\right)^2
\ln\left(1 + n\,\frac{1}{x^2}\right) \right] \,,\quad x = \frac{\varphi}{\varphi_c} \,.
\label{eq:VCM}
\end{align}
Beyond the radiative Coleman-Weinberg correction, SUGRA induces
further terms in the inflaton potential.
As stated above, these are at least of
fourth order in $\varphi$ assuming a minimal K\"ahler potential,
but also include an inflaton mass term, if the K\"ahler potential
has a noncanonical form~\cite{Linde:1997sj}.
Anyway, since we are mainly interested in the transition from inflation to
the hot thermal universe rather than in the exact inflationary dynamics,
we may disregard all SUGRA corrections.
More precisely, we can safely neglect all SUGRA effects in our analysis,
as long as the \BmL phase transition takes place at a scale, which is much lower
than the Planck scale.
Moreover, there is a curvature-induced correction to the inflaton potential,
which, however, turns out to be insignificant for all viable values of
$\lambda$~\cite{Garbrecht:2006df,Battye:2006pk}.


Finally, let us analytically estimate the parameter dependence of the
inflationary observables $A_s$ and $n_s$.
The scale $k_*$, on which these quantities are determined in CMB observations,
leaves the Hubble horizon $N_e^* \simeq 50$ e-folds before the end of
inflation~\cite{BasteroGil:2006cm}.
For intermediate values of the coupling constant, $\lambda \sim 10^{-4}$,
the inflaton field value $\varphi_*$ at this time is much larger than the critical
value, but still far below the Planck scale.
To first approximation, the effective inflaton potential
is then given as:
\begin{align}
\varphi_c \ll \varphi \ll M_P \,: \quad
V_{\textrm{eff}}\left(\varphi\right) \approx \rho_0 \left[1 +
\frac{\lambda}{16\pi^2} \ln\left(\frac{\lambda\varphi^2}{2\Lambda^2}\right)\right] \,,
\end{align}
from which one easily deduces the two slow-roll parameters $\epsilon_V$ and $\eta_V$,
\begin{align}
\epsilon_V \approx \frac{\lambda}{16\pi^2}\left|\eta_V\right| \,,\quad
\eta_V \approx - \frac{1}{2 N_e} \,,\quad
N_e \approx\frac{32\pi^3}{\lambda M_P^2}\varphi^2 \,.
\label{eq:epsVetaVNe}
\end{align}
Hence, $\epsilon_V$ is always negligibly small compared to $\left|\eta_V\right|$.
On the other hand, $\left|\eta_V\right|$ may become of order unity way
before the inflaton field reaches $\varphi_c$.
We conclude that inflation ends at $\varphi_e$, which either
corresponds to $\varphi_{\textrm{sr}}$, the field value at which
the slow-roll condition $\left|\eta_V\right| \ll 1$ becomes violated,
or to $\varphi_c$, the critical field value,
\begin{align}
\varphi_e = \textrm{max}\left[\varphi_c,\varphi_{\textrm{sr}}\right] \,,\quad
\varphi_c = v_{B-L} \,,\quad
\varphi_{\textrm{sr}} \approx \sqrt{\frac{\lambda}{\pi}}\frac{M_P}{8\pi} \,.
\label{eq:phiecsr}
\end{align}
According to Eqs.~\eqref{eq:PsPt}, \eqref{eq:Asnsrnt} and \eqref{eq:epsVetaVNe},
one finds for the amplitude of the scalar power spectrum and the scalar spectral index,
\begin{align}
A_s \approx \frac{64\pi^2}{3} N_e^* \left(\frac{v_{B-L}}{M_P}\right)^4 \,,\quad
n_s \approx 1 - \frac{1}{N_e^*} \,.
\end{align}
The measured value of $A_s$ (cf.\ Eq.~\eqref{eq:AsnsWMAP}) thus allows to estimate
the scale $v_{B-L}$,
\begin{align}
v_{B-L} \approx \left(\frac{3 A_s}{64\pi^2 N_e^*}\right)^{1/4} M_P \simeq
8 \times 10^{15} \,\textrm{GeV} \left(\frac{A_s}{2.441 \times 10^{-9}}\right)^{1/4}
\left(\frac{50}{N_e^*}\right)^{1/4} \,, \label{eq:vBLAs}
\end{align}
which is remarkably close to the GUT scale.\footnote{GUT theories are characterized
by the fact that, as a result of renormalization group running, all gauge couplings
$g_i$ unify, i.e.\ obtain the same value $g_{\textrm{GUT}}$, at some scale
$\Lambda_{\textrm{GUT}}$.
In the MSSM, the GUT coupling is expected to be
$g_{\textrm{GUT}} \simeq \sqrt{\pi/6} \simeq 0.72$,
while the GUT scale $\Lambda_{\textrm{GUT}}$ turns out to be of
$\mathcal{O}\left(10^{16}\right)\,\textrm{GeV}$.\smallskip}
Similarly, given $N_e^* \simeq 50$, we obtain
\begin{align}
n_s \simeq 0.98 \,.
\label{eq:nsRes}
\end{align}


This result for $n_s$, deviating from the best-fit value of $n_s \simeq 0.963$ by only
$1.4\,\sigma$, may be regarded as a satisfactory improvement in comparison to
Linde's original nonsupersymmetric prediction $n_s \geq 1$.
However, one can show that for very small, $\lambda \lesssim 10^{-6}$, as well
as for very large, $\lambda \gtrsim 10^{-2}$, values of the coupling constant $\lambda$,
the spectral index increases above the naive estimate in
Eq.~\eqref{eq:nsRes}~\cite{BasteroGil:2006cm}.
$n_s \simeq 0.98$ should, hence, rather be considered as a lower bound on $n_s$.
One possibility to improve on the predicted $n_s$ value is to allow
nonminimal terms in the K\"ahler potential, which, however, requires the
tuning of at least one additional dimensionless parameter to an unnaturally
small value.
Again, as we wish to focus on the \BmL phase transition instead of the inflationary dynamics
in this thesis, we shall content ourselves with the result in Eq.~\eqref{eq:nsRes} and
forget about possible SUGRA corrections in the following.


\subsection{Right-Handed Neutrinos and the Seesaw Mechanism}
\label{subsec:seesaw}


Observations of atmospheric, solar, reactor, and accelerator neutrinos clearly
indicate flavour oscillations, i.e. in-flight flavour transitions, among the three
standard model neutrino species
\cite{Fukuda:2000np,Ahmad:2001an,Eguchi:2002dm,Ahn:2006zza,Michael:2006rx}.\footnote{For
reviews on neutrino physics, cf.\ for instance
Refs.~\cite{Nakamura:2010zzi,Mohapatra:2005wg,GonzalezGarcia:2007ib}.}
These oscillations are attributed to small mass-squared differences,
$\Delta m_{\textrm{atm}}^2$ and $\Delta m_{\textrm{sol}}^2$, between
the three known neutrino states, and hence point to new physics beyond the standard model.
At the $3\,\sigma\,\textrm{CL}$, $\left|\Delta m_{\textrm{atm}}^2\right|$
and $\Delta m_{\textrm{sol}}^2$ are constrained to lie within the
following~ranges~\cite{Nakamura:2010zzi},
\begin{align}
2.07 \times 10^{-3}\,\textrm{eV}^2 \leq & \: \left|\Delta m_{\textrm{atm}}^2\right| \leq
2.75 \times 10^{-3}\,\textrm{eV}^2 \,,\\
7.05 \times 10^{-5}\,\textrm{eV}^2 \leq & \: \Delta m_{\textrm{sol}}^2 \leq
8.34 \times 10^{-5}\,\textrm{eV}^2 \,.
\label{eq:Deltam2atmsol}
\end{align}


In the standard model, neutrinos are considered to be massless
left-handed Weyl fermions.
In order to account for the small neutrino masses, we supplement
the standard model particle content by three right-handed neutrinos, so that
neutrinos can either acquire Dirac or Majorana mass terms.\footnote{As current
neutrino data is still consistent with one standard model neutrino being massless,
one is only forced to introduce two right-handed neutrinos.
We shall, however, assume that each left-handed neutrino
is complemented by a corresponding right-handed neutrino.\smallskip}
Which of these two possibilities is realized in nature is unknown at present
and represents one of the greatest questions of modern neutrino physics.
In this thesis, we shall assume that neutrinos are Majorana fermions.
A minimal mechanism capable of generating neutrino masses
of the right magnitude is then provided by the type I seesaw
mechanism ~\cite{Minkowski:1977sc,Yanagida:1979as,
Glashow:1979nm,GellMann:1980vs,Mohapatra:1979ia}\footnote{The type II and III
variants of the seesaw mechanism feature couplings of the left-handed neutrinos
to weak isospin Higgs or fermion triplets
rather than to right-handed neutrino singlets.\smallskip}.


\subsubsection{Superpotential for the Quark and Lepton Superfields}


The superpotential setting the
stage for the type~I seesaw mechanism may be cast in the following form,
\begin{align}
W_{\textrm{Seesaw}} = \frac{1}{\sqrt{2}} h_i^n n_i^c n_i^c S_1 +
h_{ij}^\nu \mathbf{5}_i^* n_j^c H_u \,.
\label{eq:WSeesaw}
\end{align}
It combines with the MSSM superpotential $W_{\textrm{MSSM}}$ to
give the total superpotential for all chiral quark and lepton superfields
present in our theoretical framework,
\begin{align}
W_{\textrm{QL}} = & \: W_{\textrm{Seesaw}} + W_{\textrm{MSSM}} \,,\\
W_{\textrm{MSSM}} =  & \: h_{ij}^u \mathbf{10}_i \mathbf{10}_j H_u +
h_{ij}^d \mathbf{5}_i^* \mathbf{10}_j H_d \,. 
\label{eq:WMSSM}
\end{align}


In Eqs.~\eqref{eq:WSeesaw} and \eqref{eq:WMSSM}, all superfields have been arranged
in $SU(5)$ multiplets,\footnote{This $SU(5)$ structure of the superpotential
implies mass unification of down-type quarks and charged leptons at the GUT scale,
which may indeed be accomplished, if nonrenormalizable terms in the Lagrangian are
arranged such that they yield appropriate mass corrections~\cite{Ellis:1979fg}.
Also, note that we do not take $SU(5)$ to be the gauge group of our model
(cf.\ Eq.~\eqref{eq:gaugegroup}).}
the indices $i,j = 1,2,3$ label the different fermion generations,
\begin{align}
S_1 \sim \mathbf{1} \,,\quad
n^c \sim \mathbf{1} \,,\quad
H_u \sim \mathbf{5} \,,\quad
H_d \sim \mathbf{5}^* \,,\quad
\mathbf{5}^* = \left(d^c,\ell\right) \,,\quad
\mathbf{10} = \left(q,u^c,e^c\right) \,.
\end{align}
The left-handed Weyl fermions $\nu_{R,i}$ contained in the
neutrino superfields $n_i^c$ are the antiparticles of the right-handed neutrinos
$\bar{\nu}_{R,i}$ required by the seesaw mechanism.
The superfields $n_i^c$ hence carry \BmL charge $q_{n^c} = 1$,
which explains why they couple to the negatively charged Higgs superfield
$S_1$ rather than to the positively charged Higgs superfield $S_2$.
Similarly, as can be seen from the superpotential in Eq.~\eqref{eq:WSeesaw},
the superfields $n_i^c$ have $R$ charge $R\left(n_i^c\right) = 1$, which
forces the $R$ charges of $S_1$ and $S_2$ to vanish (cf.\ Sec.~\ref{subsec:infltn}).
Furthermore, including three right-handed neutrinos into the particle
spectrum of our model allows us to promote the \textit{global} $U(1)_{B-L}$, which is
a symmetry of the standard model Lagrangian even at the quantum level, to an
anomaly-free \textit{local} symmetry~\cite{Babu:1989tq,Babu:1989ex}.
Before the \BmL phase transition, the full gauge group $G$ of our model
is hence given as
\begin{align}
G = G_{\textrm{SM}} \times U(1)_{B-L} \,,\quad
G_{\textrm{SM}} = SU(3)_C \times SU(2)_W \times U(1)_Y \,.
\label{eq:gaugegroup}
\end{align}
It is an intriguing possibility that, allowing for a rescaling of the \BmL coupling
strength, the group $G$ may in fact represent an intermediate stage
in the breaking of some GUT gauge group such as $SO(10)$ down to the standard model
gauge group $G_{\textrm{SM}}$~\cite{Buchmuller:1991ce}.
A~hint towards such a scenario might be that one quark-lepton family of the standard model
plus one right-handed neutrino exactly fits into the smallest complex representation of $SO(10)$.


The Higgs superfields $H_u$ and $H_d$ in Eqs.~\eqref{eq:WSeesaw} and \eqref{eq:WMSSM}
transform in the fundamental and anti-fundamental representation of $SU(5)$, respectively,
and thus both consistent of a colour triplet as well as a weak isospin doublet.
While the Higgs doublets are supposed to obtain masses of order the electroweak scale,
the masses of the Higgs triplets have to exceed the GUT scale in order to avoid too rapid
proton decay.
This dilemma is known as the \textit{doublet-triplet splitting problem}~\cite{Witten:1981nf}.
We assume one of the many mechanisms proposed in the literature and capable of
solving this problem (cf.\ Ref.~\cite{Yamashita:2011an} and references therein)
to be at work and thus regard the Higgs triplets as being
projected out of the superpotential $W_{\textrm{QL}}$.
Correspondingly, we shall denote with $H_u$ and $H_d$ the doublet components of the
respective Higgs multiplets from now on.
During the electroweak phase transition, the electrically neutral components of
these two doublets acquire nonzero
VEVs, $v_u = \left<H_u^0\right>$ and $v_d = \left<H_d^0\right>$,
which leads to the spontaneous breakdown of the electroweak
symmetry.\footnote{The description of the electroweak phase transition in the MSSM
is conceptually similar to the standard model description, which we outlined in Sec.~\ref{subsec:transitions}.
Also in the MSSM, the phase transition is typically governed by the evolution of a
single light scalar degree of freedom, so that the same effective finite-temperature
theory is applicable as in the standard case~\cite{Farrar:1996cp}.}
Adding $v_u$ and $v_d$ in quadrature yields the electroweak scale,
$v_{\textrm{EW}} = \left(v_u^2 + v_d^2\right)^{1/2}$, while the ratio $v_u / v_d$
defines the mixing angle $\beta$ via $\tan \beta = v_u / v_d$.
In the following, we will assume large $\tan\beta$, implying that
$v_d \ll v_u \simeq v_{\textrm{EW}}$.



Finally, $h^n$, $h^\nu$, $h^u$, and $h^d$ denote Yukawa matrices.
In general, $h^n$ is an arbitrary complex symmetric matrix.
By means of a Takagi diagonalization it can, however, always be brought into a diagonal
form with real and nonnegative entries $h_i^n$ on the diagonal such that
$h_i^n \leq h_{i+1}^n$~\cite{Takagi:1925aa}.
Eq.~\eqref{eq:WSeesaw} presents $W_{\textrm{Seesaw}}$ in a form which
assumes that such a diagonalization has previously been performed.


\subsubsection{Neutrino Masses from \boldmath{$B$}\boldmath{$-$}\boldmath{$L$} and Electroweak Symmetry Breaking}


During the \BmL phase transition at the end of hybrid inflation
(cf.\ Sec.~\ref{subsec:infltn}), the Higgs superfield $S_1$ acquires
a nonzero VEV, which spontaneously breaks \BmL
and generates a Majorana masses for the neutrino superfields $n_i^c$,
\begin{align}
S_1 \rightarrow \frac{v_{B-L}}{\sqrt{2}} + S_1 \,: \quad \frac{1}{\sqrt{2}} h_i^n n_i^c n_i^c S_1
\rightarrow \frac{1}{2} M_i n_i^c n_i^c + \frac{1}{\sqrt{2}} h_i^n n_i^c n_i^c S_1 \,.
\label{eq:S1shift}
\end{align}
Consequently, the massless right-handed Weyl neutrinos $\bar{\nu}_{R,i}$ combine with
the massless left-handed Weyl antineutrinos $\nu_{R,i}$ to form heavy Majorana
neutrinos,\footnote{Here, $\nu_{R,i}$ and $\bar{\nu}_{R,i}$ denote two-component spinors,
while the $N_i$ are four-component spinors.}
\begin{align}
N_i = \begin{pmatrix} \nu_{R,i} \\ \bar{\nu}_{R,i}\end{pmatrix} \,.
\label{eq:NinuRi}
\end{align}
From this relation between the Majorana neutrinos $N_i$ and the Weyl fermions
$\nu_{R,i}$ and $\bar{\nu}_{R,i}$ it is evident that the $N_i$, rather than being
uncharged under $B$$-$$L$, do not carry any definite \BmL charge at all.
As we shall see in the next section, this property of the heavy Majorana
neutrinos is a crucial ingredient for the generation of the primordial
\BmL asymmetry in the early universe.
Each of the heavy Majorana neutrinos $N_i$ is accompanied by a superpartner,
with which it shares a common mass $M_i$.
These \textit{sneutrinos}~$\tilde{N}_i$ directly correspond to the
complex scalars contained in the neutrino superfields $n_i^c$.
In the true vacuum reached at the end of the \BmL phase transition,
the heavy-(s)neutrino mass matrix is given as
\begin{align}
M = v_{B-L} h^n = \textrm{diag}\left(M_1,M_2,M_3\right) \,,\quad M_1 \leq M_2 \leq M_3 \,,
\label{eq:NiMM}
\end{align}
and the seesaw superpotential contains, next to the Yukawa interaction with
the Higgs superfield $S_1$, the following two terms,
\begin{align}
W_{\textrm{Seesaw}}  \supset \frac{1}{2} M_i n_i^c n_i^c + h_{ij}^\nu \ell_i n_j^c H_u\,.
\label{eq:WSeesawM}
\end{align}


The fermionic Lagrangian derived from Eq.~\eqref{eq:WSeesawM} accounts for
the interaction of the heavy Majorana neutrinos with the standard model
lepton-Higgs pairs $\ell_i H_u$.
Considering processes at energies far below the neutrino masses $M_i$, it is
at our discretion to integrate the heavy neutrinos out of the Lagrangian.
In the low-energy effective theory below the neutrino mass threshold,
the $\ell_i \nu_{R,j}H_u$ interaction term then mutates into
the dimension-5 Weinberg operator~\cite{Weinberg:1979sa},
\begin{align}
\mathcal{L}_{(5)} = \frac{1}{2}\left[h^\nu M^{-1} \left(h^\nu\right)^T\right]_{ij}
\left(H_u \ell\right)_i\left(H_u \ell\right)_j + \textrm{h.c.} \,.
\end{align}
In the course of electroweak symmetry breaking, the Higgs doublet
$H_u$ acquires a nonzero VEV, $\left<H_u\right> = \left(0,v_u\right)^T$, so
that $\mathcal{L}_{(5)}$ assumes the form of a Majorana mass term for the
standard model neutrinos $\nu_{L,i}$ contained in $\ell_i = \left(\nu_{L,i},e_{L,i}\right)^T$,
\begin{align}
H_u \rightarrow \left(0,v_u\right)^T \,: \quad
\mathcal{L}_{(5)} \rightarrow - \frac{1}{2} \left(m_{\nu}\right)_{ij} \nu_{L,i} \nu_{L,j} +
\textrm{h.c.} \,,
\end{align}
where the light-neutrino mass matrix $m_\nu$ is given by the \textit{seesaw formula}
\begin{align}
m_\nu = - m_D M^{-1} m_D^T \,,\quad m_D = v_u h^\nu \,,
\label{eq:seesawformula}
\end{align}
with $m_D$ denoting the neutrino Dirac mass matrix.
$m_\nu$ is again a complex symmetric matrix that is
Takagi-diagonalized by a unitary transformation $\Omega$,
\begin{align}
\Omega^T m_\nu \Omega = m_\nu^{\textrm{diag}} = \textrm{diag}\left(m_1,m_2,m_3\right) \,.
\label{eq:mnudiag}
\end{align}
According to the measured differences of the masses-squared, $\Delta m_{ij}^2 = m_i^2 - m_j^2$,
the light-neutrino mass spectrum may either exhibit a \textit{normal} (NH) or an
\textit{inverted hierarchy} (IH).
For definiteness, i.e.\ in order to fix the column ordering of $\Omega$,
one conventionally numbers the light-neutrino mass eigenstates such that
\begin{align}
\textrm{normal hierarchy:} & \quad 0 \leq m_1 < m_2 \ll m_3 \,, \\
\textrm{inverted hierarchy:} & \quad 0 \leq m_3 \ll m_1 < m_2 \,.
\label{eq:nuhierarchies}
\end{align}
Since the atmospheric neutrino oscillations are predominantly sourced by the largest
$\Delta m_{ij}^2$ in the neutrino mass spectrum, while the solar neutrino oscillations
are driven by the smallest $\Delta m_{ij}^2$, we may now identify $\Delta m_{\textrm{sol}}^2$
and $\Delta m_{\textrm{atm}}^2$ as
\begin{align}
\Delta m_{\textrm{sol}}^2 = \Delta m_{21}^2 \,,\quad
\Delta m_{\textrm{atm}}^2 = \Delta m_{31}^2 \,\textrm{(NH)}\,, \Delta m_{32}^2 \,\textrm{(IH)} \,.
\end{align}
For very small $m_1$ (NH) or $m_3$ (IH), the light-neutrino mass spectrum
features one (NH) or two (IH) neutrinos with a mass of roughly
$\left(\Delta m_{\textrm{atm}}^2\right)^{1/2} \simeq 0.05 \,\textrm{eV}$.
Assuming Dirac masses of order the electroweak scale,
$\left(m_D\right)_{ij} \sim v_{\textrm{EW}}$, such small light-neutrino masses
readily follow from the seesaw formula (cf.\ Eq.~\eqref{eq:seesawformula}),
given a heavy-neutrino mass scale of $\mathcal{O}\left(10^{15}\right)\,\textrm{GeV}$.
We hence conclude that the observed neutrino oscillations strongly suggest the
existence of heavy Majorana neutrinos whose masses are generated in the course
of \BmL breaking close to the GUT scale.


The weak isospin partners of the standard model neutrinos, the charged leptons $e_{L,i}$,
acquire Dirac masses due to their Yukawa coupling to the down-type Higgs doublet $H_d$
(cf.\ the second term of $W_{\textrm{MSSM}}$ in Eq.~\eqref{eq:WMSSM}).
During the electroweak phase transition $H_d$ obtains a nonzero VEV,
$\left<H_d\right> = \left(v_d,0\right)^T$, which gives rise to the charged-lepton
Dirac mass matrix $m_{\textrm{cl}} = v_d h^d$.
In general, $m_{\textrm{cl}}$ is an arbitrary complex matrix that can be diagonalized by means
of a singular value decomposition.
For appropriate unitary matrices $L$ and $R$, one has
\begin{align}
L^T m_{\textrm{cl}} R = m_{\textrm{cl}}^{\textrm{diag}} =
\textrm{diag}\left(m_e,m_\mu,m_\tau\right) \,, \quad m_{\textrm{cl}} = v_d h^d \,.
\label{eq:mcldiag}
\end{align}


The fact that neutrinos are massive entails flavour-changing charged-current
interactions in the lepton sector, accounted for by the following Lagrangian,
\begin{align}
\mathcal{L}_{\textrm{CC}} \supset - \frac{g}{\sqrt{2}} U_{ij}
\hat{\bar{e}}_{L,i} \bar{\sigma}^\mu \hat{\nu}_{L,j} W_\mu^- + \textrm{h.c.} \,,\quad
U = L^\dagger \Omega \,.
\label{eq:LCCUPMNS}
\end{align}
Here, $U$ denotes the Pontecorvo-Maki-Nakagawa-Sakata (PMNS) or lepton mixing
matrix \cite{Pontecorvo:1957cp,Pontecorvo:1957qd,Maki:1962mu} and $\hat{e}_{L,i}$ and
$\hat{\nu}_{L,i}$ represent the charged-lepton and light-neutrino fields in the mass
eigenbasis.
A priori, the PMNS matrix is an arbitrary unitary matrix, which may be parametrized in
terms of three mixing angles $\theta_{12},\theta_{13},\theta_{23} \in \left[0,\pi/2\right]$
and six complex phases.
However, due to the freedom of redefining the phases of the charged-lepton fields,
the number of physical complex phases in $U$ reduces to three:
$\delta,\alpha_{21},\alpha_{31} \in \left[0,2\pi\right)$, where $\delta$ is referred to
as the \textit{Dirac phase} and $\alpha_{21}$ and $\alpha_{31}$ are the two
\textit{Majorana phases}.\footnote{If neutrinos were Dirac fermions, $\alpha_{21}$
and $\alpha_{31}$ could be absorbed in the light-neutrino fields and would hence be unphysical
as well, i.e. $\delta$ would remain as the sole physical phase in $U$.\smallskip}
In the standard parametrization $U$ is given as~\cite{Nakamura:2010zzi}
\begin{align}
U =
\begin{pmatrix} c_{12}c_{13} & s_{12}c_{13}e^{i\frac{\alpha_{21}}{2}}
 & s_{13}e^{i(\frac{\alpha_{31}}{2}-\delta)} \\
-s_{12}c_{23}-c_{12}s_{23}s_{13}e^{i\delta} &
\left(c_{12}c_{23}
-s_{12}s_{23}s_{13}e^{i\delta}\right)e^{i\frac{\alpha_{21}}{2}} &
s_{23}c_{13}e^{i\frac{\alpha_{31}}{2}} \\
s_{12}s_{23}-c_{12}c_{23}s_{13}e^{i\delta} &
\left(-c_{12}s_{23}
-s_{12}c_{23}s_{13}e^{i\delta}\right)e^{i\frac{\alpha_{21}}{2}} &
 c_{23}c_{13}e^{i\frac{\alpha_{31}}{2}}\end{pmatrix} \,,
\label{eq:UPMNS}
\end{align}
where $c_{ij} = \cos\theta_{ij}$ and $s_{ij} = \sin\theta_{ij}$.
Note that owing to its complex phases $\delta$, $\alpha_{21}$ and $\alpha_{31}$, the PMNS
matrix is in general not real, $U \neq U^*$, which implies the nonconservation of $CP$ in
the charged-current interactions governed by the Lagrangian in Eq.~\eqref{eq:LCCUPMNS}.
The phases $\delta$, $\alpha_{21}$ and $\alpha_{31}$ are consequently also known as
the $CP$ violation phases of the PMNS matrix.


\subsection{Baryogenesis}
\label{subsec:baryogenesis}


According to the seminal work of Andrei Sakharov in 1967~\cite{Sakharov:1967dj},
any process in the early universe which might come into question for the dynamical generation
of the primordial baryon asymmetry, i.e.\ which might provide a viable scenario of
\textit{baryogenesis}, has to satisfy three necessary conditions.
It has to
(i) violate the conservation of baryon number $B$,
(ii) violate $C$ as well as $CP$ invariance, and
(iii) involve a departure from thermal equilibrium.\footnote{As an obvious
`zeroth' condition, we have to require the process in question
to take place between the end of inflation and the freeze-out of baryons and antibaryons
at $T \sim 10..100\,\textrm{MeV}$.}
Interestingly, all ingredients for successful baryogenesis are therefore in principle
inherent in the standard model~\cite{Kuzmin:1985mm}:
$B$ is violated in $SU(2)_W$ sphaleron processes;
due to their chiral nature, all electroweak interactions maximally violate
$C$;
the presence of a complex phase in the quark mixing matrix implies $CP$ violation; and
the out-of-equilibrium condition is satisfied during the EWPT,
if it is of first order.
However, as it turns out, the amount of $CP$ violation in flavour-changing interactions
in the quark sector is too small to account for the observed baryon
asymmetry~\cite{Shaposhnikov:1987pf} and, more importantly, given a Higgs mass of
roughly $125\,\textrm{GeV}$, the EWPT is a smooth crossover rather than a first-order
phase transition~\cite{Kajantie:1996mn} (cf.\ Sec.~\ref{subsec:transitions}).
The standard scenario of \textit{electroweak baryogenesis} is hence ruled out
and the observed BAU calls for new physics beyond the standard model.


\subsubsection{Baryogenesis in Extensions of the Standard Model}


One might presume that supersymmetry could rescue electroweak baryogenesis.
The MSSM contains additional sources of $CP$ violation and the contributions
to the effective potential from scalar top quarks could render the EWPT first
order after all~\cite{Carena:2008vj}.
However, the combined data from the Higgs searches, which have been carried out
by the LHC experiments ATLAS and CMS up to now, exclude almost completely the fraction of
the MSSM parameter space, which is consistent with electroweak baryogenesis~\cite{Curtin:2012aa}.
Besides including supersymmetry, there are several alternative attempts to successfully
generate the BAU at the electroweak scale.
A viable option, for instance, is to extend the particle content of the standard model
by one real scalar gauge singlet~\cite{Espinosa:2011ax}, which might appear
as a pseudo-Nambu-Goldstone boson in the low-energy spectrum of composite-Higgs
models~\cite{Espinosa:2011eu}.
Possible drawbacks of such models are that they are often quite speculative
or might lack a connection to other observable phenomena, which can presently
be tested in experiments.


From the historical perspective, the first baryogenesis scenarios
were developed in the context of theories of grand
unification~\cite{Yoshimura:1978ex,Ignatiev:1978uf,Weinberg:1979bt}.
In \textit{GUT baryogenesis}, the primordial baryon asymmetry is generated in
the out-of-equilibrium decay of superheavy GUT particles.
But, as the GUT interactions preserve $B$$-$$L$, notably when the $U(1)_{B-L}$ is part
of the gauge group, no net \BmL asymmetry is produced during GUT baryogenesis.
Consequently, subsequent to its generation, the primordial baryon asymmetry is completely
washed out in electroweak sphaleron processes
(cf.\ Sec.~\ref{subsec:transitions}).\footnote{The fact that
the actual success of GUT baryogenesis is spoiled by the influence of sphaleron processes
shows that the first Sakharov condition should, in fact, be replaced by the requirement
that any potential baryogenesis process has to violate \BmL rather than $B$.}
Moreover, models of GUT baryogenesis typically predict too rapid proton
decay, which represents the second major reason why GUT baryogenesis is nowadays considered
less attractive.


A third scenario for the generation of the baryon asymmetry is provided by the
\textit{Affleck-Dine mechanism}~\cite{Affleck:1984fy}, which
is based on the dynamics of flat directions in the scalar potential of supersymmetric
theories.
If these flat directions have large initial values, they begin to oscillate
in the early universe due to the soft masses, which they receive from spontaneous
supersymmetry breaking.
Furthermore, if the flat directions contain scalar quarks and thus carry baryon number,
the primordial baryon asymmetry can be created in the course of their coherent oscillations.


\subsubsection{Baryogenesis via Leptogenesis}


One of the currently most favoured scenarios for the generation of the baryon
asymmetry is \textit{leptogenesis}~\cite{Fukugita:1986hr},\footnote{For reviews on leptogenesis,
cf.\ for instance Refs.~\cite{Buchmuller:2005eh,Davidson:2008bu}.\smallskip} which traces the observed
BAU back to a primordial \textit{lepton asymmetry} and which emerges as a
direct consequence of the seesaw explanation for the observed neutrino masses and mixings.
Recall that the key ingredient of the seesaw mechanism are the heavy Majorana neutrinos~$N_i$,
whose large masses lead to a suppression of the light-neutrino masses in the
low-energy effective theory.
Because the $N_i$ neutrinos do not carry definite \BmL charges (cf.\ Eq.~\eqref{eq:NinuRi}),
they can decay into final states containing leptons as well as into final states
containing antileptons.
Hence, heavy Majorana neutrinos decaying in the early universe allow for
the generation of a primordial \BmL asymmetry, in the form of a pure lepton asymmetry,
if their decays occur out of thermal equilibrium and violate $CP$
invariance.\footnote{Whether the first condition can be satisfied depends on the
production mechanism for the $N_i$ neutrinos and on the strength of their coupling to
the thermal bath.
The second condition merely requires that the Yukawa matrix $h^\nu$ has to contain
$CP$-violating complex phases.\smallskip}
Electroweak sphalerons, rather than washing out the primordial asymmetry as in the case
of GUT baryogenesis, then process the initial lepton asymmetry into the observed
BAU (cf.\ Eq.~\eqref{eq:etaB0}).


Leptogenesis represents a very attractive and well motivated scenario of baryogenesis,
as it identifies the masses and mixings in the neutrino sector on the one hand
and the BAU on the other hand as two related phenomena.
This connection implies, in particular, that the baryon asymmetry generated during
leptogenesis is sensitive to the light-neutrino mass
spectrum~\cite{Giudice:2003jh,Buchmuller:2004nz}.
Thus, leptogenesis is falsifiable in neutrino experiments aiming at determining
the absolute neutrino mass scale.
Furthermore, leptogenesis heavily relies on the assumption that neutrinos
are Majorana fermions.
It hence predicts the presence of three physical complex phases in the PMNS
matrix~\cite{Bilenky:1980cx} and the existence of neutrinoless double-beta
decay~\cite{Rodejohann:2011mu}.
The detection of neutrinoless double-beta decay would therefore provide further evidence
for the validity of the seesaw mechanism and leptogenesis.


One distinguishes between several variants of the central leptogenesis paradigm
outlined above, differing from each other, for instance, in terms of the mechanism
responsible for the generation of the heavy Majorana neutrinos in the early universe.
In the original and most extensively studied scenario, known as \textit{thermal
leptogenesis} \cite{Fukugita:1986hr,Plumacher:1997ru,Buchmuller:2004nz},\footnote{In
the following, we shall refer to thermal leptogenesis also as \textit{standard leptogenesis}.}
the $N_i$ neutrinos are simply thermally produced at temperatures exceeding the
heavy-neutrino masses, $T \gg M_i$.
To guarantee the successful generation of the baryon asymmetry,
thermal leptogenesis requires the reheating temperature after inflation $T_{\textrm{RH}}$
to be at least of $\mathcal{O}\left(10^9\right)\,\textrm{GeV}$.
In locally supersymmetric theories such a high reheating temperature may, however,
lead to a dangerously large abundance of thermally produced gravitinos,
which, depending on the superparticle mass spectrum, potentially entail various
cosmological problems (cf.\ Sec.~\ref{subsec:candidates}).
As a possible way out of the \textit{gravitino problems}, nonthermal leptogenesis
via inflaton decay~\cite{Lazarides:1991wu,Asaka:1999yd,Asaka:1999jb,HahnWoernle:2008pq}
has been proposed.
In this scenario, the $N_i$ neutrinos are nonthermally produced in the decay of the inflaton
after inflation.
The inflaton lifetime determines the temperature scale of reheating and may be adjusted
such that leptogenesis takes place at a significantly smaller temperature than in the
standard case.
Furthermore, in supersymmetric scenarios with global \BmL symmetry, the
$\tilde{N}_1$ sneutrino can play the role of the inflaton in models of
chaotic \cite{Murayama:1992ua,Ellis:2003sq} or hybrid \cite{Antusch:2004hd,Antusch:2010mv}
inflation.
After the end of inflation, the $\tilde{N}_1$ sneutrino field coherently oscillates
around the minimum of its potential\footnote{The $\tilde{N}_1$ oscillations
may even dominate the energy density of the universe after
inflation~\cite{Hamaguchi:2001gw}.\smallskip}
and leptogenesis directly proceeds via its decay.
Finally, leptogenesis may be based on the Affleck-Dine mechanism, if the scalar
potential features a flat direction which carries lepton number.
The most interesting candidate for such a flat direction is related to
the scalar component of the chiral superfield $\ell H_u$, which represents a flat
direction of the MSSM scalar potential~\cite{Murayama:1993em,Dine:1995kz}.


Since we rely on the seesaw mechanism to account for neutrino oscillations,
leptogenesis appears as a natural feature of our cosmological framework.
In our scenario, as we will discuss in more detail in Sec.~\ref{sec:orgnunvrs},
thermal as well as nonthermal processes contribute to the production of the heavy (s)neutrinos.
For now, let us summarize the most important quantities characterizing the
decay of the heavy (s)neutrinos.
The total decay rates of the $N_i$ neutrinos and $\tilde{N}_i$
sneutrinos,\footnote{The superscript $0$ indicates that the respective decay rates
are evaluated at zero temperature.}
\begin{align}
\Gamma_{N_i}^0 = \Gamma^0\big(N_i \rightarrow \ell H_u, \bar{\ell} H_u^*,
\tilde{\ell} \tilde{H}_u, \tilde{\ell}^* \bar{\tilde{H}}_u\big) \,,\quad
\Gamma_{\tilde{N}_i}^0 = \Gamma^0\big(\tilde{N}_i \rightarrow
\tilde{\ell} H_u, \bar{\ell} \bar{\tilde{H}}_u\big) \,,
\label{eq:Gamma0NiNti}
\end{align}
are readily calculated from $W_{\textrm{Seesaw}}$ (cf.\ Eq.~\eqref{eq:WSeesawM}).
At tree-level and summing over all final-state lepton flavours and weak isospin
doublet components, one finds
\begin{align}
\Gamma_{N_i}^0 = \Gamma_{\tilde{N}_i}^0 =
\frac{1}{4\pi}\big[\left(h^\nu\right)^\dagger h^\nu\big]_{ii} M_i
= \frac{1}{4\pi} \frac{\widetilde{m}_i M_i}{v_u^2} M_i \,,
\label{eq:Gamma0Ni}
\end{align}
with $\widetilde{m}_i$ denoting the effective neutrino mass of the $i^\textrm{th}$
neutrino generation,
\begin{align}
\widetilde{m}_i = \frac{\big[m_D^\dagger m_D\big]_{ii}}{M_i} =
\big[\left(h^\nu\right)^\dagger h^\nu\big]_{ii} \frac{v_u^2}{M_i} \,.
\label{eq:mitilde}
\end{align}
These effective neutrino masses determine the coupling strengths
of the heavy (s)neutrinos to the thermal bath and thus
control, \textit{inter alia}, the significance of washout effects.
As an important detail, we note that $\widetilde{m}_1$ is bounded from below by the
lightest neutrino mass, $m_1 < \widetilde{m}_1$~\cite{Fujii:2002jw}.
Consequently, constraints on $\widetilde{m}_1$ directly translate into
constraints on the light-neutrino mass spectrum.
Moreover, if the Yukawa matrix $h^\nu$ contains complex phases,
the heavy (s)neutrino decays violate $CP$ invariance, resulting
in a violation of lepton number $L$ at the quantum level.
As a convenient measure for the amount of $CP$ violation in
the heavy (s)neutrino decays, one introduces the $CP$ violation parameters
$\epsilon_i$, which indicate the lepton asymmetries produced \textit{per} decay of
a (s)neutrino from the $i^{\textrm{th}}$ generation,
\begin{align}
\epsilon_i = \frac{\Gamma^0\big(N_i \rightarrow L\big) -
\Gamma^0\big(N_i \rightarrow \bar{L}\big)}{\Gamma_{N_i}^0} =
\frac{\Gamma^0\big(\tilde{N}_i \rightarrow L\big) -
\Gamma^0\big(\tilde{N}_i \rightarrow \bar{L}\big)}{\Gamma_{\tilde{N}_i}^0} \,,
\end{align}
where we have symbolically subsumed all partial rates for decays into
final states with positive and negative lepton number, respectively.
To lowest order, one obtains the $CP$ violation parameter $\epsilon_i$ from
the interference of the respective tree-level decay amplitude with the one-loop
vertex as well as with the one-loop neutrino self-energy
correction~\cite{Covi:1996wh,Buchmuller:1997yu},
\begin{align}
\epsilon_i = - \sum_{j\neq i} \frac{\textrm{Im}\big\{\big[\left(h^\nu\right)^\dagger h^\nu
\big]_{ij}^2\big\}}{8\pi\big[\left(h^\nu\right)^\dagger h^\nu\big]_{ii}}
F\left(\frac{M_j^2}{M_i^2}\right) \,.
\label{eq:epsiloni}
\end{align}
In the MSSM supplemented by three heavy (s)neutrino generations, $F$ is given as
\begin{align}
F(x) = \sqrt{x}\left[\ln\left(\frac{1+x}{x}\right) + \frac{2}{x-1}\right] \,.
\end{align}
Throughout this thesis, we shall assume a hierarchical heavy-(s)neutrino mass spectrum,
$M_1 \ll M_2 \lesssim M_3$.
Therefore, the first (s)neutrino generation has the smallest decay rate
and hence decays last (cf.\ Eq.~\eqref{eq:Gamma0Ni}).
Typically, the lepton asymmetry generated in the decay of the two heavier
generations then gets washed out before the first generation begins to decay,
or it is eventually outweighed by the asymmetry generated in the decay of the first generation.
In either case, the final baryon asymmetry ends up being solely sensitive to the
parameter~$\epsilon_1$.
One can show that the absolute value of $\epsilon_1$ is bounded from
above~\cite{Hamaguchi:2001gw,Davidson:2002qv},
\begin{align}
\epsilon_1 = \delta_1^{\textrm{eff}} \epsilon_1^{\textrm{max}} \,,\quad
\left|\delta_1^{\textrm{eff}}\right| \leq 1 \,,
\end{align}
where $\delta_1^{\textrm{eff}}$ denotes an effective $CP$-violating phase
of the Yukawa matrix $h^\nu$ \cite{Hamaguchi:2001gw}.
Barring a conspiracy among the Yukawa couplings $h_{ij}^\nu$,
we expect $\delta_1^{\textrm{eff}}$ to be of $\mathcal{O}(1)$.
Meanwhile, the maximal $CP$ violation parameter $\epsilon_1^{\textrm{max}}$ is given as
\begin{align}
\epsilon_1^{\textrm{max}}\approx\frac{3}{8\pi}\frac{\left|\Delta m_{\textrm{atm}}^2\right|^{1/2}M_1}
{v_{\textrm{EW}}^2 \sin^2\beta} \simeq 2.1 \times 10^{-6}\left(\frac{1}{\sin^2\beta}\right)
\left(\frac{M_1}{10^{10}\,\textrm{GeV}}\right) \,.
\label{eq:eps1max}
\end{align}


\subsection{Particle Candidates for Dark Matter}
\label{subsec:candidates}


In our cosmological framework, \BmL as well as $R$ symmetry are spontaneously broken
in the vacuum that we presently live in.\footnote{Recall that \BmL is spontaneously broken
during the \BmL phase transition.
The spontaneous breaking of local supersymmetry in some hidden sector
implies the breaking of $R$ symmetry.}
However, both symmetries leave behind a residual discrete $\mathbb{Z}_2$ subgroup,
respectively referred to as matter parity $P_M$~\cite{Dimopoulos:1981zb} and
$R$ parity $P_R$~\cite{Farrar:1978xj}, under which the Lagrangian is still invariant.
%
One can show that matter and $R$ parity are exactly equivalent to each
other~\cite{Martin:1997ns}, so that we may restrict our discussion to $R$ parity
in the following.
All ordinary particles have even $R$ parity, $P_R = 1$,
while all superparticles have odd $R$ parity, $P_R = -1$.
The fact that $R$ parity remains as a residual symmetry in the low-energy effective theory
prevents the appearance of $B$- and $L$-violating terms in the MSSM superpotential
and hence guarantees the stability of the proton.
A~further, similarly important phenomenological consequence of $R$ parity is
that it renders the lightest superparticle (LSP) stable.
If the LSP does not participate in electromagnetic or strong interactions,
it is thus an excellent candidate for dark matter \cite{Pagels:1981ke,Goldberg:1983nd,Ellis:1983ew}.
Meanwhile, all heavier superparticles are unstable, so that they eventually decay
into final states featuring odd numbers of LSPs.
Most often, the decay of a heavy superparticle yields one LSP and several
ordinary particles.


The stability of the LSP in supersymmetric theories with conserved $R$ parity
and its ability to form dark matter represent the main phenomenological arguments
in favour of supersymmetry.
Also in this thesis, we shall adopt the notion that the LSP accounts for dark matter.
The nature of the LSP depends on the superparticle mass spectrum
and hence on the mechanism for the spontaneous breaking of supersymmetry.
In many cases the lightest among the MSSM neutralinos ends up being the LSP.
Neutralinos are the archetypes of weakly interacting massive particles (WIMPs)
and so WIMP dark matter in the form of neutralinos is a popular and thoroughly
studied scenario.
However, if supersymmetry is assumed to be local, the particle spectrum
also features the gravitino,\footnote{The gravitino is the spin-$3/2$ superpartner
of the spin-$2$ graviton and may be regarded as the 'gauge' field of local
supersymmetry transformations.
Its mass represents the order parameter of the super-Higgs mechanism responsible
for the spontaneous breaking of local supersymmetry.}
whose decays may render the WIMP dark matter scenario
inconsistent with, for instance, leptogenesis or BBN.
This problem can potentially be solved by promoting the gravitino itself to the LSP.
We are thus led to consider two different, in a sense quite opposite scenarios in this thesis:
For the most part (cf.\ Ch.~\ref{ch:reheating}), we will assume that the gravitino is
the LSP and hence constitutes dark matter.
By contrast, in Ch.~\ref{ch:wimp} we will take it to be the heaviest superparticle
and consider dark matter in the form of neutralinos.
Both scenarios are only consistent within certain parameter bounds,
since stable as well as unstable gravitinos can lead to various cosmological problems.
As a preparation for the further investigation, let us now discuss
these \textit{cosmological gravitino problems} in turn and highlight for which parameter choices
they are respectively circumvented.


\subsubsection{Stable Gravitino and Gravitino Dark Matter}


Any gravitino abundance existing prior to the \BmL phase transition
is completely diluted in the course of the exponential expansion during
inflation.
Primordial gravitinos, hence, do not cause any cosmological problems~\cite{Ellis:1982yb}.
After inflation, the gravitino abundance is, however, regenerated by various
mechanisms,\footnote{Notice that the production of gravitinos subsequent to inflation
is also indicated in Fig.~\ref{fig:history}.\smallskip}
which gives rise to several restrictions on the cosmological
scenario ~\cite{Nanopoulos:1983up,Khlopov:1984pf}.


Typically, the dominant mechanism for the production of gravitinos after inflation
are inelastic 2-to-2 scatterings in the thermal bath.
Assuming the gluino to be the heaviest gaugino, the thermal gravitino production
is in turn dominated by QCD processes, while electroweak processes give
only subleading contributions.
The thermal gravitino abundance $\Omega_{\widetilde{G}}^{\textrm{th}} h^2$
is then solely controlled by the reheating temperature $T_{\textrm{RH}}$,
the gravitino mass $m_{\widetilde{G}}$ and the gluino mass $m_{\tilde{g}}$.
In the limit $m_{\widetilde{G}} \ll m_{\tilde{g}}$, mainly the longitudinal, i.e.
\textit{goldstino} degrees of freedom of the gravitino with helicity $\pm 1/2$
are excited and one finds \cite{Bolz:2000fu,Pradler:2006qh},
\begin{align}
\Omega_{\widetilde{G}}^{\textrm{th}} h^2 = C
\left(\frac{T_{RH}}{10^9~\textrm{GeV}}\right)
\left(\frac{10~\textrm{GeV}}{m_{\widetilde{G}}}\right)
\bigg(\frac{m_{\tilde{g}}}{1~\textrm{TeV}}\bigg)^2 \,,
\label{eq:OmegaGth2Approx}
\end{align}
where the coefficient $C \simeq 0.26$ to leading order in the strong
coupling constant.\footnote{$C$ has an $\mathcal{O}(1)$ uncertainty
due to unknown higher-order contributions and nonperturbative effects
\cite{Bolz:2000fu}.
Resummation of thermal masses increases $C$ by about a factor
of two \cite{Rychkov:2007uq}.\smallskip}
For fixed superparticle masses, a too high reheating temperature
results in the overproduction of gravitinos, such that
$\Omega_{\widetilde{G}}^{\textrm{th}}$
exceeds the presently observed abundance of dark matter,
$\Omega_{\widetilde{G}}^{\textrm{th}}> \Omega_{\textrm{DM}}^0$,
or, even worse, gravitinos \textit{overclose}\footnote{If a particle species $i$ is
produced so abundantly that it overcloses the universe, $\Omega_i^0 > 1$,
the spatial curvature of the universe remains, of course, unchanged.
Overclosure merely refers to the fact that the presumed production mechanism
in combination with the presumed parameter values
is inconsistent with the measured value of the expansion rate $H_0$ and thus
physically not viable.\smallskip} the universe, $\Omega_{\widetilde{G}}^{\textrm{th}} > 1$.
Hence, for given $m_{\widetilde{G}}$ and $m_{\widetilde{g}}$, the requirement that
$\Omega_{\widetilde{G}}^{\textrm{th}}\leq \Omega_{\textrm{DM}}^0$
provides us with an upper bound on the reheating temperature.
In particular, the reheating temperature has to be rather low,
$T_{\textrm{RH}} \ll 10^{9}\,\textrm{GeV}$,
for a very light gravitino, $m_{\widetilde{G}} \ll 10\,\textrm{GeV}$,
which excludes thermal leptogenesis as the possible origin of the BAU
(cf.\ Sec.~\ref{subsec:baryogenesis}).
On the other hand, it is well known that the high temperatures characteristic for
thermal leptogenesis, $T_{\textrm{RH}} \sim 10^9 .. 10^{10}\,\textrm{GeV}$, can become
a virtue, if the gravitino is the LSP and has a mass of
$\mathcal{O}\left(10...100\right)\,\textrm{GeV}$.
Given a gluino mass in the TeV range, thermally produced gravitinos
can then successfully explain the observed amount of dark matter
\cite{Bolz:1998ek}.\footnote{Note that superparticle mass spectra of the required form,
i.e. containing a gravitino LSP with $m_{\widetilde{G}} \sim 10..100\,\textrm{GeV}$ alongside
a heavy gluino with $m_{\tilde{g}} \sim 1 \,\textrm{TeV}$, naturally arise
in gravity-~\cite{Chamseddine:1982jx,Barbieri:1982eh,Ibanez:1982ee}
and gaugino-mediated~\cite{Kaplan:1999ac,Chacko:1999mi} scenarios
of supersymmetry breaking.}
It is exactly this scenario, featuring a gravitino LSP with a mass around
the electroweak scale, which we will further investigate in Ch.~\ref{ch:reheating}.


In addition to the scatterings in the thermal bath, gravitinos may be produced by a
multitude of nonthermal mechanisms.
For instance, the decays of heavier superparticles, in particular, the decay of
the next-to-lightest superparticle (NLSP) may yield a sizable contribution
$\Omega_{\widetilde{G}}^{\textrm{NLSP}}$ to the gravitino abundance~\cite{Covi:1999ty,Feng:2003xh}.
The relative importance of $\Omega_{\widetilde{G}}^{\textrm{NLSP}}$ compared
to $\Omega_{\widetilde{G}}^{\textrm{th}}$ depends on the nature of the NLSP as well as the
details of the superparticle mass spectrum.
For a variety of popular NLSP candidates, the superparticle masses, that we shall consider
in this thesis, however imply that $\Omega_{\widetilde{G}}^{\textrm{NLSP}}$ is negligibly small \cite{Feng:2004mt,Steffen:2008qp}.
Furthermore, the particles from the symmetry breaking sector
\cite{Kallosh:1999jj,Giudice:1999yt} decay into the gravitino, thereby yielding
a contribution $\Omega_{\widetilde{G}}^S$ to the total gravitino abundance.
But as these particles also couple to the heavy (s)neutrinos, the decay into
the gravitino is vastly outweighed by the much faster decay into $N_i$ neutrinos
and $\tilde{N}_i$ sneutrinos.
Besides that, $\Omega_{\widetilde{G}}^S$ scales inversely proportional to
the reheating temperature~\cite{Nakayama:2010xf}.
In view of the high temperatures, which we will eventually encounter after the \BmL
phase transition, we thus expect $\Omega_{\widetilde{G}}^S$ to be negligible as well.
We conclude that for our purposes it will suffice to
approximate the total gravitino abundance $\Omega_{\widetilde{G}}$ by its
thermal contribution $\Omega_{\widetilde{G}}^{\textrm{th}}$,
\begin{align}
\Omega_{\widetilde{G}} =
\Omega_{\widetilde{G}}^{\textrm{th}}   +
\Omega_{\widetilde{G}}^{\textrm{NLSP}} +
\Omega_{\widetilde{G}}^S  + ... \approx \Omega_{\widetilde{G}}^{\textrm{th}} \,.
\end{align}


Independently of its exact production mechanism and forgetting about possible constraints
from leptogenesis for a moment, we may ask whether or not the gravitino
is ever able to reach thermal equilibrium in the early universe.
As it turns out, the gravitino equilibrium abundance always overcloses the universe,
as long as the gravitino has a mass above the keV scale~\cite{Pagels:1981ke}.
Conversely, a thermalized gravitino with $m_{\widetilde{G}} \sim 1 \,\textrm{keV}$
does not imply overclosure and thus allows for arbitrarily high reheating temperatures.
However, it is excluded nonetheless by constraints on warm dark matter~\cite{Viel:2005qj}.
Only for a gravitino mass $m_{\widetilde{G}} \lesssim 1\,\textrm{eV}$,
a thermalized gravitino becomes cosmologically viable.
But this scenario is less attractive, since then all of the MSSM superparticles
would decay into the gravitino and none of them nor the gravitino could form dark matter.
All in all, it hence seems rather unlikely that the gravitino was ever in thermal
equilibrium after inflation.


Identifying the gravitino as the LSP may provide an attractive
scenario for the nature of dark matter as well as the origin of the baryon asymmetry.
Yet, it does not come without any further restrictions.
Most notably, the decay of the NLSP into the gravitino and standard model
particles potentially spoils the success of BBN (cf.\ Sec.~\ref{subsec:BBN})
\cite{Moroi:1993mb,deGouvea:1997tn}.\footnote{This threat to the gravitino dark matter
scenario is known as the \textit{NLSP decay problem}.}
Gravitino masses compatible with thermal leptogenesis,
$m_{\widetilde{G}} \gtrsim 10\,\textrm{GeV}$, imply a long NLSP lifetime, so that
the NLSP decays way after BBN, thereby changing the theoretical predictions for the
primordial abundances of the light elements.
Requiring these changes to be insignificant, one is able to derive constraints on
the abundance of the NLSP at the time of its decay as well as on its lifetime.
These constraints may partly be translated into bounds on the masses of the NLSP and the gravitino.
Assuming a gravitino mass $m_{\widetilde{G}} \sim 10..100\,\textrm{GeV}$,
a neutralino NLSP, for instance, must definitely be heavier than
$1\,\textrm{TeV}$~\cite{Covi:2009bk}.
Similarly restrictive constraints apply to a scalar top
quark NLSP~\cite{Berger:2008ti,Kusakabe:2009jt} as well as to a scalar tau lepton
NLSP~\cite{Pospelov:2006sc,Cyburt:2006uv}.
Meanwhile, a scalar neutrino NLSP with a mass much below $1\,\textrm{TeV}$
is not in conflict with BBN, thanks to the large branching ratio of its
invisible decay mode, $\tilde{\nu} \rightarrow \widetilde{G}\nu$~\cite{Kanzaki:2006hm}.


Several solutions to the NLSP decay problem have been proposed.
The NLSP abundance could, for instance, be sufficiently diluted prior to BBN in the
course of late-time entropy production~\cite{Buchmuller:2006tt,Hasenkamp:2010if}.
Alternatively, the lifetime of the NLSP could be shortened due to additional
decay channels into particles of some hidden sector~\cite{DeSimone:2010tr,Cheung:2010qf}.
A third option, which we consider particularly attractive, is that $R$ parity is slightly
broken after all \cite{Buchmuller:2007ui,Takayama:2000uz}.
In such a scenario, given a sufficient strength of the $R$ parity-violating interactions,
the NLSP decays into standard model particles before the onset of BBN, hence
resolving all tension with the BBN predictions.
At the same time, the NLSP hardly decays into gravitinos anymore, so that the
total gravitino abundance ceases to receive contributions from NLSP decays,
$\Omega_{\widetilde{G}}^{\textrm{NLSP}} \approx 0$.
Meanwhile, the violation of $R$ parity also renders the gravitino unstable.
The decay rate of the gravitino, however, ends up being suppressed by the Planck scale
as well as by the tiny coupling strength of the $R$ parity-violating interactions.
This results in very long-lived gravitino dark matter with a lifetime exceeding the present
age of the universe by several orders of magnitude~\cite{Takayama:2000uz}.
Interestingly, the possibility that dark matter in the form of gravitinos is, in fact,
unstable leads to a rich phenomenology, which might be tested in experiments aiming
at the indirect detection of dark
matter~\cite{Bertone:2007aw,Ishiwata:2008cu,Ibarra:2008qg,Covi:2008jy}.
Similarly, the $R$ parity-violating couplings of the NLSP can be probed
in collider experiments~\cite{Buchmuller:2004rq,Ellis:2006vu,Bobrovskyi:2010ps,Bobrovskyi:2011vx}.


A slight violation of $R$ parity hence represents an appealing solution to
the NLSP decay problem.
For clarity, we however emphasize that our further investigation will not depend on
how exactly the NLSP decay problem is solved.
When discussing gravitino dark matter in the following,
the decay of the NLSP will not play any further role.


\subsubsection{Unstable Gravitino and WIMP Dark Matter}


While gravitino dark matter faces the NLSP decay problem, it is the
decay of the gravitino itself, which might spoil the success of BBN in
other dark matter scenarios
\cite{Weinberg:1982zq,Ellis:1984er,Kawasaki:2004yh,Kawasaki:2004qu,Jedamzik:2006xz}.\footnote{In
analogy to the NLSP decay problem, we may now speak of the \textit{gravitino decay problem}.}
Since its couplings are gravitationally suppressed, the gravitino typically
decays very late, i.e.\ during or after BBN.
In particular, for small gravitino masses $m_{\widetilde{G}}$
it has a long lifetime $\tau_{\widetilde{G}}$,
\begin{align}
\tau_{\widetilde{G}} =
\left[\frac{1}{4}\left(n_v + \frac{n_m}{12}\right) \frac{m_{\widetilde{G}}^3}{M_P^2}\right]^{-1}
\simeq 280\,\textrm{d} \left(\frac{100\,\textrm{GeV}}{m_{\widetilde{G}}}\right)^3 \,.
\label{eq:Glife280d}
\end{align}
Here, $n_v = 12$ and $n_m = 49$ respectively denote the number of vector and chiral multiplets
the gravitino can decay into.
In most cases, the decay of a gravitino yields one LSP and several standard model
particles.
Decaying gravitinos hence induce electromagnetic and hadronic showers,
increase the entropy of the thermal bath and give rise to a nonthermal contribution
$\Omega_{\textrm{LSP}}^{\widetilde{G}}$ to the LSP abundance.
Again, requiring the theoretical BBN predictions not to change too drastically
allows one to derive constraints on the abundance of the gravitino at the time
of its decay as well as on its lifetime.
On top of that, one has to ensure that the LSP is not overproduced
in gravitino decays, $\Omega_{\textrm{LSP}}^{\widetilde{G}} \leq
\Omega_{\textrm{DM}}^0$.
In the case of thermally produced gravitinos, these constraints can then be
translated into bounds on the gravitino mass and the reheating temperature~\cite{Kawasaki:2008qe}.
For a gravitino mass below $1\,\textrm{TeV}$,
the reheating temperature must, for instance, not be higher than
$\mathcal{O}\left(10^5..10^6\right)\,\textrm{GeV}$,
which rules out baryogenesis via thermal leptogenesis.\footnote{This bound
on the reheating temperature is significantly alleviated, if the gravitino predominantly
decays into a neutrino-sneutrino pair \cite{Kawasaki:1994bs} or into particles that are
completely decoupled from the thermal bath \cite{Hasenkamp:2011em}.
Similarly, it is also relaxed if late-time entropy production leads to a dilution
of the gravitino abundance subsequent to its generation, but prior to its decay.\smallskip}
On the other hand, a gravitino heavier than about $10~\mathrm{TeV}$ can
be consistent with primordial nucleosynthesis and leptogenesis
\cite{Kawasaki:2008qe,Weinberg:1982zq,Gherghetta:1999sw,Ibe:2004tg}.
In such a scenario, the gravitino then represents the heaviest superparticle---a
possibility, which is in fact realized in models of anomaly-mediated supersymmetry
breaking \cite{Giudice:1998xp,Randall:1998uk} and which has recently
been reconsidered in the case of wino \cite{Ibe:2011aa}, higgsino \cite{Jeong:2011sg}
and bino \cite{Krippendorf:2012ir} LSP, because it nicely fits together with a Higgs boson mass
of about $125\,\textrm{GeV}$.
In Ch.~\ref{ch:wimp}, we shall thus consider WIMP dark matter in the form
of a neutralino LSP, which is partly produced in the decay of a very heavy gravitino.


\section[SSB of \texorpdfstring{$B$$-$$L$}{B\textminus L} as the Origin of the Hot Early Universe]
{Spontaneous Breaking of \BmL \newline as the Origin of the Hot Early Universe}
\label{sec:orgnunvrs}


A multitude of observed phenomena calls for new physics beyond the standard model.
In the previous section, we have in particular addressed the apparent flatness of
our universe, the high isotropy of the CMB temperature and the statistical
properties of its tiny fluctuations, neutrino flavour oscillations, the cosmic baryon
asymmetry as well as dark matter.
We have also argued that the supersymmetric standard model supplemented by three
generations of right-handed (s)neutrinos and spontaneously broken \BmL
provides the necessary ingredients to account for all these
phenomena.\footnote{Note that our framework can be naturally embedded in supersymmetric GUT
models~\cite{Raby:2008gh}.}
The dynamical breaking of \BmL requires an extended scalar sector, which
automatically yields hybrid inflation, thereby resolving the flatness and horizon
problems as well as explaining the origin of the primordial metric fluctuations.
Likewise, \BmL breaking at the GUT scale leads to an elegant explanation of
the small neutrino masses via the seesaw mechanism and implies baryogenesis
via leptogenesis.
Finally, assuming $R$ parity to be exactly conserved or at most only slightly
violated, the LSP represents an excellent candidate for dark matter.


The main goal of this thesis now is to demonstrate that all these pieces of the
puzzle naturally fit together, yielding a consistent cosmology, which is in
accord with cosmological and astrophysical observations
as well as with the data from neutrino and collider experiments.
Our key insight is that the \BmL phase transition at the end of hybrid inflation,
i.e.\ the \textit{cosmological realization} of \BmL breaking,
successfully generates the initial conditions of the hot early universe.
After inflation, nonthermal and thermal processes produce an abundance
of heavy (s)neutrinos, whose decays generate the primordial entropy of the thermal
bath and the primordial \BmL asymmetry.
At the same time, gravitinos are produced through inelastic
scatterings in the thermal bath.
If the gravitino is the LSP, it ends up being the dominant component of dark matter.
Conversely, if the gravitino is the heaviest superparticle and one of the neutralinos
plays the role of the LSP, the decay of the gravitino into the LSP gives rise to WIMP dark matter.


In this section, we will first outline how the energy of the false vacuum of unbroken \BmL
is successively transferred into the energy of thermal radiation.\footnote{We will present
a much more detailed discussion of this process in Chs.~\ref{ch:phasetransition} and
\ref{ch:reheating}.}
In doing so, we will pay special attention to the fact after inflation the universe
is reheated through the decays of the heavy (s)neutrinos.
As we will see, this characteristic feature of our scenario
implies a nontrivial connection between the neutrino parameters
$\widetilde{m}_1$ and $M_1$
on the one hand and the superparticle masses $m_{\widetilde{G}}$
and $m_{\widetilde{g}}$ on the other hand.
In the second part of this section, we will then introduce the Froggatt-Nielsen flavour
structure, on which we will base our analysis.


\subsection{Reheating through Heavy \texorpdfstring{(S)neutrino}{Neutrino and Sneutrino} Decays}
\label{subsec:neutrinoreheating}


\subsubsection{Particle Production during (P)reheating}


The \BmL phase transition at the end of hybrid inflation is triggered by
a tachyonic instability in the scalar potential (cf.\ Sec.~\ref{subsec:infltn}).
Once the inflaton $\varphi$ reaches the critical field value $\varphi_c = v_{B-L}$,
the mass-squared of the waterfall field $\sigma$ turns negative, so that its
long-wavelength modes begin to grow exponentially fast.
This leads to the breaking of \BmL as well as to the massive production of
nonrelativistic \BmL Higgs bosons.
In fact, almost the entire energy initially stored in the false vacuum is converted
into Higgs bosons during the \BmL phase transition.
This nonperturbative energy transfer into Higgs bosons may be regarded as the first step
towards reheating the universe after inflation and is thus referred to as
\textit{tachyonic preheating}~\cite{Felder:2000hj,Felder:2001kt} (cf.\ Sec.~\ref{sec:preheating}).
It is so efficient that symmetry breaking typically completes within a single oscillation of
the scalar field distribution as it rolls towards the true vacuum.
Meanwhile, particles coupled to the \BmL Higgs field are nonadiabatically produced
as well~\cite{GarciaBellido:2001cb}.
In the context of our theoretical framework, these are in particular
the \BmL gauge DOFs, the particles from the symmetry breaking sector
as well as all three generations of heavy (s)neutrinos.
However, due their strong coupling and their large masses, the gauge particles immediately
decay after their production into heavy (s)neutrinos and the MSSM (s)quarks and (s)leptons.
The initial conditions for the further cosmic evolution after inflation
are hence determined by tachyonic preheating \textit{and} the decay of the gauge multiplet.


Since tachyonic preheating transfers most of the false vacuum energy into
nonrelativistic Higgs bosons, the universe first undergoes a phase of matter-dominated
expansion after inflation.\footnote{In Ch.~\ref{ch:reheating}, we will track the
cosmic evolution during this era by means of Boltzmann equations.\smallskip}
During this phase, the particles from the symmetry breaking sector
decay in principle into all three generations of heavy (s)neutrinos.
In our further analysis, we will however restrict ourselves to a mass spectrum
that only allows for decays into the lightest heavy (s)neutrinos.
While this special case promises to be less cluttered than the most general scenario,
hence permitting a clear and unobstructed presentation of our findings,
we expect it to fully feature all phenomenological aspects that we are interested in.
Furthermore, in addition to the decay of the particles from the symmetry breaking sector,
the $N_1$ neutrinos and $\tilde{N}_1$ sneutrino are also produced through
scattering processes in the thermal bath.
In view of the temperatures typically reached after inflation, the thermal
production of the two heavier (s)neutrino generations is by contrast always negligible.


All heavy (s)neutrinos decay into the components of the MSSM
superfields $\ell_i H_u$ as well as into the corresponding antiparticles
(cf.\ Eq.~\eqref{eq:Gamma0NiNti}).
In consequence of the strong standard model gauge interactions, the heavy-(s)neutrino
decay products immediately thermalize, so that the energy of a decaying (s)neutrino
is always quickly distributed among all MSSM degrees of freedom.
This energy transfer from the heavy (s)neutrinos, most notably
from the the heavy (s)neutrinos of the first generation, to the thermal bath is thus
responsible for the reheating of the universe.
On top of that, it also gives rise to the primordial lepton asymmetry, which
is processed by the electroweak sphalerons into the observed baryon asymmetry.
Finally, 2-to-2 scatterings in the thermal bath
generate an abundance of gravitinos.


\subsubsection{Connection Between the Neutrino Sector and Supergravity}


Let us assume for now that the gravitino is the LSP, having a mass
as it typically arises in gravity or gaugino mediation,
$m_{\widetilde{G}} \sim 10..100\,\textrm{GeV}$.
According to Eq.~\eqref{eq:OmegaGth2Approx}, the reheating temperature $T_{\textrm{RH}}$
then has to be of $\mathcal{O}\left(10^9..10^{10}\right)\,\textrm{GeV}$, so that
the thermal production of gravitinos just yields the observed relic density of dark matter.
Remarkably, temperatures of the required magnitude are indeed realized,
if the universe is reheated through the decays of the heavy
(s)neutrinos.\footnote{This observation initially stimulated our interest in the
\BmL phase transition and prompted us to carry out the research program,
the results of which are presented in this thesis.
We reported on it for the first time in Ref.~\cite{Buchmuller:2010yy}
and further expanded on it in Refs.~\cite{Buchmuller:2011mw,Buchmuller:2012wn}.}
Due to their dominant abundance, the reheating process is mainly driven by the
decay of the lightest heavy (s)neutrinos.
The reheating temperature is consequently reached, once the Hubble rate $H$ has
dropped to the value of the $N_1$ decay rate.
Because of their relativistic motion with respect to the thermal plasma,
the heavy (s)neutrinos actually decay at a lower rate as they would do at rest.
Omitting this fact for a moment, we may approximate the $N_1$ decay rate by the
vacuum decay rate $\Gamma_{N_1}^0$,
\begin{align}
T = T_{\textrm{RH}} \,:\quad H \approx \Gamma_{N_1}^0 \,.
\label{eq:TRHHGamma}
\end{align}
For values of the neutrino mass parameters $\widetilde{m}_1$ and $M_1$
compatible with thermal leptogenesis, $\widetilde{m}_1 \sim 0.01~\mathrm{eV}$ and
$M_1 \sim 10^{10}~\mathrm{GeV}$, one obtains (cf.\ Eq.~\eqref{eq:Gamma0Ni})
\begin{equation}
\Gamma_{N_1}^0 = \frac{1}{4\pi}
\frac{\widetilde{m}_1 M_1}{v_u^2} M_1 \sim 10^3\,\textrm{GeV}\,,\quad
v_u \simeq v_{\textrm{EW}} \simeq 174 \,\textrm{GeV}\,.
\label{eq:Gamma0N1Est}
\end{equation}
Furthermore, assuming that at $T = T_{\textrm{RH}}$ the entire available energy has already
been converted into radiation, the Friedmann equation tells us that,
\begin{align}
T = T_{\textrm{RH}} \,:\quad
\rho_{\textrm{tot}} \approx \rho_R \,,\quad
H^2 = \frac{8\pi}{3M_P^2} \rho_{\textrm{tot}} \approx
\frac{8\pi}{3M_P^2} \frac{\pi^2}{30} g_{*,\rho} T_{\textrm{RH}}^4 \,.
\end{align}
In combination with Eqs.~\eqref{eq:TRHHGamma} and \eqref{eq:Gamma0N1Est},
this relation provides us with the following rough estimate for the reheating
temperature,\footnote{As we shall see in Ch.~\ref{ch:reheating}, including all
effects neglected in the above discussion, the actual
reheating temperature obtained for $\widetilde{m}_1 \sim 0.01\,\textrm{eV}$
and $M_1 \sim 10^{10}\,\textrm{GeV}$ turns out to be a bit lower than this
estimate.
Our main point, however, remains valid:
reheating through the decays of heavy (s)neutrinos results in a reheating
temperature, which is controlled by neutrino parameters.}
\begin{equation}
T_{\textrm{RH}} \approx \left(\frac{90}{8\pi^3 g_{*,\rho}}\right)^{1/4}
\sqrt{\Gamma_{N_1}^0 M_P} \sim 10^{10}\,\textrm{GeV} \,,
\end{equation}
where we have used the MSSM value for the effective number of relativistic degrees
of freedom contributing to the total radiation energy density, $g_{*,\rho} = 915/4$.
This result confirms our claim that the decay of the heavy (s)neutrinos
itself might account for the high temperatures required for leptogenesis and gravitino
dark matter.
Together with the observation that the dynamics of the symmetry breaking sector
can give rise to a sizable abundance of heavy (s)neutrinos after inflation, it
serves us as a key motivation for our study of the \BmL phase transition.


Since the reheating temperature depends on the $N_1$ decay rate
and hence on $\widetilde{m}_1$ and $M_1$,
the requirement that gravitinos be the constituents of dark matter
yields a connection between neutrino and superparticle mass parameters
(cf.\ Eq.~\eqref{eq:OmegaGth2Approx}),
\begin{align}
\Omega_{\widetilde{G}}^0 h^2 =
\Omega_{\widetilde{G}}^0 h^2 \left(\widetilde{m}_1,M_1,m_{\widetilde{G}},m_{\tilde{g}}\right) =
\Omega_{\mathrm{DM}}^0 h^2 \simeq 0.11 \,.
\end{align}
The neutrino masses $\widetilde{m}_1$ and $M_1$ are in turn constrained
by the condition that the maximal baryon asymmetry, generated in the decay
of the heavy (s)neutrinos if $\epsilon_1 = \epsilon_1^{\textrm{max}}$
(cf.\ Sec.~\ref{subsec:baryogenesis}), must not be smaller than the observed one,
$\eta_B^{\textrm{max}} \geq \eta_B^{\textrm{obs}} = 6.2\times 10^{-10}$.
As we will show in Sec.~\ref{sec:scan}, this condition directly implies
a lower bound on the gravitino mass $m_{\widetilde{G}}$ as a function of
the effective neutrino mass $\widetilde{m}_1$.


In conclusion, we emphasize that the decay of the heavy (s)neutrinos after the
\BmL phase transition entirely fixes the initial conditions of the hot early
universe.
The initial baryon asymmetry, dark matter density and reheating temperature cannot
be freely chosen, but are determined by the parameters of the underlying Lagrangian,
which can in principle be measured by particle physics experiments and astrophysical
observations.


\subsection{Froggatt-Nielsen Flavour Structure}
\label{subsec:flavour}


Eventually, we wish to study the physical processes outlined in the
previous section by means of Boltzmann equations.
But before we are ready to carry out any quantitative calculations, we
need to find a way to estimate the values of the dimensionless couplings
$\lambda$ and $h_{ij}^{n,\nu}$ in the superpotential (cf.\ Secs.~\ref{subsec:infltn}
and \ref{subsec:seesaw}).
More precisely, we have to specify a flavour model, which correctly describes
the masses and mixings of quarks, charged leptons and neutrinos, and which can
at the same time be consistently extended to the symmetry-breaking sector as
well as to the sector of the heavy (s)neutrinos.
Since we would like to study our scenario for a broad spectrum of reheating
temperatures, we seek a model which is flexible enough to allow
the neutrino masses $\widetilde{m}_1$ and $M_1$ to vary within large ranges.


As of today, it poses a major theoretical challenge of particle physics
to explain the observed patterns of quark and lepton masses and mixings, in
particular the striking differences between the quark sector and the neutrino sector.
Promising elements of a theory of flavour are grand unification based on
the groups $SU(5)$, $SO(10)$ or $E_6$, supersymmetry, the seesaw mechanism
and additional flavour symmetries \cite{Raby:2008gh}.
A successful example is the  Froggatt-Nielsen mechanism~\cite{Froggatt:1978nt}
based on spontaneously broken Abelian symmetries, which parametrizes quark and lepton
mass ratios and mixings by powers of a small \textit{hierarchy parameter}~$\eta$.
Interestingly, the resulting structure of mass matrices also arises in compactifications of
higher-dimensional field and string theories, where the parameter $\eta$
is related to the location of matter fields in the compact dimensions
or to VEVs of moduli fields~\cite{Grossman:1999ra,Gherghetta:2000qt,Buchmuller:2006ik}.
In this thesis, we shall consider a Froggatt-Nielsen symmetry which
commutes with the GUT group $SU(5)$.
Our model is a variant of the model discussed in Ref.~\cite{Buchmuller:1998zf}.
It naturally explains the large atmospheric neutrino mixing
angle~\cite{Sato:1997hv,Irges:1998ax}, satisfies all constraints from flavour-changing
processes~\cite{Buchmuller:1999yg} and predicts Dirac and Majorana neutrino mass matrices
which are consistent with thermal leptogenesis~\cite{Buchmuller:1998zf}.


\subsubsection{Yukawa Couplings from Nonrenormalizable Interactions}


\begin{table}
\centering
\begin{tabular}{c|cccccccccccccccc}\hline \hline
$\psi_i$ & $\mathbf{10}_3$ & $\mathbf{10}_2$ & $\mathbf{10}_1$ &
$\mathbf{5}^*_3$ & $\mathbf{5}^*_2$ & $\mathbf{5}^*_1$ &
$n^c_3$ & $n^c_2$ & $n^c_1$ & $H_u$ & $H_d$ & $S_1$ & $S_2$ & $\Phi$  \\ \hline
$Q_i$ & 0 & 1 & 2 & $a$ & $a$ & $a+1$ & $b$ & $c$ & $d$ & 0 & 0 & 0 & 0 & $e$
\\ \hline\hline
\end{tabular}
\medskip
\caption{Chiral flavour charges $Q_i$ of the $SU(5)$ matter multiplets $\psi_i$.}
\label{tab:charges}
\end{table}


Evolving the masses of quarks and charged leptons to the scale of grand unification,
$\Lambda_{\textrm{GUT}} \sim 10^{16}\,\textrm{GeV}$, they approximately satisfy the
following relations,
\begin{equation}
m_t:m_c:m_u \sim 1:\eta^2 : \eta^4 \,, \quad
m_b : m_s : m_d \sim m_{\tau} : m_{\mu} : m_e \sim 1: \eta : \eta^3 \,,
\label{eq:hierarchies}
\end{equation}
with $\eta^2 \simeq 1/300$.
This hierarchy pattern can be well reproduced by a simple Abelian flavour symmetry,
which we will denote by $U(1)_{\textrm{FN}}$ in the following.
Let us assign nonnegative flavour charges $Q_i$ to all $SU(5)$ multiplets $\psi_i$
which are contained in our superpotential and let us also introduce an extra
$SU(5)$ singlet field $\Sigma$, the \textit{flavon}, carrying negative flavour charge,
$Q_\Sigma = -1$.
In the effective theory below some high energy scale $\Lambda > \Lambda_{\textrm{GUT}}$,
the interactions of the matter fields $\psi_i$ with the flavon $\Sigma$ are described
by nonrenormalizable terms in the superpotential,
\begin{align}
W_{\textrm{nr}} \supset C_{ijk} \left(\frac{\Sigma}{\Lambda}\right)^{Q_i + Q_j + Q_k}
\psi_i \, \psi_j \, \psi_k \,,
\label{eq:Wnr}
\end{align}
where the coefficients $C_{ijk}$ are unknown parameters, which are expected to
be of $\mathcal{O}(1)$.
Assigning a nonzero VEV $\left<\Sigma\right>$ to the scalar component of the flavon field spontaneously breaks the $U(1)_{\textrm{FN}}$ flavour symmetry and turns the
nonrenormalizable terms in Eq.~\eqref{eq:Wnr} into the Yukawa interactions of our superpotential.
We shall assume that the Higgs fields $S_{1,2}$ and $H_{u,d}$ carry no flavour
charge.
Suppressing the $\mathcal{O}(1)$ coefficients in Eq.~\eqref{eq:Wnr},
we obtain
\begin{align}
h_{ij} \sim \eta^{Q_i + Q_j} \,,\quad \sqrt{\lambda} \sim \eta^{Q_\Phi} \,,\quad
\eta = \frac{\left<\Sigma\right>}{\Lambda}\,.
\label{eq:hijlambda}
\end{align}
The hierarchies of the quark and charged-lepton masses given in Eq.~\eqref{eq:hierarchies}
are then naturally obtained for $\eta^2 \simeq 1/300$ and the chiral charges $Q_i$
listed in Tab.~\ref{tab:charges}.
Note that Eq.~\eqref{eq:hijlambda} captures the key essence of our flavour model:
up to $\mathcal{O}(1)$ factors, all dimensionless couplings in the superpotential
are given as certain powers of a common hierarchy parameter $\eta$.


\subsubsection{Parametrization of our Model}


Eq.~\eqref{eq:hijlambda} in combination with Tab.~\ref{tab:charges} allows us to
parametrize our entire model in terms of flavour charges and Higgs VEVs.
The heavy-(s)neutrino mass matrix $M$, for instance, can now be estimated as
(cf.\ Eq.~\eqref{eq:NiMM}),
\begin{align}
M = v_{B-L} h^n \,\sim v_{B-L} \,\textrm{diag}\left(\eta^{2d},\eta^{2c},\eta^{2b}\right) \,.
\label{eq:M}
\end{align}
We shall restrict our analysis to the case of a hierarchical heavy-(s)neutrino
mass spectrum, $M_1 \ll M_2 \lesssim M_3$,
which we readily obtain by imposing the following condition on the flavour charges
of the heavy (s)neutrinos,\footnote{Notice that, although we require $b=c$,
the $\mathcal{O}(1)$ uncertainties in $M$ still allow the mass ratio $M_3/M_2$ to vary
within one order of magnitude, $1 \lesssim M_3/M_2 \lesssim 10$.}
\begin{align}
b = c = d -1 \,.
\label{eq:bcd}
\end{align}
While this restriction facilitates our investigation, it still preserves all
characteristic features of our scenario, in particular the anticipated
connection between neutrino parameters and superparticle masses.
One of the simplifications implied by Eq.~\eqref{eq:bcd}, for instance, is
that the lepton asymmetry will be mostly generated by the decay
of the lightest heavy-(s)neutrino generation (cf.\ Sec.~\ref{subsec:baryogenesis}).
Having eliminated $b$ and $c$, the remaining free charges are $a$, $d$ and $e$.
As we will shortly see, these can be related to
the physical parameters $v_{B-L}$, $M_1$ and $m_S$, where the last quantity
denotes the common mass of all particles from the symmetry-breaking sector
in the true vacuum (cf.\ Ch.~\ref{ch:model}).


The neutrino Dirac mass matrix $m_D$ may now be written as (cf.\ Eq.~\eqref{eq:seesawformula})
\begin{equation}
m_D = v_u \,h^{\nu} \sim v_u\, \eta^a
\begin{pmatrix} \eta^{d+1} & \eta^{c+1} & \eta^{b+1} \\
\eta^{d}& \eta^{c} & \eta^{b} \\ \eta^{d} & \eta^{c} & \eta^{b}\end{pmatrix} \,.
\end{equation}
Together with the heavy-(s)neutrino mass matrix $M$, it yields the light-neutrino
Majorana mass matrix $m_\nu$ via the seesaw formula (cf.\ Eq.~\eqref{eq:seesawformula}),
\begin{equation}
 m_{\nu} \sim
\frac{v_u^2}{v_{B-L}} \, \eta^{2a} \,
\begin{pmatrix} \eta^{2} & \eta & \eta \\ \eta & 1 & 1 \\
\eta & 1 & 1 \end{pmatrix} \,.
\label{eq:mnu}
\end{equation}
Interestingly, the dependence on the heavy-(s)neutrino charges drops out in
the calculation of $m_\nu$, so that it ends up being solely controlled by the charge $a$.
In fact, taking into account that $v_u = v_{\textrm{EW}} \sin\beta$
and apart from the unspecified $\mathcal{O}(1)$ coefficients, the light-neutrino
mass matrix only depends one specific combination of parameters,
namely the effective mass scale $\bar{v}_{B-L}$,
\begin{align}
\bar{v}_{B-L} = \frac{v_{B-L}}{\eta^{2a} \sin^2\beta} \,.
\label{eq:vbarBL}
\end{align}
We point out that the specific hierarchy pattern inherent in $m_\nu$
directly feeds into the lepton mixing matrix $U$ and hence has a large impact on
the low-energy observables in the neutrino sector.
In Ch.~\ref{ch:neutrinos}, we will further elaborate on the neutrino phenomenology
implied by the hierarchy structure of $m_\nu$.
Finally, diagonalizing the light-neutrino mass matrix yields the light-neutrino masses $m_i$,
\begin{align}
m_1 \sim \eta^{2a+2} \frac{v_u^2}{v_{B-L}} \,,\quad
m_2 \sim m_3 \sim \eta^{2a}\frac{v_u^2}{v_{B-L}} \,.
\label{eq:mi}
\end{align}


The second matrix which we have to diagonalize in order to
obtain the PMNS matrix $U$ is the charged-lepton mass matrix
$m_{\textrm{cl}}$ (cf.\ Eq.~\eqref{eq:LCCUPMNS}).
It is given by
\begin{equation}
m_{\textrm{cl}} = v_d\,h^d \sim v_d \, \eta^a
\begin{pmatrix} \eta^{3} & \eta^{2 } & \eta \\
\eta^{2}& \eta & 1 \\ \eta^{2} & \eta & 1\end{pmatrix}\ .
\end{equation}
As an important detail, we note that the second and the third row of
the matrix $m_{\textrm{cl}}$ have the same hierarchy pattern.
This is a consequence of the same flavour charges for the second and
the third charged-lepton generation, which is in turn the origin of the
large atmospheric mixing angle.
Hence, diagonalizing $m_{\textrm{cl}}$  can in principle give a
sizable contribution to the mixing in the lepton sector.


The light-neutrino mass scale is conveniently characterized by
$\overline{m}_\nu = \sqrt{m_2 m_3}$, the geometric mean of the two light-neutrino
masses $m_2$ and $m_3$.
In the case of a normal mass ordering of the light neutrinos,
we may estimate it as (cf.\ Eq.~\eqref{eq:Deltam2atmsol})
\begin{align}
\overline{m}_\nu \approx \left[\left(\Delta m_{\textrm{sol}}^2\right)^{1/2}
\left|\Delta m_{\textrm{atm}}^2\right|^{1/2}\right]^{1/2} \sim 3 \times 10^{-2}\,\textrm{eV}\,.
\end{align}
On the other hand, estimating $\overline{m}_\nu$ with
the aid of Eq.~\eqref{eq:mi} provides us with
\begin{align}
\overline{m}_\nu \sim \eta^{2a}\frac{v_u^2}{v_{B-L}} \,,\quad
v_{B-L} \sim \eta^{2a}\frac{v_u^2}{\overline{m}_\nu} = \eta^{2a} M_0 \,,
\label{eq:vBLa}
\end{align}
where we have introduced the heavy-neutrino mass scale $M_0$ through the relation
\begin{align}
M_0 = \frac{v_u^2}{\overline{m}_\nu} \sim
\frac{\left(174\,\textrm{GeV}\right)^2}{3 \times 10^{-2}\,\textrm{eV}} \sim
1 \times 10^{15}\,\textrm{GeV} \,.
\label{eq:M0}
\end{align}
Note that, as $M_0$ follows from the seesaw formula, triple products of $\mathcal{O}(1)$
coefficients enter into its calculation.
Hence, it may in fact be easily as large as the GUT scale,
$\Lambda_{\textrm{GUT}} \sim 10^{16}\,\textrm{GeV}$.
Meanwhile, Eq.~\eqref{eq:vBLa} illustrates that there is a one-to-one
correspondence between the \BmL scale $v_{B-L}$ and the charge $a$.
Consequently, once $v_{B-L}$ is fixed, $M_1$ is directly  related to the charge
$d$ through Eq.~\eqref{eq:M},
\begin{align}
v_{B-L} \sim \eta^{2a} M_0 \,,\quad M_1 \sim \eta^{2d} v_{B-L} \,.
\label{eq:vBLaM1d}
\end{align}
Furthermore, cosmology may impose a relation between $a$ and $d$.
Thermal leptogenesis, for instance, requires $M_1 \sim 10^{10}\,\textrm{GeV}$,
which translates into $a + d = 2$~\cite{Buchmuller:1998zf}.


The results obtained so far enable us to estimate the effective
neutrino masses $\widetilde{m}_i$ as well as the $CP$ violation
parameters $\epsilon_i$.
First of all, we find (cf.\ Eq.~\eqref{eq:mitilde})
\begin{align}
\widetilde m_i = \frac{\big[m_D^\dagger m_D\big]_{ii}}{M_i}
\sim \eta^{2a}\frac{v_u^2}{v_{B-L}} \sim \overline{m}_\nu \, .
\end{align}
Since the light-neutrino mass matrix is not hierarchical,
the $\mathcal{O}(1)$ uncertainties in the Yukawa matrix $h^\nu$
can lead to large deviations from the relation between $\widetilde{m}_1$
and $\overline{m}_\nu$.
The only rigorous inequality is $\widetilde{m}_1 > m_1$~\cite{Fujii:2002jw}.
We take these uncertainties into account by varying the effective neutrino
mass $\widetilde{m}_1$ within the range
\begin{align}
10^{-5}\,\textrm{eV} \leq \widetilde{m}_1 \leq & \: 1 \,\textrm{eV} \,.
\label{eq:m1trange}
\end{align}
Because the two heavier (s)neutrino generations will turn out to play a
less prominent role in our scenario, we decide to ignore possible
deviations of $\widetilde{m}_{2,3}$ from $\overline{m}_\nu$ and
simply set $\widetilde m_{2,3} = \overline m_\nu$.
Assuming that the complex phases in the Yukawa matrix $h^\nu$ are
not accidentally suppressed, the $CP$ violation parameter $\epsilon_1$
is expected to have a value close to its upper bound (cf.\ Eq.~\eqref{eq:eps1max}),
\begin{align}
\left|\epsilon_1\right| \sim \frac{3}{8\pi} \eta^{2(a+d)} \sim \frac{3}{8\pi} \frac{M_1}{M_0}
\sim 1 \times 10^{-6} \left(\frac{M_1}{10^{10}\,\textrm{GeV}}\right) \sim
\epsilon_1^{\textrm{max}}
\label{eq:epsilon1}
\end{align}
The $CP$ violation parameters $\epsilon_{2,3}$ are enhanced compared to
$\epsilon_1$ by a factor $\eta^{-2}$, so that our results for the
parameters $\epsilon_i$ can be brought into the following neat form,
\begin{align}
\left|\epsilon_{2,3}\right| \sim \eta^{-2} \left|\epsilon_1\right| \,,\quad
\left|\epsilon_i\right| \sim \frac{1}{10} \frac{M_i}{M_0} \,.
\label{eq:epsiloniest}
\end{align}


Last but not least, let us turn to the implications of our estimate for
the coupling constant~$\lambda$ (cf.\ Eq.~\eqref{eq:hijlambda}).
As we will see in the next chapter, $\lambda$ determines the mass of the
particles from the symmetry-breaking sector $m_S$.
At the same time, it also controls the initial false vacuum energy density $\rho_0$
(cf.\ Eq.~\ref{eq:Fterm}),
\begin{align}
m_S = \sqrt{\lambda} v_{B-L} \,, \quad  \rho_0 = \frac{1}{4} \lambda v_{B-L}^4 \,.
\end{align}
Within our flavour model these two quantities are thus estimated as
\begin{align}
m_S \sim \eta^{2a+e} M_0 \,,\quad \rho_0 \sim \frac{1}{4} \eta^{8a+2e} M_0^4 \,,
\end{align}
where we have used that $Q_\Phi = e$.
We shall assume $m_S$ to be of the same order of magnitude as
the heavy-(s)neutrino masses $M_2$ and $M_3$.
This leads us to imposing the condition $e = 2b = 2c$,
which, together with Eq.~\eqref{eq:bcd}, results in
\begin{align}
e = 2b = 2c = 2(d-1) \,.
\label{eq:ebc}
\end{align}
In Chs.~\ref{ch:reheating} and \ref{ch:wimp}, we will set $m_S = M_3 = M_2$
for definiteness.
Given such a mass spectrum, the particles from the symmetry-breaking sector
only decay into the first heavy (s)neutrino-generation.
Again, this restriction simplifies our analysis, but still preserves all aspects
that we are interested in (cf.\ Sec.~\ref{subsec:neutrinoreheating}).


Owing to the two conditions in Eqs.~\eqref{eq:bcd} and Eq.~\eqref{eq:ebc},
we have left over only two free flavour charges, $a$ and $d$.
According to Eq.~\eqref{eq:vBLaM1d}, these can be traded for
the more physical quantities $v_{B-L}$ and $M_1$.
The ranges over which $a$ and $d$, and hence $v_{B-L}$ and $M_1$, are allowed
to vary are restricted by the requirement of perturbativity of all coupling constant
as well as the lower bound on $\tan\beta$.
First, to ensure that no coupling constant significantly exceeds
the top-Yukawa coupling, we require that $a\geq 0$ and $d\geq 1$.
Second, $\tan\beta > \mathcal{O}(1)$ implies $a\leq1$, whereas
there is no corresponding upper bound on $d$.
On top of that, in Sec.~\ref{sec:strings} we will discuss further
restrictions on $a$ and $d$ that follow from requiring a viable realization of
hybrid inflation as well as a not too strong production of cosmic strings
during the \BmL phase transition.
Meanwhile, as the leptogenesis process after the \BmL phase transition
is mainly driven by nonthermally produced (s)neutrinos, we do not
have to worry about the constraint from thermal leptogenesis, $a+d = 2$.
We also remark that bounds on $a$ and $d$ directly correspond to extremal
values for $v_{B-L}$ and $M_1$.
In the following, we will assume that $v_{B-L}$ and $M_1$
can continuously vary within the ranges bounded by their extremal values.
With respect to $a$ and $d$, such a variation may be effectively realized
in terms of fractional flavour charges.


To sum up, our model is parametrized by five dimensionful parameters,
the \BmL breaking scale $v_{B-L}$, the heavy (s)neutrino mass $M_1$,
the effective neutrino mass $\widetilde{m}_1$, the gravitino mass $m_{\widetilde{G}}$,
and the gluino mass $m_{\tilde{g}}$, as well as by the dimensionless $\mathcal{O}(1)$
coefficients in the Yukawa matrices, which we have left out of consideration up to now.
We will turn our attention to the $\mathcal{O}(1)$ uncertainties
of our flavour model in the next chapter, in which we will perform a
numerical Monte-Carlo study to assess the impact of the unspecified $\mathcal{O}(1)$
factors in the lepton mass matrices on the various observables of the neutrino sector.
This analysis will demonstrate that our Froggatt-Nielsen flavour model
has a rather strong predictive power after all and that it in fact implies
parameter predictions, which are in many cases much more precise than
the rough order-of-magnitude estimates derived in the present section.
In the subsequent chapters, we will then proceed employing the best-guess estimates
obtained in Ch.~\ref{ch:neutrinos}, while henceforth ignoring the $\mathcal{O}(1)$
uncertainties of our flavour model.
This means in particular that in Chs.~\ref{ch:reheating} and \ref{ch:wimp}
we will simply set all $\mathcal{O}(1)$ factors to one.

\cleardoublepage


\chapter{Neutrino Phenomenology}
\label{ch:neutrinos}


In Sec.~\ref{subsec:flavour}, we motivated the Froggatt-Nielsen mechanism
as a promising prototype for a fundamental theory of flavour, allowing us
to parametrize our entire model in terms of flavour charges and Higgs VEVs.
Generally speaking, flavour symmetries of the Froggatt-Nielsen type provide
a natural means to reconcile the large quark and charged-lepton mass hierarchies
and small quark mixing angles on the one hand with the observed small
neutrino mass hierarchies and large neutrino mixing angles on the other hand.
But despite these virtues, the predictive power of the Froggatt-Nielsen mechanism
is understood to be rather limited due to unknown coefficients of $\mathcal{O}(1)$
in the superpotential (cf.\ Eq.~\eqref{eq:Wnr}), directly implying $\mathcal{O}(1)$
uncertainties in all entries of the mass matrices.
The explicit model introduced in Sec.~\ref{subsec:flavour}, for instance, is not able
to make a precise prediction of the solar neutrino mixing angle $\theta_{12}$,
as it can accommodate both a small as well as a large value for this
observable~\cite{Sato:1997hv,Irges:1998ax,Vissani:1998xg}.


To get an idea of the range of possible predictions for a given flavour structure,
it is instructive to  treat the $\mathcal{O}(1)$ parameters as random
variables~\cite{Hall:1999sn,Sato:2000kj,Vissani:2001im}.
In this chapter, we shall therefore employ Monte-Carlo techniques
(cf.\ Sec.~\ref{sec:montecarlo}) to quantitatively study the dependence
of yet undetermined, but soon testable parameters of the neutrino sector
on the unknown $\mathcal{O}(1)$ factors. 
Using the already measured neutrino masses and mixings as input,
we find surprisingly sharp predictions (cf.\ Sec.~\ref{sec:predictions}),
which we are even able to reproduce analytically in the case of
one observable, \textit{viz.} the Majorana $CP$ violation
phase $\alpha_{21}$ (cf.\ Sec.~\ref{sec:majorana}).


The results presented in this chapter were first published in
Ref.~\cite{Buchmuller:2011tm}.


\newpage


\section[Monte-Carlo Sampling of \texorpdfstring{$\mathcal{O}(1)$}{Order One} Factors]
{Monte-Carlo Sampling of $\mathcal{O}(1)$ Factors}
\label{sec:montecarlo}


In view of the $\mathcal{O}(1)$ factors in all mass matrices,
one might expect the predictions for the observables of the
neutrino sector to be quite uncertain.
The light-neutrino mass matrix $m_\nu$, for instance, is calculated as
the product of three other matrices, the entries of which are all determined
only up to coefficients of $\mathcal{O}(1)$ (cf.\ Eq.~\eqref{eq:seesawformula}).
Hence, one may be led to the conclusion that the predictions for all observables
deriving from $m_\nu$ should have an uncertainty of roughly three orders of magnitude.
In principle, the Froggatt-Nielsen model allows, of course, for such large variations.
This is the reason why we stated in Sec.~\ref{subsec:flavour} that the heavy-neutrino
mass scale $M_0$ may be as large as the GUT scale $\Lambda_{\textrm{GUT}}$
(cf.\ Eq.~\eqref{eq:M0}) as well as the reason why we allow the effective neutrino
mass $\widetilde{m}_1$ to vary over as much as five orders of magnitude
(cf.\ Eq.~\eqref{eq:m1trange}).
As we shall demonstrate now by means of a numerical Monte-Carlo study,
such a large dispersion is, however, not characteristic for the Froggatt-Nielsen model.
In fact, quite the opposite is the case.
For many observables, the predicted values turn out to be mostly confined to narrow ranges,
extending over less than an order of magnitude.


\subsubsection{Random Mass Matrices Compatible with all Experimental Constraints}


All observables of the neutrino sector eventually derive from the mass matrices
$M$, $m_D$ and $m_{\textrm{cl}}$ (cf.\ Eqs.~\eqref{eq:NiMM}, \eqref{eq:seesawformula} and
\eqref{eq:mcldiag}) or equivalently from the Yukawa matrices $h^n$, $h^\nu$ and $h^d$
(cf.\ Eqs.~\eqref{eq:WSeesaw} and \eqref{eq:WMSSM}).
Postponing the discussion of the individual observables to Sec.~\ref{sec:predictions},
let us now focus on our numerical method to find mass matrices
which are compatible with all experimental constraints.


The unknown $\mathcal{O}(1)$ coefficients of the Yukawa matrices $h^d$,
$h^\nu$ and $h^n$ are constrained by the experimental data on neutrino
masses and mixings, with the $3\,\sigma$ confidence ranges of the respective
measurable quantities being given by~\cite{Nakamura:2010zzi}
\begin{align}
& 2.07 \times 10^{-3} \,\textrm{eV}^2 \leq |\Delta m^2_{\text{atm}}| \leq 2.75 \times 10^{-3}
\,\textrm{eV}^2 \,, \label{eq:nudata} \\
& 7.05 \times 10^{-5} \,\textrm{eV}^2 \leq  \Delta m^2_{\text{sol}} \leq 8.34 \times 10^{-5}
\,\textrm{eV}^2 \,, \nonumber \\
& 0.75 \leq \sin^2(2 \theta_{12})  \leq 0.93 \,,\nonumber\\
& 0.88 \leq  \sin^2(2 \theta_{23}) \leq 1 \,.\nonumber
\end{align}
We explicitly do not use the current bound on the smallest mixing angle,
$\theta_{13} < 0.21$ at $3\,\sigma$ \cite{Nakamura:2010zzi}, to constrain the
$\mathcal{O}(1)$ factors.
This allows us to demonstrate that nearly all values we obtain for
$\theta_{13}$ automatically obey the experimental bound,
cf.~Fig.~\ref{fig:mixing}.


Each of the Yukawa matrices $h^\nu$ and $h^d$ contains nine complex
$\mathcal{O}(1)$ factors, $C_{ij}^\nu$ and $C_{ij}^d$, respectively, 
while the Yukawa matrix $h^n$ features three real $\mathcal{O}(1)$ factors $C_i^n$.
In a numerical Monte-Carlo study, we now generate random numbers to model these
in total~39 real parameters.\footnote{Note that it is the the low-energy Yukawa
couplings which we treat as random variables.
These are related to the couplings at higher energy
scales, i.e.\ the couplings the Froggatt-Nielsen model
is actually concerned with, via renormalization group equations.
We however expect that the effect of renormalization group running
can be absorbed into a redefinition of the effective scale
$\bar{v}_{B-L}$ (cf.\ Eq.~\eqref{eq:vbarBL}), hence leaving the results
presented in the following unchanged.}
The absolute values $\left|C_{ij}\right|$
are taken to be uniformly distributed on a logarithmic scale,
while the phases in $h^\nu$ and $h^d$ are chosen to be uniformly
distributed on a linear scale,
\begin{align}
- \frac{1}{2} \leq \log_{10}\left|C_{ij}\right| \leq \frac{1}{2} \,, \quad
0 \leq \arg{C_{ij}} < 2\pi \,. \label{eq:Cranges}
\end{align}
In the following, we shall refer to those sets of coefficients which yield
mass matrices that are consistent with the experimental constraints in
Eq.~\eqref{eq:nudata} as \textit{hits}.


In a preliminary run of our Monte-Carlo code, we solely take into account
the neutrino mixing matrix $\Omega$ (cf.\ Eq.~\eqref{eq:mnudiag}) to
calculate the PMNS matrix (cf.\ Eq.~\eqref{eq:LCCUPMNS})
as well as the set of observables encoded in it.
Meanwhile, we treat the effective scale $\bar{v}_{B-L}$
(cf.\ Eq.~\eqref{eq:vbarBL}) as a random variable in the interval
$[1/\sqrt{10},\sqrt{10}]\times10^{15}\,\textrm{GeV}$.
We find that the percentage of hits strongly peaks at
$\bar{v}_{B-L} \simeq 1 \times 10^{15}\,\textrm{GeV}$.
This is interesting for two reasons.
First, it entails that given $0 \leq a \leq 1$ (cf.\ Sec.~\ref{subsec:flavour}),
the \BmL breaking scale $v_{B-L}$ lies in the range
\begin{align}
3\times 10^{12}~\text{GeV} \, \sin^2\beta \lesssim v_{B-L}
\lesssim 1\times 10^{15}~\text{GeV} \,\sin^2\beta \,.
\label{eq:vBLMCrange}
\end{align}
Recall that we assume large $\tan\beta$ (cf.\ Sec.~\ref{subsec:seesaw}).
The upper boundary of this mass range is hence close to the GUT scale.
In particular, it deviates by less than an order of magnitude from
the rough estimate of $v_{B-L}$ based on the amplitude of the scalar power spectrum
$A_s$ (cf.\ Eq.~\eqref{eq:vBLAs}).
With regard to the great flexibility of the Froggatt-Nielsen model, we
may thus regard the upper bound on $v_{B-L}$ in Eq.~\eqref{eq:vBLMCrange}
as being consistent with the estimate implied by
hybrid inflation.
Second, the fact that the $\bar{v}_{B-L}$ distribution exhibits a strong peak
allows us to fix $\bar{v}_{B-L}$ to $10^{15}\,\textrm{GeV}$
in the following computations without introducing a significant bias.


In the main run of our code, in which $\bar{v}_{B-L}$ is now fixed, we include
the charged-lepton mixing matrix $L$ (cf.\ Eq.~\eqref{eq:mcldiag})
in the calculation of the PMNS matrix.
We require the mass ratios of the charged leptons to fulfill the corresponding
experimental constraints up to an accuracy of $5\,\%$ and allow for
${1\leq\tan\beta \leq 60}$ to achieve the correct normalization of the
charged-lepton mass spectrum.
Imposing the $3\,\sigma$ constraints on the two large mixing angles
inferred from the full PMNS matrix, $U = L^\dagger\Omega$, along with the constraints on
$|\Delta m^2_{\text{atm}}|$ and $\Delta m^2_{\text{sol}}$, we then find
the searched-for samples of $\mathcal{O}(1)$ factors, which yield
random mass matrices that are compatible with the neutrino data in Eq.~\eqref{eq:nudata}.
Our final results are based on roughly $20\,000$ such hits.
For each hit, we calculate the observables in the neutrino sector as well as two
parameters relevant for leptogenesis, \textit{viz.} $\widetilde{m}_1$ and $\epsilon_1$.
This provides us with distributions for the possible values of the respective
observables.
Before presenting our results, let us first elaborate on
the statistical method with which we shall analyze these distributions.


\subsubsection{Statistical Analysis}


The relative frequency with which we encounter a certain value for an observable
might indicate the probability that this value is actually realized within the
large class of concrete flavour models covered by our analysis.
In the following, we shall therefore treat the distributions for the various
observables as \textit{probability densities} for continuous random variables,
i.e.\ our predictions for the respective observables represent best-guess
estimates according to a probabilistic interpretation of the relative
frequencies.
For each observable, we would like to deduce measures
for its central tendency and statistical dispersion from
the respective probability distribution.
Unfortunately, it is infeasible to fit all obtained distributions
with one common template distribution.
Such a procedure would lack a clear statistical justification, and it also
appears impractical, as the distributions that we obtain differ substantially
in their shapes.
We therefore choose a different approach.


We consider the \textit{median} of a distribution as its center
and we use the $68\ \%$ \textit{confidence interval} around it
as a measure for its spread.
Of course, this range of the confidence interval is reminiscent of
the $1\,\sigma$ range of a normal distribution.
More precisely, for an observable $x$ with probability density $f$, we will
summarize its central tendency and variability in the following form \cite{Cowan:1998ji},
\begin{align}
x = \hat{x}_{\Delta_-}^{\Delta_+}
\,,\quad
\Delta_\pm = x_\pm - \hat{x}
\,.
\end{align}
Here, $x_-$ and $x_+$ denote the $16\,\%$ and $84\,\%$ quantiles
with respect to the density function $f$.
The central value $\hat{x}$ is the median of $f$ and thus
corresponds to its $50\,\%$ quantile.
All three values can be calculated from the quantile function $Q$,
\begin{align}
Q(p) = \textrm{inf} \left\{x \in
\left[x_\textrm{min},x_\textrm{\text{max}}\right] : p \leq F(x)\right\}
\,,\quad
F(x) = \int_{x_\textrm{min}}^x dt \: f(t)
\,,
\end{align}
where $F$ stands for the cumulative distribution function of $x$.
We then have
\begin{align}
x_- = Q(0.16)\,,\quad
\hat{x} = Q(0.50)\,,\quad
x_+ = Q(0.84)\,.
\end{align}
Intuitively, the intervals from $x_{\textrm{min}}$ to $x_-$, $\hat{x}$, and $x_+$,
respectively, correspond to the $x$ ranges into which $16\,\%$, $50\,\%$ or $84\,\%$
of all hits fall.
This is also illustrated in the histogram
for $\sin^2 2\theta_{13}$ in Fig.~1.
Moreover, we have included vertical lines into each plot,
indicating the respective positions of $x_-$, $\hat{x}$, and $x_+$.


In our case, the median is a particularly useful measure of location.
First of all, it is resistant against outliers and hence an appropriate
statistic for such skewed distributions, as we observe them.
But more importantly, the average absolute deviation from
the median is minimal in comparison to any other reference point.
The median is thus the best guess for the outcome of a
measurement, if one is interested in being as close as possible
to the actual result, irrespectively of the sign of the error.
On the technical side, the definition of the median fits nicely
together with our method of assessing statistical dispersion.
The $68\,\%$ confidence interval as introduced above
is just constructed in such a way, that equal numbers of hits
lie in the intervals from $x_-$ to $\hat{x}$ and from $\hat{x}$
to $x_+$, respectively.
In this sense, our confidence interval represents a symmetric
error with respect to the median.


To test the robustness of our results, we check the dependence
of our distributions on the precise choice of the experimental
error intervals.
Our results, however, turn out to be insensitive to small variations
of the error margins.
For definiteness, we therefore stick to the $3\,\sigma$ ranges listed
in Eq.~\eqref{eq:nudata}.
Furthermore, we also check the effect of taking the absolute values
$\left|C_{ij}\right|$ of the $\mathcal{O}(1)$ factors to be
uniformly distributed on a linear instead of a logarithmic scale
(cf.\ Eq.~\eqref{eq:Cranges}).
Again, our results prove to be robust.


\section{Predictions for Neutrino Observables}
\label{sec:predictions}


Finally, we present the results of our numerical Monte-Carlo study.
Out of the findings which we obtain three particularly interesting ones
deserve to be highlighted:
(i) a large value for the smallest mixing angle $\theta_{13}$ in accordance
with recent results from the T2K~\cite{Abe:2011sj}, Minos~\cite{Adamson:2011qu},
Double Chooz~\cite{Abe:2011fz}, Daya Bay~\cite{An:2012eh}, and Reno~\cite{Ahn:2012nd}
experiments,
(ii) a value for the lightest neutrino mass of $\mathcal{O}\left(10^{-3}\right)\,\textrm{eV}$,
and (iii) one Majorana $CP$-violating phase in the PMNS matrix peaked at $\alpha_{21} \simeq \pi$.


\subsubsection{Mass Hierarchy}


An important unsolved puzzle of modern neutrino physics, which is closely connected to
the flavour physics of the neutrino sector, is the question as to the hierarchy
of the light-neutrino mass eigenstates.
Since up to now the sign of $\Delta m^2_{\textrm{atm}}$ has remained undetermined,
the current experimental data is still consistent with a normal
as well as with an inverted hierarchy (cf.\ Eq.~\eqref{eq:nuhierarchies}).\footnote{A
measurement of the Mikheyev-Smirnov-Wolfenstein (MSW) effect of the earth on the neutrino
oscillation probabilities could resolve this ambiguity (cf.\ Ref.~\cite{Nakamura:2010zzi}
and references therein).}
As a first result of our Monte-Carlo study, we find that all hits come with
a normal mass ordering.
We obtain no hits at all corresponding to an inverted hierarchy.
It is, however, notable that merely imposing the hierarchy pattern of the
neutrino mass matrix $m_\nu$ (cf.\ Eq.~\eqref{eq:mnu}) does \textit{not} exclude
the inverted mass ordering.
Only if the bounds on the mixing angles are taken into account as well,
this possibility is ruled out.


\subsubsection{Mixing Angles}


The flavour composition of the three neutrino mass eigenstates is characterized
the three mixing angles $\theta_{12}$, $\theta_{13}$, and $\theta_{23}$ of
the PMNS matrix (cf.\ Eq.~\eqref{eq:UPMNS}).
Two of these angles are solely bounded from one side by experiment:
for the largest mixing angle $\theta_{23}$ there merely exists a lower bound,
whereas the smallest mixing angle $\theta_{13}$ is so far only bounded from above.
Recent results from the T2K~\cite{Abe:2011sj}, Minos~\cite{Adamson:2011qu},
Double Chooz~\cite{Abe:2011fz}, Daya Bay~\cite{An:2012eh}, and Reno~\cite{Ahn:2012nd}
experiments point to a value of $\theta_{13}$ just below the current experimental
bound.\footnote{The Daya Bay and Reno experiments, whose results are the newest ones,
claim that their measurements are indicative of $\theta_{13}\neq 0$
with a significance of $5.2\,\sigma$ and $4.9\,\sigma$, respectively.}
The respective best-fit points, assuming a normal hierarchy, are
$\sin^2 2\theta_{13} = 0.11$ (T2K),
$2 \sin^2 \theta_{23} \sin^2 2 \theta_{13} = 0.041$ (Minos),
$\sin^2 2 \theta_{13} = 0.085$ (Double Chooz),
$\sin^2 2\theta_{13} = 0.092$ (Daya Bay), and
$\sin^2 2\theta_{13} = 0.113$ (Reno).
The $90\,\%$ or $68\,\%$ confidence intervals respectively read
\begin{align}
&0.03 < \sin^2 2\theta_{13} < 0.28   &&\text{T2K, 90 $\%$ CL}, \, \delta_{CP} = 0, \\
&2 \sin^2 \theta_{23} \sin^2 2 \theta_{13} < 0.12
&&\text{Minos, 90 $\%$ CL}, \, \delta_{CP} = 0,  \nonumber\\
& 0.01 < \sin^2 2\theta_{13} < 0.16  &&\text{Double Chooz, 68 $\%$ CL}, \nonumber \\
& 0.07 < \sin^2 2\theta_{13} < 0.11  &&\text{Daya Bay, 68 $\%$ CL}, \nonumber \\
& 0.09 < \sin^2 2\theta_{13} < 0.14  &&\text{Reno, 68 $\%$ CL}. \nonumber
\end{align}


\begin{figure}
\includegraphics[width=0.48\textwidth]{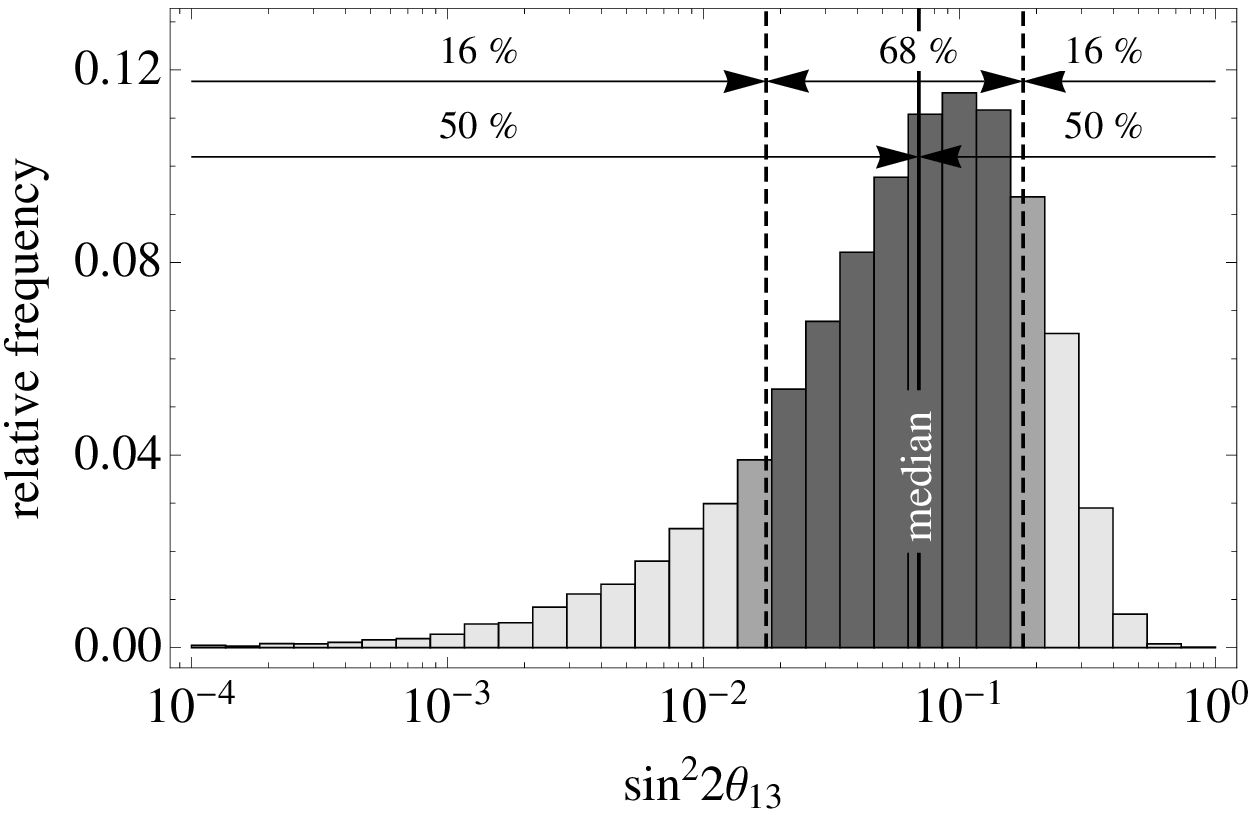}
\includegraphics[width=0.48\textwidth]{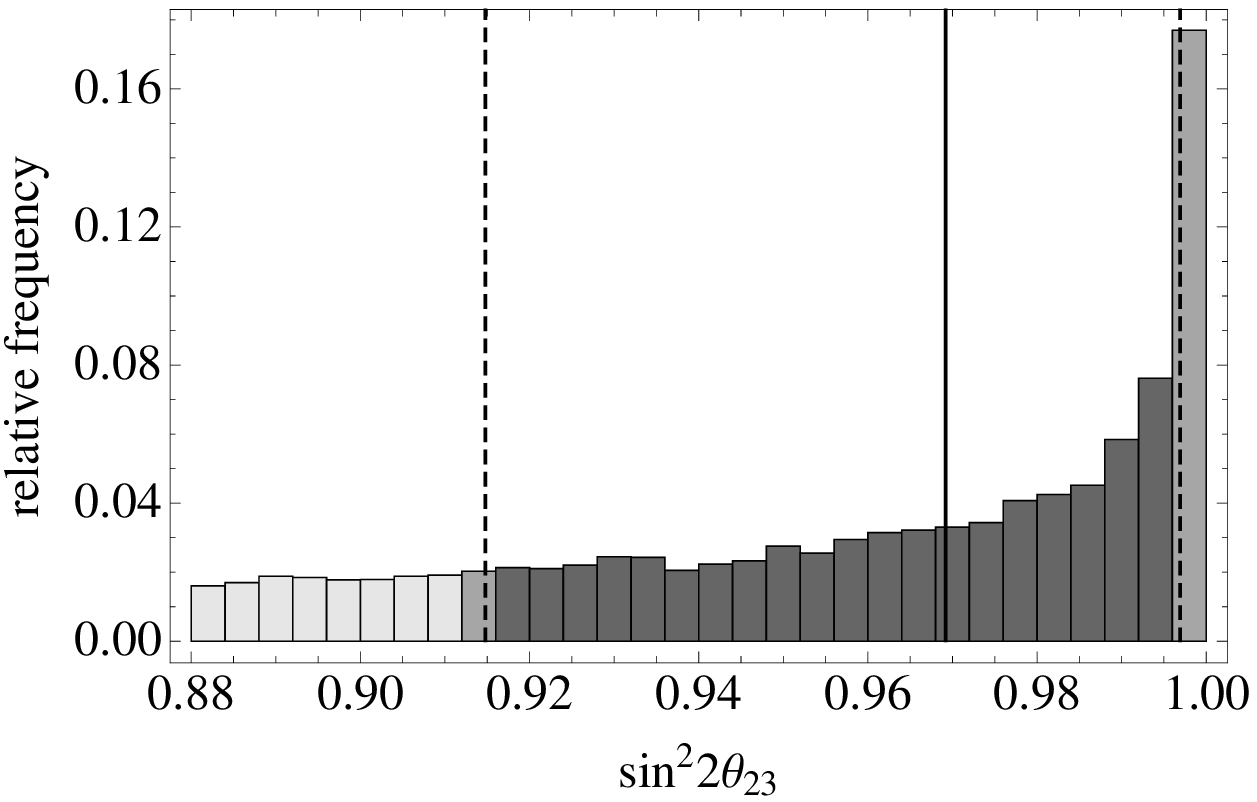}

\medskip
\caption[Neutrino mixing angles $\theta_{13}$ and $\theta_{23}$.]
{Neutrino mixing angles $\theta_{13}$ and $\theta_{23}$.
The vertical lines denote the positions of the medians (solid lines) and
the boundaries of the $68\,\%$ confidence regions (dashed lines) of the
respective distributions.}
\label{fig:mixing}
\end{figure}


We find sharp predictions for $\theta_{13}$ and $\theta_{23}$
within the experimental bounds,
\begin{align}
\sin^2 2\theta_{13} = 0.07^{+0.11}_{-0.05}  \,,
\quad  \sin^2 2\theta_{23} = 0.97^{+0.03}_{-0.05} \,,
\end{align}
with the corresponding distributions being shown in Fig.~\ref{fig:mixing}.
These results are quite remarkable: our prediction for $\theta_{23}$
points to maximal mixing of atmospheric neutrinos, while the rather
large value for $\theta_{13}$ is consistent with the recent T2K, Minos,
Double Chooz, Daya Bay and Reno results.


Finally, we remark that the strong mixing in the lepton sector apparently
derives for the most part from the peculiar hierarchy pattern of the neutrino
mass matrix $m_{\nu}$ (cf.\ Eq.~\eqref{eq:mnu}).
In our Monte-Carlo study, we observe that our results for the mixing
angles are not much affected when omitting the charged-lepton
mixing matrix $L$ in the calculation of the PMNS matrix.
We hence conclude that the PMNS matrix is approximately given by
$\Omega$, the matrix which diagonalizes $m_{\nu}$.


\subsubsection{Absolute Mass Scale}


\begin{figure}
\includegraphics[width=0.48\textwidth]{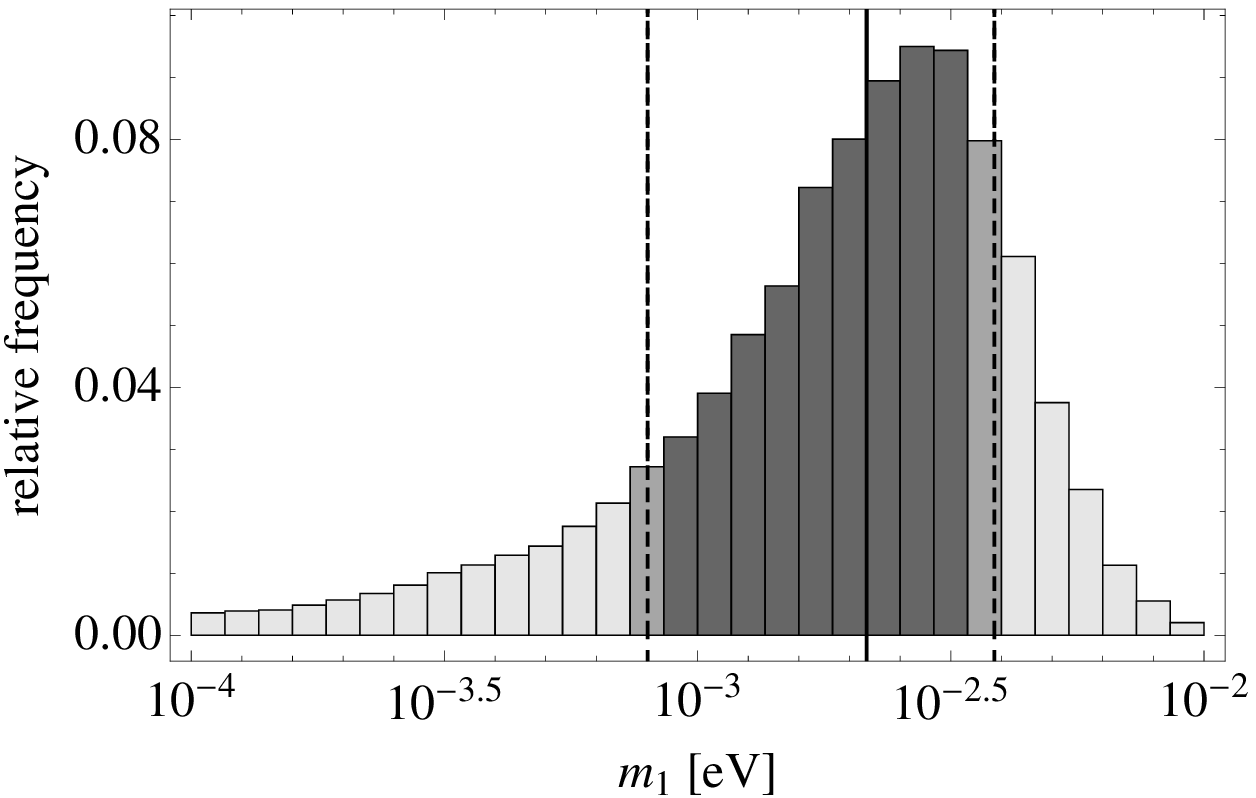}
\includegraphics[width=0.48\textwidth]{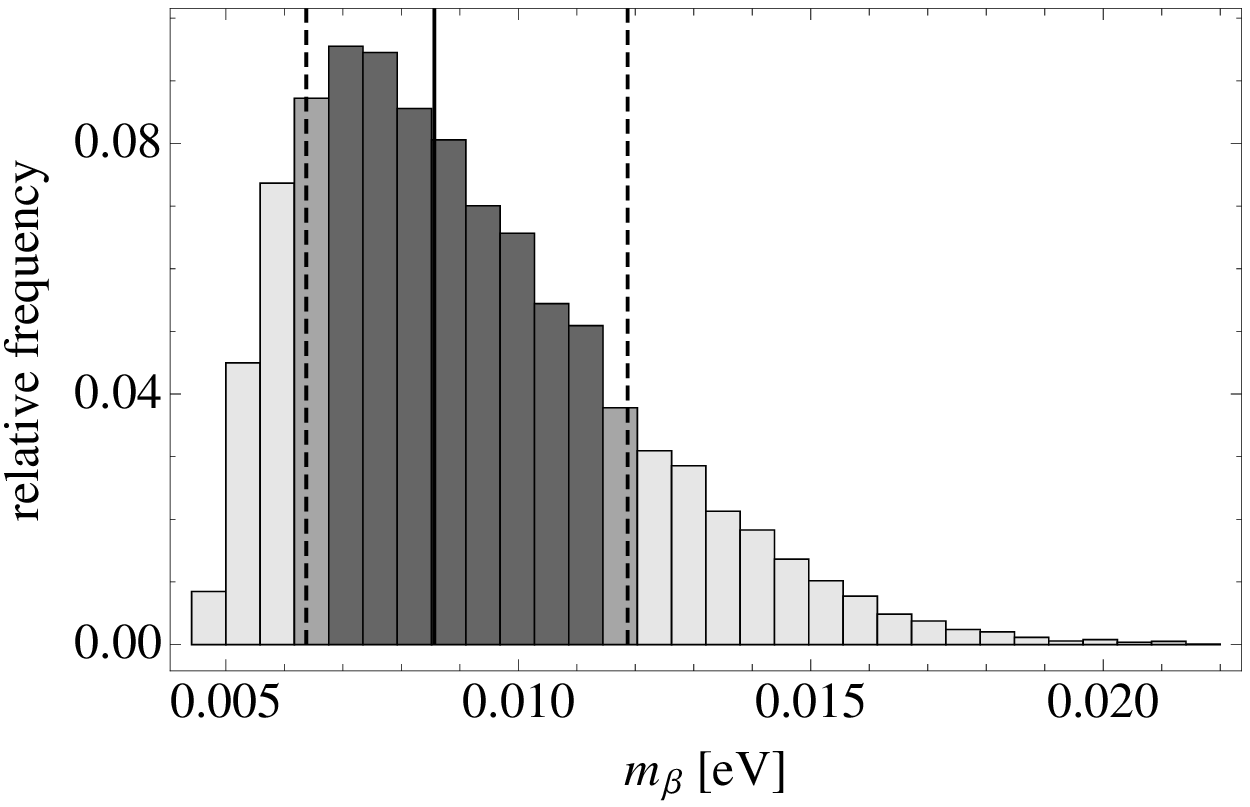}

\medskip
\caption[Lightest neutrino mass $m_1$ and effective neutrino mass in tritium decay $m_\beta$]
{Lightest neutrino mass $m_1$ and effective neutrino mass in tritium decay $m_\beta$.
Vertical lines and shadings as in Fig.~\ref{fig:mixing}.
\label{fig:m1mtritium}}
\end{figure}


The absolute neutrino mass scale determines the impact of cosmic neutrinos
on the formation of matter structure in the early universe (cf.\ Sec.~\ref{subsec:neutrinos}),
represents a crucial parameter of leptogenesis (cf.\ Sec.~\ref{subsec:baryogenesis})
and enters into the description of various low-energy neutrino phenomena.
Neutrino oscillation experiments are unfortunately insensitive to
the absolute neutrino mass scale.
But thanks to its influence on cosmology and low-energy neutrino processes,
it is experimentally accessible nonetheless.
Recall that the combination of several cosmological data sets allows
to put an upper bound on $m_{\textrm{tot}}$ (cf.\ Eq.~\eqref{eq:mnutotrange}),
\begin{align}
 m_{\text{tot}} = \sum_i m_i \lesssim 0.28 \,\textrm{eV} \,.
\label{eq:mtot}
\end{align}
The Planck satellite is expected to be sensitive to values of $m_{\text{tot}}$
as low as roughly $0.1\,\textrm{eV}$~\cite{Planck:2006aa}.
A further constraint on the absolute neutrino mass scale arises from
measurements of the electron spectrum induced by the $\beta^-$ decay of tritium.
Such experiments are able to provide information on the effective
neutrino mass $m_\beta$, for which there only exists an upper bound
at present~\cite{Nakamura:2010zzi},
\begin{align}
m_{\beta}^2  = \sum_i \left|U_{ei}\right|^2 m_i^2 < 4\,\textrm{eV}^2 \,.
\label{eq:mbeta}
\end{align}
By comparison, the Katrin experiment, which will start taking data soon, aims
at reaching a sensitivity of $0.04 \,\text{eV}^2$~\cite{Beck:2010zzb}.
Finally, the absolute neutrino mass scale also enters into the decay amplitude of
neutrinoless double-beta decay, \textit{viz.} through the effective neutrino mass
$m_{0\nu\beta\beta}$,
\begin{align}
 m_{0\nu\beta\beta} = \Big|\sum_i U_{ei}^2 \, m_i \Big| \,.
\label{eq:m0nubb}
\end{align}
The authors of Ref.~\cite{KlapdorKleingrothaus:2001ke} claim to have a measured
a value of $0.11..0.56\,\textrm{eV}$ for this effective mass.
Dedicated experiments searching for neutrinoless double-beta decay
and capable of scrutinizing this claim, such as Gerda~\cite{Meierhofer:2011zz}
with a design sensitivity of $0.09..0.20\,\textrm{eV}$, are on the way.
Note that $m_{0\nu\beta\beta}$ does not only depend on the absolute neutrino mass
scale and the mixing angles $\theta_{12}$ and $\theta_{13}$, but also on the
$CP$-violating phases $(\alpha_{31} - 2 \delta)$ and $\alpha_{21}$.


Again, our Monte-Carlo study provides us with sharp predictions.
Our best-guess estimates for the neutrino masses listed
in Eqs.~\eqref{eq:mtot}, \eqref{eq:mbeta} and \eqref{eq:m0nubb} as well
as for the lightest neutrino mass $m_1$ are (cf.\ Figs.~\ref{fig:m1mtritium} and
\ref{fig:masses})
\begin{align}
m_{\text{tot}} = & \: 6.0^{+0.3}_{-0.3} \times 10^{-2} \,\textrm{eV} \,, \quad
m_{\beta} = 8.6^{+3.3}_{-2.2} \times 10^{-3}  \,\textrm{eV} \,, \\
m_{0\nu\beta\beta} = & \: 1.5^{+0.9}_{-0.8} \times 10^{-3}  \,\textrm{eV} \,, \quad
m_1 = 2.2^{+1.7}_{-1.4} \times 10^{-3} \, \textrm{eV} \,. \nonumber
\end{align}
The fact that $m_1$ turns out to be merely of $\mathcal{O}\left(10^{-3}\right)\,\textrm{eV}$
implies a low neutrino mass scale, unfortunately beyond the reach of current and
upcoming experiments.


\subsubsection{\boldmath{$CP$} Violation Phases}


\begin{figure}
\includegraphics[width=0.48\textwidth]{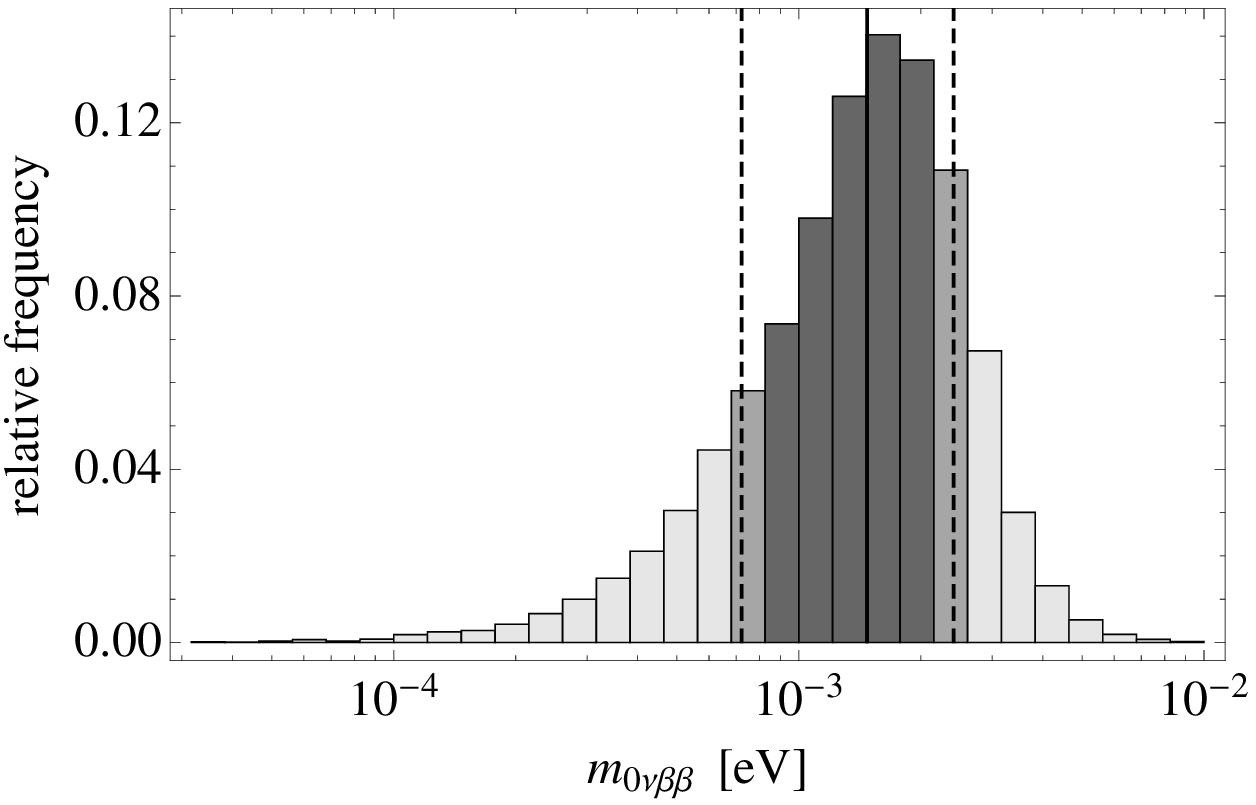}
\includegraphics[width=0.48\textwidth]{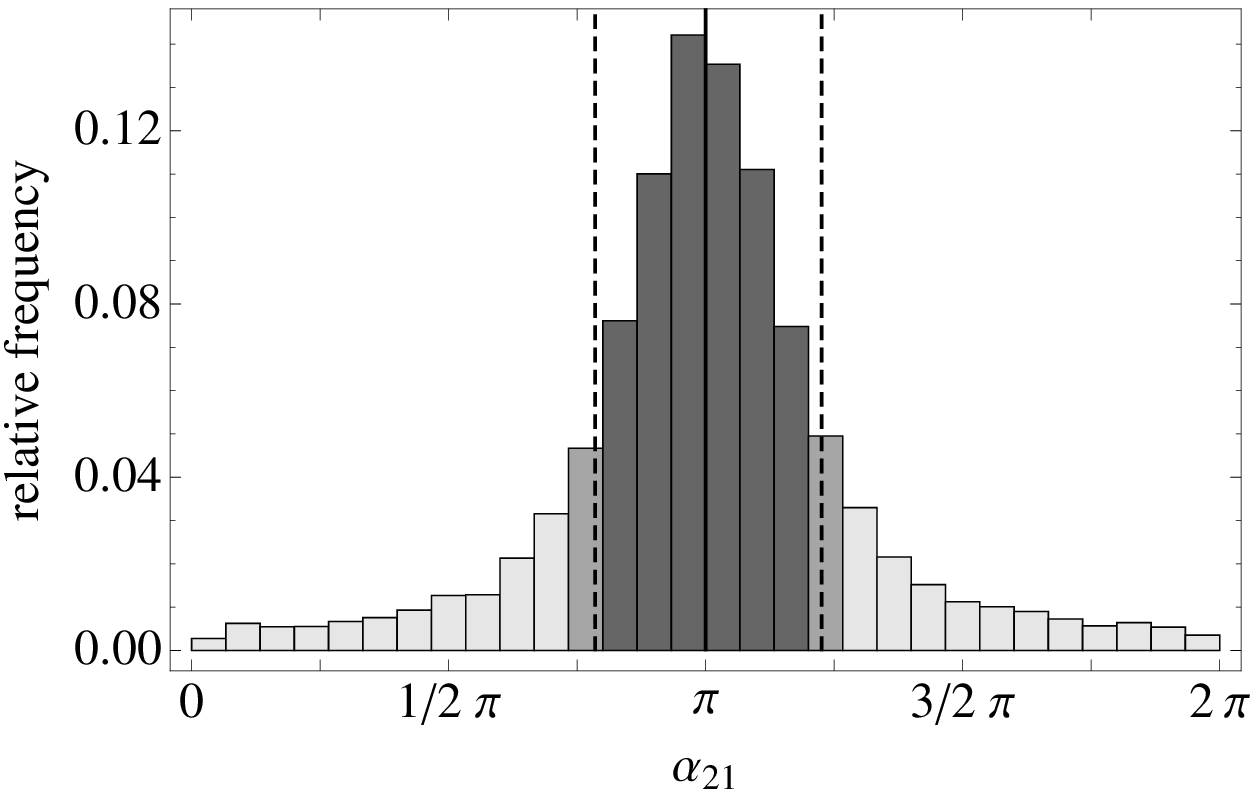}

\medskip
\caption[Effective neutrino mass in $0\nu\beta\beta$ decay $m_{0\nu\beta\beta}$ and
Majorana phase $\alpha_{21}$]
{Effective neutrino mass in neutrinoless double-beta decay $m_{0\nu\beta\beta}$ and
$CP$-violating Majorana phase $\alpha_{21}$.
Vertical lines and shadings as in Fig.~\ref{fig:mixing}.
\label{fig:masses}}
\end{figure}


The small value of our prediction for $m_{0\nu\beta\beta}$ can be traced back to
a relative minus sign between the $m_1$ and $m_2$ terms in Eq.~\eqref{eq:m0nubb},
which is caused by a strong peak in the distribution of the Majorana phase
$\alpha_{21}$ at $\alpha_{21} \simeq \pi$ (cf.\ Fig.~\ref{fig:masses}),
\begin{align}
\frac{\alpha_{21}}{\pi} = 1.0^{+0.2}_{-0.2}  \,.
\end{align}
In the next section, we will demonstrate by means of a simplified
analytic calculation how this preference for $\alpha_{21}$ values close
to $\pi$ directly emerges as a consequence of the hierarchy structure
of the the neutrino mass matrix $m_\nu$.
For the other phases of the PMNS matrix, the Majorana phase $\alpha_{31}$
as well as the Dirac phase $\delta$, we find no such distinct behaviour,
but approximately flat distributions.


\subsubsection{Leptogenesis Parameters}


\begin{figure}
\includegraphics[width=0.48\textwidth]{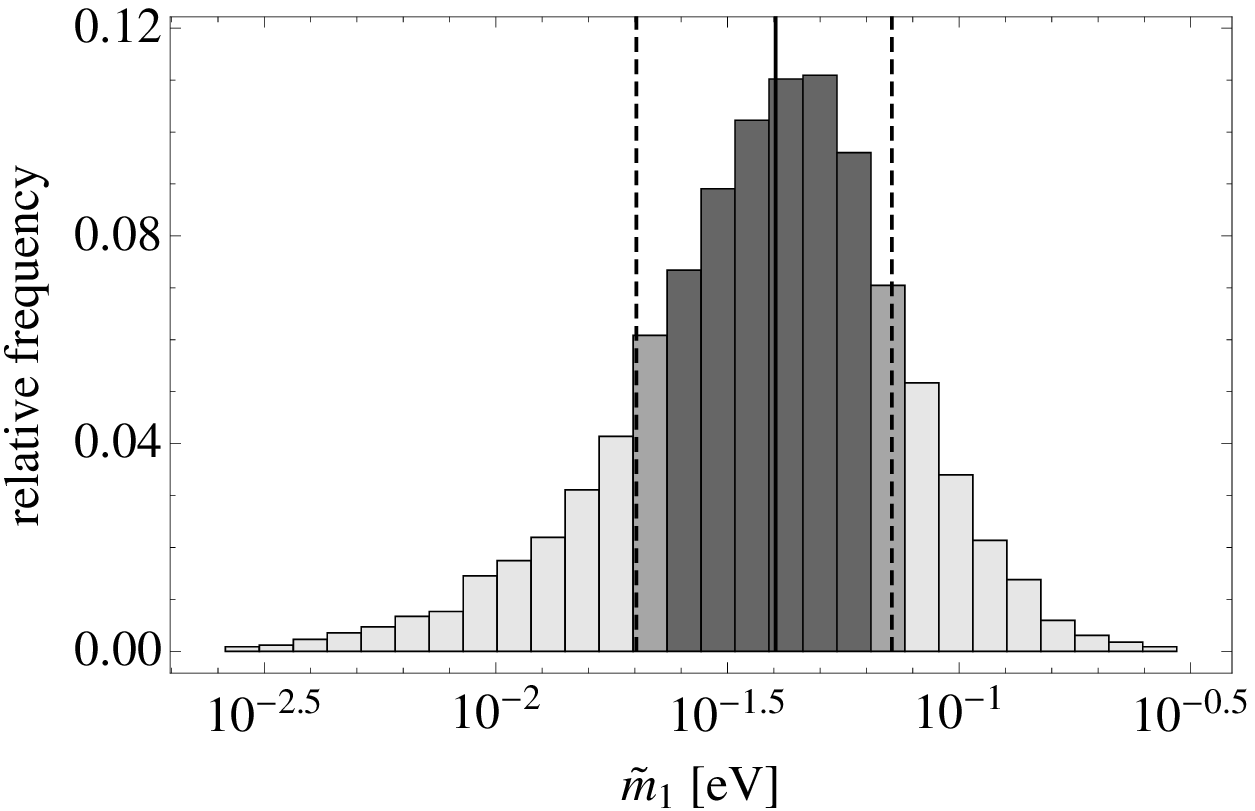}
\includegraphics[width=0.48\textwidth]{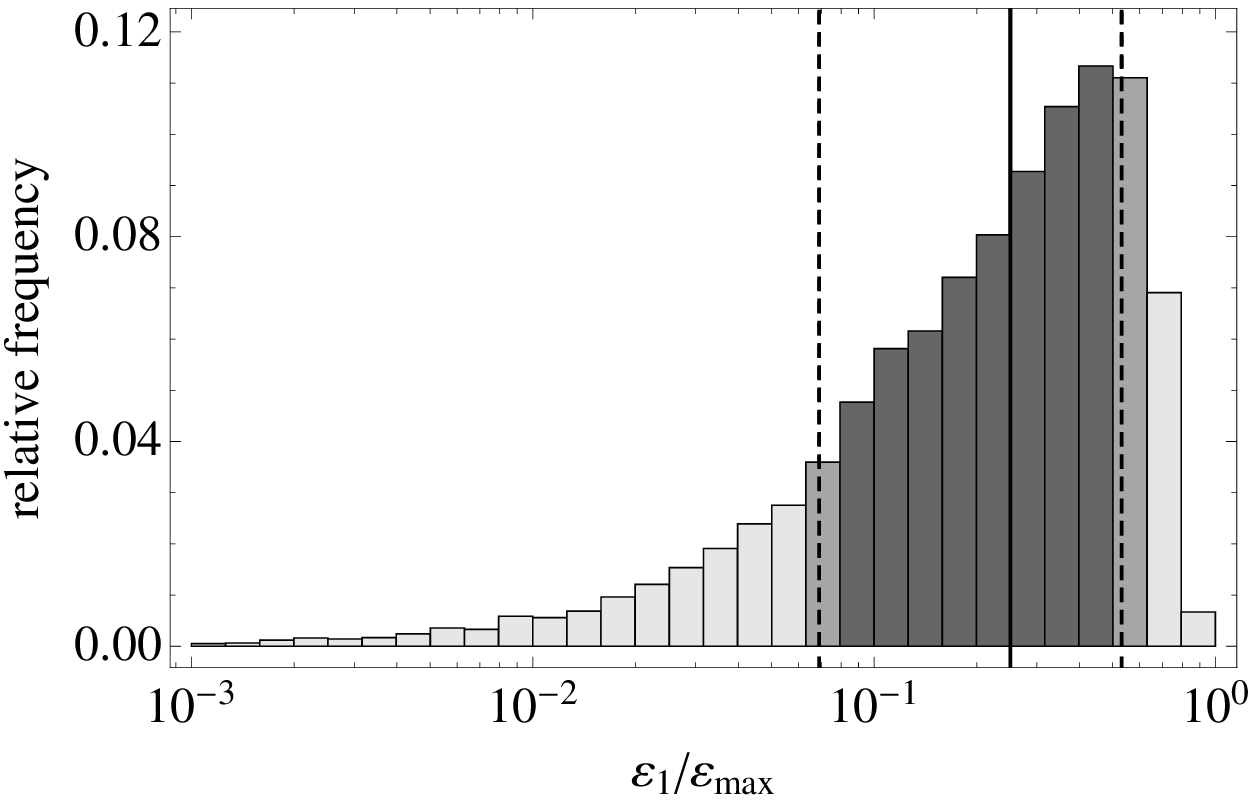}

\medskip
\caption[Effective neutrino mass $\widetilde{m}_1$ and $CP$ violation parameter $\epsilon_1$]
{Effective neutrino mass of the first generation $\widetilde{m}_1$
and $CP$ violation parameter $\epsilon_1$. Vertical lines and shadings as in Fig.~\ref{fig:mixing}.
\label{fig:CP}}
\end{figure}


Leptogenesis links the low-energy neutrino physics to the high-energy
physics of the early universe (cf.\ Sec.~\ref{subsec:baryogenesis}).
The parameters which capture this connection are the effective neutrino
mass $\widetilde{m}_1$ (cf.\ Eq.~\eqref{eq:mitilde}) and
the $CP$ violation parameter $\epsilon_1$ (cf.\ Eq.~\eqref{eq:epsiloni}).
Our best-guess estimates are (cf.\ Fig.~\ref{fig:CP})
\begin{align}
\widetilde{m}_1 = 4.0^{+3.1}_{-2.0} \times 10^{-2} \, \text{eV} \,, \quad
\frac{\epsilon_1}{\epsilon_1^{\text{max}}} = 0.25^{+0.28}_{-0.18}\,.
\label{eq:m1tepsilon1}
\end{align}


The large value for $\widetilde{m}_1$ indicates
a clear preference for the strong washout regime \cite{Buchmuller:2005eh,Davidson:2008bu}.
Note that there is typically a hierarchy between $\widetilde{m}_1$ and the lightest
neutrino mass $m_1$ of about one order of magnitude.
Generally speaking, we observe that all $\widetilde{m}_1$ values
generated in our Monte-Carlo study fall into a range extending
over roughly two orders of magnitude (cf.\ Fig.~\ref{fig:CP}).
This result renders our decision to allow for a variation of $\widetilde{m}_1$
over \textit{five} orders of magnitude quite conservative.
In other words, it assures us that varying $\widetilde{m}_1$ as indicated in
Eq.~\eqref{eq:m1trange} will certainly suffice to cover all $\widetilde{m}_1$ values
compatible with the Froggatt-Nielsen model.
In particular, the chosen range of $\widetilde{m}_1$ values easily covers our best-guess
estimate for $\widetilde{m}_1$ (cf.\ Eq.~\eqref{eq:m1tepsilon1}).


The relative frequency of possible $\epsilon_1$ values peaks close to
the upper bound $\epsilon_1^{\text{max}}$, with most of the hits lying
one order of magnitude or less below $\epsilon_1^{\text{max}}$.
This confirms our earlier expectation that the effective
$CP$-violating phase $\delta_1^{\textrm{eff}} = \epsilon_1 / \epsilon_1^{\textrm{max}}$
should be of $\mathcal{O}(1)$.
Hence, when using $\epsilon_1^{\text{max}}$ to estimate the produced lepton asymmetry
in leptogenesis, we should expect that the \textit{actually} produced asymmetry,
corresponding to the \textit{actual} value of $\epsilon_1$, is only slightly smaller than
our estimate, i.e.\ the maximum possible asymmetry.


\subsubsection{Theoretical versus Experimental Input}


The results presented in this section are obtained from the combination of
two conceptually different inputs:
the hierarchy pattern of the neutrino mass matrix $m_\nu$
(cf.\ Eq.~\eqref{eq:mnu}) on the one hand and the experimental
constraints listed in Eq.~\eqref{eq:nudata} on the other hand.
For most neutrino observables, the distributions indeed arise from
the interplay between both of these ingredients.
To give an example, the hierarchy structure alone does \textit{not} favour
a large solar mixing angle $\theta_{12}$, nor does it typically yield a ratio
$r = \Delta m^2_{\text{sol}} / \Delta m^2_{\text{atm}}$ of about $1/30$.
Given the hierarchy structure in Eq.~\eqref{eq:mnu}, $r$ rather tends to be too large instead
\cite{Altarelli:2002sg,Masina:2005am,Plentinger:2007px,Plentinger:2006nb}.\footnote{In
fact, in our case, this discrepancy is not as severe as we do not directly generate
random coefficients for the entries of $m_\nu$, but rather calculate $m_\nu$ from the
seesaw formula (cf.\ Eq.~\eqref{eq:seesawformula}) after generating random coefficients
for the entries of the matrices $m_D$ and $M$.}
Only the requirement of consistency with the experimental data eventually
singles out the subset of random mass matrices which we are able to use for our analysis.
As another example, consider the smallest mixing angle $\theta_{13}$
and the smallest neutrino mass $m_1$.
In the case of these observables, the hierarchy structure of the neutrino
mass matrix automatically implies small values, similar to those in
Figs.~\ref{fig:mixing} and \ref{fig:m1mtritium}.
However, the exact distributions, including the precise position of
the peaks, only arise after implementing the experimental constraints.
A notable exception to this scheme is the Majorana phase $\alpha_{21}$.
As we shall see in the next section, the peak in the $\alpha_{21}$ distribution
at $\alpha_{21} \simeq \pi$ is a result of the hierarchy structure of the neutrino
matrix alone.


In conclusion, we remark that we expect our results also to hold beyond
flavour models of the Froggatt-Nielsen type.
An obvious example are extradimensional models which lead to the same type
of light neutrino mass matrix~\cite{Asaka:2003iy}.
On the other hand, quark-lepton mass hierarchies and the presently available neutrino
data cannot determine the remaining observables in a model-independent way.
This is, for instance, illustrated by the fact that our present knowledge about
quark and lepton masses and mixings is still consistent with an inverted neutrino
mass hierarchy.
As a consequence, further measurements of neutrino parameters will be able to
falsify certain patterns of flavour mixing and thereby provide valuable guidance
for the theoretical origin of quark and lepton mass matrices.


\section[Analytic Derivation of the Majorana Phase \texorpdfstring{$\alpha_{21}$}{alpha\textunderscore21}]
{Analytic Derivation of the Majorana Phase \texorpdfstring{\boldmath{$\alpha_{21}$}}{alpha\textunderscore21}}
\label{sec:majorana}


The complex phases of the $\mathcal{O}(1)$ coefficients in the neutrino mass matrix
$m_\nu$ and the charged-lepton mass matrix $m_{\textrm{cl}}$ are randomly distributed.
One would thus naively expect that also the Majorana phases
$\alpha_{21}$ and $\alpha_{31}$ in the PMNS matrix can take arbitrary values.
By contrast, the distribution of values for $\alpha_{21}$ that we obtain
from our numerical Monte-Carlo study (cf. Fig.~\ref{fig:masses}) clearly features a
prominent peak at $\alpha_{21} \simeq \pi$.
In this section we shall demonstrate by means of a simplified example
how the structure of the neutrino mass matrix $m_\nu$ may partly
fix the phases of the corresponding mixing matrix $U$.


Consider the following simplified light-neutrino Majorana mass matrix $m_\nu$,
\begin{align}
m_\nu = \hat{v} \begin{pmatrix}
\eta^2 & \eta e^{i\beta}  & \eta \\ \eta e^{i\beta} & 1 & 1 \\ \eta & 1 & 1\end{pmatrix}
\,,\quad
\hat{v} = \frac{v^2_{\textrm{EW}} }{\bar{v}_{B-L}}
\,,
\label{eq:toym}
\end{align}
where $\beta\in\left[0,2\pi\right)$ is an arbitrary complex phase.
For simplicity, let us neglect any effects on the mixing matrix $U$
from the diagonalization of $m_{\textrm{cl}}$.
That is, we define $U$ such that $U^T m_\nu U = \textrm{diag}\left(m_i\right)$,
with $m_i^2$ denoting the eigenvalues of $m_\nu^\dagger m_\nu$,
\begin{align}
\frac{m_{1,2}^2}{\hat{v}^2}= & \: \eta^2 \sin^2\left(\beta/2\right)
\left[2\mp\eta \left(5+3\cos\left(\beta\right)\right)^{1/2}\right]
+ \mathcal{O}\left(\eta^4\right)\,,\\
\frac{m_3^2}{\hat{v}^2}= & \: 4\left(1 + \eta^2\left[1-\sin^2\left(\beta/2\right)\right]\right)
+ \mathcal{O}\left(\eta^4\right)\nonumber
\,.
\end{align}
The first two mass eigenvalues are nearly degenerate.
This is a consequence of the particular hierarchy pattern of
the matrix $m_\nu$, which originally stems from the equal flavour
charges of the $\textbf{5}_2^*$ and $\textbf{5}_3^*$ multiplets.
The relative sign of the $\mathcal{O}\left(\eta^3\right)$ contributions
to $m_1^2$ and $m_2^2$ eventually shows up again in entries of $U$, for instance,
\begin{align}
U_{11,12} = \mp\frac{2\left(5+3\cos\left(\beta\right)\right)^{1/2}}{3 + e^{i\beta}}
\,\exp\left(-\frac{i}{2}\textrm{Arg}\left[\mp z\right]\right)
+ \mathcal{O}\left(\eta\right)
\,.
\end{align}
with $z = 1-\cos\left(\beta\right)-2i\sin\left(\beta\right).$
The phase $\alpha_{21} = 2 \left(\textrm{Arg}\left[U_{12}/U_{11}\right]
\:\textrm{mod}\:\pi\right)$ in the matrix $U$ represents the analog
of the Majorana phase $\alpha_{21}$ in the PMNS matrix, cf. Eq.~\eqref{eq:UPMNS}.
According to our explicit results for $U_{11}$ and $U_{12}$, it is independent
of the arbitrary phase $\beta$ to leading order in $\eta$,
\begin{align}
\alpha_{21} \simeq 2 \left(\textrm{Arg}\left[- \exp\left(-\frac{i}{2}
\textrm{Arg}\left[+z\right]
+\frac{i}{2} \textrm{Arg}
\left[-z\right]
\right)\right] \:\textrm{mod}\:\pi\right) = \pi
\,.
\end{align}
In a similar way, we may determine the phase analogous
to the Majorana phase $\alpha_{31}$.
However, due to the hierarchy between the mass eigenvalues
$m_1$ and $m_3$, the first and third column of the matrix $U$
differ significantly from each other, thus  leading to a phase
that depends on $\beta$ at all orders of $\eta$.


Including corrections to all orders in $\eta$ and scanning over
the phase $\beta$ numerically shows that the maximum possible deviation
of $\alpha_{21}$ from $\pi$ is, in fact, of $\mathcal{O}\left(\eta^4\right)$.
Adding more complex phases to the matrix $m_\nu$ in Eq.~\eqref{eq:toym}
gradually smears out the peak in the distribution of $\alpha_{21}$ values.
The distribution which is reached in the case of six different phases
is already very similar to the one in Fig.~\ref{fig:masses}.
We conclude that, despite the need for corrections,
the rough picture sketched in this section remains valid:
the hierarchy pattern of the neutrino mass matrix directly
implies that $\alpha_{21}$ tends to be close to $\alpha_{21} = \pi$.



\chapter[Supersymmetric Abelian Higgs Model]
{Supersymmetric\newline Abelian Higgs Model}
\label{ch:model}


In order to incorporate hybrid inflation as well as the \BmL phase transition
into our cosmological scenario, we introduced the superpotential $W_{B-L}$
(cf.\ Eq.~\eqref{eq:WBL}) in Ch.~\ref{ch:framework}.
The chiral superfields contained in this superpotential, the inflaton field $\Phi$
and the two Higgs fields $S_{1,2}$, partly carry \BmL charge,
but all transform as standard model gauge singlets.
Their dynamics are hence fully accounted for
by a supersymmetric Abelian gauge theory with gauge group $U(1)_{B-L}$.
The field-theoretic description of \BmL breaking at the end of inflation
represents in particular a variant of the \textit{supersymmetric Abelian Higgs model}.


The goal of this chapter now is to derive the full supersymmetric
Lagrangian for the Abelian Higgs model describing the \BmL phase transition
in unitary gauge.
We will first consider \BmL to be unbroken
and compute the Lagrangian of a general supersymmetric Abelian gauge theory
in arbitrary gauge (cf.\ Sec.~\ref{sec:beforeSSB}).
Then, after going to unitary gauge, we will evaluate this general Lagrangian
for the specific field content of our model in the broken phase
(cf.\ Sec.~\ref{sec:afterSSB}).
This will allow us to calculate the masses, decay rates and branching ratios
of all particles coupling to the \BmL Higgs boson.
One important result of our analysis at this point will be that
during the \BmL phase transition the mass eigenvalues under study
are time-dependent, which gives rise to nonperturbative particle production.


\newpage


\section{Before Spontaneous Symmetry Breaking}
\label{sec:beforeSSB}


As a first step towards the supersymmetric Abelian Higgs model of the
\BmL phase transition, we derive the Lagrangian of a general $U(1)$
gauge theory featuring $N$ chiral superfields $\Phi_i$ and one massless
vector superfield $V$.
By virtue of its particular transformation law, the vector superfield $V$
ensures the invariance of the Lagrangian under super-gauge transformations.
In addition to the fields $\Phi_i$ and $V$, the theory also contains the gravity
multiplet, consisting of the graviton $G$ and the gravitino $\widetilde{G}$, which
gives rise to further, Planck-suppressed operators in the Lagrangian.
As the \BmL phase transition takes place around the GUT scale,
$v_{B-L} \sim \Lambda_{\textrm{GUT}} \ll M_P$, these SUGRA corrections are however
irrelevant, so that we may neglect them in the following (cf.\ Sec.~\ref{subsec:infltn}).
More precisely, for the purposes of this thesis, it will suffice to compute
the Lagrangian of the Abelian Higgs model for the case of global supersymmetry.
By contrast, we are not allowed to facilitate our calculation by choosing
a specific gauge.\footnote{The Lagrangian of the Abelian Higgs model in \textit{Wess-Zumino gauge},
for instance, is well known and listed in all standard textbooks on supersymmetry and supergravity
\cite{Wess:1992cp,Bailin:1994qt,Weinberg:2000cr}.}
During the \BmL phase transition, the vector superfield turns massive,
which is best described in unitary gauge, where the physical DOFs are manifest.
To be able to evaluate the Lagrangian in unitary gauge later on,
we now first have to calculate it in arbitrary gauge.
In doing so, we will closely follow the notation of Ref.~\cite{Wess:1992cp}.


\subsection{From Superspace to the Component Lagrangian}
\label{subsec:lagrangian}


The total Lagrangian $\mathcal{L}$ of our Abelian gauge theory splits into three pieces,
\begin{align}
\mathcal{L} = \mathcal{L}_G + \mathcal{L}_K + \mathcal{L}_W\,.
\end{align}
Here, $\mathcal{L}_G$ encompasses the gauge-kinetic terms, $\mathcal{L}_K$ takes care
of the kinetic terms and gauge interactions of the chiral superfields $\Phi_i$ and
$\mathcal{L}_W$ contains all interactions stemming from the superpotential $W$.
Let us now calculate each of these contributions to $\mathcal{L}$ in terms of
the components of the superfields $\Phi_i$ and $V$.


\subsubsection{Expansion of the Superfields in Superspace Coordinates}


The supersymmetric Abelian Higgs model may be formulated as a theory on \textit{superspace}.
In four spacetime dimensions and assuming the number of generators of supersymmetry
transformations to be minimal, i.e.\ $\mathcal{N}=1$ supersymmetry, superspace is
spanned by four bosonic, commuting coordinates $x^\mu$, where $\mu=0,1,2,3$, as well as
four fermionic, anticommuting coordinates $\theta_\alpha$ and $\bar{\theta}^{\dot{\alpha}}$,
where $\alpha,\dot{\alpha} = 1,2$.
The superfields $\Phi_i$ and $V$ are nothing but functions of these superspace
coordinates and should be understood in terms of their power series expansion
in $\theta$ and $\bar{\theta}$~\cite{Wess:1992cp}.
The chiral superfields $\Phi_i$ are expanded as follows,
\begin{align}
\Phi_i = \phi_i + i \theta \sigma^\mu \bar{\theta} \partial_\mu \phi_i +
\frac{1}{4}\theta\theta\bar{\theta}\bar{\theta}\Box\phi_i + \sqrt{2}\theta \psi_i
-\frac{i}{\sqrt{2}}\theta\theta\partial_\mu \psi_i \sigma^\mu\bar{\theta} + \theta\theta F_i\,,
\label{eq:Phii}
\end{align}
where we have introduced $\phi_i$, $\psi_i$ and $F_i$ as the components of $\Phi_i$,
all of which are fields on spacetime.
$\phi_i$ and $F_i$ are complex scalars, having mass dimension $1$ and $2$, respectively,
while $\psi_i$ is an ordinary left-chiral Weyl fermion.
As we will shortly see, $F_i$ is not dynamical, i.e.\ an auxiliary field,
and can therefore be integrated out.
Under hermitian conjugation, the superfield in Eq.~\eqref{eq:Phii} turns into
\begin{align}
\Phi_i^\dagger = \phi_i^* - i \theta \sigma^\mu \bar{\theta} \partial_\mu \phi_i^* +
\frac{1}{4}\theta\theta\bar{\theta}\bar{\theta}\Box\phi_i^* + \sqrt{2}\bar{\theta}\bar{\psi}_i
+\frac{i}{\sqrt{2}}\bar{\theta}\bar{\theta}\theta \sigma^\mu \partial_\mu \bar{\psi}_i + \bar{\theta}\bar{\theta} F_i^*\,.
\label{eq:Phiidagger}
\end{align}


The vector superfield $V$ satisfies the reality condition $V^\dagger = V$
and is given by
\begin{align}
& \textrm{Arbitrary gauge:}\qquad
V = C + i\theta\chi -i\bar{\theta}\bar{\chi} + \frac{i}{2}\theta\theta\left(M + i N\right)
-\frac{i}{2}\bar{\theta}\bar{\theta}\left(M - iN\right)
\label{eq:Varbi}\\
& - \theta \sigma^\mu\bar{\theta} A_\mu
+ i\theta\theta\bar{\theta}\left(\bar{\xi} + \frac{i}{2}\bar{\sigma}^\mu\partial_\mu \chi\right)
-i\bar{\theta}\bar{\theta}\theta\left(\xi + \frac{i}{2}\sigma^\mu\partial_\mu \bar{\chi}\right)
+ \frac{1}{2}\theta\theta\bar{\theta}\bar{\theta}\left(D + \frac{1}{2}\Box C\right)\,.\nonumber
\end{align}
$C$, $\chi$, $M$, $N$, $A^\mu$, $\xi$, and $D$ are the spacetime-dependent component
fields of~$V$, where $C$, $M$, $N$, and $D$ are real scalars of mass dimension $0$, $1$, $1$
and $2$, respectively, while $\chi$ and $\xi$ are left-chiral Weyl fermions of mass
dimension $1/2$ and $3/2$, respectively.
$A^\mu$ is an ordinary real four-vector.
$C$, $\chi$, $M$, $N$, and $D$ are auxiliary fields, the latter three of which
we will eventually integrate out, whereas $C$ and $\chi$ will become dynamical
during the \BmL phase transition.
By performing an appropriate super-gauge transformation on $V$, one can eliminate
all auxiliary fields in $V$ except for $D$, such that $V$ reduces to
\begin{align}
\textrm{Wess-Zumino gauge:}\qquad
V = - \theta \sigma^\mu\bar{\theta} A_\mu + i\theta\theta\bar{\theta}\bar{\xi}
-i\bar{\theta}\bar{\theta}\theta\xi + \frac{1}{2}\theta\theta\bar{\theta}\bar{\theta}D\,.
\label{eq:VWZgauge}
\end{align}
The gauge in which $V$ takes this form is referred to as the Wess-Zumino gauge.
Compared to other gauge choices, it allows for the technically simplest treatment
of the vector superfield $V$.
After gauge-fixing to Wess-Zumino gauge, the Lagrangian is no longer manifestly
invariant under supersymmetry transformations, but still gauge-invariant
under ordinary $U(1)$ transformations.
A slight relaxation of the Wess-Zumino gauge consists in merely gauging away
$\chi$, $M$ and $N$, which corresponds to fixing the gauge only up to complexified
$U(1)$ transformations, i.e.\ transformations featuring a complex
gauge transformation parameter.
In this \textit{complex gauge}, $C$ remains as an auxiliary field in $V$,
\begin{align}
\textrm{Complex gauge:}\qquad
V = C - \theta \sigma^\mu\bar{\theta} A_\mu + i\theta\theta\bar{\theta}\bar{\xi}
-i\bar{\theta}\bar{\theta}\theta\xi + \frac{1}{2}\theta\theta\bar{\theta}\bar{\theta}D\,.
\end{align}


It is easy to construct the Lagrangian of the Abelian Higgs model from
the chiral superfields $\Phi_i$ and $\Phi_i^\dagger$ as well as the vector
superfield $V$ in Wess-Zumino gauge (cf.\ Eq.~\eqref{eq:VWZgauge})
\cite{Wess:1992cp,Bailin:1994qt,Weinberg:2000cr}.
However, since we wish to eventually describe the \BmL phase transition
in unitary gauge, we now face the task to generalize the result
commonly quoted in the literature to arbitrary gauge.
Instead of $V$ as given in Eq.~\eqref{eq:VWZgauge}, we now have to calculate
the Lagrangian of the Abelian Higgs model using $V$ as given in Eq.~\eqref{eq:Varbi},
thereby taking into account \textit{all} auxiliary fields.


\subsubsection{Gauge-Kinetic Terms and Terms Deriving from the Superpotential}


$\mathcal{L}_G$ and $\mathcal{L}_W$ are solely constructed from gauge-invariant quantities.
While the Lagrangian $\mathcal{L}_G$ is obtained from the gauge-invariant chiral superfield
$\mathcal{W}_\alpha$, the supersymmetric generalization of the
Abelian field strength tensor $F_{\mu\nu} = \partial_\mu A_\nu - \partial_\nu A_\mu$,
the Lagrangian $\mathcal{L}_W$ follows straightforwardly from the gauge-invariant
superpotential $W$.
These two contributions to the total Lagrangian $\mathcal{L}$ hence look the same
in all gauges, so that we may simply adopt the standard expressions
for $\mathcal{L}_G$ and $\mathcal{L}_W$, which one readily finds in Wess-Zumino gauge,
\begin{align}
\mathcal{L}_G = & \: -\frac{1}{4} F_{\mu\nu} F^{\mu\nu}
- i\bar{\xi}\bar{\sigma}^\mu \partial_\mu \xi
+ \frac{1}{2}D^2 \,, \label{eq:LG}\\
\mathcal{L}_W = & -\frac{1}{2} \sum_{i,j} W_{ij} \psi_i\psi_j + \sum_i W_i F_i + \textrm{h.c.} \,.
\label{eq:LW}
\end{align}
Of course, the superpotential is actually a holomorphic function of the chiral
superfields, $W = W\left(\left\{\Phi_k\right\}\right)$.
In Eq.~\eqref{eq:LW}, to the benefit of a convenient notation,
it is, however, interpreted as a function of the corresponding complex scalars,
$W = W\left(\left\{\phi_k\right\}\right)$.
The functions $W_i$ and $W_{ij}$ then stand for
\begin{align}
W_i = \frac{\partial}{\partial \phi_i} W\left(\left\{\phi_k\right\}\right)\,,\quad
W_{ij} = \frac{\partial^2}{\partial \phi_i \partial \phi_j} W\left(\left\{\phi_k\right\}\right)\,.
\end{align}


\subsubsection{Kinetic Terms and Gauge Interactions of the Chiral Superfields}


$\mathcal{L}_K$ receives a contribution $\mathcal{L}_K^i$ for each chiral superfield $\Phi_i$.
Assuming canonical kinetic terms for the $\Phi_i$ component fields, $\mathcal{L}_K$
is uniquely given as
\begin{align}
\mathcal{L}_K = \sum_i \mathcal{L}_K^i \,,\quad
\mathcal{L}_K^i = \left[\Phi_i^\dagger e^{p_iV} \Phi_i\right]_D
\,,\quad p_i = 2g q_i\,.
\label{eq:LagKi}
\end{align}
Here, the subscript $D$ indicates that only
the $\theta\theta\bar{\theta}\bar{\theta}$ component, i.e.\ the $D$-term, of the field
product $\Phi_i^\dagger e^{p_iV} \Phi_i$ is to be included in $\mathcal{L}_K^i$.
Meanwhile, $g$ is the $U(1)$ gauge coupling and $q_i$ denotes the $U(1)$ gauge charge
of the chiral superfield $\Phi_i$.
In Chs.~\ref{ch:phasetransition}, \ref{ch:reheating} and \ref{ch:wimp}, we will
take $g$ to be the GUT gauge coupling,
$g = g_{\textrm{GUT}} \simeq \sqrt{\pi/6}$.
Expanding the exponential in Eq.~\eqref{eq:LagKi} in
powers of $\left(V-C\right)$, the Lagrangian $\mathcal{L}_K^i$ turns into
\begin{align}
\mathcal{L}_K^i = & \: e^{p_i C} \bigg[\Phi_i^\dagger
\bigg\{1 + p_i\left(V-C\right) + \frac{p_i^2}{2}\left(V-C\right)^2 \label{eq:LKi}\\
+ & \: \frac{p_i^3}{6}\left(V-C\right)^3
+ \frac{p_i^4}{24}\left(V-C\right)^4\bigg\}\Phi_i\bigg]_D \,.
\nonumber
\end{align}
All higher powers of $\left(V-C\right)$ vanish, as they only involve products of at least
five Grassmannian superspace coordinates, which are all zero.


In a first step towards $\mathcal{L}_K^i$, we calculate $\left(V-C\right)^n$
for $n = 2,3,4$.
Given the component expansion of $V$ in Eq.~\eqref{eq:Varbi} and
making heavy use of the following spinor identities~\cite{Wess:1992cp},
\begin{align}
\left(\chi\theta\right)\left(\psi\theta\right) = & \: -\frac{1}{2}\left(\chi\psi\right)\left(\theta\theta\right)\,,&
\left(\bar{\chi}\bar{\theta}\right)\left(\bar{\psi}\bar{\theta}\right) = & \:
-\frac{1}{2}\left(\bar{\chi}\bar{\psi}\right)\left(\bar{\theta}\bar{\theta}\right)\,,\\
\left(\chi\theta\right)\bar{\theta}\bar{\sigma}^\mu\theta = & \: -\frac{1}{2}
\bar{\theta}\bar{\sigma}^\mu\chi\left(\theta\theta\right)\,,&
\left(\bar{\chi}\bar{\theta}\right)\bar{\theta}\bar{\sigma}^\mu\theta = & \:
-\frac{1}{2}\bar{\chi}\bar{\sigma}^\mu\theta\left(\bar{\theta}\bar{\theta}\right)\,,\nonumber\\
\chi\sigma^\mu\bar{\psi} = & \: - \bar{\psi}\bar{\sigma}^\mu \chi\,,\phantom{\frac{1}{2}}&
\left(\chi\sigma^\mu\bar{\psi}\right)^\dagger = & \: \psi \sigma^\mu \bar{\chi}\,,\nonumber\\
\bar{\theta}\bar{\sigma}^\mu\theta \bar{\theta}\bar{\sigma}^\nu\theta = & \: -\frac{1}{2}
\eta^{\mu\nu} \left(\theta\theta\right)\left(\bar{\theta}\bar{\theta}\right)\,,&
\theta^\alpha\theta^\beta\theta^\gamma = & \:
\bar{\theta}_{\dot{\alpha}}\bar{\theta}_{\dot{\beta}}\bar{\theta}_{\dot{\gamma}} = 0\,,\nonumber
\end{align}
we find
\begin{align}
\left(V-C\right)^2 = & \: \frac{1}{2}\chi^2 \theta\theta + \frac{1}{2}\bar{\chi}^2\bar{\theta}\bar{\theta}
+ 2\chi\theta\bar{\chi}\bar{\theta} + \chi\theta\bar{\theta}\bar{\theta}\left(M-iN\right)
+ \bar{\chi}\bar{\theta}\theta\theta\left(M+iN\right) \\
+ & \: 2i\left(\chi\theta-\bar{\chi}\bar{\theta}\right) \bar{\theta}\bar{\sigma}^\mu\theta A_\mu
+ \bigg[-\frac{1}{2}A_\mu A^\mu + \frac{1}{2}\left(M^2 + N^2\right) \nonumber\\
-& \: \chi\xi -\bar{\chi}\bar{\xi} - \frac{i}{2}\bar{\chi}\bar{\sigma}^\mu\partial_\mu \chi
+ \frac{i}{2}\partial_\mu\bar{\chi}\bar{\sigma}^\mu\chi \bigg]\theta\theta\bar{\theta}\bar{\theta}\nonumber \\
\left(V-C\right)^3 = & \: \frac{3i}{2}\bar{\chi}^2\chi\theta\bar{\theta}\bar{\theta}
- \frac{3i}{2}\chi^2\bar{\chi}\bar{\theta}\theta\theta + \bigg[\frac{3i}{4}\bar{\chi}^2\left(M+iN\right)
-\frac{3i}{4}\chi^2\left(M-iN\right) \\
+ & \: \frac{3}{2}\bar{\chi}\bar{\sigma}^\mu\chi A_\mu\bigg]
\theta\theta\bar{\theta}\bar{\theta} \nonumber\\
\left(V-C\right)^4 = & \: \frac{3}{2}\chi^2\bar{\chi}^2\theta\theta\bar{\theta}\bar{\theta}\,.
\end{align}
These expressions allow us to work out the field products
$\left[\Phi_i^\dagger\left(V-C\right)^n\Phi_i\right]_D$, where $n = 0,\textrm{...},4$,
constituting the Lagrangian $\mathcal{L}_K^i$ in Eq.~\eqref{eq:LKi},
\begin{align}
\left[\Phi_i^\dagger\Phi_i\right]_D = & \: -\frac{1}{2}\partial_\mu \phi_i^* \partial^\mu \phi_i
+\frac{1}{4}\phi_i^*\Box \phi_i +\frac{1}{4}\phi_i\Box \phi_i^*
+ \frac{i}{2}\partial_\mu\bar{\psi}_i\bar{\sigma}^\mu \psi_i \label{eq:PV0P}\\
- &\: \frac{i}{2}\bar{\psi}_i\bar{\sigma}^\mu \partial_\mu \psi_i + F_i^*F_i \,,
\phantom{\bigg\{}\nonumber\\
\left[\Phi_i^\dagger\left(V-C\right)\Phi_i\right]_D = & \: \frac{1}{2}\phi_i^*\phi_i\left(D+\frac{1}{2}\Box C\right)
+ \frac{1}{2}\bar{\psi}_i\bar{\sigma}^\mu\psi_i A_\mu + \bigg\{\frac{i}{2}\phi_i^*\partial^\mu\phi_i A_\mu   \label{eq:PV1P}\\
+ & \: \frac{i}{\sqrt{2}}\phi_i^*\psi_i\left(\xi + \frac{i}{2}\sigma^\mu\partial_\mu\bar{\chi}\right)
- \frac{1}{2\sqrt{2}} \phi_i^* \bar{\chi}\bar{\sigma}^\mu\partial_\mu\psi_i
\phantom{\bigg\{}\nonumber\\
- & \: \frac{i}{2}\phi_i^* F_i\left(M-iN\right)
+ \frac{1}{2\sqrt{2}}\bar{\chi}\bar{\sigma}^\mu\psi_i\partial_\mu\phi_i^*
+ \frac{i}{\sqrt{2}} \bar{\psi}_i\bar{\chi} F_i + \textrm{h.c.}\bigg\}\,, \nonumber \\
\left[\Phi_i^\dagger\left(V-C\right)^2\Phi_i\right]_D = & \: \phi_i^*\phi_i
\bigg(-\frac{1}{2}A_\mu A^\mu
+\frac{1}{2}\left(M^2 + N^2\right) - \chi\xi - \bar{\chi}\bar{\xi} \label{eq:PV2P}\\
- & \: \frac{i}{2}\bar{\chi}\bar{\sigma}^\mu\partial_\mu \chi
+ \frac{i}{2}\partial_\mu\bar{\chi}\bar{\sigma}^\mu\chi\bigg)
+ \chi\psi_i \bar{\chi}\bar{\psi}_i
+ \bigg\{ -\frac{i}{2}\phi_i^*\bar{\chi}\bar{\sigma}^\mu\chi \partial_\mu \phi_i \nonumber\\
- & \: \frac{1}{\sqrt{2}}\phi_i^*\psi_i\chi\left(M-iN\right) - \frac{i}{\sqrt{2}}\phi_i^*\bar{\chi}\bar{\sigma}^\mu \psi_i A_\mu
+ \frac{1}{2}\phi_i^* \bar{\chi}^2 F_i +\textrm{h.c.}\bigg\}\,, \nonumber \\
\left[\Phi_i^\dagger\left(V-C\right)^3\Phi_i\right]_D = & \: \frac{3}{2}\phi_i^*\phi_i \bar{\chi}\bar{\sigma}^\mu \chi A_\mu
+ \bigg\{\frac{3i}{4}\phi_i^*\phi_i\bar{\chi}^2\left(M+iN\right) \label{eq:PV3P}\\
+ &\: \frac{3i}{2\sqrt{2}}\bar{\chi}\bar{\psi}_i\chi^2\phi_i
+ \textrm{h.c}\bigg\}\,,\nonumber\\
\left[\Phi_i^\dagger\left(V-C\right)^4\Phi_i\right]_D = & \: \frac{3}{2} \phi_i^* \phi_i \chi^2\bar{\chi}^2 \,. \label{eq:PV4P}
\end{align}
By shifting around spacetime derivatives with the aid of integrations by parts,
we are able to combine several terms in the above field products.
We shall use that
\begin{align}
e^{p_i C} \left(\frac{1}{4}\phi_i^*\Box \phi_i +\frac{1}{4}\phi_i\Box \phi_i^*\right)
= & \: e^{p_i C} \bigg(-\frac{1}{2}\partial_\mu\phi_i^*\partial^\mu\phi_i
-\frac{p_i}{4}\phi_i^*\partial_\mu C \partial^\mu\phi_i \\
- & \: \frac{p_i}{4}\phi_i\partial_\mu C \partial^\mu\phi_i^*\bigg)
+ d \,,\nonumber\\
\frac{i}{2} e^{p_i C} \partial_\mu\bar{\psi}_i\bar{\sigma}^\mu \psi_i
= & \:  -\frac{i}{2} e^{p_i C} \left(\bar{\psi}_i\bar{\sigma}^\mu \partial_\mu\psi_i
+ p_i \bar{\psi}_i\bar{\sigma}^\mu\psi_i\partial_\mu C\right) + d \,,
\phantom{\bigg\{}\\
\frac{i p_i^2}{4}e^{p_i C} \phi_i^* \phi_i \partial_\mu\bar{\chi}\bar{\sigma}^\mu \chi
= & \: \frac{i p_i^2}{4}e^{p_i C} (-\phi_i^* \phi_i \bar{\chi}\bar{\sigma}^\mu \partial_\mu \chi
- \phi_i^* \partial_\mu \phi_i \bar{\chi}\bar{\sigma}^\mu \chi \phantom{\bigg\{}\\
- & \: \phi_i \partial_\mu \phi_i^* \bar{\chi}\bar{\sigma}^\mu \chi
- p_i \phi_i^* \phi_i \partial_\mu C \bar{\chi}\bar{\sigma}^\mu \chi)
+ d \,, \phantom{\bigg\{}\nonumber\\
- \frac{p_i}{2\sqrt{2}}e^{p_i C} \phi_i^* \psi_i \sigma^\mu \partial_\mu \bar{\chi}
= & \: - \frac{p_i}{2\sqrt{2}}e^{p_i C} (\phi_i^* \bar{\chi} \bar{\sigma}^\mu \partial_\mu\psi_i
+ \partial_\mu \phi_i^* \bar{\chi} \bar{\sigma}^\mu\psi_i \phantom{\bigg\{}\\
+ & \: p_i \phi_i^* \partial_\mu C \bar{\chi} \bar{\sigma}^\mu\psi_i)
+ d \,.\nonumber
\end{align}
Here, $d$ denotes total derivatives, which we do not need
to include into the Lagrangian.
Inserting these relations into Eqs.~\eqref{eq:PV0P}, \eqref{eq:PV1P} and \eqref{eq:PV2P},
we obtain
\begin{align}
\left[\Phi_i^\dagger\Phi_i\right]_D = & \: -\partial_\mu \phi_i^* \partial^\mu \phi_i
- i\bar{\psi}_i\bar{\sigma}^\mu \partial_\mu \psi_i + F_i^*F_i \label{eq:PV0P2}\\
- & \: \frac{p_i}{4}\phi_i^*\partial_\mu C \partial^\mu\phi_i
- \frac{p_i}{4}\phi_i\partial_\mu C \partial^\mu\phi_i^*
- \frac{i p_i}{2}\bar{\psi}_i\bar{\sigma}^\mu\psi_i\partial_\mu C\,,\phantom{\bigg\{}\nonumber\\
\left[\Phi_i^\dagger\left(V-C\right)\Phi_i\right]_D = & \: \frac{1}{2}\phi_i^*\phi_i\left(D+\frac{1}{2}\Box C\right)
+ \frac{1}{2}\bar{\psi}_i\bar{\sigma}^\mu\psi_i A_\mu + \bigg\{\frac{i}{2}\phi_i^*\partial^\mu\phi_i A_\mu \label{eq:PV1P2}\\
+ & \: \frac{i}{\sqrt{2}}\phi_i^*\psi_i\xi
- \frac{1}{\sqrt{2}} \phi_i^* \bar{\chi}\bar{\sigma}^\mu\partial_\mu\psi_i
- \frac{p_i}{2\sqrt{2}} \phi_i^* \bar{\chi} \bar{\sigma}^\mu\psi_i \partial_\mu C
\phantom{\bigg\{}\nonumber \\
- & \: \frac{i}{2}\phi_i^* F_i\left(M-iN\right)
+ \frac{i}{\sqrt{2}} \bar{\psi}_i\bar{\chi} F_i + \textrm{h.c.}\bigg\}\,, \nonumber \\
\left[\Phi_i^\dagger\left(V-C\right)^2\Phi_i\right]_D = & \:
\phi_i^*\phi_i \bigg(-\frac{1}{2}A_\mu A^\mu
+\frac{1}{2}\left(M^2 + N^2\right) - \chi\xi - \bar{\chi}\bar{\xi} \label{eq:PV2P2}\\
- & \: \frac{i p_i}{2}\bar{\chi}\bar{\sigma}^\mu\chi\partial_\mu C\bigg)
- i\phi_i^*\bar{\chi}\bar{\sigma}^\mu \partial_\mu \left(\phi_i \chi\right)
+ \chi\psi_i \bar{\chi}\bar{\psi}_i \nonumber\\
+ & \: \bigg\{-\frac{1}{\sqrt{2}}\phi_i^*\psi_i\chi\left(M-iN\right) - \frac{i}{\sqrt{2}}\phi_i^*\bar{\chi}\bar{\sigma}^\mu \psi_i A_\mu \nonumber\\
+ & \: \frac{1}{2}\phi_i^* \bar{\chi}^2 F_i +\textrm{h.c.}\bigg\}\,. \nonumber
\end{align}
Note the newly emerged couplings to $\partial_\mu C$, which have been generated
by the various integrations by parts.
Our results in Eqs.~\eqref{eq:PV3P} and \eqref{eq:PV4P} remain unchanged,
so that we are able to construct a preliminary expression for $\mathcal{L}_K^i$
(cf.\ Eq.~\eqref{eq:LKi}) by assembling the field products in
Eqs.~\eqref{eq:PV3P}, \eqref{eq:PV4P}, \eqref{eq:PV0P2}, \eqref{eq:PV1P2}, and \eqref{eq:PV2P2}.


\subsubsection{Eliminating the Auxiliary Fields}


Our calculation thus far has provided us with kinetic terms for the gauge fields
$A_\mu$ and $\xi$ (cf.\ Eq.~\eqref{eq:LG}) as well as for the scalar and fermionic components
$\phi_i$ and $\psi_i$ of the chiral superfields $\Phi_i$ (cf.\ Eq.~\eqref{eq:PV0P2}).
The fields $F_i$, $C$, $\chi$, $M$, $N$, and $D$ are, by contrast, not dynamical and may
hence be integrated out.
An important observation, however, is that the Lagrangian $\mathcal{L}_K^i$ contains
two terms (cf.\ Eq.~\eqref{eq:PV1P2}, and \eqref{eq:PV2P2}) which turn into
kinetic terms for $C$ and $\chi$, once the scalar field $\phi_i$ acquires a nonzero VEV.
As exactly this happens during the \BmL phase transition, we shall keep
the auxiliary fields $C$ and $\chi$ in the Lagrangian, anticipating them to become dynamical
in the course of \BmL breaking.
Meanwhile, no term in the total Lagrangian features a derivative of $F_i$, $M$, $N$,
or $D$, i.e.\ no term could possibly give rise to a kinetic term for any of these fields,
and hence all of them always remain auxiliary.
Let us now integrate them out of the Lagrangian.


The auxiliary gauge field $D$ only appears in two terms in the total
Lagrangian (cf.\ Eqs.~\eqref{eq:LG} and \eqref{eq:PV1P2}),
which we collect in the Lagrangian $\mathcal{L}_D$,
\begin{align}
\mathcal{L}_D = \frac{1}{2} D^2 + \frac{1}{2} \sum_i p_i e^{p_i C} \phi_i^* \phi_i D\,.
\label{eq:LagD}
\end{align}
The equation of motion of the field $D$ is hence given as
\begin{align}
\frac{\partial \mathcal{L}}{\partial D} =
D + \frac{1}{2} \sum_i p_i e^{p_i C} \phi_i^* \phi_i = 0\,,\quad
D = - \frac{1}{2} \sum_i p_i e^{p_i C} \phi_i^* \phi_i \,.
\end{align}
Substituting the solution for $D$ back into Eq.~\eqref{eq:LagD} yields
\begin{align}
\mathcal{L}_D  = \frac{1}{2} D^2 - D^2 = - \frac{1}{2} D^2 = - V_D \,,\quad
V_D  = \frac{1}{8} \sum_{ij} p_i p_j e^{\left(p_i+p_j\right)C} \phi_i^*\phi_i \phi_j^* \phi_j\,,
\label{eq:LDremain}
\end{align}
where we will refer to $V_D$ as the $D$-term scalar potential in the following.


Next, we compile all terms featuring $F_i$ and $F_i^*$ in one common Lagrangian $\mathcal{L}_F^i$,
\begin{align}
\mathcal{L}_F^i = & \: \bigg\{W_i F_i + e^{p_i C} \bigg[F_i^* F_i + p_i\left(-\frac{i}{2}\phi_i^*\left(M-iN\right)F_i
+ \frac{i}{\sqrt{2}}\bar{\psi}_i\bar{\chi}F_i\right) \label{eq:LFi}\\
+ & \: \frac{p_i^2}{2}\frac{1}{2}\phi_i^*\bar{\chi}^2 F_i\bigg]
+ \textrm{h.c.}\bigg\} - e^{p_i C} F_i^* F_i \nonumber\,.
\end{align}
The last term in Eq.~\eqref{eq:LFi} prevents us from double-counting
$e^{p_i C}F_i^*F_i$, which appears twice within the curly brackets.
$\mathcal{L}_F^i$ can be  simplified by observing that
\begin{align}
\mathcal{L}_F^i = \frac{\partial \mathcal{L}_F^i}{\partial F_i} F_i +
\frac{\partial \mathcal{L}_F^i}{\partial F_i^*} F_i^* - e^{p_i C} F_i^* F_i\,.
\label{eq:LFiobs}
\end{align}
The first two terms on the right-hand side of this relation
vanish due the equations of motions of $F_i$ and $F_i^*$, which provides us
with a compact expression for $\mathcal{L}_F^i$,
\begin{align}
\frac{\partial \mathcal{L}_F^i}{\partial F_i} =
\frac{\partial \mathcal{L}_F^i}{\partial F_i^*} = 0 \,,\quad
\mathcal{L}_F^i =  - e^{p_i C} F_i^* F_i\,.
\label{eq:LFicompact}
\end{align}
Written out explicitly, the equation of motion for $F_i^*$ reads
\begin{align}
\frac{\partial \mathcal{L}_F^i}{\partial F_i} = 
W_i + e^{p_i C} \left[F_i^* + p_i\left(-\frac{i}{2}\phi_i^*\left(M-iN\right)
+ \frac{i}{\sqrt{2}}\bar{\psi}_i\bar{\chi}\right) +
\frac{p_i^2}{2}\frac{1}{2}\phi_i^*\bar{\chi}^2 \right] = 0\,.
\end{align}
From this we obtain $F_i^*$ as a function of $W_i$, $M-iN$ as well as fermionic expressions,
\begin{align}
F_i^* = - \left[e^{-p_i C} W_i + p_i\left(-\frac{i}{2}\phi_i^*\left(M-iN\right)
+ \frac{i}{\sqrt{2}}\bar{\psi}_i\bar{\chi}\right) + \frac{p_i^2}{2}\frac{1}{2}\phi_i^*\bar{\chi}^2\right] \,.
\label{eq:Fis}
\end{align}
We may distinguish terms in $\mathcal{L}_F^i$ that are either proportional to
$e^{-p_iC}$, 1, or $e^{p_iC}$,
\begin{align}
\mathcal{L}_F^i = - e^{-p_i C} L_-^i - L_0^i - e^{p_i C} L_+^i\,.
\label{eq:LFisplit}
\end{align}
Combining our results in Eq.~\eqref{eq:LFicompact} and \eqref{eq:Fis}, we find
\begin{align}
L_-^i = & \: W_i W_i^* \,,\\
L_0^i = & \: W_i \left[p_i\left(\frac{i}{2}\phi_i\left(M+iN\right)
- \frac{i}{\sqrt{2}}\psi_i\chi\right) + \frac{p_i^2}{2}\frac{1}{2}\phi_i\chi^2\right] + \textrm{h.c.} \,,\\
L_+^i = & \: \frac{p_i^4}{16}\phi_i^*\phi_i \chi^2\bar{\chi}^2 + \frac{p_i^2}{2}\psi_i\chi \bar{\psi}_i\chi
+ \frac{p_i^2}{4}\phi_i^*\phi_i\left(M^2 + N^2\right) \label{eq:Lpi}\\
+ & \:\bigg\{-\frac{p_i^2}{2\sqrt{2}}\phi_i^*\psi_i\chi\left(M-iN\right)
+  \frac{ip_i^3}{8}\phi_i^*\phi_i\bar{\chi}^2\left(M+iN\right) - \frac{i p_i^3}{4\sqrt{2}}\phi_i^*\bar{\chi}^2\psi_i\chi
+ \textrm{h.c.}\bigg\} \,. \nonumber
\end{align}
All terms appearing in $L_+^i$ already exist in $\mathcal{L}_K^i$.
Due to the additional negative sign in Eq.~\eqref{eq:LFisplit}, all of these terms
drop out of the total Lagrangian!
Eventually, we are therefore left with only one contribution $\mathcal{L}_{MN}$
to the total Lagrangian which still contains auxiliary fields other than $C$ and $\chi$,
i.e.\ the auxiliary gauge fields $M$ and $N$ to be exact,
\begin{align}
\mathcal{L}_{MN} = - \sum_i \left\{\frac{i}{2} W_i p_i\phi_i\left(M+iN\right) + \textrm{h.c.} \right\}\,.
\end{align}
Thinking of the superpotential $W$ as a function of the scalar fields $\phi_i$,
its variation $\delta W$ under a gauge transformation, characterized by a rotation
angle $\lambda$, can be directly related to the corresponding variations
$\delta \phi_i$ of the scalar fields.
The gauge invariance of the superpotential, $\delta W = 0$, then implies
the vanishing of $\mathcal{L}_{MN}$,
\begin{align}
0 = \delta W = \sum_i W_i \delta \phi_i =
- \frac{i}{2} \lambda \sum_i W_i p_i\phi_i\,,\quad \mathcal{L}_{MN} = 0 \,.
\end{align}
The only remaining contributions to the total Lagrangian contained in
$\mathcal{L}_F^i$ are thus those terms in $L_-^i$ and $L_0^i$ which
do not involve any auxiliary fields except for $C$ and $\chi$,
\begin{align}
\sum_i \mathcal{L}_F^i \supset - V_F - \sum_i \left\{ W_i \left[
\frac{p_i^2}{4}\phi_i\chi^2 -
\frac{i p_i}{\sqrt{2}}\psi_i\chi\right] + \textrm{h.c.} \right\} \,,
\label{eq:LFiremain}
\end{align}
where we have introduced $V_F$ to refer to the $F$-term scalar potential,
\begin{align}
V_F = \sum_i e^{-p_i C}W_i^* W_i\,.
\end{align}


\subsubsection{Field Redefinitions}


Having integrated out $F_i$, $M$, $N$, and $D$, there is only one step left
that separates us from writing down our final result for the total Lagrangian.
As noted above, the auxiliary fields $C$ and $\chi$ have mass dimension $0$ and $1/2$,
respectively.
To promote them to fields with canonical mass dimension,
we rescale them as follows,\footnote{In the following, we will again omit the primes
on $C$ and $\chi$ and implicitly understand that from now on $C$ and $\chi$ refer to the
fields of the correct mass dimension.}
\begin{align}
C \rightarrow C' = \frac{p\tilde{v}}{\sqrt{2}}C \,,\quad
\chi \rightarrow \chi' = \frac{p\tilde{v}}{\sqrt{2}}\chi \,.
\label{eq:rescalings}
\end{align}
Here, $p$ denotes an arbitrary real constant, $p \in \mathbb{R}$,
and $\tilde{v}$ is a spacetime-dependent auxiliary scalar field,
$\tilde{v} = \tilde{v}\left(t,\vec{x}\right)$, of mass dimension $1$.
The rescalings in Eq.~\eqref{eq:rescalings} are such that $C$ and $\chi$
acquire canonical kinetic terms, once a subset of scalar fields obtains
nonzero VEVs $\left<\phi_i^*\phi_i\right>$
and given that the product $p\tilde{v}$ is then identified as
\begin{align}
p\tilde{v}(t) = \left(\sum_i p_i^2 \left<\phi_i^*\phi_i\right>\right)^{1/2} \,.
\label{eq:pviden}
\end{align}
This implies that, assuming the scalar VEVs not to vary over space, the mass scale $\tilde{v}$ is
merely a function of time, $\tilde{v} = \tilde{v}(t)$, rather than a full-fledged scalar field.
Furthermore, in the special case of only one scalar field $\phi_0$ acquiring a nonvanishing VEV,
which corresponds to the physical situation during the \BmL phase transition,
Eq.~\eqref{eq:pviden} reduces to $p\tilde{v} = p_0 \left<\phi_0^*\phi_0\right>^{1/2}$,
which suggests to identify $p$ and $\tilde{v}$ with $p_0$ and $\left<\phi_0^*\phi_0\right>^{1/2}$,
respectively.


\subsection{General Lagrangian in Arbitrary Gauge}


We are now ready to piece together the results which we have obtained
so far in this section.
The following five steps lead us to our final expression for $\mathcal{L}$,
the total Lagrangian of the supersymmetric $U(1)$ gauge theory: we
(i) add our results for $\mathcal{L}_G$ (cf.\ Eq.~\eqref{eq:LG}),
$\mathcal{L}_K$ (cf.\ Eqs.~\eqref{eq:LagKi}, \eqref{eq:LKi}, \eqref{eq:PV3P},
\eqref{eq:PV4P}, \eqref{eq:PV0P2}, \eqref{eq:PV1P2}, and \eqref{eq:PV2P2}),
and $\mathcal{L}_W$ (cf.\ Eq.~\eqref{eq:LW});
(ii) remove all terms from this sum, which are contained in $L_+^i$ (cf.\ Eq.~\eqref{eq:Lpi});
(iii) set the auxiliary fields $F_i$, $M$, $N$, and $D$ to zero;
(iv) include by hand the terms in Eqs.~\eqref{eq:LDremain} and \eqref{eq:LFiremain}; and
(v) rescale the auxiliary fields $C$ and $\chi$ according to Eq.~\eqref{eq:rescalings}.
In the resultant Lagrangian, we of course recover all terms
which one usually obtains in Wess-Zumino or complex gauge.
But more importantly, besides that we obtain a further contribution $\mathcal{L}_{C\chi}$
to the total Lagrangian, featuring nonstandard couplings to $C$, $\partial_\mu C$ and $\chi$,
\begin{align}
\mathcal{L} = \mathcal{L}_{\textrm{WZ}} + \mathcal{L}_{C\chi} \,,\quad
\mathcal{L}_{\textrm{WZ}} = \mathcal{L}_{\textrm{WZ}}^{\textrm{kin}} +
\mathcal{L}_{\textrm{WZ}}^{\textrm{gauge}} +
\mathcal{L}_{\textrm{WZ}}^{\textrm{ferm}} - V_F - V_D \,. \label{eq:Ltotres}
\end{align}
We distinguish five different contributions to $\mathcal{L}_{\textrm{WZ}}$,
which are respectively given as
\begin{align}
\mathcal{L}_{\textrm{WZ}}^{\textrm{kin}} = & \: - \frac{1}{4} F_{\mu\nu}F^{\mu\nu}
- i \bar{\xi}\bar{\sigma}^\mu \partial_\mu \xi \\
- & \: \sum_i \exp\left(p_i\sqrt{2}C/(p\tilde{v})\right)
\left(\partial_\mu \phi_i^* \partial^\mu \phi_i
+ i\bar{\psi}_i\bar{\sigma}^\mu \partial_\mu \psi_i\right)\,, \nonumber\phantom{\bigg[}\\
\mathcal{L}_{\textrm{WZ}}^{\textrm{gauge}} = & \: \sum_i \exp\left(p_i\sqrt{2}C/(p\tilde{v})\right)
\bigg[\frac{p_i}{2}\left(i\phi_i^*\partial^\mu\phi_i - i\phi_i\partial^\mu\phi_i^*
+ \bar{\psi}_i\bar{\sigma}^\mu\psi_i\right) A_\mu \label{eq:Lgauge}\\
- & \: \frac{p_i^2}{4} \phi_i^*\phi_i A_\mu A^\mu \bigg]\,, \nonumber\\
\mathcal{L}_{\textrm{WZ}}^{\textrm{ferm}} = & \:\sum_i \exp\left(p_i\sqrt{2}C/(p\tilde{v})\right)
\frac{i p_i}{\sqrt{2}}\phi_i^* \psi_i \xi -\frac{1}{2} \sum_{i,j} W_{ij} \psi_i\psi_j
+ \textrm{h.c.}\,, \phantom{\bigg[} \label{eq:Lferm}\\
V_F = & \: \sum_i \exp\left(-p_i\sqrt{2}C/(p\tilde{v})\right) W_i^* W_i\,, \phantom{\bigg[} \label{eq:VF}\\
V_D = & \: \frac{1}{8} \sum_{ij} p_i p_j \exp\left((p_i + p_j)\sqrt{2}C/(p\tilde{v})\right)
\phi_i^*\phi_i \phi_j^* \phi_j\,. \phantom{\bigg[} \label{eq:VD}
\end{align}
Expanding the exponential functions in $\mathcal{L}_{\textrm{WZ}}$ to leading
order in the auxiliary field $C$, which is in fact equivalent to setting $C$ to zero,
yields the familiar Lagrangian in Wess-Zumino gauge.
All terms of higher order in $C$, i.e.\ all terms involving at least one power of $C$
at all, correspond to additional couplings that arise
when performing a super-gauge transformation from Wess-Zumino to complex gauge.
Meanwhile, the Lagrangian $\mathcal{L}_{C\chi}$ assumes the following form,
\begin{align}
\mathcal{L}_{C\chi} = & \: \sum_i
\exp\left(p_i\sqrt{2}C/(p\tilde{v})\right)
\bigg[ \frac{p_i}{2\sqrt{2}} \phi_i^*\phi_i\Box \frac{C}{p\tilde{v}}
- \frac{i p_i^2}{p \tilde{v}} \phi_i^*\bar{\chi}\bar{\sigma}^\mu\partial_\mu 
\frac{\phi_i\chi}{p\tilde{v}} \\
+ & \: \frac{p_i^3}{2 (p\tilde{v})^2} \phi_i^* \phi_i \bar{\chi}\bar{\sigma}^\mu
\chi A_\mu
- \frac{ p_i^2}{\sqrt{2}p\tilde{v}} \phi_i^* \phi_i \left(\chi\xi
+ \bar{\chi}\bar{\xi}\right)
 \nonumber \\
+ & \: \frac{i p_i}{\sqrt{2}}\left(\frac{i}{2}\phi_i^*\partial_\mu \phi_i 
+ \frac{i}{2}\phi_i\partial_\mu \phi_i^*
- \bar{\psi}_i \bar{\sigma}_\mu \psi_i
- \frac{p_i^2}{(p\tilde{v})^2}\phi_i^*\phi_i\bar{\chi}\bar{\sigma}_\mu\chi\right)
\partial^\mu \frac{C}{p\tilde{v}} \nonumber \\
- & \: \left\{\frac{ p_i^2}{\sqrt{2}p\tilde{v}} \phi_i^*\bar{\chi}\bar{\sigma}^\mu
\psi_i\partial_\mu \frac{C}{p\tilde{v}}
 + \frac{ p_i}{p\tilde{v}} \phi_i^* \bar{\chi}\bar{\sigma}^\mu \partial_\mu \psi_i
+ \frac{i p_i^2}{2p\tilde{v}} \phi_i^* \bar{\chi}\bar{\sigma}^\mu\psi_i A_\mu
+ \textrm{h.c.}\right\}\bigg] \nonumber \\
- &  \: \sum_i \left\{W_i \left(\frac{p_i^2}{2(p\tilde{v})^2}\phi_i\chi^2
- \frac{i p_i}{p \tilde{v}}\psi_i\chi\right) + \textrm{h.c.} \right\} \,.\nonumber
\end{align}
In Wess-Zumino gauge the auxiliary fields $C$ and $\chi$ are zero,
implying that the Lagrangian $\mathcal{L}_{C\chi}$ vanishes in this gauge as well.


\subsection{Gauge and Mass Eigenstates}


The Abelian Higgs model of the \BmL phase transition
corresponds to the supersymmetric $U(1)$ gauge theory featuring the
chiral superfields $\Phi$, $S_1$ and $S_2$ in combination with
the superpotential $W_{B-L}$ (cf.\ Eq.~\eqref{eq:WBL}),
\begin{align}
W_{B-L} = \frac{\sqrt{\lambda}}{2}\Phi\left(v_{B-L}^2 - 2 S_1 S_2\right) \,. \nonumber
\end{align}
We thus readily obtain the Lagrangian governing the dynamics of the \BmL
phase transition by applying the general result, which we computed in
the previous section (cf.\ Eq.~\eqref{eq:Ltotres}),
to the special case of $N = 3$ chiral superfields $\Phi_i = \Phi,S_1,S_2$
whose interactions are determined by the superpotential in Eq.~\eqref{eq:WBL}.
Before actually writing down the Lagrangian of the Abelian Higgs
model, we shall however discuss in more detail the \BmL Higgs superfields
$S_1$ and $S_2$, in particular their relationship to each other in unitary gauge.


\subsubsection{Gauge Eigenstates}


$S_1$ and $S_2$ carry definite \BmL charges, $q_S = q_{S_2} = -q_{S_1} = 2$
(cf.\ Sec.~\ref{subsec:infltn}).
In the following we will therefore refer to their scalar and fermionic
components, $s_{1,2}$ and $\tilde{s}_{1,2}$, as the Higgs fields in the
\textit{gauge basis} or as the \textit{gauge eigenstates}.
Let us now calculate the scalar potential $V$ for the scalar fields $s_{1,2}$.
Before the spontaneous breaking of $B$$-$$L$, the gauge fields $C$ and $\chi$
do not posses kinetic terms and are thus not dynamical.
Prior to the \BmL phase transition, the \textit{physical gauge}
hence corresponds to the Wess-Zumino gauge, in which both $C$ and $\chi$ vanish.
In Wess-Zumino gauge, $V$ is given as the the sum of $V_F$ and $V_D$
(cf.\ Eqs.~\eqref{eq:VF} and \eqref{eq:VD}) after setting $C$ to zero,
\begin{align}
V = & \: V_F + V_D \,, \quad
V_F = V_F^{(0)} + V_F^{(1)} + V_F^{(2)} + V_F^{(1,2)} \,,\label{eq:Vphis12}\\
V_F^{(0)} = & \: \frac{\lambda}{4} v_{B-L}^4 \,, \quad
V_F^{(1)} = \lambda \left|s_1\right|^2 \left|\phi\right|^2 \,, \quad
V_F^{(2)} = \lambda \left|s_2\right|^2 \left|\phi\right|^2 \,,
\phantom{\frac{p_S^2}{8}}\nonumber\\
V_F^{(1,2)} = & \: \lambda \left|s_1\right|^2 \left|s_2\right|^2
- \frac{\lambda}{2}v_{B-L}^2\left(s_1 s_2 + s_1^* s_2^*\right) \,, \quad
V_D = \frac{p_S^2}{8} \left(\left|s_1\right|^2 - \left|s_2\right|^2\right)^2 \,.\nonumber
\end{align}
Here, $\phi \in \Phi$ denotes the complex scalar contained in the inflaton superfield $\Phi$
and $p_S$ is given as $p_S = 2 g q_S$.
This result for $V$ as a function of $\phi$, $s_1$ and $s_2$ illustrates two important
aspects.
(i) Due to the mass mixing term in $V_F^{(1,2)}$, the scalar fields $s_{1,2}$, and hence
the chiral superfields $S_{1,2}$ as well, do not correspond to the physical mass eigenstates.
(ii) In the supersymmetric true vacuum, $V$ has to vanish, which, given our result
for $V_D$, enforces $\left|s_1\right| = \left|s_2\right|$ at the end of the \BmL phase transition.


\subsubsection{Mass Eigenstates}


The scalar mass matrix of the \BmL Higgs sector is diagonalized
by performing a unitary transformation on the scalar Higgs fields
$s_1$ and $s_2^*$,
\begin{align}
s_{\pm} = \frac{1}{\sqrt{2}} \left(s_1 \pm s_2^*\right) \,.
\label{eq:spm}
\end{align}
As $s_1$ and $s_2^*$ are equally charged under the $U(1)_{B-L}$,
the two superpositions $s_\pm$ also have definite and, in fact,
equal \BmL charges.
This has the particular virtue that all of the product operators $s_\pm^*s_\pm$
are gauge-invariant.
In passing, we also mention that it is not feasible to perform
the transformation in Eq.~\eqref{eq:spm} directly on the level of the
superfields $S_{1,2}$ in the superpotential.
To see this, note that $s_2^*$ is the scalar component of the
conjugate superfield $S_2^\dagger$.
But as the superpotential is supposed to be a holomorphic function,
it must not contain $S_2^\dagger$ nor any other conjugate field.
Applying now Eq.~\eqref{eq:spm} to our result
in Eq.~\eqref{eq:Vphis12} provides us with the scalar potential $V$
as a function of the scalar fields $\phi$ and $s_\pm$,
\begin{align}
V = & \: V_F + V_D \,, \quad
V_F = V_F^{(+)} + V_F^{(-)} + V_F^{(\pm)} \,,\label{eq:Vphispm}\\
V_F^{(+)} = & \: \frac{\lambda}{4}\left(\left|s_+\right|^2 - v_{B-L}^2\right)^2
+ \lambda \left|s_+\right|^2 \left|\phi\right|^2\,, \nonumber\\
V_F^{(-)} = & \: \frac{\lambda}{4}\left(\left|s_-\right|^4 + 2v_{B-L}^2\left|s_-\right|^2\right)
+ \lambda \left|s_-\right|^2 \left|\phi\right|^2\,, \nonumber\\
V_F^{(\pm)} = & \: -\frac{\lambda}{4}\left(s_+^2 s_-^{*2} + s_+^{*2} s_-^2 \right) \,,\quad
V_D = \frac{p_S^2}{8} \left(s_+s_-^* + s_+^*s_-\right)^2 \,.\nonumber
\end{align}
Evidently, the scalar fields $s_\pm$ represent indeed the physical mass eigenstates.
As anticipated in Sec.~\ref{subsec:infltn}, the scalar mass eigenvalues
squared $m_{s_\pm}^2$ turn out to be,
\begin{align}
m_{s_\pm}^2 = \frac{\lambda}{2}\left(\varphi^2 \mp v_{B-L}^2\right) \,,
\label{eq:mspm}
\end{align}
where we have used that $\phi = \varphi/\sqrt{2}e^{i\theta}$, with $\varphi$ being
the inflaton field.
The scalars $s_\pm$ are accompanied by two massive higgsinos
$\tilde{s}_{\pm}$ with Majorana masses $m_{\tilde{s}_\pm}$,
\begin{align}
\tilde{s}_+ = \frac{i}{\sqrt{2}}e^{i\theta/2}\left(\tilde{s}_1 + \tilde{s}_2\right) \,,\quad
\tilde{s}_- = \frac{1}{\sqrt{2}}e^{i\theta/2}\left(\tilde{s}_1 - \tilde{s}_2\right) \,,\quad
m_{\tilde{s}_{\pm}} = \sqrt{\frac{\lambda}{2}} \varphi \,.
\end{align}
Together, the scalar and fermionic fields $s_\pm$ and $\tilde{s}_\pm$ constitute
the Higgs fields in the \textit{mass basis}.
In contrast to $s_\pm$, the higgsino fields $\tilde{s}_\pm$ are constructed from oppositely
charged gauge eigenstates, which implies that they do not carry definite \BmL charges.
Because of that, it is not sensible to combine the scalar and the fermionic Higgs mass eigenstates
in common chiral superfields $S_\pm$.
Finally, we recall that in Sec.~\ref{subsec:infltn} it was the mass
splitting between the fields $s_\pm$ and $\tilde{s}_\pm$, which was
after all responsible for the emergence of a nonvanishing contribution to the
Colemann-Weinberg potential (cf.\ Eq.~\eqref{eq:VCM}).


The transition from the gauge to the mass basis allows
us to identify which scalar DOF contained in $S_{1,2}$ actually corresponds
to the real \BmL Higgs boson or waterfall field $\sigma$.
As can be seen from Eq.~\eqref{eq:Vphispm}, for inflaton field
values below the critical point, $\varphi < \varphi_c$, the potential of the
complex scalar $s_+ = \sigma_+/\sqrt{2}e^{i\zeta_+}$ has the shape of a \textit{Mexican hat}.
Its radial component $\sigma_+$ hence plays the role of the waterfall field of the
\BmL phase transition.%
\footnote{As we intend to reserve the field name $\sigma$ for the waterfall
field in unitary gauge, we shall refer to the radial component of $s_+$ as $\sigma_+$.
The real scalar $\sigma_+$ may then be regarded as the generalization of $\sigma$ to arbitrary gauge.}
Meanwhile, its angular component $\zeta_+$ is one of the Goldstone bosons, which
are absorbed into the vector multiplet in the course of \BmL breaking.
Once the inflaton field $\varphi$ drops below $\varphi_c$,
the complex Higgs field $s_+$ acquires a nonzero VEV $v$,
approaching $v_{B-L}$ at large times,
\begin{align}
v(t) = \left<s_+^*s_+\right>^{1/2} \,,\quad
\lim_{t\rightarrow\infty} v(t) = v_{B-L} \,.
\label{eq:vt}
\end{align}
The fact that $v$ goes precisely to $v_{B-L}$ rather than to $v_{B-L}$ times some numerical
factor is due our specific normalization of the parameters in the superpotential
(cf.\ Eq.~\eqref{eq:WBL}).
In the literature~\cite{Nakayama:2010xf,Copeland:1994vg,Dvali:1994ms,Battye:2006pk,BasteroGil:2006cm},
$W_{B-L}$ is often defined as $W_{B-L} = \sqrt{\lambda}\Phi\left(v_{B-L}^2 - S_1 S_2\right)$,
i.e.\ without any additional factors of $2$, which results in the complex Higgs boson $s_+$
obtaining a VEV of $\sqrt{2}v_{B-L}$ or equivalently in the waterfall field $\sigma_+$
obtaining a VEV of $2v_{B-L}$.
In this case, the actual scale of \BmL breaking, $\sqrt{2}v_{B-L}$, is
larger than the dimensionful parameter in the superpotential, $v_{B-L}$, by a factor of $\sqrt{2}$.


\subsubsection{Higgs Fields in Unitary Gauge}


As soon as $s_+$ develops a nonvanishing VEV, \BmL is spontaneously broken,
which is, \textit{inter alia}, reflected in the vector multiplet $V$ turning massive.
Before \BmL breaking, the only physical gauge DOFs are one massless
vector boson as well as one left-chiral gaugino.
Now, during the \BmL phase transition, two further bosonic and two further fermionic DOFs,
i.e.\ the particle content of one chiral multiplet, are absorbed into the vector multiplet,
so that it henceforth consists of one massive vector boson, one real gauge scalar and one
Dirac gaugino (cf.\ Sec.~\ref{subsec:physicalDOFs}).
In the physical gauge after \BmL breaking, i.e.\ in \textit{unitary gauge},
the number of chiral multiplets is hence reduced by one.
Figuratively speaking, we may say that one chiral multiplet is \textit{eaten} by
the massless vector multiplet for the purpose of rendering it massive.
In the context of our Abelian Higgs model of the \BmL phase transition, we
are able to eliminate one chiral multiplet by performing a super-gauge transformation,
which maps $S_1$ and $S_2$ to the same chiral superfield $S$.
At the same time, such a super-gauge transformation relates the vector
superfield $V$ in arbitrary gauge to its counterpart in unitary gauge $Z$,
\begin{equation}
S_{1,2} = \frac{1}{\sqrt{2}} S \exp\left(\pm i \Lambda \right) \,,\quad
V = Z + \frac{i}{p_S}\left(\Lambda - \Lambda^\dagger\right) \,,
\label{eq:sgaugetrafo}
\end{equation}
for some appropriate chiral superfield $\Lambda$.
For clarity, we iterate once more:
$S$ and $Z$ are the Higgs and the vector superfield in unitary gauge,
in which the physical DOFs are manifest, $S_{1,2}$ and $V$ are the respective
fields in arbitrary gauge and $\Lambda$ is the corresponding super-gauge transformation
parameter, relating these two sets of fields to each other.
After symmetry breaking, the Wess-Zumino gauge is no longer of any use, since
it is only able to account for the dynamics of a massless vector multiplet.
The special choice $\Lambda = 0$ corresponds to performing no super-gauge
transformation at all and hence staying in unitary gauge,
\begin{align}
\textrm{$S_{1,2}$ and $V$ in unitary gauge:}\qquad
S_1 = S_2 = \frac{1}{\sqrt{2}} S \,,\quad V = Z \,.
\label{eq:S1S2VUniG}
\end{align}


The chiral superfields $S$ and $\Lambda$ contain the complex scalar fields
$s = \left(\sigma + i \tau\right)/\sqrt{2}$ and $\lambda = \left(a + i b\right)/\sqrt{2}$.
According to Eqs.~\eqref{eq:spm} and \eqref{eq:sgaugetrafo}, these are related
to the scalar Higgs fields in the gauge and in the mass basis in the following way,
\begin{align}
s_{1,2} = \frac{1}{\sqrt{2}} \,s \,e^{\pm i\lambda} \,,\quad
s_{\pm} = \frac{1}{2} \left(s \,e^{i\lambda} \pm s^* e^{i\lambda^*}\right) \,.
\label{eq:s12spm}
\end{align}
In unitary gauge, $\lambda$ is zero, which is reflected in
the disappearance of the complex phases of the scalar fields in the mass basis,
$s_+ = \sigma_+ /\sqrt{2}e^{i\zeta_+}$ and $s_- = \sigma_- /\sqrt{2}e^{i\zeta_-}$,
\begin{align}
\lambda = 0 \:: \quad
s_+ = \frac{1}{2}\left(s + s^*\right) = \textrm{Re}\left\{s\right\} = \frac{\sigma}{\sqrt{2}}\,,\quad
s_- = \frac{1}{2}\left(s - s^*\right) = i\,\textrm{Im}\left\{s\right\} = \frac{i\tau}{\sqrt{2}} \,.
\label{eq:spmsigmatau}
\end{align}
Here, the result for $s_+$ directly implies that the real scalar $\sigma$
has to be identified as the physical waterfall field in unitary gauge.
Its relation to $\sigma_+$, the waterfall field in arbitrary gauge,
also follows from Eq.~\eqref{eq:s12spm}.
Expanding $s_\pm$, $s$ and $\lambda$
into their real components, the second identity in Eq.~\eqref{eq:s12spm}
allows us to express the real scalar Higgs DOFs in arbitrary gauge, $\sigma_\pm$
and $\zeta_\pm$, as functions of the real scalar Higgs DOFs in unitary
gauge, $\sigma$ and $\tau$, as well as of the two super-gauge transformation
parameters $a$ and $b$,
\begin{align}
\sigma_+ = & \: \left[\sigma^2 \cosh^2\left(\frac{b}{\sqrt{2}}\right) +
\tau^2 \sinh^2\left(\frac{b}{\sqrt{2}}\right)\right]^{1/2} \,, \\
\sigma_- = & \: \left[\sigma^2 \sinh^2\left(\frac{b}{\sqrt{2}}\right) +
\tau^2 \cosh^2\left(\frac{b}{\sqrt{2}}\right)\right]^{1/2} \,, \nonumber\\
\tan\zeta_+ = & \: \frac{\sigma\sin\left(a/\sqrt{2}\right)\cosh\left(b/\sqrt{2}\right)
-\tau\cos\left(a/\sqrt{2}\right)\sinh\left(b/\sqrt{2}\right)}
{\sigma\cos\left(a/\sqrt{2}\right)\cosh\left(b/\sqrt{2}\right)
+\tau\sin\left(a/\sqrt{2}\right)\sinh\left(b/\sqrt{2}\right)} \,, \nonumber\\
\tan\zeta_- = & \: \frac{\sigma\sin\left(a/\sqrt{2}\right)\sinh\left(b/\sqrt{2}\right)
-\tau\cos\left(a/\sqrt{2}\right)\cosh\left(b/\sqrt{2}\right)}
{\sigma\cos\left(a/\sqrt{2}\right)\sinh\left(b/\sqrt{2}\right)
+\tau\sin\left(a/\sqrt{2}\right)\cosh\left(b/\sqrt{2}\right)} \,. \nonumber
\end{align}
Restricting ourselves to ordinary gauge transformations, i.e.\
discarding the possibility of complex gauge transformations by
setting the parameter $b$ to $0$, we find in particular,
\begin{align}
\sigma_+ = \left|\sigma\right| \,, \quad \sigma_- = \left|\tau\right| \,, \quad
\zeta_+ = \frac{a}{\sqrt{2}} + \left(1 - \textrm{sgn}\left(\sigma\right)\right)\frac{\pi}{2} \,, \quad
\zeta_- = \frac{a}{\sqrt{2}} + \textrm{sgn}\left(\tau\right)\frac{\pi}{2} \,,
\end{align}
where the shifts of $\zeta_+$ and $\zeta_-$ relative to $a/\sqrt{2}$ ensure that,
in unitary gauge, $s_+$ and $s_-/i$ have the same sign as $\sigma$ and $\tau$,
respectively.
Acting with some ordinary gauge transformation on
the complex Higgs fields $s_+$ and $s_-$ in unitary gauge therefore results in
\begin{align}
s_+ = & \:\frac{\sigma}{\sqrt{2}} \rightarrow
\frac{\left(-1\right)^{\left(1 - \textrm{sgn}\left(\sigma\right)\right)/2}\left|\sigma\right|}
{\sqrt{2}}\,e^{i a/\sqrt{2}} = \frac{\sigma}{\sqrt{2}} \,e^{i a/\sqrt{2}} \,, \\
s_- = & \: \frac{i\tau}{\sqrt{2}} \rightarrow
\frac{i^{\textrm{sgn}\left(\tau\right)}\left|\tau\right|}{\sqrt{2}}\,
e^{i a/\sqrt{2}} = \frac{i\tau}{\sqrt{2}} \,e^{i a/\sqrt{2}} \,, \nonumber
\end{align}
which illustrates that $s_\pm$ carry indeed equal \BmL charges
(cf.\ the comment below Eq.~\eqref{eq:spm}).


The complex scalar $s$ is the only remaining Higgs boson in unitary gauge.
It acquires the same VEV as $s_+$, one of the two complex
Higgs bosons in arbitrary gauge  (cf.\ Eq.~\eqref{eq:vt}),
\begin{align}
s = \frac{1}{\sqrt{2}}\left(\sigma + i \tau\right) \,,\quad
\left<s^*s\right>^{1/2} = \frac{1}{2}\left<\sigma^2\right>^{1/2} = 
\left<s_+^*s_+\right>^{1/2} = v(t) \,.
\label{eq:sVEVss}
\end{align}
This demonstrates once more that $\sigma$, the real component of $s$, is the
physical symmetry-breaking Higgs boson or waterfall field of the \BmL phase transition.
$\tau$, the imaginary component of $s$, remains by contrast massive
and hence stabilized at $\tau = 0$ at all times.
With the aid of Eqs.~\eqref{eq:Vphispm} and \eqref{eq:spmsigmatau}, we now readily
obtain the scalar potential $V$ as a function of $\sigma$ and $\tau$ in the special
case, in which the auxiliary field $C$ vanishes,\footnote{As $\sigma$ and $\tau$ are
fields in unitary gauge, one actually also has to take into account the auxiliary gauge scalar
$C$ in the calculation of their scalar potential (cf.\ Sec.~\ref{eq:LagUniG}).}
\begin{align}
V = & \: V_F + V_D \,, \:\:
V_F = V_F^{(\sigma)} + V_F^{(\tau)} + V_F^{(\sigma\tau)} \,,\:\:
V_F^{(\sigma\tau)} = \frac{\lambda}{8}\sigma^2\tau^2 \,,\:\: V_D = 0 \,, \label{eq:Vsigmatau}\\
V_F^{(\sigma)} = & \: \frac{\lambda}{16}\left(\sigma^2 - 2v_{B-L}^2\right)^2
+ \frac{1}{2}\lambda \sigma^2 \left|\phi\right|^2\,, \nonumber\\
V_F^{(\tau)} = & \: \frac{\lambda}{16}\left(\tau^4 + 4v_{B-L}^2\tau^2\right)
+ \frac{1}{2}\lambda \tau^2 \left|\phi\right|^2\,. \nonumber
\end{align}
Restoring the complex scalar $s$, this result can be written in a more compact way,
\begin{align}
V = & \: \frac{\lambda}{4}\left|v_{B-L}^2 - s^2\right|^2 +
\lambda\left|s\right|^2\left|\phi\right|^2 \label{eq:Vstau}\\
= & \: \frac{\lambda}{4}\left(v_{B-L}^2 - \left|s\right|^2\right)^2
+ \frac{1}{2}\lambda v_{B-L}^2\tau^2
+ \lambda\left|s\right|^2\left|\phi\right|^2 \nonumber \,,
\end{align}
which nicely illustrates how the complex field $s$ is
stabilized in the direction of its imaginary
component $\tau$ by means of an additional mass term for this component.


\section{During and After Spontaneous Symmetry Breaking}
\label{sec:afterSSB}


Building upon the results of the previous section, we are now able to
(i) derive the Lagrangian of the Abelian Higgs model of the \BmL phase transition,
(ii) identify the physical DOFs of our model and
(iii) calculate all relevant decay rates.


\subsection{Lagrangian of the SSB Sector in Unitary Gauge}
\label{eq:LagUniG}


We evaluate the Lagrangian in Eq.~\eqref{eq:Ltotres} for the special case of $N = 3$
chiral superfields $\Phi_i = \Phi,S_1,S_2$,
interacting with each other via the superpotential $W_{B-L}$ (cf.\ Eq.~\eqref{eq:WBL}),
and perform a super-gauge transformation to unitary gauge (cf.\ Eq.~\eqref{eq:S1S2VUniG}).
Identifying $p$ with $p_S = 2gq_S$ and denoting the scalar and fermionic
components of $S$ and $\Phi$ by $\big(s,\tilde{s}\big)$ and $\big(\phi,\tilde{\phi}\big)$,
respectively, we find
\begin{align}
\mathcal{L}_{\textrm{WZ}}^{\textrm{kin}} = & \:
- \frac{1}{4} F_{\mu\nu}F^{\mu\nu}
- i \bar{\xi}\bar{\sigma}^\mu \partial_\mu \xi
- \partial_\mu \phi^* \partial^\mu \phi
- i\bar{\tilde \phi}\bar{\sigma}^\mu \partial_\mu \tilde{\phi} \label{eq:Lkinres}\\
- & \: \cosh\left(\sqrt{2}C/\tilde{v}\right) \left(\partial_\mu s^* \partial^\mu s
+ i\bar{\tilde s}\bar{\sigma}^\mu \partial_\mu \tilde s \right) \,, \nonumber \\
\mathcal{L}_{\textrm{WZ}}^{\textrm{gauge}} = & \: \sinh\left(\sqrt{2}C/\tilde{v}\right)
\left[\frac{p_S}{2}\left(i s^*\partial^\mu s - i s\partial^\mu s^*
+ \bar{\tilde s}\bar{\sigma}^\mu\tilde s \right) A_\mu \right] \label{eq:Lgaugeres} \\
- & \: \cosh\left(\sqrt{2}C/\tilde{v}\right) \frac{p_S^2}{4} \left|s\right|^2 A_\mu A^\mu \,, \nonumber \\
\mathcal{L}_{\textrm{WZ}}^{\textrm{ferm}} = & \:\sinh\left(\sqrt{2}C/\tilde{v}\right)
\frac{i p_S}{\sqrt{2}}s^* \tilde s \xi + \frac{1}{2} \sqrt{\lambda} \phi \tilde s \tilde s
+ \sqrt{\lambda} s \tilde \phi \tilde s + \textrm{h.c.}\,,  \label{eq:Lfermres} \\
V_F = & \: \frac{\lambda}{4} |v^2_{B-L} - s^2|^2
+ \cosh\left(\sqrt{2}C/\tilde{v}\right) \lambda \left|s\right|^2 \left|\phi\right|^2 \,,
\label{eq:VFres}\\
V_D = & \: \frac{p_S^2}{8} \sinh^2\left(\sqrt{2}C/\tilde{v}\right) \left|s\right|^4 \,.
\label{eq:VDres}
\end{align}
The Lagrangian taking care of the couplings of the gauge fields $C$ and $\chi$ now reads
\begin{align}
\mathcal{L}_{C\chi} = & \: \sinh\left(\sqrt{2}C/\tilde{v}\right)
\bigg[\frac{\left|s\right|^2}{2\sqrt{2}}\Box\frac{C}{\tilde{v}}
+ \frac{p_S\left|s\right|^2}{2 \tilde{v}^2} \bar{\chi}\bar{\sigma}^\mu \chi A_\mu
- \left\{\frac{s^*}{\tilde{v}}\bar{\chi}\bar{\sigma}^\mu \partial_\mu \tilde s + \textrm{h.c.} \right\} \label{eq:LCchi} \\
+ & \:\frac{i}{\sqrt{2}} \left(\frac{i}{2} s^*\partial_\mu s + \frac{i}{2} s \partial_\mu s^*
- \bar{\tilde s} \bar{\sigma}_\mu \tilde s - \frac{\left|s\right|^2}{\tilde{v}^2} \bar{\chi}\bar{\sigma}_\mu\chi\right)\partial^\mu \frac{C}{ \tilde{v}} \bigg]  \nonumber \\
- & \: \cosh\left(\sqrt{2}C/\tilde{v}\right) \bigg[\frac{is^*}{\tilde{v}}\bar{\chi}\bar{\sigma}^\mu\partial_\mu \frac{s \, \chi}{\tilde{v}} + \frac{p_S \left|s\right|^2}{ \sqrt{2} \tilde{v}}\left(\chi\xi + \bar{\chi}\bar{\xi}\right) \nonumber \\
+ & \: \left\{\frac{s^*}{\sqrt{2} \tilde{v}} \bar{\chi}\bar{\sigma}^\mu \tilde s \partial_\mu \frac{C}{\tilde{v}}
+ \frac{i p_S s^*}{2\tilde{v}}\bar{\chi}\bar{\sigma}^\mu \tilde s A_\mu + \textrm{h.c.} \right\}\bigg]
+ \bigg\{\frac{\sqrt{\lambda}s^2}{2 \tilde{v}^2}\phi\chi^2 + \textrm{h.c.} \bigg\}  \nonumber \,.
\end{align}


We account for the spontaneous breaking of \BmL by shifting the complex Higgs field
$s$ around its time-dependent expectation value (cf.\ Eqs.~\eqref{eq:vt} and \eqref{eq:sVEVss}),
\begin{align}
s \rightarrow v(t) + s = v(t) + \frac{1}{\sqrt{2}}\left(\sigma + i\tau \right) \,, \quad
\lim_{t\rightarrow\infty}v(t) = v_{B-L} \,.
\label{eq:shift}
\end{align}
Up to now, we denoted with $s$ and $\sigma$ the fluctuations of the complex
\BmL Higgs boson and of its real component around the origin,
i.e.\ around the false vacuum.
From now on, $s$ and $\sigma$ shall, however, refer to the respective fluctuations
around the homogeneous Higgs background, i.e.\ eventually, once the \BmL phase
transition is completed, around the true vacuum.
%
%
The replacement in Eq.~\eqref{eq:shift}
induces mass terms for all particles, which are coupled to the Higgs boson $s$
(cf.\ Sec.~\ref{subsec:physicalDOFs}), and gives rise to kinetic terms for
the gauge fields $C$ and $\chi$,
\begin{align}
\mathcal{L}_{C\chi} \supset - \frac{1}{2}\frac{v^2}{\tilde{v}^2} \partial_\mu C \partial^\mu C
- \frac{v^2}{\tilde{v}^2} i \bar{\chi}\bar{\sigma}^\mu\partial_\mu \chi \,.
\end{align}
As anticipated in Sec.~\ref{subsec:lagrangian}, the mass scale $\tilde{v}$
has to be identified with the VEV $v$, in order to
obtain canonically normalized kinetic terms (cf.\ Eq.~\eqref{eq:pviden}),
\begin{align}
\tilde{v}(t) = \,v(t) = \left<s_+^* s_+\right> \,.
\end{align}


\subsection{Physical Degrees of Freedom and Time-Dependent Masses}
\label{subsec:physicalDOFs}


\begin{figure}
\begin{center}
\includegraphics[width=1.0\textwidth]{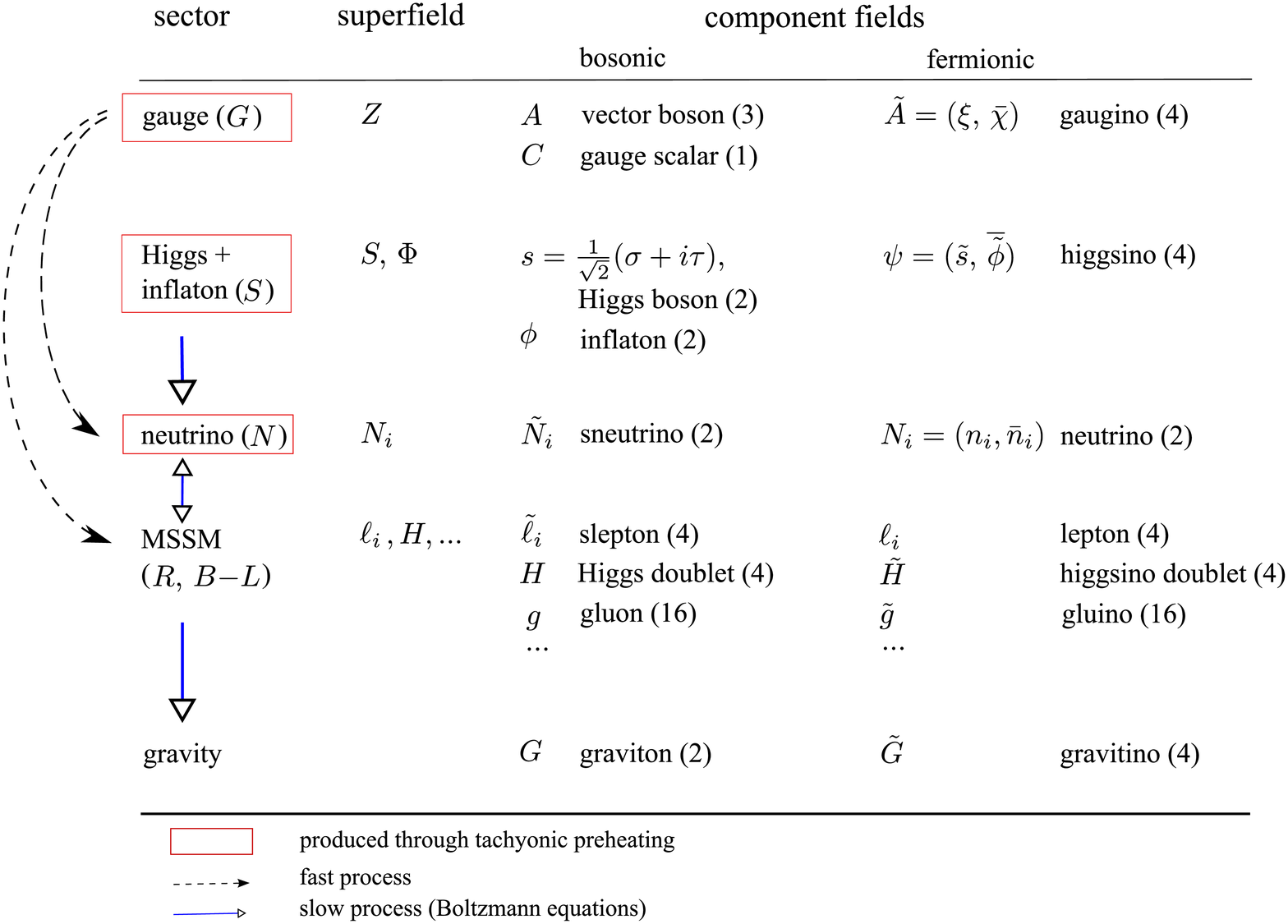}\bigskip
\caption[Particle spectrum and interactions after spontaneous \BmL breaking]
{Physical particle spectrum as well as all relevant
production and decay processes after the spontaneous breaking of \BmL symmetry.
The Higgs field $\sigma$ and all particles coupled to it are produced
during tachyonic preheating (red boxes).
The gauge DOFs then decay nearly instantaneously (black, dashed arrows),
whereas the decay and production of the other particles can be described
by Boltzmann equations (blue, solid arrows).
The numbers in parentheses denote the respective internal DOFs.
\label{fig:overview}}
\end{center}
\end{figure}


With the full Lagrangian of the Abelian Higgs model at our disposal,
we are now ready to evaluate how the bosonic and fermionic DOFs, which we
initially introduced through the superfields $\Phi$, $S_{1,2}$ and $V$,
are eventually distributed among the physical particles in the broken phase.
In Fig.~\ref{fig:overview}, we give an overview of the particle spectrum and indicate
how the various particle species are respectively produced as well as how
they respectively interact with each other.


\subsubsection{Symmetry-Breaking Sector}


Out of the four real scalar DOFs initially contained in $S_{1,2}$,
only two DOFs remain as independent real scalar fields after the \BmL phase
transition---the waterfall field $\sigma$ as well as its partner $\tau$.
We can read off the masses of these two particles, $m_\sigma$ and $m_\tau$,
from the $F$-term scalar potential (cf.\ Eqs.~\eqref{eq:Vsigmatau} and \eqref{eq:VFres}),
\begin{align}
V_F \supset & \:
\frac{\lambda}{16}\Big[\big(\sqrt{2}v+ \sigma\big)^2 - 2v_{B-L}^2\Big]^2
+ \frac{1}{2}\lambda \big(\sqrt{2}v+ \sigma\big)^2 \left|\phi\right|^2
\supset \frac{1}{2}m_\sigma^2\sigma^2 \,, \\
V_F \supset & \:
\frac{\lambda}{16}\left(\tau^4 + 4v_{B-L}^2\tau^2\right) +
\frac{\lambda}{8}\big(\sqrt{2}v+ \sigma\big)^2\tau^2
+ \frac{1}{2}\lambda \tau^2 \left|\phi\right|^2
\supset \frac{1}{2}m_\tau^2\tau^2 \,,
\end{align}
where $m_\sigma$ and $m_\tau$ are given as
\begin{align}
m_\sigma^2(t) = & \: \frac{\lambda}{2}\left(3v^2(t) - v_{B-L}^2 + 2 \left|\phi(t)\right|^2\right) \,,
\label{eq:msigma2}\\
m_\tau^2(t) = & \; \frac{\lambda}{2}\left(v^2(t) + v_{B-L}^2 + 2 \left|\phi(t)\right|^2\right) \,.
\label{eq:mtau2}
\end{align}
In the supersymmetric true vacuum, $\phi$ vanishes and $v$ has reached its final
value,\footnote{In Sec.~\ref{sec:preheating}, when discussing the nonperturbative
production of particles during tachyonic preheating, we will work in the
\textit{quench approximation}, in which the inflaton field $\phi$ is straight away
set to zero as soon as the waterfall field becomes unstable.
In the quench approximation, the terms proportional to $\left|\phi\right|^2$
in Eqs.~\eqref{eq:msigma2} and \eqref{eq:mtau2} may hence be omitted from the beginning.}
\begin{align}
t\rightarrow \infty \:: \qquad v(t) = v_{B-L} \,,\quad \phi = 0 \,,\quad
m_\sigma^2(t) = m_\sigma^2(t) = \lambda v_{B-L}^2 \,.
\end{align}
The real scalars $\sigma$ and $\tau$ are accompanied by the complex scalar $\phi$.
Its mass $m_\phi$ can also be read off from the $F$-term scalar potential
(cf.\ Eq.~\eqref{eq:VFres}),
\begin{align}
V_F \supset
\cosh\big(\sqrt{2}C/v\big) \lambda \left|v+s\right|^2 \left|\phi\right|^2
\supset m_\phi^2 \left|\phi\right|^2 \,,\quad m_\phi^2(t) = \lambda v^2(t) \,.
\label{eq:mphi}
\end{align}
Finally, the fermionic component $\tilde{s}$ of the superfield $S$ pairs up
with the fermionic component $\tilde{\phi}$ of the inflaton superfield $\Phi$
to form a Dirac fermion $\psi = \big(\tilde{s},\bar{\tilde{\phi}}\big)^T$, the higgsino,
which acquires a mass $m_\psi$ in the course of \BmL breaking (cf.\ Eq.~\eqref{eq:Lfermres}),
\begin{align}
\mathcal{L}_{\textrm{WZ}}^{\textrm{ferm}} \supset & \:
\sqrt{\lambda}(v+s) \, \tilde \phi \tilde s + \textrm{h.c.} \supset
m_\psi \, \tilde\phi \tilde s + \textrm{h.c.} \,,\quad
m_\psi(t) = \sqrt{\lambda} v(t) \,.
\end{align}
As required by supersymmetry, which is restored at the end of the \BmL phase transition,
all particles originating from the chiral superfields $\Phi$ and $S_{1,2}$ end up having
same mass in the true vacuum,
\begin{align}
t\rightarrow \infty \:: \qquad
m_\sigma(t) = m_\tau(t) = m_\phi(t) = m_\psi(t) = m_S = \sqrt{\lambda}v_{B-L} \,.
\end{align}


\subsubsection{Gauge Sector}


Next, we turn to the gauge sector.
One key implication of the spontaneous breaking of \BmL is that it turns
the massless vector multiplet into a full massive vector multiplet with four scalar
and four fermionic DOFs.
The mass $m_A$ of the vector boson $A$ can be read off from the Lagrangian
$\mathcal{L}_{\textrm{WZ}}^{\textrm{gauge}} $ (cf. Eq.~\eqref{eq:Lgaugeres}),
\begin{align}
\mathcal{L}_{\textrm{WZ}}^{\textrm{gauge}} \supset
- \cosh\big(\sqrt{2}C/v\big) \frac{p_S^2}{4} \left|v + s\right|^2 A_\mu A^\mu
\supset - \frac{1}{2} m_A^2 A_\mu A^\mu \,.
\end{align}
where $m_A$ now depends on $p_S = 2g q_S = 4g$ rather than the coupling constant $\lambda$,
\begin{align}
m_A^2(t) = \frac{1}{2}p_S^2v^2(t) = 2g^2q_S^2v^2(t)= 8g^2v^2(t) \,.
\label{eq:mA}
\end{align}
Similarly to $\tilde{s}$ and $\tilde{\phi}$,
the Weyl fermion $\xi$ and the former auxiliary field $\chi$ consort with each other
to form a Dirac fermion $\tilde{A} = \big(\xi,\bar{\chi}\big)^T$, the gaugino,
which obtains a mass $m_{\tilde{A}}$ during the \BmL phase transition.
The mass term for the gaugino is contained in $\mathcal{L}_{C\chi}$,
which also features the kinetic terms for $C$ and $\chi$ (cf.\ Eq.~\eqref{eq:LCchi}),
\begin{align}
\mathcal{L}_{C\chi} \supset - \cosh\big(\sqrt{2}C/v\big)
\frac{p_S \left|v+s\right|^2}{ \sqrt{2} v}\chi\xi + \textrm{h.c.} \supset
-m_{\tilde{A}}\,\chi\xi + \textrm{h.c.} \,,
\end{align}
from which we infer that $m_{\tilde{A}} = m_A$ at all times (cf.\ Eq.~\eqref{eq:mA}).
As previously mentioned, the vector multiplet absorbs two real scalar DOFs,
which initially belong to the Higgs superfields $S_{1,2}$---one ends up being
the longitudinal component of the massive vector boson $A$, the other is to be
identified with the gauge field $C$, which becomes dynamical as soon as \BmL is broken.
During the \BmL phase transition, $C$ acquires a mass $m_C$, which receives
contributions from the $D$-term scalar potential (cf.\ Eq.~\eqref{eq:VDres}) as well
as from the Lagrangian $\mathcal{L}_{C\chi}$ (cf.\ Eq.~\eqref{eq:LCchi}),
\begin{align}
\mathcal{L}_{C\chi} - V_D \supset & \:
\sinh\big(\sqrt{2}C/v\big)\frac{\left|v+s\right|^2}{2\sqrt{2}}\Box\frac{C}{v}
- \frac{p_S^2}{8} \sinh^2\big(\sqrt{2}C/v\big) \left|v+s\right|^4
\label{eq:LmC}\\
\supset & \: \frac{1}{2} m_C^2 C^2 \,, \quad
m_C^2 = \frac{1}{2}p_S^2v^2 + v\frac{\partial^2}{\partial t^2}\frac{1}{v} =
\frac{1}{2}p_S^2v^2 + 2\frac{\dot{v}^2}{v^2} - \frac{\ddot{v}}{v} \,. \nonumber
\end{align}
The mass term in the Lagrangian $\mathcal{L}_{C\chi}$ relies on the fact
that during the \BmL phase transition $v$ is a function of time.
Once $v$ has reached its final value, it vanishes and $m_C$ is solely accounted
for by the mass term in $V_D$.


In order to assess the relative importance of the two contributions to the
mass of the scalar $C$ in the course of \BmL breaking, we need to
know the time dependence of the VEV $v$.
To this end, we expand the complex Higgs boson $s_+$ into momentum eigenfunctions
$s_+(k,t)e^{-i\mathbf{k}\mathbf{x}}$, where $k = \left|\mathbf{k}\right|$,
and write $v$ as an integral over the Higgs mode functions $s_+(k,t)$
(cf.\ Eq.~\eqref{eq:vt}),
\begin{align}
v^2(t) = \left<s_+^*s_+\right> = \int\frac{d^3k}{\left(2\pi\right)^3} \left|s_+(k,t)\right|^2
\,. \label{eq:v2tmodes}
\end{align}
Given the solutions to the mode equations for all wavenumbers $k$,
this relation yields $v$ as a function of time.
Due to the quartic self-interaction of the field $s_+$ (cf.\ Eq.~\eqref{eq:Vphispm}),
the equations of motion of the respective field modes are, however, nonlinear and
thus require a numerical treatment.
The authors of Ref.~\cite{GarciaBellido:2001cb} perform a fully nonlinear lattice simulation
to study the evolution of the Higgs VEV in the quench approximation and conclude that, apart from
strongly damped oscillations after symmetry breaking, it is well approximated
by a smooth step function interpolating between the false, $v = 0$, and
the true vacuum, $v = v_{B-L}$,
\begin{align}
v(t) \approx \frac{v_{B-L}}{2}\left[1 + \tanh\frac{m_S\left(t - t_{\textrm{PH}}\right)}{2}\right]
\,,\quad m_S = \sqrt{\lambda}v_{B-L} \,.
\label{eq:vtapprox}
\end{align}
Here, $t_{\textrm{PH}}$ denotes the time at the end of tachyonic preheating and
we fix the origin of the time axis by setting $t_{\textrm{PH}} = 0$.
Now, inserting the approximate expression for $v$ in Eq.~\eqref{eq:vtapprox} into
our result for $m_C$ (cf.\ Eq.~\eqref{eq:LmC}), we obtain
\begin{align}
m_C^2(t) \approx \left(\frac{1}{2}p_S^2 + \frac{\lambda}{1 + \exp\left(m_S t\right)}\right)v^2(t)
\approx \frac{1}{2}p_S^2 v^2(t) \,.
\end{align}
The contribution to $m_C$ from the Lagrangian $\mathcal{L}_{C\chi}$ falls off
exponentially fast and is hence always negligibly small at sufficiently late times.
But even at early times, $t < 0$, it is outweighed by the contribution from the $D$-term
scalar potential since, owing to perturbativity, $\lambda \ll \frac{1}{2}p_S^2 \simeq 4\pi/3$
(cf.\ Sec.~\ref{subsec:flavour}).
On top of that, the extra mass term in $\mathcal{L}_{C\chi}$ significantly
complicates the calculation describing the nonperturbative production of $C$ quanta
during tachyonic preheating (cf.\ Sec.~\ref{sec:preheating}).
As our final results concerning the generation of entropy, baryon asymmetry and
dark matter prove to be rather insensitive to the dynamics of the gauge sector
in any case (cf.\ Ch.~\ref{ch:reheating}), we shall therefore ignore the contribution
to $m_C$ from the Lagrangian $\mathcal{L}_{C\chi}$ in the following
and treat the gauge particles as if they had equal masses at all times,
$m_C \approx m_{\tilde{A}} = m_A$.
In the supersymmetric true vacuum, this approximation turns into an exact statement,
\begin{align}
t \rightarrow\infty \::\qquad
m_A(t) = m_{\tilde{A}}(t) = m_C(t) = m_G = \frac{1}{\sqrt{2}}p_S v_{B-L} \,.
\end{align}


\subsubsection{Neutrino Sector}


So far, we have seen how the spontaneous breaking of \BmL causes the bosonic and fermionic
DOFs initially contained in the superfields $\Phi$, $S_{1,2}$ and $V$ to assemble in new
physical particles.
For completeness, let us now also list all further particle species present in the broken phase.


The neutrino superfields $n_i^c$ are coupled to the Higgs superfield $S_1$ via
a common term in the seesaw superpotential (cf.\ Eq.~\eqref{eq:WSeesaw}).
In unitary gauge and after \BmL breaking, we have to replace $S_1$ by
$\frac{1}{\sqrt{2}}\left(v+S\right)$ in this term (cf.\ Eq.~\eqref{eq:S1shift}),
\begin{align}
S_1 \rightarrow \frac{1}{\sqrt{2}}\left(v+S\right) \,,\quad
W_{\textrm{Seesaw}} \supset \frac{1}{\sqrt{2}} h_i^n n_i^c n_i^c S_1 \rightarrow
\frac{1}{2} h_i^n n_i^c n_i^c \left(v+S\right) \,.
\label{eq:WSeesawShift}
\end{align}
In consequence of the new coupling to the homogeneous Higgs background,
the fermionic components of the fields $n_i^c$ and $n_i$ combine into three
heavy Majorana neutrinos $N_i$ (cf.\ Eq.~\eqref{eq:NinuRi}).
The superpartners of these neutrinos, the heavy sneutrinos $\tilde{N}_i$,
are identified with the scalar components of the fields $n_i^c$.
From Eq.~\eqref{eq:WSeesawShift} it directly follows that the heavy neutrinos
share common masses $M_i$ with their superpartners at all times,
\begin{align}
M_i(t) = h_i^n v(t) \,.
\end{align}


\subsubsection{MSSM and Gravitino}


Finally, our model features the entire MSSM particle content as well as the
gravity multiplet consisting of the graviton and the gravitino.
We assume that supersymmetry is spontaneously broken before the end of inflation
in some hidden sector.
The mediation of supersymmetry breaking to the visible sector via, for instance,
gravitational or loop-suppressed interactions then induces soft masses for all
MSSM superparticles as well as for the gravitino (cf.\ Sec.~\ref{subsec:candidates}).
The soft masses, which are partly generated for the particles coupling to the \BmL
Higgs boson $s$, are negligibly small compared to their masses generated
in the course of \BmL breaking.
Furthermore, as soon as a primordial thermal bath has emerged, all MSSM particles
obtain effective thermal masses due to their rapid gauge and Yukawa interactions.
At the high temperatures reached during reheating after the \BmL phase transition,
these thermal masses always exceed by far the corresponding soft masses.
By contrast, all other particles interact too feebly with the thermal bath
and thus do not obtain sizable thermal masses.


\subsection{Decay Rates and Branching Ratios}
\label{subsec:rates}


A central result of the previous section is that the masses of all particles
coupled to the Higgs boson $s$ rapidly grow while the Higgs VEV $v$ evolves from
$0$ to $v_{B-L}$ (cf.\ Eq.~\eqref{eq:vtapprox}).
This sudden change in inertia leads to the nonadiabatic
production of these particles~\cite{GarciaBellido:2001cb}.
In Sec.~\ref{sec:preheating}, we will discuss the particle abundances generated
during tachyonic preheating in more detail.
For now, we merely state that all particles from the SSB, gauge and neutrino sectors
are produced during the \BmL phase transition, \textit{viz.} predominantly with
momenta $k$ much smaller than their respective masses, $k \ll m$.
Next to a dominating abundance of nonrelativistic Higgs bosons, tachyonic preheating
thus also gives rise to a gas of nonrelativistic higgsinos, inflatons, gauge particles,
and heavy (s)neutrinos.


All of these particles are unstable and decay after preheating into lighter
DOFs (cf.\ Fig.~\ref{fig:overview}).
To determine the relevant decay channels, we have to identify all renormalizable
operators in the Lagrangian, which entail kinematically allowed two-body decays.
Due to our particular choice of Froggatt-Nielsen flavour charges
(cf.\ Eq.~\eqref{eq:ebc}), the mass spectrum of our model exhibits
the following hierarchy,\footnote{For definiteness, we will set $m_S = M_3 = M_2$
in Chs.~\ref{ch:reheating} and \ref{ch:wimp}.}
\begin{align}
m_G \sim v_{B-L} \gg m_S \sim M_3 \sim M_2 \gg M_1 \gg v_{\textrm{EW}}\,.
\label{eq:spectrum}
\end{align}
When identifying the relevant operators in the Lagrangian, we therefore have to look for
terms that couple (i) one gauge particle to two particles from any other sector,
(ii) one Higgs boson, higgsino, inflaton or (s)neutrino of the second or third
generation to (s)neutrinos of the first generation or MSSM particles, or (iii)
one (s)neutrino of the first generation to MSSM particles.


\subsubsection{Gauge Sector}


All operators which couple fields of the gauge sector to fields of the SSB sector
contain at least two gauge fields.
Real gauge particles, i.e.\ gauge particles on the mass shell, can thus not decay into
Higgs bosons, higgsinos or inflatons.
Meanwhile, the interactions of the heavy (s)neutrinos as well as of the MSSM particles with the
\BmL vector boson $A$ are accounted for by (cf.\ Eq.~\eqref{eq:Lgauge}),
\begin{align}
\mathcal{L}_{\textrm{WZ}}^{\textrm{gauge}} \supset \sum_i \exp\left(p_i\sqrt{2}C/(p_Sv)\right)
\left[\frac{p_i}{2}\left(i\phi_i^*\partial^\mu\phi_i
- i\phi_i\partial^\mu\phi_i^* + \bar{\psi}_i\bar{\sigma}^\mu\psi_i\right) A_\mu \right] \,,
\label{eq:Adec}
\end{align}
from which one obtains the actual operator describing the two-body decays of the vector $A$
as the lowest-order term when expanding the exponential in powers of $C$.
The interactions with the gaugino are governed by (cf.\ Eq.~\eqref{eq:Lferm})
\begin{align}
\mathcal{L}_{\textrm{WZ}}^{\textrm{ferm}} \supset \sum_i \exp\left(p_i\sqrt{2}C/(p_Sv)\right)
\frac{i p_i}{\sqrt{2}}\phi_i^* \psi_i \xi + \textrm{h.c.}
\supset \sum_i \frac{i p_i}{\sqrt{2}}\phi_i^* \psi_i \xi + \textrm{h.c.} \,.
\label{eq:Atdec}
\end{align}
As far as we restrict ourselves to renormalizable interactions, the gauge
field $C$ only decays into the scalar components $\phi_i$ of the heavy (s)neutrino
and MSSM superfields.
The strongest interaction between one $C$ particle and two scalars derives
from the $D$-term scalar potential (cf.\ Eqs.~\eqref{eq:VD} and \eqref{eq:VDres}),
\begin{align}
V_D = & \: \frac{1}{8}\Big[p_S \sinh\big(\sqrt{2}C/v\big)\left|v+s\right|^2 +
\sum_i p_i \exp\left(p_i\sqrt{2}C/(p_Sv)\right)\left|\phi_i\right|^2\Big]^2 \label{eq:Cdec1} \\
\supset & \: \frac{p_S}{4}\sinh\big(\sqrt{2}C/v\big)\left|v+s\right|^2
\sum_i p_i \exp\left(p_i\sqrt{2}C/(p_Sv)\right)\left|\phi_i\right|^2 \nonumber \\
\supset & \: \frac{p_S}{4v} \sqrt{2}C \left|v+s\right|^2
\sum_i p_i \left|\phi_i\right|^2
\supset \frac{v}{\sqrt{2}} C \sum_i \frac{1}{2}p_S p_i \left|\phi_i\right|^2 \,. \nonumber
\end{align}
Beyond that, the $F$-term scalar potential contains an operator
coupling the scalar $C$ to two heavy sneutrinos $\tilde{N}_i$ (cf.\ Eq.~\eqref{eq:VF}),
\begin{align}
V_F \supset & \: \exp\left(-p_{n^c}\sqrt{2}C/(p_Sv)\right) \sum_i W_{n_i^c}^* W_{n_i^c} \,,\:
W_{n_i^c} = h_i^n \tilde{N}_i (v+s) + h_{ji}^\nu \tilde{\ell}_j H_u  \,, \label{eq:Cdec2} \\
\supset & \:  -\frac{p_{n^c}}{p_Sv}\sqrt{2}C \left|v+s\right|^2
\sum_i \left(h_i^n\right)^2 \big|\tilde{N}_i \big|^2
\supset - \frac{v}{\sqrt{2}} C \sum_i \left(h_i^n\right)^2 \big|\tilde{N}_i \big|^2 \,, \nonumber
\end{align}
where we have used that the neutrino superfields $n_i^c$ carry \BmL charge $q_{n^c} = 1$,
such that $p_{n^c} = 2gq_{n_c} = p_S/2$.
The strength of the interactions in Eqs.~\eqref{eq:Cdec1} and \eqref{eq:Cdec2}
is determined by the effective gauge couplings $\frac{1}{2}p_Sp_i$ as well as
by the Yukawa couplings $\left(h_i^n\right)^2$, respectively.
Since $\left(h_i^n\right)^2 \ll \frac{1}{2}p_S\left|p_i\right|$ (cf.\ Sec.~\ref{subsec:flavour}),
we shall neglect the contribution from Eq.~\eqref{eq:Cdec2} to the decay rate for the process
$C \rightarrow \tilde{N}_i \tilde{N}_i^*$ in the following.


Given the operators in Eqs.~\eqref{eq:Adec}, \eqref{eq:Atdec} and \eqref{eq:Cdec1},
we are now able to calculate the tree-level decay rates and branching ratios of the particles
$A$, $\tilde{A}$ and $C$ by means of standard methods~\cite{Dreiner:2008tw}.\footnote{Note
that the conventions for the spacetime metric $\eta_{\mu\nu}$ as well as for the sigma
matrices $\sigma^\mu$ and $\bar{\sigma}^\mu$ employed in Ref.~\cite{Dreiner:2008tw}
differ from the conventions of Ref.~\cite{Wess:1992cp}.
Appendix A of Ref.~\cite{Dreiner:2008tw}, however, provides a manual for how to translate
between these two different conventions.}
As it turns out, all three particles decay at the same total rate $\Gamma_G^0$,
\begin{align}
\Gamma_A^0 = \Gamma_{\tilde{A}}^0 = \Gamma_C^0 = \Gamma_G^0 =
\frac{g^2}{16\pi} m_G \sum_i q_i^2 \left[1-\left(2m_i/m_G\right)^2\right]^{1/2} \,,
\label{eq:GammaG0}
\end{align}
with the sum running over all heavy (s)neutrino and MSSM multiplets $i$ carrying \BmL
charges $q_i$.
As $m_G$ is much larger than all supersymmetry-breaking soft masses,
we may treat the MSSM particles as massless.
$\Gamma_G^0$ can then be written as
\begin{align}
\Gamma_G^0 =
\frac{g^2}{16\pi} m_G \left(\sum_{\textrm{MSSM}}q_i^2 + \sum_i
\left[1-\left(2M_i/m_G\right)^2\right]^{1/2}\right) \,,\quad \sum_{\textrm{MSSM}}q_i^2 = 13\,,
\end{align}
where the second sum now runs over the three heavy (s)neutrino multiplets $\big(N_i,\tilde{N}_i\big)$.
Next, introducing the function $R$ through the relation
\begin{align}
R\left(q,m\right) = \frac{q^2\left[1-\left(2 m/m_G\right)^2\right]^{1/2}}
{13 + \sum_i \left[1-\left(2M_i/m_G\right)^2\right]^{1/2}} \,,
\end{align}
allows us to state the branching ratios for the various final states into which the gauge
particles $A$, $\tilde{A}$ and $C$ are able to decay in a particularly convenient form,
\begin{align}
\textrm{Br}\big(A\rightarrow N_iN_i\big)=&\:\frac{2}{3}\,R(1,M_i)\,,\quad \label{eq:AAtCBr}
\textrm{Br}\big(A\rightarrow\psi_i\bar{\psi}_i\big)=\frac{2}{3}\,R(q_i,0)\,,\\
\textrm{Br}\big(A\rightarrow\tilde{N}_i\tilde{N}_i^*\big)=&\:\frac{1}{3}\,R(1,M_i)\,,\quad
\textrm{Br}\big(A\rightarrow\phi_i\phi_i^*\big)=\frac{1}{3}\,R(q_i,0)\,,\\
\textrm{Br}\big(\tilde{A}\rightarrow N_i\tilde{N}_i\big)=&\:\phantom{\frac{2}{3}\,}R(1,M_i)\,,\quad
\textrm{Br}\big(\tilde{A}\rightarrow\bar{\psi}_i\phi_i\big)=\phantom{\frac{2}{3}\,}R(q_i,0)\,,\\
\textrm{Br}\big(C\rightarrow\tilde{N}_i\tilde{N}_i^*\big)=&\:\phantom{\frac{2}{3}\,}R(1,M_i)\,,\quad
\textrm{Br}\big(C\rightarrow\phi_i\phi_i^*\big)=\phantom{\frac{2}{3}\,}R(q_i,0)\,.
\end{align}
Here, $\phi_i$ and $\psi_i$ denote the components of an arbitrary
MSSM matter multiplet.


\subsubsection{Symmetry-Breaking Sector}


The SSB sector only interacts with the gauge as well as with the
heavy (s)neutrino sector.
In addition, as a result of our specific mass spectrum (cf.\ Eq.~\eqref{eq:spectrum}),
the particles of the SSB sector are only allowed to decay into the
(s)neutrinos of the first generation.
All operators accounting for the decay of the Higgs boson,
higgsino and inflaton hence derive from the following terms in the
superpotential,\footnote{If the decays into the second and third (s)neutrino generations
were kinematically allowed, the calculation of the respective partial decay rates
would be, of course, analogous.}
\begin{align}
W \supset & \:
\frac{\sqrt{\lambda}}{2} \Phi \left[v_{B-L}^2 - \left(v+S\right)^2\right]
+ \frac{1}{2} h_1^n n_1^c n_1^c (v+S) \label{eq:WSdecay}\\
\supset & \: - \sqrt{\lambda} v \,\Phi S + \frac{1}{2} h_1^n v\, n_1^c n_1^c
+ \frac{1}{2} h_1^n n_1^c n_1^c S \,. \nonumber
\end{align}
Inserting these terms into Eq.~\eqref{eq:Lferm}, we obtain the following Yukawa interactions,
\begin{align}
\mathcal{L}_{\textrm{WZ}}^{\textrm{ferm}} \supset
-\frac{1}{2} \sum_{i,j} W_{ij} \,\psi_i\psi_j + \textrm{h.c.}
\supset -\frac{1}{2} h_1^n s\,\nu_{R,1}\nu_{R,1} - h_1^n \tilde{N}_1 \, \nu_{R,1} \tilde{s}
+ \textrm{h.c.} \label{eq:LfermSdec}
\end{align}
The former of these two operators describes the decay of the scalars $\sigma$
and $\tau$ into a pair of $N_1$ neutrinos, while the latter governs the decay
of the higgsino $\psi$ into an $N_1$ neutrino and an $\tilde{N}_1^*$ sneutrino.
Moreover, the superpotential in Eq.~\eqref{eq:WSdecay} implies the following
trilinear couplings in the $F$-term scalar potential (cf.\ Eq.~\eqref{eq:VF}),
\begin{align}
V_F = & \: \sum_i \exp\left(-p_i\sqrt{2}C/(p_S v)\right) W_i^* W_i \supset
W_{n_1^c}^* W_{n_1^c} + W_S^* W_S \label{eq:VFSdec}\\
\supset & \: \left(h_1^n\right)^2 v \big|\tilde{N}_i\big|^2 s -
\frac{1}{2} h_1^n \sqrt{\lambda} v \tilde{N}_1^* \tilde{N}_1^*\phi + \textrm{h.c.} \,,
\nonumber
\end{align}
which respectively describe the decay of the real component of $s$, i.e.\ of the scalar $\sigma$,
into a $\tilde{N}_1\tilde{N}_1^*$ pair as well as the decay of the inflaton $\phi$ into
two $\tilde{N}_1$ sneutrinos.
Note that the coupling of the scalar $\tau$ to the sneutrino field product
$\big|\tilde{N}_i\big|^2$ drops out of the $F$-term scalar potential,
when adding the hermitian conjugate of the two terms explicitly stated
in Eq.~\eqref{eq:VFSdec}.\footnote{In order to obtain a trilinear coupling
of $\tau$ to two sneutrinos in the potential $V_F$, we would have to include an
explicit mass term for the superfield $S$ into the superpotential,
$W \supset \frac{1}{2}mS^2$.}


With the operators in Eqs.~\eqref{eq:LfermSdec} and \eqref{eq:VFSdec}
at hand, we are ready to calculate the tree-level decay rates and branching ratios
of the particles $\sigma$, $\tau$, $\psi$, and $\phi$.
Just as in the case of the gauge multiplet, all particles decay
at the same total rate $\Gamma_S^0$,
\begin{align}
\Gamma_\sigma^0 = \Gamma_\tau^0 = \Gamma_\psi^0 = \Gamma_\phi^0 = \Gamma_S^0 =
\frac{1}{32\pi}\frac{M_1^2}{v_{B-L}^2} m_S \left[1-\left(2 M_1^2/m_S\right)^2 \right]^{1/2} \,.
\label{eq:GammaS0}
\end{align}
The branching ratios for the respective final states turn out to be mostly trivial,
\begin{align}
\textrm{Br}\big(\sigma \rightarrow N_1N_1\big) = & \: 1-\left(2 M_1^2/m_S\right)^2 \,, \quad
\textrm{Br}\big(\sigma \rightarrow \tilde{N}_1 \tilde{N}_1^*\big) = \left(2 M_1^2/m_S\right)^2
\,, \label{eq:sigmatauphipsiBr}\\
\textrm{Br}\big(\tau \rightarrow N_1N_1\big) = & \:1 \,,\quad
\textrm{Br}\big(\psi \rightarrow N_1 \tilde{N}_1^*\big) = 1 \,,\quad
\textrm{Br}\big(\phi \rightarrow \tilde{N}_1 \tilde{N}_1\big) = 1 \,.
\end{align}


\subsubsection{Neutrino Sector}


By now, we have encountered three different production
mechanisms for heavy (s)neutrinos: tachyonic preheating $(\textrm{PH})$, the decay of
the \BmL vector boson and its superpartners $(G)$ and the decay of the particles
from the SSB sector $(S)$.
Beyond that, heavy (s)neutrinos may also be thermally produced $(\textrm{th})$,
i.e.\ through inverse decay processes in the thermal bath (cf.\ Sec.~\ref{sec:boltzmann}).
All of these mechanisms yield heavy (s)neutrinos components
$\big(N_i^{\textrm{x}},\tilde{N}_i^{\textrm{x}}\big)$, where
$\textrm{x} = \textrm{PH},G,S,\textrm{th}$,
with different characteristic energies.
Due to the effect of relativistic time dilatation, they thus all decay at different
rates $\Gamma_{N_i}^{\textrm{x}}$,
\begin{align}
\Gamma_{N_i}^{\textrm{x}} = \left<\frac{M_i}{E_{N_i}}\right>_{\textrm{x}} \Gamma_{N_i}^0 \,,
\label{eq:GammaxNi}
\end{align}
Here, $\Gamma_{N_i}^0$ denotes the zero-temperature (s)neutrino decay rate.
It directly follows from the seesaw superpotential (cf.\ Eq.~\eqref{eq:WSeesawM})
and is given by (cf.\ Eq.~\eqref{eq:Gamma0Ni}),
\begin{align}
\Gamma_{N_i}^0 = \frac{1}{4\pi} \frac{\widetilde{m}_i M_i}{v_u^2} M_i \,, \nonumber
\end{align}
For completeness, let us also mention in passing the branching ratios
for the various final states, into which the heavy (s)neutrinos can decay
(cf.\ Eq.~\ref{eq:Gamma0NiNti}),
\begin{align}
\textrm{Br}\big(N_i \rightarrow \ell H_u\big) = & \: \frac{1}{4} \,,\quad \label{eq:NNtBr}
\textrm{Br}\big(N_i \rightarrow \tilde{\ell} \tilde{H}_u\big) = \frac{1}{4} \,, \quad
\textrm{Br}\big(\tilde{N}_i \rightarrow \tilde{\ell} H_u\big) = \frac{1}{2} \,,\\
\textrm{Br}\big(N_i \rightarrow \bar{\ell} H_u^*\big) = & \: \frac{1}{4} \,,\quad
\textrm{Br}\big(N_i \rightarrow \tilde{\ell}^* \bar{\tilde{H}}_u\big) = \frac{1}{4} \,, \quad
\textrm{Br}\big(\tilde{N}_i \rightarrow \bar{\ell} \bar{\tilde{H}}_u\big) = \frac{1}{2} \,.
\end{align}
The prefactor of $\Gamma_{N_i}^0$ in Eq.~\eqref{eq:GammaxNi} is the inverse time
dilatation factor averaged over the momenta of all (s)neutrinos of a given component,
\begin{align}
\left<\frac{M_i}{E_{N_i}}\right>_{\textrm{x}} = \frac{1}{n_{N_i}^\textrm{x}}
\frac{g_{N_i}}{\left(2\pi\right)^3} \int d^3p \frac{M_i}{E_{N_i}} f_{N_i}^\textrm{x}(t,p) \,, \quad
E_{N_i} = \sqrt{p^2 + M_i^2} \,, \label{eq:dilatation}
\end{align}
with $n_{N_i}^\textrm{x}$ and $f_{N_i}^\textrm{x}$ being the number density and
the phase space distribution function of the (s)neutrino species
$\big(N_i^{\textrm{x}},\tilde{N}_i^{\textrm{x}}\big)$, respectively
(cf.\ Sec.~\ref{sec:boltzmann}).
Typically, these two quantities exhibit a nontrivial time dependence and thus
need to be determined by means of the Boltzmann equation for the (s)neutrinos
under study.

\cleardoublepage


\chapter{Nonperturbative Dynamics}
\label{ch:phasetransition}

The total decay rates and branching ratios which we derived from
the Lagrangian of the supersymmetric Abelian Higgs model in the previous chapter
are important ingredients to the study of the reheating process after the \BmL
phase transition.
Reheating is a \textit{perturbative} process, which we will investigate by means of
semiclassical Boltzmann equations in the next chapter.
For now, we shall focus on the \textit{nonperturbative} dynamics of the
\BmL phase transition.


As the symmetry breaking at the end of hybrid inflation proceeds very rapidly
and abruptly, it represents what is often referred to as a \textit{waterfall}
phase transition (cf.\ Sec.~\ref{subsec:infltn}).
It is accompanied by the production of local topological defects
in the form of cosmic strings as well as the nonadiabatic production
of particles coupled to the Higgs field $s$ during tachyonic preheating.
In this chapter, we will discuss these two nonperturbative processes in turn.
First, we will estimate the amount of cosmic strings produced during
the \BmL phase transition and use the current bound on the string tension
to constrain the parameters of hybrid inflation, \textit{viz.}\ the
\BmL breaking scale $v_{B-L}$ and the coupling constant
$\lambda$ (cf.\ Sec.~\ref{sec:strings}).
Subsequent to that, we will sketch the computation of the particle abundances
generated during tachyonic preheating (cf.\ Sec.~\ref{sec:preheating}),
which we will require as initial data for our study of the reheating process
in the next chapter.


\newpage


\section{Production of Cosmic Strings}
\label{sec:strings}


In the true vacuum of the Abelian Higgs model, the expectation value of the
Higgs field $s_+ = \sigma_+/\sqrt{2}e^{i\zeta_+}$ equals the \BmL breaking scale,
$\langle s_+^*s_+\rangle^{1/2} = v_{B-L}$.
The vacuum manifold of the Abelian Higgs model is thus isomorphic to the circle
and may be parametrized in terms of the Goldstone phase $\zeta_+ \in \left[0,2\pi\right)$.
Since the circle is not a simply connected manifold, i.e.\ since it exhibits a
\textit{nontrivial topology}, the classical field equations of the Abelian Higgs
model admit solitonic solutions, which describe one-dimensional \textit{topological defects}
known as cosmic strings.\footnote{For reviews on cosmic strings,
cf.\ for instance Refs.~\cite{Vilenkin:1984ib,Hindmarsh:1994re,Hindmarsh:2011qj}.}


In our case, the tension of a cosmic string equals its energy per unit length $\mu$,
\begin{align}
\mu = 2 \pi  v_{B-L}^2 B(\beta) \,, \quad
B\left(\beta\right) \simeq
\begin{cases}
1.04 \,\beta^{0.195}\,, & 10^{-2} \lesssim \beta \ll 1  \,, \\
2.4/\ln\left(2/\beta\right)\,, & \beta \lesssim 10^{-2} \,,
\end{cases}
\label{eq:mu}
\end{align}
where $\beta = \lambda / (8 g^2)$.
The characteristic distance $\xi$ separating two strings at the moment of their formation
is approximately given as~\cite{Copeland:2002ku},
\begin{align}
\xi \approx (- \lambda \, v_{B-L} \, \dot{\varphi}_c)^{-1/3} \,,
\label{eq:xi}
\end{align}
with $\dot{\varphi}_c$ denoting the velocity of the inflaton field $\varphi$
at the time when it reaches its critical value $\varphi_c = v_{B-L}$.
We obtain $\dot{\varphi}_c$ directly from the Klein-Gordon equation, i.e.\
the equation of motion for the inflaton field (cf.\ Eq.~\eqref{eq:KleinGordon}),
\begin{align}
- \dot{\varphi}_c = \frac{1}{3H}\left(V'+ \ddot{\varphi}\right) \,.
\end{align}
For small and intermediate values of the coupling constant, $\lambda \lesssim 10^{-3}$,
inflation ends because of the tachyonic instability in the scalar potential
(cf.\ Eq.~\eqref{eq:phiecsr}).
The slow-roll approximation is then valid all the way until
the onset of the \BmL phase transition (cf.\ Eq.~\eqref{eq:KleinGordonSR}),
so that we may neglect $\ddot{\varphi}$ in the calculation of $\dot{\varphi}_c$
and approximate the Hubble rate $H$ by $H_I$ (cf.\ Eq.~\eqref{eq:HINe}),
\begin{align}
\lambda \lesssim 10^{-3} \::\qquad
- \dot{\varphi}_c \approx \frac{V'}{3 H_I} \,.
\label{eq:phidotsr}
\end{align}
By contrast, for large values of the coupling constant, $\lambda \gtrsim 10^{-3}$,
inflation ends because the slow-roll condition $\left|\eta_V\right| \ll 1$
becomes violated at some inflaton field value $\varphi_{\textrm{sr}} \gg \varphi_c$.
In this case, we are hence not allowed to compute $\dot{\varphi}_c$ within the
slow-roll approximation.
The expression for $\dot{\varphi}_c$ in Eq.~\eqref{eq:phidotsr} remains, however,
useful nonetheless, as it now provides us with an upper estimate of $\dot{\varphi}_c$,
\begin{align}
\lambda \gtrsim 10^{-3} \::\qquad
- \dot{\varphi}_c < \frac{V'}{3H} <  \frac{V'}{3 H_I} \,,
\label{eq:phidotcbound}
\end{align}
where we have used that $\ddot{\varphi} < 0$ and $H > H_I$.
Note that for the range of $v_{B-L}$ and $\lambda$ values that we are interested in
the only relevant contribution to the inflaton potential is the radiative
Coleman-Weinberg correction $V_{\textrm{CW}}$, while the terms
induced by SUGRA are negligible~\cite{Buchmuller:2000zm,Battye:2006pk}.
In summary, given the expressions for $H_I$ and $V_{\textrm{CW}}$ in
Eqs.~\eqref{eq:HINe} and \eqref{eq:VCM} and irrespectively of whether
$\left|\eta_V\right| \ll 1$ or not, we find that the absolute value
of $\dot{\varphi}_c$ cannot be larger than
\begin{align}
\frac{V'}{3 H_I} = \frac{\ln4}{64\sqrt{6}\pi^{5/2}}\,
\lambda^{3/2} \, M_P \, v_{B-L} \simeq \lambda^{3/2}\, v_{B-L}^2
\left(\frac{6\times 10^{15}\,\textrm{GeV}}{v_{B-L}}\right) \,.
\label{eq:Vp3HI}
\end{align}


With the aid of Eqs.~\eqref{eq:mu} and \eqref{eq:xi}, the energy density $\rho_{\text{string}}$
stored in cosmic strings just after the end of the \BmL phase transition can be calculated as
\begin{align}
\rho_{\text{string}} = \frac{\mu}{\xi^2} \,.
\end{align}
In the following, we will always employ the slow-roll expression for
$\dot{\varphi}_c$ (cf.\ Eq.~\eqref{eq:Vp3HI}) in the calculation of $\rho_{\text{string}}$.
For $\lambda \lesssim 10^{-3}$, our results for $\rho_{\text{string}}$ will hence represent
actual estimates of the string energy density, while for $\lambda \gtrsim 10^{-3}$,
given Eq.~\eqref{eq:phidotcbound} and the fact that $\rho_{\text{string}}$ scales like
$\left(-\dot{\varphi}_c\right)^{2/3}$, we will merely obtain upper bounds on the
string energy density.
As a first observation, we note that the fraction $\rho_{\text{string}} / \rho_0$
of the total energy stored in cosmic strings directly after the \BmL phase transition
monotonically increases with $\lambda$.
This is because larger values of $\lambda$ entail higher string tensions
as well as shorter average distances between two strings.
Some numerical results, illustrating the increasing efficiency of the
production of cosmic strings for larger $\lambda$ values, are listed in
Tab.~\ref{tab:strings}.\footnote{Including SUGRA corrections into the inflaton
potential on condition of a canonical K\"ahler potential~\cite{Linde:1997sj},
these results remain practically unchanged.
Given a nonminimal K\"ahler potential, inflation can no longer be successfully
realized for $\lambda < \lambda_{\textrm{min}}$, with $\lambda_{\textrm{min}}$ depending on
the coefficients of the noncanonical terms in the K\"ahler potential~\cite{Nakayama:2010xf}.
In this case, our results agree with the outcome of the full calculation,
based on the complete potential, as long as $\lambda \gtrsim \lambda_{\textrm{min}}$.}


\begin{table}
\centering
\begin{tabular}{c|ccccccc}\hline \hline
$\lambda$ & $10^{-8}$ & $10^{-7}$ & $10^{-6}$ & $10^{-5}$ & $10^{-4}$ & $10^{-3}$ & $10^{-2}$\\\hline
$H^{-1}/\xi$ & $5$ & $10$ & $20$ & $50$ & $100$ & $200$ & $\lesssim 500$ \\
$\rho_{\text{string}}/\rho_0 \,\left[\%\right]$  &
$0.002$ & $0.01$ & $0.07$ & $0.4$ & $2$ & $10$ & $\lesssim 80\phantom{0}$ \\\hline\hline
\end{tabular}
\medskip
\caption[Efficiency of the production of cosmic strings]
{Efficiency of the production of cosmic strings.
$H^{-1}/\xi$, the ratio of the Hubble radius to the string separation scale, is a measure
for the abundance of cosmic strings at the end of the \BmL phase transition, while
$\rho_{\text{string}}/\rho_0$, the string energy density relative to the initial false
vacuum energy density, indicates the amount of energy stored in cosmic strings directly after
symmetry breaking.
The numbers shown in this table correspond to $v_{B-L}$ kept
fixed at $5 \times 10^{15}\,\textrm{GeV}$.}
\label{tab:strings}
\end{table}


Subsequent to their production, the cosmic strings may intercommute
with each other, which leads to the formation of closed string loops next
to the abundance of infinitely extended strings.
These string loops oscillate, thereby loosing energy into gravitational waves
as well as into the Higgs and gauge DOFs~\cite{Hindmarsh:2008dw,Dufaux:2010cf}.
After a relaxation time $t_{\text{string}}$, roughly given by
the distance scale $\xi$~\cite{Hindmarsh:2008dw}, the cosmic string network
enters the \textit{scaling regime}, which is characterized by an abundance of
only $\mathcal{O}(1)$ cosmic strings per Hubble volume and a string energy
density $\rho_{\text{string}}$ scaling like $H^2 M_P^2$.
Relic cosmic strings are a possible source for primordial density
fluctuations, gravitational lensing as well as gravitational waves.
Hence, if cosmic strings are indeed generated during some phase transition in
the early universe, they should reveal their existence in a variety of present-day
cosmological observations.
The fact that no effects related to cosmic strings have
been observed so far implies an upper bound on the string
tension~\cite{Battye:2010xz,Dunkley:2010ge,Urrestilla:2011gr,Dvorkin:2011aj}.
The actual values quoted in the literature partly differ from each other.
In this thesis, we will work with the following representative value,
\begin{align}
G \mu \lesssim 5 \times 10^{-7} \,,
\end{align}
with $G =  M_P^{-2}$ being Newton's constant.
This constraint directly translates into an upper bound on $v_{B-L}$,
which weakly depends on $\lambda$ (cf.\ Eq.~\eqref{eq:mu}),
\begin{align}
v_{B-L} \lesssim 1.8 \times 10^{-4}  \left( \ln \frac{16 g^2}{\lambda} \right)^{1/2} M_P \,.
\end{align}
We conclude that, for reasonable values of the coupling constant
$\lambda$, the nonobservation of cosmic strings alone already
excludes the possibility of \BmL breaking taking place \textit{above} the GUT
scale, i.e.\ $v_{B-L} \lesssim 1 \times 10^{16}\,\textrm{GeV}$ for
$\lambda > 10^{-20}$.


Similarly to us, the authors of Ref.~\cite{Nakayama:2010xf} also discuss
the production of cosmic strings during the \BmL phase transition at the end
of supersymmetric $F$-term hybrid inflation.\footnote{Cf.\ also the analyses in
Refs.~\cite{Battye:2006pk,Battye:2010hg,Jeannerot:2005mc}.}
Combining the requirement of successful inflation with bounds on the parameter
space inferred from the amplitude of the CMB power spectrum $A_s$ (cf.\ Eq.~\eqref{eq:AsnsWMAP})
as well as the nonobservation of cosmic strings~\cite{Battye:2010xz},
they find consistency among all observations for
\begin{align}
3 \times 10^{15}\,\textrm{GeV} \lesssim v_{B-L} \lesssim 7 \times 10^{15}\,\textrm{GeV}
\,, \quad 10^{-4} \lesssim  \sqrt{\lambda}  \lesssim 10^{-1} \,.
\label{eq:vBLlambdabounds}
\end{align}
Note in particular that, up to a factor of $\mathcal{O}(1..10)$, the range of
\textit{viable} $v_{B-L}$ values is compatible with the upper boundary of
the range of \textit{preferred} $v_{B-L}$ values, which we found in our
Monte-Carlo study of the Froggatt-Nielsen model (cf.\ Eq.~\eqref{eq:vBLMCrange}).
The synopsis of our results in Eqs.~\eqref{eq:vBLAs}, \eqref{eq:vBLMCrange}
and \eqref{eq:vBLlambdabounds} hence leads to the conclusion that the flavour
charge $a$, which controls the magnitude of $v_{B-L}$ relative to mass scale
$M_0$ (cf.\ Eq.~\eqref{eq:vBLaM1d}), should be zero.
In addition to that, the results of Ref.~\cite{Nakayama:2010xf} in
Eq.~\eqref{eq:vBLlambdabounds} imply a proportionality factor of about $5$
in Eq.~\eqref{eq:vBLaM1d},
\begin{align}
a = 0 \,,\quad v_{B-L} \simeq 5 \times \eta^{2a} M_0 \simeq 5 \times 10^{15}\,\textrm{GeV}\,.
\label{eq:vBLrange}
\end{align}
A factor of $5$ can still be accounted for by the Froggatt-Nielsen model, since
it is \textit{triple} products of $\mathcal{O}(1)$ factors which enter into the calculation
of $v_{B-L}$ from the neutrino mass matrices $M$ and $m_D$.
Likewise, the viable range of $\lambda$ values can be translated into constraints
on the flavour charge $d$ (cf.\ Eqs.~\eqref{eq:hijlambda} and \eqref{eq:ebc}),
\begin{align}
\sqrt{\lambda} \sim \eta^e \,,\quad e = 2(d-1) \,,\quad 1.4 \lesssim d \lesssim 2.6 \,,
\label{eq:drange}
\end{align}
and therefore, according to Eq.~\eqref{eq:vBLaM1d}, also into constraints on the heavy-(s)neutrino
mass $M_1$.
The parameter space, which we shall investigate in this thesis, is hence defined by
(cf.\ Eqs.~\eqref{eq:m1trange}, \eqref{eq:vBLrange} and \eqref{eq:drange}),
\begin{align}
& \: 10^{-5}\,\textrm{eV} \leq \widetilde{m}_1 \leq 1 \,\textrm{eV} \,,\label{eq:parameterspace} \\
& \: v_{B-L} = 5 \times 10^{15}\,\textrm{GeV} \,, \nonumber \\
& \: 10^9\,\textrm{GeV} \leq M_1 \leq 3 \times 10^{12}\,\textrm{GeV} \,. \nonumber
\end{align}


The production and decay of cosmic strings can in principle have a large influence
on the further evolution of the universe after the \BmL phase transition.
However, as we will argue in the following, for our purposes it is not
necessary to consider any processes related to cosmic strings in more detail
as long as we restrict ourselves to the parameter space in Eq.~\eqref{eq:parameterspace}.
By means of Eq.~\eqref{eq:xi}, we find that for $\lambda \sim 10^{-3}..10^{-2}$
the number of cosmic strings per Hubble volume $N_s = \left(H^{-1}/\xi\right)^2$ is
as large as $\mathcal{O}\left(10^5\right)$.
For $\lambda \gtrsim 6 \times 10^{-3}$, the upper bound on the fractional
string energy density exceeds $50\,\%$.
In the case of such large $\lambda$ values, the cosmic string relaxation time,
$t_{\text{string}} \sim \xi \sim 10^{-3} H^{-1}$,
is however much shorter than the Hubble time $H^{-1}$.
Most of the string energy is hence converted back into Higgs and gauge DOFs before the
reheating processes has really begun.
The exact mechanism of energy loss of cosmic strings is not yet fully understood,
which prompts us to refrain from attempting to precisely describe it.
As an important result, we should, however, keep in mind that \textit{at the very most}
about half of the initial false vacuum energy density may be processed via an intermediate
population of cosmic strings into the particles of the Higgs and gauge multiplets.
Generically, the effects of cosmic strings are much less important.
For $\lambda \lesssim 10^{-4}$, for instance, their relative energy contribution
is at the level of at most $\mathcal{O}\left(1\,\%\right)$.
Due to supersymmetry, the additional higgsinos produced in the decay of cosmic
strings decay into the same supermultiplet, \textit{viz.}\ $n_1^c$, as the Higgs bosons
produced during tachyonic preheating (cf.\ Sec.~\ref{subsec:rates}).
We thus expect that it should not make a qualitative difference, whether the extra
higgsinos produced in string decays are taken into account in the description
of the reheating process or not.
Meanwhile, the extra gauge particles produced in string decays predominantly decay
into radiation, which is subsequently quickly diluted during the phase of matter
domination after preheating.
We hence claim that the reheating process after the \BmL phase transition is, in fact,
mostly insensitive to all string-induced modifications of its initial conditions.
In a numerical study of the reheating process similar to the one discussed
in Ch.~\ref{ch:reheating}, we are able to confirm this claim.
Considering the case of extremal string production, we shift half of the energy
initially stored in the Higgs bosons at the end of preheating into the gauge
DOFs and calculate the resulting abundances of entropy, baryon asymmetry and
gravitino dark matter.
We find no deviations from the results presented in
Section~\ref{sec:evolution} above the percent level.
These findings serve us a justification to neglect all effects related to
the production and decay of cosmic strings in the remainder of this thesis.


\section{Tachyonic Preheating}
\label{sec:preheating}


Next to the production of cosmic strings, the decay of the false vacuum entails a second
nonperturbative process, \textit{viz.}\ tachyonic preheating~\cite{Felder:2000hj,Felder:2001kt},
which also results in particle production~\cite{GarciaBellido:2001cb}.
In this section, we shall first discuss tachyonic preheating, i.e.\ the transfer of
the false vacuum energy into a gas of nonrelativistic Higgs bosons, and then the
nonadiabatic particle production associated with it.


\subsubsection{Energy Transfer from the False Vacuum to Nonrelativistic Higgs Bosons}


When the inflaton field $\varphi$ reaches its critical value $\varphi_c = v_{B-L}$,
the scalar potential develops a tachyonic instability in the direction of the waterfall
field $\sigma$.
As we shall demonstrate now, this triggers the growth of the long-wavelength modes
of the Higgs field $s_+ = \sigma_+/\sqrt{2}e^{i\zeta_+}$ at an exponential rate.
Neglecting the expansion of the universe during the \BmL phase transition,
the linearized equations of motion of the Higgs mode functions $s_+(k,t)$
are given by (cf.\ Eqs.~\eqref{eq:Vphispm} and \eqref{eq:mspm}),
\begin{align}
\ddot{s}_+(k,t) + \left(k^2 + m_{s_+}^2 \right) s_+(k,t) = 0 \,,\quad
m_{s_+}^2 = \frac{\lambda}{2}\left(\varphi^2 - \varphi_c^2\right) \,, \quad
\varphi \leq \varphi_c \,.
\label{eq:sEOMs}
\end{align}
The solutions of these mode equations corresponding to momenta $k$ which
satisfy $k^2 < -m_{s_+}^2$ are indeed superpositions of a growing and a decaying
exponential.
For a fixed value of the inflaton field $\varphi$, for instance, we may explicitly write
\begin{align}
s_+(k,t) = A(k) \exp\left(\omega_k t\right) + B(k) \exp\left(- \omega_k t\right) \,,\quad
\omega_k = \sqrt{-m_{s_+}^2 - k^2} \,. \label{eq:modesol}
\end{align}
This spinodal growth of the long-wavelength modes directly translates into
an exponential growth of the variance $v^2$  of the Higgs field (cf.\ Eq.~\eqref{eq:v2tmodes})
as well as of the occupation numbers $n_k$ of the respective modes~\cite{GarciaBellido:2001cb},
\begin{align}
v^2(t) = \left<s_+^*s_+\right> = \int\frac{d^3k}{\left(2\pi\right)^3} \left|s_+(k,t)\right|^2
\,,\quad n_k(t) = \left|s_+^*(k,t)\dot{s}_+(k,t)\right| - \frac{1}{2} \,,
\end{align}
Tachyonic preheating continues, until the nonlinear term in the full mode
equations, which is induced by the quartic Higgs self-coupling, begins to
compensate for the negative mass squared, i.e.\ when the curvature
of the Higgs potential vanishes.
As an important detail, note that, given the initial conditions $s_+ = \dot{s}_+ = 0$,
the classical component of the Higgs field $\left<s_+\right>$ never acquires a VEV.
This illustrates that the decay of the false vacuum is a purely quantum
mechanical process, which is solely driven by the exponentially growing quantum
fluctuations of the Higgs field.
As anticipated in Sec.~\ref{subsec:transitions}, the fact that $\left<s_+\right> = 0$
also implies that \BmL actually never becomes broken during the phase transition, but merely hidden.


At the end of preheating, the occupation numbers of the long-wavelength
Higgs modes are exponentially large, which allows for a treatment of these modes as
an ensemble of colliding \textit{semiclassical waves}~\cite{GarciaBellido:2002aj}.
Another interpretation of the large occupation numbers at low momenta, the one which we
will adopt in the following, is that after preheating the universe is filled by
a dominating abundance of nonrelativistic Higgs bosons, i.e.\ \textit{particles in coherent
quantum states}.
Finally, we recapitulate that the \BmL phase transition comes to an end once the Higgs
VEV has become as large as the \BmL breaking scale, $\langle s_+^*s_+\rangle^{1/2} = v_{B-L}$.
To study the evolution of $v$ at late times, when the nonlinear Higgs dynamics can no
longer be neglected, one has to resort to numerical lattice simulations.
In the quench approximation, for instance, matching the exponentially growing,
analytic solution for $v$ at early times with
the numerical solution for $v$ at late times, one finds that the time evolution
of $v$ is well described by the smooth step function in Eq.~\eqref{eq:vtapprox}.
This implies in particular that the energy transfer from the false vacuum into
Higgs bosons is a very fast process, which typically completes within a single oscillation
of the scalar field distribution.


\subsubsection{Waterfall Conditions and Quench Approximation}


For a broad range of parameters, the false vacuum decays almost instantaneously as soon
as the inflaton field reaches its critical value.
By definition, this is equivalent to the statement that the \BmL phase transition
generically takes place in the \textit{waterfall regime}.
The realization of the waterfall regime is subject to
two conditions, which were first formulated by Linde with respect to
its original nonsupersymmetric version of hybrid inflation~\cite{Linde:1993cn}.
To ensure a rapid and abrupt vacuum decay, once the tachyonic instability
in the scalar potential has appeared, two time scales,
$\Delta t_\sigma$ and $\Delta t_\phi$, must be much shorter than the Hubble time,
\begin{align}
\Delta t_\sigma,\,\Delta t_\phi \ll H_I^{-1} \,.
\label{eq:waterfallcond}
\end{align}
$\Delta t_\sigma$ denotes the time it takes until the negative mass squared
of the waterfall field has become sizable, so that the exponential growth of the
Higgs quantum fluctuations sets in.
Meanwhile, $\Delta t_\phi$ characterizes the time scale, on which the inflaton
field value changes after the onset of symmetry breaking.
It is hence also a measure for the time it takes for the inflaton to reach
the minimum of the scalar potential.
In Ref.~\cite{Linde:1993cn}, Linde translates the two requirements in
Eq.~\eqref{eq:waterfallcond} into relations among the parameters of his model,
to which he refers as the two \textit{waterfall conditions}.
We shall now extend Linde's results to the supersymmetric case and derive the two
waterfall conditions for supersymmetric $F$-term hybrid inflation.


The relevant DOFs driving tachyonic preheating are the semiclassical
infrared modes of the Higgs field $s_+$ (cf.\ Eq.~\eqref{eq:modesol}).
Among all long-wavelength modes, the $k=0$ mode grows in particular
at the largest rate, $\omega_0 = \left|m_{s_+}\right|$.
We may thus define $\Delta t_\sigma$ as the time scale, on which the growth of
this mode sets in, $\Delta t_\sigma = \omega_0^{-1}$.
Expanding $\omega_0^2$ in a Taylor series up to first order in $\Delta t_\sigma$ around
$\Delta t_\sigma = 0$ yields
\begin{align}
\omega_0^2 = - m_{s_+}^2 = - \frac{\lambda}{2}\left(\varphi^2 - \varphi_c^2\right) \approx
- \frac{\lambda}{2} \left.\frac{d}{dt}\right|_{t_c} \varphi^2 \Delta t_\sigma \,, \quad
\varphi\left(t_c\right) = \varphi_c \,.
\end{align}
Identifying $\omega_0^2$ on the left-hand side of this relation with $\Delta t_\sigma^{-2}$
and inserting the slow-roll expression for $\dot{\varphi}_c$ (cf.\ Eq.~\eqref{eq:Vp3HI}),
we obtain the first waterfall condition,
\begin{align}
\Delta t_\sigma^{-3} \approx - \lambda \varphi_c \dot{\varphi}_c \gg H_I^3 \,,\quad
\sqrt[4]{\lambda} M_P \gg \frac{4\pi}{\sqrt[4]{3\ln4}} v_{B-L} \,,
\end{align}
which may be cast into the form of a lower bound on the coupling constant $\lambda$,
\begin{align}
\lambda \gg \frac{256\pi^4}{3\ln3}\left(\frac{v_{B-L}}{M_P}\right)^4 \simeq
2 \times 10^{-10} \left(\frac{v_{B-L}}{5 \times 10^{15}\,\textrm{GeV}}\right)^4\,.
\end{align}
In the parameter space under study (cf.\ Eq.~\eqref{eq:parameterspace}),
it is certainly fulfilled.


The time scale governing the inflaton dynamics is given by the time-dependent mass
of the inflaton field which it acquires in the course of symmetry breaking,
$\Delta t_\phi = m_\phi^{-1} = \big(\sqrt{\lambda}v\big)^{-1}$ (cf.\ Eq.~\eqref{eq:mphi}).
Shortly after the onset of tachyonic preheating, the Higgs VEV $v$ grows
exponentially (cf. Eqs.~\eqref{eq:v2tmodes} and \eqref{eq:modesol}), so that
the condition $\Delta t_\phi = m_\phi^{-1}$ is typically satisfied at rather early
times, $\Delta t_\phi = C_\phi m_S^{-1}$, with $C_\phi$ being of
$\mathcal{O}(10)$~\cite{GarciaBellido:2001cb}.
The second waterfall condition then assumes the form of an upper bound on the
possible values of the \BmL scale,
\begin{align}
\Delta t_\phi^{-1} = \frac{m_S}{C_\phi} \gg H_I \,,\quad
v_{B-L} \ll \sqrt{\frac{3}{2\pi}}\frac{M_P}{C_\phi}
\simeq \frac{8}{C_\phi} \times 10^{18}\,\textrm{GeV} \,.
\end{align}
Also the second waterfall condition is clearly fulfilled in our parameter space.
Moreover, the fact that $C_\phi \gg 1$ illustrates
that the inflaton reaches the minimum of the scalar potential on a time
scale much shorter than the actual duration of symmetry breaking, which only
comes to an end once $m_\phi = m_S$.


In conclusion, we find that for the values of the parameters $v_{B-L}$ and
$\lambda$, which we shall consider in this thesis, (i) the negative mass squared of
the waterfall field essentially pops up, once $\varphi$ reaches its critical value,
and that (ii) the inflaton field rapidly rolls down towards the minimum of the scalar potential.
To simplify our investigation of tachyonic preheating, we may therefore henceforth
work in the \textit{quench approximation}, in which the inflaton field $\varphi$
is straight away set to zero as soon as the waterfall field becomes unstable.
This is to say that we do not slowly turn on the negative mass squared $m_{s_+}^2$
of the waterfall field, but instead instantly introduce it with its largest possible
absolute value,
\begin{align}
\textrm{Quench approximation:}\qquad
\varphi \leq \varphi_c \quad\rightarrow\quad
m_{s_+}^2 = - \frac{\lambda}{2} \, v_{B-L} \,.
\end{align}
Physically, the quench approximation corresponds to the limiting case of a
particularly rapid and abrupt waterfall transition.
It is applicable, if $v_{B-L}$ and $\lambda$ clearly satisfy the two waterfall
conditions, which is definitely guaranteed in our case.
The fact that the quench approximation holds for the parameter values
that we are interested in also serves as an \textit{a posteriori} justification for
the omission of the Hubble expansion in our discussion of tachyonic
preheating (cf.\ Eq.~\eqref{eq:sEOMs}).


\subsubsection{Nonadiabatic Particle Production During Tachyonic Preheating}


The sudden change in the masses of the particles coupled to the Higgs
field $s_+$ leads to the nonadiabatic production of these particles
during tachyonic preheating, which can be studied using the formalism
of quantum fields in strong backgrounds.
In the quench approximation, employing the smooth step function in Eq.~\eqref{eq:vtapprox}
to describe the time evolution of the Higgs VEV $v$, it is even feasible to derive
analytic expressions for the occupation numbers $n_k^B$ and $n_k^F$ of the produced
bosons and fermions~\cite{GarciaBellido:2001cb}.
One obtains $n_k^B$ and $n_k^F$ in three steps.
First of all, one has to rewrite the mode equations of the bosonic and fermionic fields
coupled to the Higgs field $s_+$ as oscillator equations with time-dependent and partly
complex frequencies.
Then, one needs to solve these oscillator equations in terms of hypergeometric functions.
In the third and last step, one can use the solutions of the mode equations to compute
the Bogoliubov coefficients, relating the mode functions in the asymptotic past to the
mode functions in the asymptotic future.
The Bogoliubov coefficients then directly yield the desired occupation numbers.
For all bosons and fermions the masses of which increase linearly with the Higgs VEV,
$m_B  = h_B v$ and $m_F = h_F v$, one finds
\begin{align}
n_k^B = & \: \frac{\cosh\left[\pi\sqrt{4\alpha^2-1}\right]-
\cosh\left[2\pi\left(\omega_+-\omega_-\right)/m_S\right]}{
\sinh\left[2\pi\omega_-/m_S\right]
\sinh\left[2\pi\omega_+/m_S\right]} \,, \label{eq:nkBnkF}\\[6pt]
n_k^F = & \: \frac{\cosh\left[2\pi\alpha\right]-
\cosh\left[2\pi\left(\omega_+-\omega_-\right)/m_S\right]}{
2\sinh\left[2\pi\omega_-/m_S\right]
\sinh\left[2\pi\omega_+/m_S\right]} \,. \nonumber
\end{align}
Here, $\alpha = m_i/m_S$ denotes the ratio of the mass $m_i$ of the respective
particle $i$ in the true vacuum to the Higgs boson mass $m_S$.
Meanwhile, $\omega_-$ and $\omega_+$ are the asymptotic \textit{in} and
\textit{out frequencies}, $\omega_- = k$ and $\omega_+ = \sqrt{k^2 + m_i^2}$.
Apparently, $n_k^B$ and $n_k^F$ are largest for the low-momentum modes, $k \ll m_i$.
We hence conclude that tachyonic preheating primarily entails the production
of \textit{nonrelativistic} particles.


According to our analysis in Sec.~\ref{subsec:physicalDOFs}, the masses of
the inflaton $\phi$, the higgsino $\psi$ as well as all particles from the
gauge and neutrino sector increase linearly with the Higgs VEV $v$,
such that the formulae in Eq.~\eqref{eq:nkBnkF} can be readily used to calculate
the occupation numbers of these particles after preheating.
The real Higgs scalar $\tau$, however, represents an exception, which is not covered
by the results in Eq.~\eqref{eq:nkBnkF}.
As compared to, for instance, $\phi$ and $\psi$, the scalar $\tau$ comes with
a constant contribution to its mass $m_\tau$ (cf.\ Eq.~\eqref{eq:mtau2}), which directly
derives from $m_{s_-}$ in the quench approximation, $m_{s_-} = \sqrt{\lambda}/2 \,v_{B-L}$
(cf.\ Eq.~\eqref{eq:mspm}).
Note in particular that it is this constant mass term, which stabilizes the
Higgs field $s$ in the $\tau$ direction, so that the role of the waterfall
direction is solely due to the Higgs field $\sigma$ (cf.\ Eq.~\eqref{eq:Vstau}).
Neglecting the expansion of the universe, the linearized mode equation for
the Higgs field $\tau$ takes the same form as the mode equations for all other
scalars coupled to $s_+$,
\begin{align}
\ddot{\tau}(t,k) + \left(k^2 + m^2_\tau(t)\right) \tau(t,k) = 0 \,,
\end{align}
the only difference being the constant contribution to the mass squared, which
is absent in the case of the other scalars.
To restore the standard form of the mode equation,
we simply absorb the constant mass term in the momentum $k$.
In the language of Ref.~\cite{GarciaBellido:2001cb}, this is equivalent to a
shift in the asymptotic \textit{in} frequency,
\begin{align}
k^2 \rightarrow k^2 + \lambda/2\, v_{B-L}^2 \,,\quad
\omega_-(k) = k \rightarrow \sqrt{k^2 + \lambda/2\, v_{B-L}^2} \,.
\end{align}
Performing this shift in the expression for $n_k^B$ in Eq.~\eqref{eq:nkBnkF}
results in a drastic suppression of the $\tau$ occupation numbers.
We find that, due to the large initial $\tau$ mass, the production of $\tau$
quanta during tachyonic preheating is less efficient by as much as four orders
of magnitude for the smallest and hence most important momenta.
This is physically intuitive, as the presence of a constant mass term already
at the onset of symmetry breaking implies that a larger amount of energy is necessary
to excite a given mode.
On top of that, the only tree-level decay channel of the Higgs boson $\tau$
is into a pair of neutrinos $N_1$, whose production is vastly dominated by
the decays of the much more abundant Higgs boson $\sigma$ (cf.\ Sec.~\ref{subsec:rates}).
Because of their low abundance as well as their inferiority to the $\sigma$ bosons,
we will completely neglect the $\tau$ particles in our analysis in
Chs.~\ref{ch:reheating} and \ref{ch:wimp}.


For all other particles produced during tachyonic preheating,
we will employ the following formulae for the final number and energy
densities at $t = t_{\textrm{PH}}$, which have been obtained by the
authors of Ref.~\cite{GarciaBellido:2001cb} by fitting the results of
their nonlinear numerical lattice simulations to the function
$f(\alpha, \gamma) = \sqrt{\alpha^2 + \gamma^2} - \gamma$,\footnote{The
nonadiabatic production of particles during the \BmL phase
transition can be significantly enhanced by quantum effects
\cite{Berges:2010zv}, which, however, require further investigations.}
\begin{align}
n_B\left(t_{\textrm{PH}}\right) \simeq & \:
1 \times 10^{-3} g_i m_S^3 f(\alpha, 1.3)/\alpha \,, \label{eq:rhonBF} \\
n_F\left(t_{\textrm{PH}}\right) \simeq & \:
3.6 \times 10^{-4} g_i m_S^3 f(\alpha, 0.8)/\alpha \,, \nonumber \\
\rho_B\left(t_{\textrm{PH}}\right)/\rho_0 \simeq & \:
2 \times 10^{-3} \, g_i \, \lambda \, f(\alpha, 1.3) \,, \nonumber \\
\rho_F\left(t_{\textrm{PH}}\right)/\rho_0 \simeq & \:
1.5 \times 10^{-3} \, g_i \, \lambda \, f(\alpha, 0.8) \,, \nonumber
\end{align}
with $g_i$ counting the internal DOFs of the respective particles (cf. Fig.~\ref{fig:overview}).
Just as the $\sigma$ bosons themselves, these particles are mostly produced
with very low momenta, i.e.\ nonrelativistically.
Furthermore, we observe that the total energy fraction transferred from the
Higgs background into bosonic and fermionic DOFs typically ends up being quite small.
The backreaction of the produced particles with the Higgs field
is consequently expected to be insignificant, justifying \textit{a posteriori}
our procedure of first determining the time evolution of the Higgs VEV and then
using this Higgs VEV as a homogeneous background in the mode equations.
Another interesting consequence of Eq.~\eqref{eq:rhonBF} is that
the decay of the gauge DOFs produced during tachyonic preheating yields
equal amounts of scalar and fermionic particles, with regard to each matter
multiplet into which the gauge DOFs can decay.
This directly follows from combining the branching ratios in
Eq.~\eqref{eq:AAtCBr} with the fact that the number densities $n_B$
and $n_F$ are proportional to $g_i$.

\cleardoublepage


\chapter{The Reheating Process}
\label{ch:reheating}

Tachyonic preheating after hybrid inflation and the subsequent decay of
the \BmL gauge multiplet set the stage for the emergence of the hot early universe.
As outlined in Sec.~\ref{subsec:neutrinoreheating}, the reheating process after the
\BmL phase transition is driven by the decay of thermally and nonthermally produced
heavy (s)neutrinos, which entails the generation of a primordial lepton asymmetry as a byproduct.
Meanwhile, inelastic 2-to-2 scatterings in the thermal bath
generate an abundance of gravitinos, which might either give rise to gravitino
or WIMP dark matter.


In this chapter, we will now elaborate on the reheating process
and demonstrate in particular that it is indeed capable of engendering the observed BAU
$\eta_B^{\textrm{obs}}$ (cf.\ Sec.~\ref{subsec:baryncmttr})
as well as the relic density of dark matter
$\Omega_{\textrm{DM}}^{\textrm{obs}}h^2$ (cf.\ Sec.~\ref{subsec:DM}), i.e.\
successfully generating the initial conditions of the hot early universe.
Postponing the discussion of WIMP dark matter to the next chapter,
we will now take the gravitino to be the LSP and consider the possibility
of gravitino dark matter.
The appropriate tool to track the cosmic evolution after the \BmL phase transition
quantitatively are the Boltzmann equations for the various particle species under study.
After carefully deriving them (cf.\ Sec.~\ref{sec:boltzmann}),
we will first solve these Boltzmann equations for a representative
choice of parameter values (cf.\ Sec.~\ref{sec:evolution}).
This will provide us with a detailed and time-resolved description of all particle abundances.
Then, we will carry out a scan of the parameter space, which will allow
us to determine relations between neutrino and superparticle masses
(cf.\ Sec.~\ref{sec:scan}).


The results presented in this chapter were first published in
Refs.~\cite{Buchmuller:2010yy,Buchmuller:2011mw,Buchmuller:2012wn}.


\newpage


\section{Boltzmann Equations}
\label{sec:boltzmann}


The relevant physical particles remaining after the decay of the gauge DOFs are:
(i) the Higgs boson $\sigma$, the inflaton $\phi$ and the higgsino $\psi$,
(ii) all three generations of heavy (s)neutrinos $\big(N_i,\tilde{N}_i\big)$,
(iii) all MSSM particles, and (iv) the gravitino $\widetilde{G}$.
In this section, we will successively derive the Boltzmann equations
which respectively describe the evolution of these particle species in an
expanding Friedmann-Lema\^itre background.
The formalism of Boltzmann equations as well as our notational
conventions are summarized in App.~\ref{sec:kinetic}.
Depending on the particle species, we will either work with the actual
Boltzmann equation for the respective phase space distribution function $f_X$
or with the integrated Boltzmann equation for the respective comoving
number density $N_X$, which counts the number of $X$ particles in a comoving
spatial volume of size $a^3$,
\begin{align}
N_X (t) = a^3 n_X(t) =
a^3 \frac{g_X}{\left(2\pi\right)^3}\int d^3p \,f_X(t,p) \,.
\label{eq:NXa3nX}
\end{align}
Here, $n_X$ denotes the number density of particle species $X$ (cf.\ Eq.~\ref{eq:numdendef})
and $a$ is the cosmic scale factor, which accounts for the expansion of
the universe in the Robertson-Walker metric.
All physical observables are invariant under a rescaling of the scale factor.
For convenience, we thus set $a_{\textrm{PH}} = a\left(t_{\textrm{PH}}\right) = 1$
in our analysis, where $t_{\textrm{PH}} = 0$ corresponds to the time at the end of
preheating (cf.\ Eq.~\eqref{eq:vtapprox}).\footnote{The scale factor $a$
actually has the unit of a length, $[a] = \textrm{GeV}^{-1}$.
For the ease of notation, we will, however, always omit the unit symbol,
when stating values of $a$.
Alternatively, all explicit values of $a$ which we will state in the following
may be understood as indicating the ratio $a/a_{\textrm{PH}}$.}


The time-dependence of the scale factor $a$ is controlled by the Friedmann equation.
For a flat universe and a constant equation of state $\omega = \rho / p$ between
some reference time $t_0<t$ and time $t$, the Friedmann equation is solved by
\begin{align}
a(t) = a(t_0) \bigg[ 1 + \frac{3}{2} (1 + \omega)
\left( \frac{8 \pi}{3 M_P^2} \rho_{\text{tot}}(t_0) \right)^{1/2}\left(t-t_0\right)\bigg] ^{\frac{2}{3(1+\omega)}} \,.
\label{eq:scalefac}
\end{align}
After preheating, the total energy density of the
universe $\rho_{\textrm{tot}}$ is dominated by the abundance
of nonrelativistic Higgs bosons, i.e.\ $\omega = 0$.
In the course of reheating, the initial Higgs boson energy
is, however, gradually transferred into MSSM radiation.
Eventually, we thus have $\omega = 1/3$.
In the intervening time, the  equation of state parameter
$\omega$ changes continuously.
We approximate this behaviour by working with a \textit{piecewise constant},
\textit{effective} equation of state with coefficients~$\omega_i$ in the
intervals $(t_i, t_{i+1}]$, where $a_{\textrm{PH}} \leq a(t_i) < a(t_{i+1})$.
We determine the $\omega_i$ iteratively by requiring self-consistency of the
Friedmann equation, i.e.\ by numerically solving the following equation for
all time intervals $(t_i, t_{i+1}]$, until we reach $\omega_i=1/3$,
\begin{align}
\frac{\rho_{\text{tot}}(t_i)}{\rho_{\text{tot}}(t_{i+1})} = \left( \frac{a(t_{i+1})}{a(t_i)}
\right)^{3(1 + \omega_i)} \,, \quad \rho_{\text{tot}} \approx \rho_{\sigma} + \rho_{N_1}^S \,.
\end{align}
In the computation of the effective equation of state parameters $\omega_i$,
we approximate the total energy density $\rho_{\text{tot}}$ by its two dominant
components---the energy density of the Higgs bosons $\rho_{\sigma} $ and the
energy density of the neutrinos produced in Higgs, higgsino and inflaton decays
$\rho_{N_1}^S$.
For both densities we are able to derive analytical expressions
(cf.\ Eqs.~\eqref{eq:rhosigmaphipsisol} and \eqref{eq:rhoXntsol}), so that we can determine
the time evolution of the scale factor \textit{before} solving any Boltzmann equation
numerically.
In conclusion, we also note that in the following we will calculate the Hubble rate
$H$ as $\dot{a}/a$ using the piecewise defined scale factor in Eq.~\eqref{eq:scalefac}.


\subsection{Symmetry-Breaking Sector}


The particles of the SSB sector are solely produced during tachyonic preheating
and exclusively decay into the heavy (s)neutrinos of the first generation
(cf.\ Sec.~\ref{subsec:rates}).
The Boltzmann equations for $\sigma$, $\phi$ and $\psi$ read
\begin{align}
\hat{\mathcal{L}} f_\sigma = & \:
- C_{\sigma}(\sigma \rightarrow N_1 N_1)
- C_{\sigma}(\sigma \rightarrow \tilde{N}_1 \tilde{N}_1^*)
\,, \label {eq:BEsigmaphipsi} \\
\hat{\mathcal{L}} f_\phi = & \:
- C_{\phi}(\phi \rightarrow \tilde{N}_1 \tilde{N}_1) \,, \nonumber \\
\hat{\mathcal{L}} f_\psi = & \:
- C_{\psi}(\psi \rightarrow N_1 \tilde{N}_1^*) \,. \nonumber
\end{align}


In order to solve
these equations, let us consider for a moment the general case of an arbitrary
particle $X$ which is produced at some time $t_0$ with an initial distribution
function $f_X^0$, but which merely decays after its production
into other particles $ij..$, not being replenished by other processes.
The time evolution of $f_X$ is then described by the following Boltzmann equation,
\begin{align}
\hat{\mathcal{L}} f_X =  \mathcal{C}_X = - \sum_{ij} C_X(X \rightarrow ij..) \,,
\label{eq:BEdec}
\end{align}
with the total collision $\mathcal{C}_X$ being given by
\begin{align}
\mathcal{C}_X = - \sum_{ij..}
\frac{1}{2 g_X} \int d\Pi\left(X|i,j,..\right) \left(2\pi\right)^4
\delta^{(4)}f_X \left|\mathcal{M}\left(X \rightarrow i j ..\right)\right|^2 \,.
\end{align}
By definition of the total zero-temperature decay rate $\Gamma_X^0$ of the particle
$X$, this operator can be simplified to $\mathcal{C}_X = - m_X f_X \Gamma_X^0$,
such that Eq.~\eqref{eq:BEdec} turns into
\begin{align}
\frac{1}{E_X} \, \hat{\mathcal{L}} f_X(t,p) =
\left(\frac{\partial}{\partial t} - H p \frac{\partial}{\partial p}\right)
f_X(t,p) = \frac{d}{dt}f_X(t,p) = - \frac{m_X}{E_X} \Gamma_X^0 f_X(t,p) \,.
\label{eq:BEdecsimpl}
\end{align}
This is a linear homogeneous ordinary differential equation in time, which has
a unique solution for each initial distribution function $f_X^0$,
\begin{align}
f_X(t,p) = f_X^0(t,p) \exp\left[-\Gamma_X^0 \int_{t_0}^t dt'
\frac{m_X}{\mathcal{E}_X\left(E_X;t,t'\right)}\right] \,, \quad E_X = \sqrt{p^2 + m_X^2} \,.
\label{eq:fdecsol}
\end{align}
where $\mathcal{E}_X$ denotes the energy of an $X$ particle
at time $t_2$ which at time $t_1$ has an energy $E_1$.
Irrespectively of the time ordering of $t_1$ and $t_2$, we have
\begin{align}
\mathcal{E}_X\left(E_1;t_1,t_2\right) = \left[\left(\frac{a_1}{a_2}\right)^2
\left(E_1^2 - m_X^2\right) + m_X^2 \right]^{1/2} \,,\quad a_{1,2} = a\left(t_{1,2}\right) \,.
\end{align}


If the particle species $X$ is exclusively produced with a particular initial
momentum $p_*$, its initial distribution function is proportional to a delta function,
\begin{align}
f_X^0(t,p) \propto \delta(p_0-p_*) \,, \quad p = \frac{a_0}{a} \, p_0 \,,\quad a_0 = a(t_0) \,.
\label{eq:fdelta}
\end{align}
Here, $p_0$ is the momentum of an $X$ particle at time $t_0$, which evolves
into a momentum $p$ at time $t$.
The correct normalization of the delta function in Eq.~\eqref{eq:fdelta}
is obtained from matching the integral of $f_X^0$ over
$g_X d^3p/\left(2\pi\right)^3$ with the initial number density $n_X(t_0)$,
\begin{align}
f_X^0(t,p) = \frac{2\pi^2}{g_X} n_X(t_0)\,\frac{\delta\left(p_0- p_*\right)}{p_0^2} =
\frac{2\pi^2}{g_X} \left(\frac{a_0}{a}\right)^3 n_X(t_0) \,
\frac{\delta\left(p - \left(a_0/a\right)p_*\right)}{p^2} \,. \label{eq:fpstarres}
\end{align}
The delta function contained in $f_X^0$ allows us to rewrite Eq.~\eqref{eq:fdecsol}
as follows,
\begin{align}
f_X(t,p) = f_X^0(t,p) \exp\left[-\Gamma_X^0 \int_{t_0}^t dt'
\frac{m_X}{\mathcal{E}_X\left(E_*;t_0,t'\right)}\right] \,, \quad E_* = \sqrt{p_*^2 + m_X^2} \,.
\end{align}
This step has eliminated the nontrivial $p$ dependence of $f_X$, so that we
may now integrate it over $g_X d^3p/\left(2\pi\right)^3$ in order to calculate the number
density $n_X$,
\begin{align}
n_X(t) = \left(\frac{a_0}{a}\right)^3 n_X(t_0) \exp\left[-\Gamma_X^0 \int_{t_0}^t dt'
\frac{m_X}{\mathcal{E}_X\left(E_*;t_0,t'\right)}\right] \,.
\end{align}
In the case that $p_* = 0$, the expressions for $E_*$,
$\mathcal{E}_X$, $f_X^0$, $f_X$ and $n_X$ simplify to
\begin{align}
p_* = & \: 0 \,,\quad E_* = \mathcal{E}_X = m_X \,,\quad f_X^0(t,p) =
\frac{2\pi^2}{g_X} \left(\frac{a_0}{a}\right)^3 n_X(t_0) \,
\frac{\delta\left(p\right)}{p^2} \label{eq:pstar0res}\\
f_X(t,p) = & \: f_X^0(t,p) \, e^{-\Gamma_X^0 (t-t_0)} \,,\quad
n_X(t) = \left(\frac{a_0}{a}\right)^3 n_X(t_0) \, e^{-\Gamma_X^0 (t-t_0)} \,. \nonumber
\end{align}


Since $\sigma$, $\phi$ and $\psi$ are mostly produced with very low momenta
during tachyonic preheating (cf.\ Eq.~\eqref{eq:nkBnkF}), it is justified to
approximate their initial distribution functions by delta functions.
Our results in Eq.~\eqref{eq:pstar0res} thus directly provide us with
the solutions of the Boltzmann equations in Eq.~\eqref{eq:BEsigmaphipsi},
\begin{align}
f_X(t,p) = \frac{2 \pi^2}{g_X}\left(\frac{a_{\textrm{PH}}}{a}\right)^3
n_X\left(t_{\textrm{PH}}\right) \,
\frac{\delta\left(p\right)}{p^2}\, e^{-\Gamma_S^0 (t-t_{\textrm{PH}})}
\,, \qquad X = \sigma, \, \phi ,\, \psi \,, \label{eq:fsigmaphipsi}
\end{align}
where we have used the fact that $\sigma$, $\phi$ and $\psi$ all have the same
total decay rate $\Gamma_S^0$ (cf.\ Eq.~\eqref{eq:GammaS0}).
Recall that $N_X = a^3 n_X$ (cf.\ Eq.~\eqref{eq:NXa3nX}).
The comoving number densities $N_X$ corresponding to these distribution
functions $f_X$ are then given by
\begin{align}
N_X(t) = N_X(t_{\textrm{PH}}) \, e^{-\Gamma_S^0 (t-t_{\textrm{PH}})}
\,, \qquad X = \sigma, \, \phi ,\, \psi \,. \label{eq:Nsigmaphipsisol}
\end{align}
Taking into account that $a_{\textrm{PH}} = 1$, the initial comoving number
densities $N_\phi(t_{\textrm{PH}})$ and $N_\psi(t_{\textrm{PH}})$ readily
follow from Eq.~\eqref{eq:rhonBF}.
The initial comoving number density of Higgs bosons $N_\sigma(t_{\textrm{PH}})$
can be deduced from the initial Higgs energy density $\rho_\sigma(t_{\textrm{PH}})$,
which in turn is given as the difference of the initial vacuum energy density $\rho_0$
and the energy densities of all particles produced during tachyonic preheating,
\begin{align}
N_\sigma(t_{\textrm{PH}}) = & \: a_{\textrm{PH}}^3 \, n_\sigma(t_{\textrm{PH}}) =
n_\sigma(t_{\textrm{PH}}) = \frac{1}{m_S} \rho_\sigma(t_{\textrm{PH}}) \\
= & \: \frac{\rho_0}{m_S}
\left(1 - \sum_{\textrm{bosons}} \rho_B\left(t_{\textrm{PH}}\right)/\rho_0 -
\sum_{\textrm{fermions}} \rho_F\left(t_{\textrm{PH}}\right)/\rho_0\right) \,. \nonumber
\end{align}
As $\sigma$, $\phi$ and $\psi$ are all nonrelativistic, the energy densities $\rho_X$
are trivially related to the comoving number densities $N_X$ in Eq.~\eqref{eq:Nsigmaphipsisol},
\begin{align}
\rho_X(t) = \frac{m_S}{a^3} \, N_X(t) = m_S \, n_X(t_{\textrm{PH}})
\left(\frac{a_{\textrm{PH}}}{a}\right)^3 e^{-\Gamma_S^0 (t-t_{\textrm{PH}})}
\,, \qquad X = \sigma, \, \phi ,\, \psi \,. \label{eq:rhosigmaphipsisol}
\end{align}


\subsection{Neutrino Sector}
\label{subsec:BEneutrinos}


\subsubsection{Heavy (S)neutrinos from Tachyonic Preheating}


Next to the particles of the SSB sector, tachyonic preheating also
entails the production of the gauge DOFs as well as of all three
heavy (s)neutrino generations.
Again, the respective initial comoving number densities
$N_X\left(t_{\textrm{PH}}\right) = n_X(t_{\textrm{PH}})$
can be calculated employing our results in Eq.~\eqref{eq:rhonBF}.
In the following, we will label all quantities associated with the heavy
(s)neutrino components $\big(N_i^{\textrm{PH}},\tilde{N}_i^{\textrm{PH}}\big)$
produced during preheating with an upper index $\textrm{PH}$.


\subsubsection{Heavy (S)neutrinos and Radiation from the Decay of the Gauge DOFs}


Due to their strong coupling and their large mass, the gauge particles $A$, $\tilde{A}$
and $C$ have a very short lifetime $t_G = t_{\textrm{PH}} + 1/\Gamma_G^0$
(cf.\ Eq.~\eqref{eq:GammaG0}) and hence decay practically immediately after
preheating.
It is thus not necessary to explicitly resolve the time dependence of
their number densities.
Instead, we may simply approximate the comoving number densities
of the gauge particles by step functions,
\begin{align}
N_X(t) = N_X\left(t_{\textrm{PH}}\right) \,\Theta\left(t_G - t\right) \,, \quad
t \geq t_{\textrm{PH}} \,, \quad X = A, \, \tilde{A},\, C \,.
\end{align}
Interpreting the comoving number densities $N_X$ as functions of the scale factor,
we would have to replace $\Theta\left(t_G - t\right)$ by $\Theta\left(a_G - a\right)$,
where $a_G = a(t_G)$.
As a technical detail, we note that we will use $a_G$ as the initial value of
the scale factor, when solving the Boltzmann equations numerically.


The decay of the gauge particles gives rise to further heavy (s)neutrino components
$\big(N_i^G,\tilde{N}_i^G\big)$ as well as to an initial
abundance of radiation.\footnote{Similarly to the heavy (s)neutrinos
$\big(N_i^{\textrm{PH}},\tilde{N}_i^{\textrm{PH}}\big)$, we will label
all quantities associated with the heavy (s)neutrino produced in the decay
of the gauge particles with an upper index $G$.
This notational convention is consistently employed throughout this thesis.
The lower index of a quantity such as a number density, decay rate, etc.\ always
indicates the particle to which the respective quantity belongs,
whereas its upper index always refers to the origin of this particle.}
This time, the respective initial comoving number densities are determined
by $N_A\left(t_{\textrm{PH}}\right)$, $N_{\tilde{A}}\left(t_{\textrm{PH}}\right)$
and $N_C\left(t_{\textrm{PH}}\right)$ as well as the branching ratios of
the gauge particles (cf.\ Eq.~\eqref{eq:AAtCBr}),
\begin{align}
N_{N_i}^G(t_G) = & \:
2\, N_A\left(t_{\textrm{PH}}\right)\textrm{Br}\big(A\rightarrow N_iN_i^{\phantom{*}}\big)
+ N_{\tilde{A}}\left(t_{\textrm{PH}}\right) \textrm{Br}\big(\tilde{A}\rightarrow N_i\tilde{N}_i\big)
\,, \\
N_{\tilde{N}_i}^G(t_G) = & \:
2\, N_A\left(t_{\textrm{PH}}\right)\textrm{Br}\big(A\rightarrow\tilde{N}_i\tilde{N}_i^*\big)
+ N_{\tilde{A}}\left(t_{\textrm{PH}}\right) \textrm{Br}\big(\tilde{A}\rightarrow N_i\tilde{N}_i\big)
\nonumber \\
+ & \: 2\, N_C\left(t_{\textrm{PH}}\right) \textrm{Br}\big(C\rightarrow\tilde{N}_i\tilde{N}_i^*\big)
\,, \nonumber \\
N_R^G(t_G) = & \:
2 \, N_A\left(t_{\textrm{PH}}\right)\textrm{Br}\big(A\rightarrow\textrm{MSSM}\big)
+ 2 \, N_{\tilde{A}}\left(t_{\textrm{PH}}\right)
\textrm{Br}\big(\tilde{A}\rightarrow\textrm{MSSM}\big) \nonumber \\
+ & \: 2 \, N_C\left(t_{\textrm{PH}}\right)\textrm{Br}\big(C\rightarrow\textrm{MSSM}\big)
\nonumber \,,
\end{align}
where $\textrm{Br}\big(A\rightarrow\textrm{MSSM}\big)$,
$\textrm{Br}\big(\tilde{A}\rightarrow\textrm{MSSM}\big)$ and
$\textrm{Br}\big(C\rightarrow\textrm{MSSM}\big)$ are given by
\begin{align}
\textrm{Br}\big(A\rightarrow\textrm{MSSM}\big) = & \: \sum_{\textrm{MSSM}}
\left[\textrm{Br}\big(A\rightarrow\psi_i\bar{\psi}_i\big) +
\textrm{Br}\big(A\rightarrow\phi_i\phi_i^*\big)\right] \,, \\
\textrm{Br}\big(\tilde{A}\rightarrow\textrm{MSSM}\big) = & \: \sum_{\textrm{MSSM}}
\textrm{Br}\big(\tilde{A}\rightarrow\bar{\psi}_i\phi_i\big) \,, \nonumber \\
\textrm{Br}\big(C\rightarrow\textrm{MSSM}\big) = & \: \sum_{\textrm{MSSM}}
\textrm{Br}\big(C\rightarrow\phi_i\phi_i^*\big) \,. \nonumber
\end{align}
Note that $N_{\tilde{N}_i}^G$ subsumes the comoving number densities of
the $\tilde{N}_i$ sneutrinos and the $\tilde{N}_i^*$
antisneutrinos.
We emphasize that, unless explicitly indicated otherwise, all further quantities
associated with the heavy sneutrinos $\tilde{N}_i$ are meant to equally comprise
the heavy sneutrinos $\tilde{N}_i$ \textit{and} the heavy antisneutrinos $\tilde{N}_i^*$.
Moreover, as already mentioned in Sec.~\ref{sec:preheating}, combining our
results for the branching ratios of the gauge DOFs and the initial
comoving number densities produced during preheating
(cf.\ Eqs.~\eqref{eq:AAtCBr} and \eqref{eq:rhonBF}), one can easily
show that $N_{N_i}^G(t_G) = N_{\tilde{N}_i}^G(t_G)$, which in turn implies
that $N_{N_i}^G = N_{\tilde{N}_i}^G$ at all times (cf.\ Eq.~\eqref{eq:NN23sol}).


An important caveat applies to the comoving number density $N_R^G(t_G)$.
As the MSSM particles produced in the decay of the gauge DOFs quickly thermalize
after their production, $N_R^G(t_G)$ must not be used as the initial value of
the comoving radiation number density, when solving the radiation Boltzmann equation.
Instead, $N_R^G(t_G)$ merely provides us with the initial radiation energy
density $\rho_R(t_G)$.
Since the gauge particles are nonrelativistic at all times, their decay products
are always equipped with an initial energy of $m_G/2$.
Making use of the expressions for the radiation number and the radiation
energy density in Eq.~\eqref{eq:nRrhoR}, we find
\begin{align}
\rho_R(t_G) = & \: \frac{m_G}{2} \frac{N_R^G(t_G)}{a_G^3} \,, \quad
T(t_G) = \left(\frac{30}{\pi^2 g_{*,\rho}} \rho_R(t_G)\right)^{1/4} \,,\\
N_R(t_G) = & \: a_G^3 \frac{\zeta(3)}{\pi^2}g_{*,n} T^3(t_G) \,. \nonumber
\end{align}


\subsubsection{Heavy (S)neutrinos of the Second and Third Generation}


The heavy (s)neutrinos of the second and third generation are solely produced during
preheating as well as in the decay of the gauge DOF and they exclusively decay into MSSM
lepton-Higgs pairs.
Their Boltzmann equations are hence of the same type as the template Boltzmann equation
in Eq.~\eqref{eq:BEdec},
\begin{align}
\hat{\mathcal{L}} f_{N_{2,3}} = & \:
- C_{N_{2,3}}\big(N_{2,3}\rightarrow\ell H_u, \bar{\ell} H_u^*,
\tilde{\ell} \tilde{H}_u, \tilde{\ell}^* \bar{\tilde{H}}_u\big)
\,, \label{eq:BENNt23} \\
\hat{\mathcal{L}} f_{\tilde{N}_{2,3}} = & \:
-C_{\tilde{N}_{2,3}}\big(\tilde{N}_{2,3}\rightarrow\tilde{\ell} H_u,
\bar{\ell} \bar{\tilde{H}}_u\big)
-C_{\tilde{N}_{2,3}^*}\big(\tilde{N}_{2,3}^*\rightarrow\tilde{\ell}^* H_u^*,
\ell\tilde{H}_u\big)
\,. \nonumber
\end{align}
Consequently, $f_{N_{2,3}}$ and $f_{\tilde{N}_{2,3}}$, the solutions of the $N_{2,3}$ and
$\tilde{N}_{2,3}$ Boltzmann equations, are of the same form as the distribution
function in Eq.~\eqref{eq:fdecsol}.
Each of these solutions consists of two independently evolving parts, respectively accounting
for the $\big(N_i^{\textrm{PH}},\tilde{N}_i^{\textrm{PH}}\big)$ (s)neutrinos as well
as for the $\big(N_i^G,\tilde{N}_i^G\big)$ (s)neutrinos,
\begin{align}
f_X(t,p) = & \: f_X^{\textrm{PH}}(t,p) + \Theta\left(t-t_G\right) f_X^G(t,p)
\,,\qquad X = N_i,\tilde{N}_i \,,\quad i = 2,3 \,,\phantom{\bigg]} \label{eq:fN23sol}\\
f_X^{\textrm{PH}}(t,p) = & \: f_X^{\textrm{PH}}(t_{\textrm{PH}},p)
\, e^{-\Gamma_{N_i}^0 (t-t_{\textrm{PH}})} \,, \nonumber\\
f_X^G(t,p) = & \: f_X^G(t_G,p) \exp\left[-\int_{t_G}^t dt'
\frac{M_i \,\Gamma_{N_i}^0}{\mathcal{E}_X\left(m_G/2;t_G,t'\right)}\right]\,, \nonumber 
\end{align}
with the initial distribution functions $f_X^{\textrm{PH}}(t_{\textrm{PH}},p)$
and $f_X^G(t_G,p)$ being similar to those in Eqs.~\eqref{eq:pstar0res} and
\eqref{eq:fpstarres},
\begin{align}
f_X^{\textrm{PH}}(t_{\textrm{PH}},p) = & \: \frac{2\pi^2}{g_{N}}
\frac{N_X^{\textrm{PH}}(t_{\textrm{PH}})}{a^3}
\, \frac{\delta\left(p\right)}{p^2} \,,\qquad X = N_i,\tilde{N}_i \,,\quad i = 2,3 \,, \\
f_X^G(t_G,p) = & \: \frac{2\pi^2}{g_N} \frac{N_X^G(t_G)}{a^3} \,
\frac{\delta\left(p - \left(a_G/a\right)p_*\right)}{p^2} \,,\quad
p_* = \sqrt{\left(\frac{m_G}{2}\right)^2 - M_i^2} \,. \nonumber
\end{align}
Integrating the distribution functions in Eq.~\eqref{eq:fN23sol} over the heavy (s)neutrino
momentum space yields the corresponding comoving number densities,
\begin{align}
N_X(t) = & \: N_X^{\textrm{PH}}(t) + \Theta\left(t-t_G\right) N_X^G(t)
\,,\qquad X = N_i,\tilde{N}_i \,,\quad i = 2,3 \,, \phantom{\bigg]} \label{eq:NN23sol}\\
N_X^{\textrm{PH}}(t) = & \: N_X^{\textrm{PH}}(t_{\textrm{PH}})
\, e^{-\Gamma_X^0 (t-t_{\textrm{PH}})} \,, \nonumber\\
N_X^G(t) = & \: N_X^G(t_G) \exp\left[-\int_{t_G}^t dt'
\frac{M_i \,\Gamma_{N_i}^0}{\mathcal{E}_X\left(m_G/2;t_G,t'\right)}\right]\,. \nonumber 
\end{align}


\subsubsection{Nonthermally Produced Heavy (S)neutrinos of the First Generation}


Similarly to the two heavier (s)neutrino flavours, the heavy (s)neutrinos of the first
generation are also produced during tachyonic preheating and in the decay of the gauge
multiplet.
On top of that, the abundances of the $N_1$ neutrinos and $\tilde{N}_1$ sneutrinos
receive additional contributions from the decay of the Higgs boson $\sigma$, the
inflaton $\phi$ and the higgsino $\psi$ as well as from thermal production processes.
The corresponding Boltzmann equations are consequently more involved,
\begin{align}
\hat{\mathcal{L}} f_{N_1} = & \: 2 \, C_{N_1}(\sigma \rightarrow N_1 N_1)
+ C_{{N}_1} (\psi \rightarrow N_1 \tilde{N}_1^*) \label{eq:BENNt1}\\
+ & \: \phantom{2} \,
C_{N_1} \big(N_1\leftrightarrow\ell H_u, \bar{\ell} H_u^*,
\tilde{\ell} \tilde{H}_u, \tilde{\ell}^* \bar{\tilde{H}}_u\big) \,,  \nonumber \\
\hat{\mathcal{L}} f_{\tilde{N}_1} = & \: 2 \,
C_{\tilde{N}_1} (\sigma \rightarrow \tilde{N}_1 \tilde{N}_1^*)
+ 2 \, C_{\tilde{N}_1} (\phi \rightarrow \tilde{N}_1 \tilde{N}_1)
+ C_{\tilde{N}_1} (\psi \rightarrow N_1 \tilde{N}_1^*) \nonumber \\
+ & \: \phantom{2} \,
C_{\tilde{N}_1} \big(\tilde{N}_1 \leftrightarrow
\tilde{\ell} H_u, \bar{\ell} \bar{\tilde{H}}_u\big) 
+ C_{\tilde{N}_1^*}\big(\tilde{N}_1^*\leftrightarrow\tilde{\ell}^* H_u^*,
\ell\tilde{H}_u\big)\,. \nonumber
\end{align}
Again, these Boltzmann equations are best tackled by decomposing the (s)neutrino
distribution functions $f_{N_1}$ and $f_{\tilde{N}_1}$ into independently evolving parts,
\begin{align}
f_X(t,p) = & \:f_X^{\textrm{nt}}(t,p) + f_X^{\textrm{th}}(t,p)
\,,\qquad X = N_1,\tilde{N}_1 \,, \\
= & \: f_X^{\textrm{PH}}(t,p) + \Theta\left(t-t_G\right) f_X^G(t,p)
+ f_X^S(t,p) + f_X^{\textrm{th}}(t,p) \,. \nonumber
\end{align}
As the Boltzmann equations in Eq.~\eqref{eq:BENNt1} are linear in $f_{N_1}$
and $f_{\tilde{N}_1}$, they may be rewritten as a set of independent \textit{partial}
Boltzmann equations, respectively describing the time evolution of one of the
distribution functions $f_X^{\textrm{x}}$, where $X = N_1,\tilde{N}_1$ and
$\textrm{x} = \textrm{PH},G,S,\textrm{th}$.
Of course, the Boltzmann equations for $f_X^{\textrm{x}}$ and $N_X^{\textrm{x}}$, with
$X = N_1,\tilde{N}_1$ and $\textrm{x} = \textrm{PH},G$ take exactly the
same form as the Boltzmann equations for the two heavier (s)neutrino generations
in Eq.~\eqref{eq:BENNt23}.
Our results for $f_X^{\textrm{x}}$ and $N_X^{\textrm{x}}$, with
$X = N_{2,3},\tilde{N}_{2,3}$ and $\textrm{x} = \textrm{PH},G$, in Eqs.~\eqref{eq:fN23sol}
and \eqref{eq:NN23sol} can thus readily be generalized to the first heavy (s)neutrino
generation.


In order to solve the Boltzmann equations for $f_{N_1}^S$ and $f_{\tilde{N}}^S$, i.e.\
the heavy (s)neutrinos produced in the decay of particles from the SSB sector,
let us consider for a moment the general case of an arbitrary
particle $X$ which is continuously produced in the decay of nonrelativistic
particles $i$ and which itself steadily decays into other particles $ab..$.
The time evolution of $f_X$ is then described by,
\begin{align}
\hat{\mathcal{L}} f_X = \sum_{ij}(1+\delta_{Xj})\,C_X(i \rightarrow Xj)
- \sum_{ab..} C_X(X \rightarrow ab..)\,.
\label{eq:BEproddec}
\end{align}
Here, the Kronecker delta $\delta_{Xj}$ accounts for the fact that \textit{two}
$X$ particles are produced in case of $j = X$.
Just as in our discussion of the Boltzmann equation in Eq.~\eqref{eq:BEdec},
the sum over the decay operators $C_X(X \rightarrow ab..)$ can again be simplified
to $m_Xf_X\Gamma_X^0$ (cf.\ Eq.~\eqref{eq:BEdecsimpl}).
The production operators $C_X(i \rightarrow Xj)$ are given by
\begin{align}
C_X(i \rightarrow Xj) =
\frac{1}{2 g_X} \int d\Pi\left(X|i;j\right) \left(2\pi\right)^4
\delta^{(4)}f_i \left|\mathcal{M}\left(i\rightarrow X j\right)\right|^2 \,.
\label{eq:prodop}
\end{align}
We shall restrict ourselves to particles $i$ which
are instantaneously produced at time $t_0$ with vanishing initial momentum.
According to our results in Eq.~\eqref{eq:pstar0res}, we are thus allowed
to use the following expression for $f_i$,
\begin{align}
f_i(t,p) = \frac{2\pi^2}{g_i} n_i(t) \frac{\delta(p)}{p^2} \,.
\end{align}
For a matrix element squared $\left|\mathcal{M}\right|^2$ which is independent
of the momenta of the particles $i$ and $j$ the collision operator in
Eq.~\eqref{eq:prodop} can then be rewritten as
\begin{align}
C_X(i \rightarrow Xj) =
\frac{1}{2 g_X} \frac{2\pi^2}{g_i} n_i
\left|\mathcal{M}\left(i\rightarrow X j\right)\right|^2
\frac{S(i;X,j)\,\delta(p-p_*)}{8\pi \,m_i \,p_*} \,, \label{eq:prodop2}
\end{align}
with $p_* = \left[\left(m_i/2\right)^2 - m_X^2\right]^{1/2}$.
Employing the standard expression for
the decay rate of a two-body decay, we are able to make the following substitution,
\begin{align}
\frac{1}{2}\frac{S(i;X,j)}{8\pi\, m_i} \frac{1}{g_i}
\left|\mathcal{M}\left(i\rightarrow X j\right)\right|^2 =
\frac{E_*}{p_*} \, \Gamma_i^0 \left(i\rightarrow X j\right)\,, \quad E_* = \frac{m_i}{2} \,.
\end{align}
The collision operator in Eq.~\eqref{eq:prodop2} consequently turns into
\begin{align}
C_X(i \rightarrow Xj) =
E_* \, \frac{2\pi^2}{g_X} \gamma\left(i\rightarrow X j\right)
\frac{\delta(p-p_*)}{p_*^2} \,,
\end{align}
where we have used that
$\gamma\left(i\rightarrow X j\right) = n_i\,\Gamma_i^0 \left(i\rightarrow X j\right)$
for a nonrelativistic particle species $i$.
The Boltzmann equation can now be brought into the following form
\begin{align}
\frac{d}{dt}f_X(t,p) =  \frac{2\pi^2}{g_X} \sum_{ij}(1+\delta_{Xj})\,
\gamma\left(i\rightarrow X j\right)\frac{\delta(p-p_*)}{p^2}
- \frac{m_X}{E_X} \Gamma_X^0 f_X(t,p) \,.
\end{align}
Again, we end up with a linear homogeneous ordinary differential equation in time.
Starting from zero initial abundance, $f_X^0 = 0$, its unique solution is given by
\begin{align}
f_X(t,p) = & \: \frac{2\pi^2}{g_X} \sum_i \int_{t_0}^t dt'\,\gamma_{i,X}(t')
\frac{\delta(p'-p_*)}{p'^2}
\exp\left[-\int_{t'}^t dt''\frac{m_X \Gamma_X^0}{\mathcal{E}_X(E_X;t,t'')}\right] \,,\\
\gamma_{i,X}(t) = & \: \sum_j (1+\delta_{Xj})\,\gamma\left(i\rightarrow X j\right)
\,,\quad p' =  \frac{a}{a'} p \,,\quad a' = a(t') \,. \nonumber
\end{align}
Thanks to the delta function in the integral over $dt'$, the energy
$\mathcal{E}_X(E_X;t,t'')$ in the integral over $dt''$ can be replaced by
$\mathcal{E}_X(m_i/2;t',t'')$.
This eliminates the nontrivial momentum dependence of the distribution function $f_X$.
The delta function itself can be rewritten as a function of
$p$, the momentum of an $X$ particle at time $t$,
\begin{align}
\delta(p'- p_*) = \frac{a'}{a} \,\delta(p-(a'/a)\,p_*)
\end{align}
Our final result for the distribution function $f_X$ then reads,
\begin{align}
f_X(t,p) = & \: \frac{2\pi^2}{g_X} \sum_i \int_{t_0}^t dt' \left(\frac{a'}{a}\right)^3
\gamma_{i,X}(t') \frac{\delta(p-(a'/a)\,p_*)}{p^2} \\
\times & \:
\exp\left[-\int_{t'}^t dt''\frac{m_X \Gamma_X^0}{\mathcal{E}_X(m_i/2;t',t'')}\right] \,.
\nonumber
\end{align}
Owing to the trivial momentum dependence of $f_X$, we can easily deduce the
number, energy and interaction densities of the particle species $X$
(cf.\ Eqs.~\eqref{eq:numdendef}, \eqref{eq:engdendef} and \eqref{eq:gammadef}).
For the ease of notation, let us introduce $\delta n_X(t',t)$ to denote the
number density at time $t$ of those $X$ particles which are produced at time $t'$.
As we will see immediately, $\delta n_X$ is given by
\begin{align}
\delta n_X(t',t) = \sum_i \delta n_X^i (t',t)
= \sum_i \left(\frac{a'}{a}\right)^3 \gamma_{i,X}(t')
\exp\left[-\int_{t'}^t dt''\frac{m_X \Gamma_X^0}{\mathcal{E}_X(m_i/2;t',t'')}\right] \,.
\label{eq:deltan}
\end{align}
Indeed, with the aid of $\delta n_X$ and $\delta n_X^i$, we are able to summarize
our results as follows,
\begin{align}
f_X(t,p) = & \: \frac{2\pi^2}{g_X} \sum_i \int_{t_0}^t dt' \, \delta n_X^i(t',t) 
\frac{\delta(p-(a'/a)\,p_*)}{p^2} \,, \label{eq:fnrhogammaX}\\
n_X(t) = & \: \int_{t_0}^t dt'\, \delta n_X(t',t) = \sum_i \int_{t_0}^t dt'\, \delta n_X^i(t',t) =
\sum_i n_X^i(t) \,,\quad \nonumber \\
\rho_X(t) = & \: \sum_i\int_{t_0}^t dt'\, \mathcal{E}_X(m_i/2;t',t) \, \delta n_X^i(t',t) \nonumber \\
\gamma_X(t) = & \: \gamma\left(X\rightarrow ab ..\right) = \sum_i
\int_{t_0}^t dt'\, \frac{m_X \,\Gamma_X^0}{\mathcal{E}_X(m_i/2;t',t)} \, \delta n_X^i(t',t)
= \sum_i n_X^i(t) \,\Gamma_X^i(t) \,, \nonumber
\end{align}
where $\Gamma_X^i$ represents the zero-temperature decay rate of the particle species
$X$, weighted with the average inverse time dilatation factor for $X$ particles produced
in the decay of $i$ particles (cf.\ Eqs.~\eqref{eq:GammaxNi} and \eqref{eq:dilatation}),
\begin{align}
\Gamma_X^i(t) = \left<\frac{m_X}{E_X}\right>_i \Gamma_X^0 =
\int_{t_0}^t dt'\, \frac{m_X}{\mathcal{E}_X(m_i/2;t',t)}
\frac{\delta n_X^i(t',t)}{n_X^i(t)} \, \Gamma_X^0 \,. \label{eq:GammaXi}
\end{align}


The general expressions in Eqs.~\eqref{eq:deltan}, \eqref{eq:fnrhogammaX}
and \eqref{eq:GammaXi} are readily applied to the heavy (s)neutrinos of the first
generation, which are produced in the decay of particles from the SSB sector.
To obtain the corresponding expressions for the $N_1^S$ neutrinos and $\tilde{N}_1^S$
sneutrinos one merely has to perform the following substitutions,
\begin{align}
t_0 \rightarrow & \: t_{\textrm{PH}} \,,\quad i \rightarrow \sigma,\phi,\psi \,,\quad
m_i \rightarrow m_S \,,\quad X \rightarrow  N_1^S, \tilde{N}_1^S \,,\quad
g_X \rightarrow g_N \,\quad m_X \rightarrow M_1 \,,\\
\Gamma_X^0 \rightarrow & \: \Gamma_{N_1}^0  \,, \quad
\gamma_{i,X} \rightarrow \gamma_{S, N_1}, \gamma_{S, \tilde{N}_1} \,, \nonumber
\end{align}
with the interaction densities $\gamma_{S, N_1}$ and $\gamma_{S, \tilde{N}_1}$
being defined as
\begin{align}
\gamma_{S,N_1}(t) = & \: 2 \,n_\sigma(t)\,\Gamma^0(\sigma \rightarrow N_1 N_1)
+ n_\psi(t)\,\Gamma^0(\psi \rightarrow N_1 \tilde{N}_1^*)  \,, \\
\gamma_{S,\tilde{N}_1}(t) = & \: 2\,n_\sigma(t)\,
\Gamma^0(\sigma \rightarrow \tilde{N}_1 \tilde{N}_1^*)
+ 2\,n_\phi(t)\,\Gamma^0(\phi \rightarrow \tilde{N}_1 \tilde{N}_1)
+ n_\psi(t)\,\Gamma^0(\psi \rightarrow N_1 \tilde{N}_1^*)  \,.
\nonumber
\end{align}
To sum up, we find the following analytical results for the comoving number
densities of the nonthermal heavy (s)neutrinos of the first generation,
\begin{align}
N_X^{\textrm{nt}}(t) = & \: N_X^{\textrm{PH}}(t) + N_X^G(t) + N_X^S(t)
\,,\qquad X = N_1,\tilde{N}_1 \,, \label{eq:NXntsol} \phantom{\bigg\{}\\
N_X^{\textrm{PH}}(t) = & \: N_X^{\textrm{PH}}(t_{\textrm{PH}})\,
e^{-\Gamma^0_{N_1}(t-t_{\text{PH}})} \,, \:\:
N_X^G(t) = N_X^G(t_G)\,
\exp\left[-\int_{t_G}^t dt' \frac{M_1 \,\Gamma^0_{N_1} }{\mathcal{E}_X(m_G/2;t_G,t')}\right]
\,, \nonumber \\
N_X^S(t) = & \: \int_{t_{\textrm{PH}}}^t dt'\,a'^3\, \gamma_{S,X}(t')
\exp\left[-\int_{t'}^t dt''\frac{M_1 \,\Gamma_{N_1}^0}{\mathcal{E}_X(m_S/2;t',t'')}\right]
\nonumber \,.
\end{align}
Meanwhile, the corresponding energy densities look as follows,
\begin{align}
\rho_X^{\textrm{nt}}(t) = & \: \rho_X^{\textrm{PH}}(t) + \rho_X^G(t) + \rho_X^S(t)
\,,\qquad X = N_1,\tilde{N}_1 \,, \label{eq:rhoXntsol} \phantom{\bigg\{}\\
\rho_X^{\textrm{PH}}(t) = & \: \frac{M_1}{a^3} \, N_X^{\textrm{PH}}(t) \,, \:\:
\rho_X^G(t) = \frac{\mathcal{E}_X(m_G/2;t_G,t)}{a^3} \, N_X^G(t) \,, \nonumber \\
\rho_X^S(t) = & \: \int_{t_{\textrm{PH}}}^t dt'\,\frac{\mathcal{E}_X(m_S/2;t',t)}{a^3}
a'^3\, \gamma_{S,X}(t')
\exp\left[-\int_{t'}^t dt''\frac{M_1 \,\Gamma_{N_1}^0}{\mathcal{E}_X(m_S/2;t',t'')}\right]
\nonumber \,.
\end{align}


The expressions for the comoving number densities of the $N_1$ neutrinos and
the $\tilde{N}_1$ sneutrinos in Eq.~\eqref{eq:NXntsol} only differ in terms of
the interaction density $\gamma_{S,X}$.
The time dependence of $\gamma_{S,N_1}$ and $\gamma_{S,\tilde{N}_1}$ is however
the same, so that the ratio $N_{\tilde{N}_1}^S/N_{N_1}^S$ is a constant at all times,
\begin{align}
\gamma_{S,X}(t) \propto \frac{e^{-\Gamma_S^0(t-t_{\textrm{PH}})}}{a^3} \,,\quad
\frac{N_{\tilde{N}_1}^S}{N_{N_1}^S} =
\frac{2\,n_\sigma(t_{\textrm{PH}})\,
\textrm{Br}(\sigma \rightarrow \tilde{N}_1 \tilde{N}_1^*)
+ 2\,n_\phi(t_{\textrm{PH}}) + n_\psi(t_{\textrm{PH}})}
{2 \,n_\sigma(t_{\textrm{PH}})\,\textrm{Br}(\sigma \rightarrow N_1 N_1)
+ n_\psi(t_{\textrm{PH}})} \,, \label{eq:NtSNSratio}
\end{align}
where we have used that $\textrm{Br}(\psi \rightarrow N_1 \tilde{N}_1^*) =
\textrm{Br}(\phi \rightarrow \tilde{N}_1 \tilde{N}_1) = 1$ (cf.\ Eq.~\eqref{eq:sigmatauphipsiBr}).
According to the parametrization of our model, the ratio $N_{\tilde{N}_1}^S/N_{N_1}^S$
solely depends on the heavy (s)neutrino mass $M_1$.
For those values of $M_1$ that we are interested in this dependence is, however, rather weak.
Varying $M_1$ between $10^9\,\textrm{GeV}$ and $3\times10^{12}\,\textrm{GeV}$
(cf.\ Eq.~\eqref{eq:parameterspace}), the ratio $N_{\tilde{N}_1}^S/N_{N_1}^S$ only increases
from roughly $4 \times 10^{-5}$ to roughly $2 \times 10^{-4}$.


\subsubsection{Thermally Produced Heavy (S)neutrinos of the First Generation}


Unlike the two heavier (s)neutrino flavours, the (s)neutrinos of the first generation are
also thermally produced through inverse decay processes in the bath.
In the Boltzmann equations in Eq.~\eqref{eq:BENNt1},
the production and the decay of the thermal (s)neutrinos are accounted for by
the collision operators featuring an double arrow.
The partial Boltzmann equations for the distribution functions $f_{N_1}^{\textrm{th}}$
and $f_{\tilde{N}_1}^{\textrm{th}}$ are hence of the following form,
\begin{align}
\hat{\mathcal{L}} f_X = & \: \mathcal{C}_X = \sum_{ij..} C_X(X \leftrightarrow ij..) =
\frac{1}{2g_X} \sum_{ij} \int d \Pi \left(X|i,j,..\right) \left(2\pi\right)^4 \delta^{(4)}
\label{eq:BEtherm} \\
\times & \: \left[f_i f_j..\left|\mathcal{M}\left(X \leftarrow ij..\right)\right|^2
- f_X \left|\mathcal{M}\left(X \rightarrow ij..\right)\right|^2 \right]\,, \nonumber
\end{align}
with $ij..$ denoting particles in thermal equilibrium.
Neglecting any effects of $CP$ violation in the decays and inverse decays of
the $X$ particles, we are able to identify
$\left|\mathcal{M}\left(X \leftarrow ij..\right)\right|^2$
with $\left|\mathcal{M}\left(X \rightarrow ij..\right)\right|^2$.
Furthermore, approximating the distributions functions $f_i, f_j, ..$
by Maxwell-Boltzmann distributions (cf.\ Eq.~\eqref{eq:fMB}),
energy conservation implies,
\begin{align}
f_i f_j .. = f_i^{\textrm{eq}} f_j^{\textrm{eq}}..
\approx e^{-E_i/T} e^{-E_j/T}.. = e^{-E_X / T} = f_X^{\textrm{eq}} \,.
\label{eq:engcon}
\end{align}
Employing the definition of the zero-temperature decay rate $\Gamma_X^0$ of
the particle species $X$, the Boltzmann equation in Eq.~\eqref{eq:BEtherm} can
therefore be rewritten as
\begin{align}
\frac{d}{dt} f_X = & \: - \left(f_X - f_X^{\textrm{eq}}\right)
\frac{1}{2g_XE_X} \sum_{ij..} \int d \Pi \left(X|i,j,..\right) \left(2\pi\right)^4 \delta^{(4)}
\left|\mathcal{M}\left(X \rightarrow ij..\right)\right|^2 \label{eq:BEthermsimp} \\
= & \: - \left(f_X - f_X^{\textrm{eq}}\right) \frac{m_X}{E_X} \Gamma_X^0 \,. \nonumber
\end{align}
This Boltzmann equation equally describes the evolution of $f_{N_1}^{\textrm{th}}$
as well as of $f_{\tilde{N}_1}^{\textrm{th}}$.
Thus, even before attempting to solve it, we already know that
$f_{N_1}^{\textrm{th}}=f_{\tilde{N}_1}^{\textrm{th}}$ and hence
$N_{N_1}^{\textrm{th}}=N_{\tilde{N}_1}^{\textrm{th}}$ at all
times.\footnote{This statement presupposes, of course, that both
thermal neutrinos and thermal sneutrinos start out with the same initial
distribution function, which, however, is certainly the case.
At $t=t_G$, we have $f_{N_1}^{\textrm{th}} = f_{\tilde{N}_1}^{\textrm{th}} = 0.$}
The exact distribution function of the thermal
(s)neutrinos is given as the unique solution of Eq.~\eqref{eq:BEthermsimp}
for vanishing initial conditions,
\begin{align}
f_X^{\textrm{th}}(t,p) = \int_{t_G}^t dt' \frac{M_1\,\Gamma_{N_1}^0}
{\mathcal{E}_X(E_X;t,t')} f_X^{\textrm{eq}}(t',p) \,
\exp\left[-\int_{t'}^t dt''\frac{M_1\,\Gamma_{N_1}^0}{\mathcal{E}_X(E_X;t,t'')}\right]
\,,\:\: X = N_1, \tilde{N}_1 \,. \label{eq:fthexact}
\end{align}


As the thermal (s)neutrinos are produced within a broad range of energies
$\mathcal{E}_X(E_X;t,t')$, the exact solution for $f_X^{\textrm{th}}$ cannot
be as easily integrated over momentum space as the nonthermal distribution
function $f_X$ in Eq.~\eqref{eq:fnrhogammaX}.
However, since the thermal (s)neutrinos inherit their momentum distribution
from the particles in the thermal bath, it is reasonable to assume that they
are approximately in kinetic equilibrium (cf.\ Eq.~\eqref{eq:kineq}),
\begin{align}
f_X^{\textrm{th}}(t,p) \approx \frac{N_X^{\textrm{th}}(t)}{N_X^{\textrm{eq}}(t)}
f_X^{\textrm{eq}}(t,p) \,,\quad f_X^{\textrm{eq}}(t,p) = e^{-E_X/T}
\,,\qquad X = N_1,\tilde{N}_1 \,. \label{eq:fthapprox}
\end{align}
This approximation holds, if the quotient $f_X^{\textrm{th}}/f_X^{\textrm{eq}}$,
with $f_X^{\textrm{th}}$ taken from Eq.~\eqref{eq:fthexact}, is independent of
the (s)neutrino momentum $p$.
To be able to check whether this is indeed the case at all times, we need
to know the temperature of the thermal bath $T$ as a function of time.
The temperature $T$, however, is determined from the solution of the radiation Boltzmann
equation, which, as we will see further below, also contains terms involving the distribution
function $f_X^{\textrm{th}}$.
Consequently, we are only able to examine the \textit{self-consistency} of our
approximation in Eq.~\eqref{eq:fthapprox}.
Given a solution for $T$, we can check \textit{a posteriori} how well the approximate
distribution function in Eq.~\eqref{eq:fthapprox} coincides with the exact expression
in Eq.~\eqref{eq:fthexact}, i.e.\ to which extent the exact solution $f_X^{\textrm{th}}$ indeed
exhibits the same momentum dependence as $f_X^{\textrm{eq}}$.
Neglecting all effects due to supersymmetry and the decay of the gauge DOFs, we
perform such an analysis for a particular point in parameter space.
A detailed discussion of our findings is given in Appendix B of
Ref.~\cite{Buchmuller:2011mw}.
Here, we merely remark that, in the case of the investigated parameter point,
$f_X^{\textrm{th}}/f_X^{\textrm{eq}}$ turns out to depend, in general,
only very weakly on the (s)neutrino momentum $p$.
For particularly small and large momenta, the ratio $f_X^{\textrm{th}}/f_X^{\textrm{eq}}$
initially deviates from its mean value $\langle f_X^{\textrm{th}}/f_X^{\textrm{eq}}\rangle_p$
by as much as one order of magnitude.
In the course of the reheating process, the momentum dependence
of $f_X^{\textrm{th}}/f_X^{\textrm{eq}}$ then vanishes almost completely.
This is to say that the distribution function of the thermal (s)neutrinos
steadily converges towards the equilibrium distribution in Eq.~\eqref{eq:fthapprox}.
From this perspective, the approximation of kinetic equilibrium may hence be regarded
as justified.


Inelastic $2$-to-$2$ scatterings of the (s)thermal neutrinos involving MSSM
(s)quark pairs speed up the equilibration of the (s)neutrino distribution
function \cite{HahnWoernle:2009qn}.
This results in a larger abundance of thermal (s)neutrinos at earlier times.
On the other hand, these scatterings also tend to increase the efficiency
of washout processes such that, after all, their impact on the final lepton
asymmetry generated in the decay of the thermal (s)neutrinos is negligible for our purposes.
Again restricting ourselves to the nonsupersymmetric case and omitting the decay
of the gauge DOFs, we are able to numerically confirm this picture by solving
the relevant set of Boltzmann equations~\cite{Buchmuller:2011mw}.


Assuming that the thermal (s)neutrinos are in kinetic equilibrium throughout
the reheating process, Eq.~\eqref{eq:BEthermsimp} is easily integrated over momentum space.
Eventually, we obtain
\begin{align}
a H \frac{d}{da} N_X^{\textrm{th}} = - \Gamma_X^{\textrm{th}}
\left(N_X^{\textrm{th}}-N_X^{\textrm{eq}}\right)
\,,\qquad X = N_1,\tilde{N}_1 \,. \label{eq:BENth}
\end{align}
Here, $N_X^{\text{eq}}$ denotes the comoving number density of the thermal (s)neutrinos
in thermal equilibrium (cf.\ Eq.~\eqref{eq:nrhoXeq}) and $\Gamma_X^{\textrm{th}}$
stands for the zero-temperature decay rate of the thermal (s)neutrinos, weighted
with the corresponding average inverse time dilatation factor,
\begin{align}
\Gamma^{\textrm{th}}_X = \left<\frac{M_1}{E_X}\right>_{\textrm{th}} \Gamma_X^0 =
\frac{a^3}{N_X^\textrm{th}} \frac{g_N}{\left(2\pi\right)^3}
\int d^3p \frac{M_1}{E_X} f_x^\textrm{th} \, \Gamma_X^0 =
\frac{K_1(M_1/T)}{K_2(M_1/T)} \, \Gamma_X^0 \,,\quad X = N_1,\tilde{N}_1 \,,
\label{eq:GammathN1}
\end{align}
where $K_1$ and $K_2$ are the modified Bessel functions of the second kind of order $1$ and $2$.


\subsection{MSSM Degrees of Freedom}
\label{subsec:BELR}


All three generations of heavy (s)neutrinos exclusively decay into the
lepton-Higgs pairs of the MSSM.
Due the strong standard model gauge interactions, the heavy-(s)neutrino
decay products immediately thermalize, so that the energy of a decaying (s)neutrino
is always quickly distributed among all MSSM DOFs.
In this sense, the energy transfer from the heavy (s)neutrinos to the thermal
bath represents the actual reheating process after the \BmL phase transition.
The production of entropy during reheating is conveniently
described by means of the Boltzmann equation for the comoving number density $N_R$
of MSSM or radiation quanta.
Meanwhile, the decay of the heavy (s)neutrinos also entails the generation of a
primordial lepton asymmetry.
Before turning to the radiation Boltzmann equation, let us derive the Boltzmann
equation for the comoving number density $N_L = N_\ell - N_{\bar{\ell}}$,
which characterizes the excess of leptons $\ell$ over antileptons $\bar{\ell}$.


\subsubsection{Lepton Asymmetry}


The Boltzmann equation for the lepton asymmetry $L$ directly follows from the
respective equations for the lepton supermultiplet $\ell$ and the antilepton supermultiplet $\bar{\ell}$,
\begin{align}
\hat{\mathcal{L}} f_L = & \: \hat{\mathcal{L}} f_\ell - \hat{\mathcal{L}} f_{\bar{\ell}} \,,
\label{eq:BEL}\\
\hat{\mathcal{L}} f_\ell = & \: \sum_i \left[
C_\ell(\ell H_u \leftrightarrow N_i) +
C_\ell(\tilde{\ell} \tilde{H}_u \leftrightarrow N_i) +
C_\ell(\ell \tilde{H}_u \leftrightarrow \tilde{N}_i^*) +
C_\ell(\tilde{\ell} H_u \leftrightarrow \tilde{N}_i) \right] +
2 \, \mathcal{C}_\ell^{\textrm{red}}
\,,\nonumber
\end{align}
with $\mathcal{C}_\ell^{\textrm{red}}$ being the reduced collision operator
for all $2$-to-$2$ scattering processes with a heavy (s)neutrino in the intermediate state
which result in a change of the total lepton number by two units,
i.e.\ $\Delta L = 2$ scatterings processes with lepton-Higgs pairs in the external states,
\begin{align}
\mathcal{C}_\ell^{\textrm{red}} = & \:
C_\ell^{\textrm{red}}(\ell H_u \leftrightarrow \bar{\ell} H_u^*) +
C_\ell^{\textrm{red}}(\tilde{\ell} \tilde{H}_u \leftrightarrow \tilde{\ell}^* \bar{\tilde{H}}_u) +
C_\ell^{\textrm{red}}(\ell H_u \leftrightarrow \tilde{\ell}^* \bar{\tilde{H}}_u)
\label{eq:Clred} \\ + & \: \nonumber
C_\ell^{\textrm{red}}(\tilde{\ell} \tilde{H}_u \leftrightarrow \bar{\ell} H_u^*) +
C_\ell^{\textrm{red}}(\ell \tilde{H}_u \leftrightarrow \tilde{\ell}^* H_u^*) +
C_\ell^{\textrm{red}}(\tilde{\ell} H_u \leftrightarrow \bar{\ell} \bar{\tilde{H}}_u)
\,.
\end{align}
The Boltzmann equation for the antilepton multiplet $\bar{\ell}$ is readily obtained
by $CP$-conjugating each term in the Boltzmann equation for the lepton multiplet $\ell$.


The ordinary collision operators in Eq.~\eqref{eq:BEL}, accounting for the decays and
inverse decays of the heavy (s)neutrinos $\big(N_i,\tilde{N}_i\big)$, are able to
mimic $\Delta L = 2$ scatterings with on-shell (s)neutrinos in the intermediate state.
They, however, disregard off-shell scatterings, even though these equally affect
the evolution of the lepton asymmetry.
To remedy this flaw, we have to include the reduced collision operators
$\mathcal{C}_\ell^{\textrm{red}}$ and $\mathcal{C}_{\bar{\ell}}^{\textrm{red}}$
in the Boltzmann equations for $\ell$ and $\bar{\ell}$, which incorporate
the off-shell contributions to all relevant $\Delta L = 2$ scattering processes.
For $M_1 \ll 10^{14}\,\textrm{GeV}$, the $CP$-preserving parts of the reduced
collision operators are negligibly small~\cite{Buchmuller:2002rq}.
We thus discard them, keeping only their $CP$-violating contributions,
\begin{align}
\mathcal{C}_X^{\textrm{red}} = \mathcal{C}_{X,CP}^{\textrm{red}} +
\mathcal{C}_{X,\cancel{CP}}^{\textrm{red}}
\approx \mathcal{C}_{X,\cancel{CP}}^{\textrm{red}}
\,,\qquad X = \ell, \bar{\ell} \,.\label{eq:CXCPpres}
\end{align}
The operators $\mathcal{C}_{X,\cancel{CP}}^{\textrm{red}}$ can be computed
by calculating the $CP$-violating contributions $\mathcal{C}_{X,\cancel{CP}}$
to the full collision operators $\mathcal{C}_X$ and then subtracting their
on-shell parts $\mathcal{C}_{X,\cancel{CP}}^{\textrm{on}}$,
\begin{align}
\mathcal{C}_{X,\cancel{CP}}^{\textrm{red}} =
\mathcal{C}_{X,\cancel{CP}} - \mathcal{C}_{X,\cancel{CP}}^{\textrm{on}}
\,,\qquad X = \ell, \bar{\ell} \,.
\end{align}
As we prove in Appendix~\ref{ch:scatterings}, the unitarity and $CPT$ invariance
of the $S$ matrix imply that the $CP$-violating operators $\mathcal{C}_{X,\cancel{CP}}$
vanish up to corrections of $\mathcal{O}\left((h^\nu)^4\right)$.
Hence, the inclusion of the reduced collision operators
$\mathcal{C}_{X,\cancel{CP}}^{\textrm{red}}$ is practically equivalent to
the subtraction of the operators $\mathcal{C}_{X,\cancel{CP}}^{\textrm{on}}$,
which describe $\Delta L = 2$ scatterings with real intermediate states,
\begin{align}
\mathcal{C}_{X,\cancel{CP}}^{\textrm{red}} =
- \mathcal{C}_{X,\cancel{CP}}^{\textrm{on}} + \mathcal{O}\left((h^\nu)^4\right)
\,,\qquad X = \ell, \bar{\ell} \,. \label{eq:CXCPviol}
\end{align}


The individual collision operators $C_X^{\textrm{on}}$ contributing
to $\mathcal{C}_X^{\textrm{on}}$ may be rewritten as collision operators
for inverse decays of MSSM lepton-Higgs pairs into \textit{real} heavy (s)neutrinos,
\begin{align}
& \: C_X^{\textrm{on}}\left(I \leftrightarrow \bar{F} \right) =
\sum_i
\left[\textrm{Br}\left(R_i\rightarrow I\right)
C_X^{\textrm{on}}\left(\bar{F} \rightarrow R_i \right)-
\textrm{Br}\left(R_i\rightarrow \bar{F}\right)
C_X^{\textrm{on}}\left(I \rightarrow R_i \right)\right] \,, \label{eq:CXon}\\
& \: X_{\phantom{i}} = \ell, \bar{\ell} \,,\quad
R_i = N_i \,,\:\: F = I,\tilde{I} \:\:\textrm{for}\:\: I = \ell H_u , \tilde{\ell}\tilde{H}_u
\,;\quad \nonumber\\
& \: R_i = \tilde{N}_i \,, \:\: F = \tilde{I} \:\: \textrm{for} \:\:I = \tilde{\ell} H_u \,; \quad
R_i = \tilde{N}_i^* \,, \:\: F = \tilde{I} \:\:\textrm{for}\:\: I = \ell \tilde{H}_u \,. \nonumber
\end{align}
Here, $\bar{F}$ represents the pair of antiparticles corresponding to
the pair of particles $F$.
Similarly, $\tilde{I}$ is the pair of superparticles corresponding to
the pair of particles $I$.
Note that this relation equally applies to the $CP$-conserving parts
$C_{X,CP}^{\textrm{on}}$ of the collision operators $C_X^{\textrm{on}}$ as
well as to their $CP$-violating parts $C_{X,\cancel{CP}}^{\textrm{on}}$.
After rewriting all operators $C_{X,\cancel{CP}}^{\textrm{on}}$
as operators for inverse decays, the Boltzmann equations for $\ell$ and $\bar{\ell}$
end up solely containing collision operators, all of which respectively
look like one of the following two prototypes,
\begin{align}
C_X^{\textrm{on}}(Xj \rightarrow R_i) = & \:
\frac{1}{2g_X} \int d \Pi \left(X|j;R_i\right) \left(2\pi\right)^4 \delta^{(4)}
f_X f_j \left|\mathcal{M}\left(Xj \rightarrow R_i\right)\right|^2 \,, \label{eq:CXonproto}\\
C_X^{\textrm{on}}(R_i \rightarrow Xj) = & \:
\frac{1}{2g_X} \int d \Pi \left(X|R_i;j\right) \left(2\pi\right)^4 \delta^{(4)}
f_{R_i} \left|\mathcal{M}\left(R_i \rightarrow Xj\right)\right|^2 \,. \nonumber
\end{align}
The only amplitudes squared which we have to calculate are hence
those describing the decays and inverse decays of heavy (s)neutrinos.
Using the definition of the $CP$ violation parameters $\epsilon_i$ as well
as $CPT$ invariance, the various partial amplitudes squared can be related
to the tree-level amplitudes squared $|\mathcal{M}_i|^2$,
\begin{align}
|\mathcal{M}_i|^2 =
\big|\mathcal{M}\big(N_i\leftrightarrow\ell H_u, \bar{\ell} H_u^*,
\tilde{\ell} \tilde{H}_u, \tilde{\ell}^* \bar{\tilde{H}}_u\big)\big|^2 =
8 \big[\left(h^\nu\right)^\dagger h^\nu\big]_{ii} M_i^2 \,.
\end{align}
Up to first order in the $CP$ violation parameters $\epsilon_i$,
one finds~\cite{Covi:1996wh},
\begin{align}
& \: \label{eq:amplitudes}
|\mathcal{M}(N_i \rightarrow \ell H_u)|^2 =
|\mathcal{M}(\bar{\ell} H_u^* \rightarrow N_i)|^2 =
|\mathcal{M}(N_i \rightarrow \tilde{\ell} \tilde{H}_u)|^2 =
|\mathcal{M}(\tilde{\ell}^* \bar{\tilde{H}}_u \rightarrow N_i)|^2 \\
= & \: \nonumber
|\mathcal{M}(\tilde{N}_i^* \rightarrow \ell \tilde{H}_u)|^2 =
|\mathcal{M}(\bar{\ell} \bar{\tilde{H}}_u \rightarrow \tilde{N}_i)|^2 =
|\mathcal{M}(\tilde{N}_i \rightarrow \tilde{\ell} H_u)|^2 =
|\mathcal{M}(\tilde{\ell}^* H_u^* \rightarrow \tilde{N}_i^*)|^2 \\
= & \: \nonumber
\frac{1}{4} \left(1 + \epsilon_i \right) |\mathcal{M}_i|^2 \,, \\
& \: \nonumber
|\mathcal{M}(N_i \rightarrow \bar{\ell} H_u^*)|^2 =
|\mathcal{M}(\ell H_u \rightarrow N_i)|^2 =
|\mathcal{M}(N_i \rightarrow \tilde{\ell}^* \bar{\tilde{H}}_u)|^2 =
|\mathcal{M}(\tilde{\ell} \tilde{H}_u \rightarrow N_i)|^2 \\
= & \: \nonumber
|\mathcal{M}(\tilde{N}_i \rightarrow \bar{\ell} \bar{\tilde{H}}_u)|^2 =
|\mathcal{M}(\ell \tilde{H}_u \rightarrow \tilde{N}_i^*)|^2 =
|\mathcal{M}(\tilde{N}_i^* \rightarrow \tilde{\ell}^* H_u^*)|^2 =
|\mathcal{M}(\tilde{\ell} H_u \rightarrow \tilde{N}_i)|^2 \\
= & \: \nonumber
\frac{1}{4} \left(1 - \epsilon_i \right) |\mathcal{M}_i|^2 \,.
\end{align}


The relations in Eqs.~\eqref{eq:CXon}, \eqref{eq:CXonproto} and
\eqref{eq:amplitudes} enable us to compute the reduced
collision operators in the Boltzmann equations for $\ell$ and $\bar{\ell}$.
As we are only interested in the $CP$-violating contributions
$\mathcal{C}_{X,\cancel{CP}}^{\textrm{red}}$ to the operators
$\mathcal{C}_X^{\textrm{red}}$, we solely take into account those parts
of the various amplitudes squared which are proportional to $\epsilon_i$.
Working up to leading order in $\epsilon_i$, it is sufficient to employ
the tree-level results for the (s)neutrino branching ratios (cf.\ Eq.~\eqref{eq:NNtBr})
and to approximate the distribution functions of all particles in the lepton and Higgs
multiplets by Maxwell-Boltzmann distributions.
This latter simplification allows us in particular to replace the product
$f_X f_j$ in $C_X^{\textrm{on}}(Xj \rightarrow R_i)$ (cf.\ Eq.~\eqref{eq:CXonproto})
by the (s)neutrino equilibrium distribution function $f_{N_i}^{\textrm{eq}}$
(cf.\ Eq.~\eqref{eq:engcon}).
Up to corrections of $\mathcal{O}\left(\epsilon_i^2\right)$, we find
\begin{align}
\mathcal{C}_{X,\cancel{CP}}^{\textrm{red}}
\approx - \mathcal{C}_{X,\cancel{CP}}^{\textrm{on}} =
\mp\frac{1}{2g_X} \sum_i \epsilon_i\int d \Pi \left(X|j;R_i\right) \left(2\pi\right)^4
\delta^{(4)} f_{N_i}^{\textrm{eq}} \left|\mathcal{M}_i\right|^2 \,, \qquad X = \ell,\bar{\ell} \,.
\end{align}
The collision operators describing the decays and inverse decays of
the heavy (s)neutrinos are of a similar form.
In contrast to the reduced collision operators, we, however, do not take the
(s)leptons to be in thermal equilibrium, when calculating these operators.
Combining our results for all collision operators, the Boltzmann equations
for $\ell$ and $\bar{\ell}$ finally read
\begin{align}
\hat{\mathcal{L}} f_X = & \: \frac{1}{2g_X} \sum_i \int d \Pi \left(X|j;R_i\right)
\left(2\pi\right)^4 \delta^{(4)} \frac{1}{4}\left|\mathcal{M}_i\right|^2
\label{eq:BEfell}\\
\times & \: \Big[(1 \pm \epsilon_i)(2 f_{N_i} + f_{\tilde{N}_i} + f_{\tilde{N}_i^*})
\mp 8 \epsilon_i f_{N_i}^{\textrm{eq}} - 2 (1 \mp \epsilon_i) f_X f_j^{\textrm{eq}}\Big]
\,, \qquad X = \ell,\bar{\ell} \,. \nonumber
\end{align}
The difference of these two equations yields the Boltzmann equation
for the lepton asymmetry,
\begin{align}
\hat{\mathcal{L}} f_L = & \: \frac{1}{2g_L} \sum_i \int d \Pi \left(L|j;R_i\right)
\left(2\pi\right)^4 \delta^{(4)}\left|\mathcal{M}_i\right|^2 \label{eq:BEfL}\\
\times & \: \Big[\epsilon_i (f_{N_i} - f_{N_i}^{\textrm{eq}}) +
\frac{\epsilon_i}{2} (f_{\tilde{N}_i} + f_{\tilde{N}_i^*} - 2 f_{N_i}^{\textrm{eq}})
- \frac{f_{N_i}^{\textrm{eq}}}{2 N_\ell^{\textrm{eq}}} N_L \Big]
\,, \nonumber
\end{align}
where we have used the fact that the MSSM (s)leptons are in kinetic equilibrium,
\begin{align}
f_L = f_\ell - f_{\bar{\ell}} = \frac{N_\ell}{N_\ell^{\textrm{eq}}} f_\ell^{\textrm{eq}}
- \frac{N_{\bar{\ell}}}{N_{\bar{\ell}}^{\textrm{eq}}} f_{\bar{\ell}}^{\textrm{eq}}
= \frac{N_L}{N_\ell^{\textrm{eq}}} f_\ell^{\textrm{eq}}
\,, \quad f_L f_j^{\textrm{eq}} = \frac{N_L}{N_\ell^{\textrm{eq}}} f_{N_i}^{\textrm{eq}} \,.
\end{align}


Just as in the case of the Boltzmann equations for the heavy (s)neutrinos,
we again split the (s)neutrino distribution functions into independently
evolving components.
The integration of Eq.~\eqref{eq:BEfL} over the lepton number momentum space
then provides us with the Boltzmann equation for the comoving number density $N_L$.
Up to corrections of $\mathcal{O}\left(\epsilon_i^2, (h^\nu)^4\right)$, we obtain
\begin{align}
a H \frac{d}{da} N_L = a H \frac{d}{da} \left(N_L^{\textrm{nt}} + N_L^{\textrm{th}}\right)
\,,\quad
a H \frac{d}{da} N_L^{\textrm{x}} =
\hat{\Gamma}_L^{\textrm{x}} N^{\textrm{x}}_L - \hat{\Gamma}_W N_{L}^{\textrm{x}}
\,,\quad \textrm{x} = \textrm{nt}, \textrm{th} \,, \label{eq:BELint}
\end{align}
with the washout rate $\hat{\Gamma}_W$ and the effective (non)thermal
production rates $\hat{\Gamma}_L^{\textrm{nt,th}}$ given by
\begin{align}
\hat{\Gamma}_W = & \:
\sum_i \frac{N_{N_i}^{\text{eq}}}{2 N_{\ell}^{\textrm{eq}}}\Gamma_{N_i}^{\textrm{th}}
\approx \frac{N_{N_1}^{\text{eq}}}{2 N_{\ell}^{\textrm{eq}}}\Gamma_{N_1}^{\textrm{th}}\,, 
\label{eq:Lrates}\\
\hat{\Gamma}_L^{\text{nt}} = & \:  \big(N_L^{\textrm{nt}}\big)^{-1}
\Bigg[\sum_{R_i} \epsilon_i
\left(\Gamma_{R_i}^{\textrm{PH}} N_{R_i}^{\textrm{PH}} +
\Gamma_{R_i}^G N_{R_i}^G\right)
+ \epsilon_1 \big(\Gamma_{N_1}^S N_{N_1}^S +
\Gamma_{\tilde{N}_1}^S N_{\tilde{N}_1}^S \big) \Bigg]\,, \nonumber \\
\hat{\Gamma}_L^{\textrm{th}} = & \: \big(N_L^{\textrm{th}}\big)^{-1}
\sum_i \epsilon_i \,\Gamma_{N_i}^{\textrm{th}} \big(N_{N_i}^{\textrm{th}}
+ N_{\tilde{N}_i}^{\textrm{th}} - 2 N_{N_i}^{\textrm{eq}}\big) \approx
\big(N_L^{\textrm{th}}\big)^{-1} \epsilon_1 \,\Gamma_{N_1}^{\textrm{th}}
\big(N_{N_1}^{\textrm{th}} + N_{\tilde{N}_1}^{\textrm{th}} - 2 N_{N_1}^{\textrm{eq}}\big)\,.
\nonumber \phantom{\Bigg]}
\end{align}
Here, all quantities labeled with an index $\tilde{N}_i$ equally comprise
the heavy sneutrinos $\tilde{N}_i$ \textit{and} the heavy antisneutrinos $\tilde{N}_i^*$.
As for the washout rate $\hat{\Gamma}_W$ and the thermal production rate
$\hat{\Gamma}_L^{\textrm{th}}$, we only consider the contributions from the
first heavy (s)neutrino generation.
Since $M_3 \sim M_3 \gg M_1$, the thermal as well as equilibrium abundances of the
two heavier (s)neutrino generations, $N_{N_i}^{\textrm{th}}$, $N_{\tilde{N}_i}^{\textrm{th}}$
and $N_{N_i}^{\textrm{eq}}$, where $i = 2,3$, are strongly suppressed, so that we can safely
neglect the corresponding terms in the Boltzmann equations.
The various decay rates $\Gamma_{R_i}^{\textrm{x}}$ are computed according to
Eqs.~\eqref{eq:GammaxNi} and \eqref{eq:dilatation}.
Note that explicit expressions for the rates $\Gamma_{N_1}^S$,
$\Gamma_{\tilde{N}_1}^S$ and $\Gamma_{N_1}^{\textrm{th}}$ are given in
Eqs.~\eqref{eq:GammaXi} and \eqref{eq:GammathN1}, respectively.
Meanwhile, the remaining rates, i.e.\ $\Gamma_{R_i}^{\textrm{PH}}$ and $\Gamma_{R_i}^G$,
can be easily calculated using our results for the distribution functions
$f_{R_i}^{\textrm{PH}}$ and $f_{R_i}^G$ in Eq.~\eqref{eq:fN23sol}.
We remark that Eq.~\eqref{eq:Lrates} nicely illustrates the connection between
the decay rates $\Gamma_{R_i}^{\textrm{x}}$ on the one hand and the effective
production rates $\hat{\Gamma}_W$ and $\hat{\Gamma}_L^{\textrm{nt,th}}$
on the other hand.
The latter describe in particular the relative change in the lepton asymmetry
due to a given process and can hence be directly compared to the Hubble
rate $H$ in order to assess the efficiency of the respective process.\footnote{This
is a general feature of any effective production rate $\hat{\Gamma}_i$.
To see this, note that the Boltzmann equation for the number density
$n_X$ of a species $X$ can always be written as
$\dot{n}_X / n_X = \big(\hat{\Gamma}/H - 3\big) H$, where
$\hat{\Gamma} = \sum_i \hat{\Gamma}_i$.}
Furthermore, we point out that we have introduced $N_L^{\textrm{nt}}$ and
$N_L^{\textrm{th}}$ in Eq.~\eqref{eq:BELint} to denote the nonthermal
and thermal contributions to the total lepton asymmetry
$N_L = N_L^{\textrm{nt}}+ N_L^{\textrm{th}}$, respectively.
The comparison of the corresponding final baryon asymmetries $\eta_B^{\textrm{nt}}$
and $\eta_B^{\textrm{th}}$ will eventually allow us to identify the relative importance
of nonthermal and thermal leptogenesis in different regions of parameter space
(cf.\ Secs.~\ref{sec:evolution} and \ref{sec:scan}).
For the parameters $\epsilon_i$,
we employ the Froggatt-Nielsen estimates in Eq.~\eqref{eq:epsiloniest}.
The final baryon asymmetry inferred from the solution of the Boltzmann equation in
Eq.~\eqref{eq:BELint} then corresponds to the maximum possible baryon asymmetry,
i.e.\ an upper bound on the actually produced asymmetry.
As we have seen in Sec.~\ref{sec:predictions}, in the context of the Froggatt-Nielsen
model, we should, however, expect that the actually produced asymmetry,
corresponding to the actual value of $\epsilon_1$, is only slightly smaller than
this upper bound (cf.\ Eq.~\eqref{eq:m1tepsilon1}).


\subsubsection{Radiation}


The progress of the reheating process after the \BmL phase transition
is reflected in the time evolution of the temperature of the thermal bath $T$.
In principle, the behaviour of $T$ as a function of time $t$ is determined by the
following non-linear first-order differential equation, which directly follows from
the covariant energy conservation,
\begin{align}
\dot{\rho}_{\textrm{tot}} + 3H\left(\rho_{\textrm{tot}}
+ p_{\textrm{tot}}\right) = 0\,.
\label{eq:covengcon}
\end{align}
We, however, choose to pursue a different approach and infer the temperature
$T$ from the comoving number density of radiation quanta, i.e.\ MSSM particles,
$N_R$ (cf.\ Eq.~\eqref{eq:nRrhoR}),
\begin{align}
N_R = a^3 \,\frac{\zeta(3)}{\pi^2} \, g_{*,n} \, T^3 \,,\quad
T = \left(\frac{\pi^2}{g_{*,n} \, \zeta(3)} \frac{N_R}{a^3} \right)^{1/3} \,.
\label{eq:tempNR}
\end{align}
Similarly to the comoving number densities of all other species,
the time evolution of $N_R$ may be studied by means of an appropriate
Boltzmann equation.
Hence, deducing the temperature $T$ from $N_R$ has the advantage that it allows
us to consistently describe the reheating process exclusively in terms of a set of
Boltzmann equations for comoving number densities $N_X$.


While the Boltzmann equation for the lepton asymmetry $L$ corresponds to
the difference of the respective equations for $\ell$ and $\bar{\ell}$
(cf.\ Eq.~\eqref{eq:BEfell} and \eqref{eq:BEfL}), the Boltzmann equation
for the distribution function $f_R$ of MSSM particles is related to the sum
of these two equations,
\begin{align}
\hat{\mathcal{L}} f_R = \hat{r}_R \left( \hat{\mathcal{L}} f_\ell +
\hat{\mathcal{L}} f_{\bar{\ell}}\right) \,. \label{eq:BERdef}
\end{align}
Here, $\hat{r}_R$ counts the number of radiation quanta effectively
added to the thermal bath in the decay of a heavy (s)neutrino.
It has to be thought of as an operator acting on the various (s)neutrino distribution
functions $f_{R_i}^{\textrm{x}}$ contained in $\hat{\mathcal{L}} f_{\ell,\bar{\ell}}$
in the following way,
\begin{align}
\hat{r}_R f_{R_i}^{\textrm{x}} = r_{R_i}^{\textrm{x}} f_{R_i}^{\textrm{x}}
\,,\quad R_i = N_i, \tilde{N}_i \,,\quad \textrm{x} = \textrm{PH}, G, S, \textrm{th} \,,
\end{align}
with $r_{R_i}^{\textrm{x}}$ denoting the effective increase of radiation quanta
due to the decay of a heavy (s)neutrino $R_i$ which originates from a production
mechanism $\textrm{x}$.
Let us now derive an explicit expression for $r_{R_i}^{\textrm{x}}$.
We consider a spatial volume $V$ in which heavy (s)neutrinos $R_i^{\textrm{x}}$
of average energy $\varepsilon_{R_i}^{\textrm{x}}$ decay into MSSM lepton-Higgs pairs.
The (s)neutrino decay products thermalize practically instantaneously
after their production, so that we may neglect the cosmic expansion
for the moment.
Per decay, the energy density of the thermal bath is then increased by
$\varepsilon_{R_i}^{\textrm{x}}/V$ and a new thermal equilibrium at a slightly higher
temperature is established right after the decay,
\begin{align}
\rho_R \rightarrow \rho_R + \frac{\varepsilon_{R_i}^{\textrm{x}}}{V} \,,\quad
T \rightarrow T \left(1 + \frac{1}{\rho_R}\frac{\epsilon_{R_i}^{\textrm{x}}}{V}\right)^{1/4} \,.
\end{align}
This increase in $T$ corresponds to an increase
in the number density $n_R$ (cf.\ Eq.~\eqref{eq:nRrhoR}),
\begin{align}
n_R \rightarrow n_R
\left(1 + \frac{1}{\rho_R}\frac{\epsilon_{R_i}^{\textrm{x}}}{V}\right)^{3/4}
\simeq n_R + \frac{3}{4}\frac{n_R}{\rho_R}\frac{\varepsilon_{R_i}^{\textrm{x}}}{V}
= n_R + \frac{r_{R_i}^{\textrm{x}}}{V} \,, \:\:\:
r_{R_i}^{\textrm{x}} = \frac{3\,\varepsilon_{R_i}^{\textrm{x}}}{4\,\varepsilon_R}
\,,\:\:\: \varepsilon_R = \frac{\rho_R}{n_R} \,.
\end{align}
Note that, similarly to $\varepsilon_R$, the average (s)neutrino
energy $\varepsilon_{R_i}^{\textrm{x}}$ may also be obtained as the ratio of the
corresponding energy density to the corresponding number density,
$\varepsilon_{R_i}^{\textrm{x}} = \rho_{R_i}^{\textrm{x}}/n_{R_i}^{\textrm{x}}$.


With these remarks on the operator $\hat{r}_R$ in mind, we are now ready to
write down the radiation Boltzmann equation.
Neglecting all effects of $CP$ violation, we find
\begin{align}
\hat{\mathcal{L}} f_R = \frac{1}{2g_R} \sum_i \int d \Pi \left(R|j;R_i\right)
\left(2\pi\right)^4 \delta^{(4)} \left|\mathcal{M}_i\right|^2 \hat{r}_R
\Big[f_{N_i} + \frac{1}{2} f_{\tilde{N}_i} + \frac{1}{2} f_{\tilde{N}_i^*}
- 2 f_{N_i}^{\textrm{eq}}\Big] \,.
\end{align}
After decomposing the (s)neutrino distribution functions into their respective
components, the integration of this equation over momentum space yields the
Boltzmann equation for the comoving number density $N_R$ of MSSM DOFs,
\begin{align}
a H \frac{d}{da} N_R = a H \frac{d}{da} \left(N_R^{\textrm{nt}} + N_R^{\textrm{th}}\right) =
\hat{\Gamma}_R N_R \,,\quad
a H \frac{d}{da} N_R^{\textrm{x}} = \hat{\Gamma}_R^{\textrm{x}} N^{\textrm{x}}_R
\,,\quad \textrm{x} = \textrm{nt}, \textrm{th} \,,
\label{eq:BERint}
\end{align}
where the effective (non)thermal production rates
$\hat{\Gamma}_R^{\textrm{nt,th}}$ are given by
\begin{align}
\hat{\Gamma}_R^{\text{nt}} = & \:  \big(N_R^{\textrm{nt}}\big)^{-1}
\Bigg[\sum_{R_i}
\left(r_{R_i}^{\textrm{PH}}\Gamma_{R_i}^{\textrm{PH}} N_{R_i}^{\textrm{PH}} +
r_{R_i}^G \Gamma_{R_i}^G N_{R_i}^G\right)
+ \big(r_{N_1}^S \Gamma_{N_1}^S N_{N_1}^S +
r_{\tilde{N}_1}^S \Gamma_{\tilde{N}_1}^S N_{\tilde{N}_1}^S \big)\Bigg]\,, \\
\hat{\Gamma}_R^{\textrm{th}} = & \: \big(N_R^{\textrm{th}}\big)^{-1}
\sum_i r_{R_i}^{\textrm{th}} \,\Gamma_{N_i}^{\textrm{th}} \big(N_{N_i}^{\textrm{th}}
+ N_{\tilde{N}_i}^{\textrm{th}} - 2 N_{N_i}^{\textrm{eq}}\big) \approx
\big(N_R^{\textrm{th}}\big)^{-1} r_{R_1}^{\textrm{th}} \,\Gamma_{N_1}^{\textrm{th}}
\big(N_{N_1}^{\textrm{th}} + N_{\tilde{N}_1}^{\textrm{th}} - 2 N_{N_1}^{\textrm{eq}}\big)
\,. \nonumber
\end{align}
We note that this result for the radiation Boltzmann equation is very similar
in form to the Boltzmann equation for the lepton asymmetry in Eq.~\eqref{eq:BELint}.
The only differences are that Eq.~\eqref{eq:BERint} contains no washout but only
production terms and that it features the factors $r_{R_i}^{\textrm{x}}$ instead
of the $CP$ violation parameters $\epsilon_i$.
Moreover, we point out that the factor $r_{R_i}^{\textrm{th}}$ can be equally used to
count the number of radiation quanta produced in the decay of thermal neutrinos
$N_i^{\textrm{th}}$ as well as the quanta produced in the decay of thermal sneutrinos
$\tilde{N}_i^{\textrm{th}}$.
Likewise, it also applies to heavy (s)neutrinos in thermal equilibrium
(cf.\ Eq.~\eqref{eq:nrhoXeq}),
\begin{align}
r_{R_i}^{\textrm{th}} = r_{N_i}^{\textrm{th}} = r_{\tilde{N}_i}^{\textrm{th}}
= r_{N_i}^{\textrm{eq}} = r_{\tilde{N}_i}^{\textrm{eq}} =
\frac{3\,\varepsilon_{R_i}^{\textrm{th}}}{4\,\varepsilon_R} \,, \quad
\varepsilon_{R_i}^{\textrm{th}} = \varepsilon_{R_i}^{\textrm{eq}} =
3 T + \frac{K_1(M_i/T)}{K_2(M_i/T)} M_i \,.
\end{align}
We also remark that we have introduced the total radiation production
rate $\hat{\Gamma}_R$ in Eq.~\eqref{eq:BERint},
\begin{align}
\hat{\Gamma}_R = N_R^{-1} \, a H\frac{d N_R}{da} =
N_R^{-1} \, \frac{d N_R}{dt} =
\frac{N_R^{\textrm{nt}}}{N_R} \hat{\Gamma}_R^{\textrm{nt}} +
\frac{N_R^{\textrm{th}}}{N_R} \hat{\Gamma}_R^{\textrm{th}} \,.
\label{eq:GammaR}
\end{align}
$\hat{\Gamma}_R$ counts the relative increase in the comoving
radiation number density $N_R$ per unit time.
It will prove to be a useful quantity in our discussion of
the reheating process in Sec.~\ref{sec:evolution}.


\subsection{Gravitinos}


Gravitinos are predominantly produced through inelastic QCD $2$-to-$2$ scattering
processes in the thermal bath.\footnote{Cf.\ Sec.~\ref{subsec:candidates} for a
comprehensive discussion of all conceivable gravitino production mechanisms.}
The integrated Boltzmann equation governing the time evolution of the comoving
gravitino number density $N_{\widetilde{G}}$ reads
\begin{align}
a H \frac{d}{da} N_{\widetilde{G}} =  \hat{\Gamma}_{\widetilde{G}} N_{\widetilde{G}} \,.
\label{eq:BoltzGravi}
\end{align}
In supersymmetric QCD, up to leading order in the strong gauge coupling $g_s$,
one obtains the following expression for the total production rate
$\hat{\Gamma}_{\widetilde{G}}$~\cite{Bolz:2000fu},
\begin{align}
\hat{\Gamma}_{\widetilde{G}}(T) = \frac{a^3}{N_{\widetilde{G}}}
\left(1+\frac{m^2_{\tilde{g}}(T)}{3 m^2_{\widetilde{G}}}\right)
\frac{54\, \zeta(3)\, g_s^2(T)}{\pi^2 M_P^2} \, T^6
\left[\ln\left(\frac{T^2}{m_g^2(T)}\right) + 0.8846 \right] \,,
\label{eq:GammaG}
\end{align}
Here, $m_{\tilde{g}}$ denotes the energy scale-dependent gluino mass and $m_g$ is the
gluon plasma mass,
\begin{align}
m_{\tilde{g}}(t) = \frac{g_s^2(T)}{g_s^2(\mu_0)} \, m_{\tilde{g}}(\mu_0) \,,\quad
m_g(t) = \sqrt{\frac{3}{2}} \, g_s(T) \, T\,.
\label{eq:mgtmg}
\end{align}
As reference scale $\mu_0$, we choose the $Z$ boson mass $M_Z \simeq 91.18\,\textrm{GeV}$,
where the strong coupling constant is given by
$\alpha_s(\mu_0) = g_s^2(\mu_0)/\left(4\pi\right) \simeq 0.118$~\cite{Nakamura:2010zzi}.
The scale dependence of $g_s$ is controlled by the corresponding MSSM
renormalization group equation, which is solved by
\begin{align}
g_s(\mu(T)) = g_s(\mu_0) \left[1 + \frac{3}{8 \pi^2} \, g_s^2(\mu_0) \,
\ln \frac{\mu(T)}{\mu_0} \right]^{-1/2} \,.
\end{align}
with $\mu$ being the typical energy scale during reheating.
It can be estimated as the average energy per relativistic particle in the thermal bath,
$\mu(T) \simeq \varepsilon_R \simeq 3T$.
For instance at temperatures $T= 10^8,10^{10},10^{12}\,\textrm{GeV}$,
the strong coupling constant therefore takes the values $g_s = 0.90, 0.84,0.80$.
The gravitino mass $m_{\tilde G}$ and the gluino mass at the electroweak scale
$m_{\tilde{g}} = m_{\tilde{g}}(\mu_0)$ remain as free parameters.


\section{Time Evolution of the Particle Abundances}
\label{sec:evolution}


Combining the initial conditions set by the \BmL phase transition
with the Boltzmann equations derived in the previous
section poses an initial-value problem.
Its solution allows us to quantitatively describe the generation of entropy,
matter and dark matter due to the production and decay of heavy (s)neutrinos.
We have numerically solved this problem for all values of the input
parameters within the ranges specified in Eq.~\eqref{eq:parameterspace}.
In this section, we will first illustrate our findings for a representative
choice of parameter values.
In Sec.~\ref{sec:scan}, we will then turn to the investigation of the parameter space.


The results presented in this section were first published in Ref.~\cite{Buchmuller:2012wn}.
In this paper, in contrast to our earlier studies~\cite{Buchmuller:2010yy, Buchmuller:2011mw},
we take into account all (super)particles involved in the reheating process, in particular
the gauge DOFs.
This allows us to give a realistic, time-resolved description of the reheating process.
Furthermore, compared to Refs.~\cite{Buchmuller:2010yy, Buchmuller:2011mw},
we consider a higher \BmL scale,
$v_{B-L} = 5 \times 10^{15} \,\textrm{GeV}$, in Ref.~\cite{Buchmuller:2012wn},
which renders reheating after the \BmL phase transition compatible with
hybrid inflation and cosmic strings (cf.\ Sec.~\ref{sec:strings}).


\subsection{Particle Masses and Couplings}


Let us study the evolution of the universe after inflation for
\begin{align}
M_1 = 5.4 \times 10^{10}\,\textrm{GeV}\,,\quad
\widetilde{m}_1 = 4.0\times 10^{-2}\,\textrm{eV}\,,\quad
m_{\widetilde{G}} = 100\,\textrm{GeV}\,,\quad
m_{\tilde{g}} = 1\,\textrm{TeV}\,.
\label{eq:exampleparameters}
\end{align}
As we will see later in Sec.~\ref{subsec:gravitinoDM},
requiring successful leptogenesis as well as the right gravitino abundance
to explain dark matter typically forces
$M_1$ to be close to $10^{11}\,\textrm{GeV}$.
Here, we adjust its explicit numerical value such that, given the values for
$\widetilde{m}_1$ and $m_{\widetilde{G}}$, the gravitino abundance comes out
right in order to account for dark matter.
The choice for $\widetilde{m}_1$ represents the best-guess estimate in the context
of the Froggatt-Nielsen flavour model, which we obtained in our Monte-Carlo study
in Ch.~\ref{ch:neutrinos} (cf.\ Eq.~\eqref{eq:m1tepsilon1}).
In scenarios of gravity- or gaugino-mediated supersymmetry breaking, the gravitino
often acquires a soft mass of $\mathcal{O}(100)\,\textrm{GeV}$, which is why we
set $m_{\widetilde{G}}$ to $100\,\textrm{GeV}$.
A gluino mass of $1\,\textrm{TeV}$ is close to the current lower bounds from
ATLAS \cite{ATLAS:2011ad} and CMS \cite{Chatrchyan:2011zy}.
The values in Eq.~\eqref{eq:exampleparameters} readily determine
several further important model parameters:
\begin{align}
m_S     & = 1.6 \times 10^{13} \,\textrm{GeV} \,, &
M_{2,3} & = 1.6 \times 10^{13} \,\textrm{GeV} \,, &
\label{eq:secondaryparameters}\\
\Gamma_S^0         & =  1.9 \times 10 \,\textrm{GeV} \,, &
\Gamma_{N_{2,3}}^0 & =  2.1 \times 10^{10} \,\textrm{GeV} \,, &
\Gamma_{N_1}^0     & =  3.0 \times 10^5 \,\textrm{GeV} \,,
\nonumber\\
\lambda        & =  1.0 \times 10^{-5} \,, &
\epsilon_{2,3} & =  - 1.6 \times 10^{-3} \,, &
\epsilon_1     & =  5.3 \times 10^{-6} \,. \nonumber
\end{align}
Here, we have chosen opposite signs for the $CP$ parameters $\epsilon_1$ and
$\epsilon_{2,3}$, so that the sign of the total lepton asymmetry
always indicates which contribution from the various (s)neutrino decays is
the dominant one.

\begin{figure}
\begin{center}
\includegraphics[width=11.25cm]{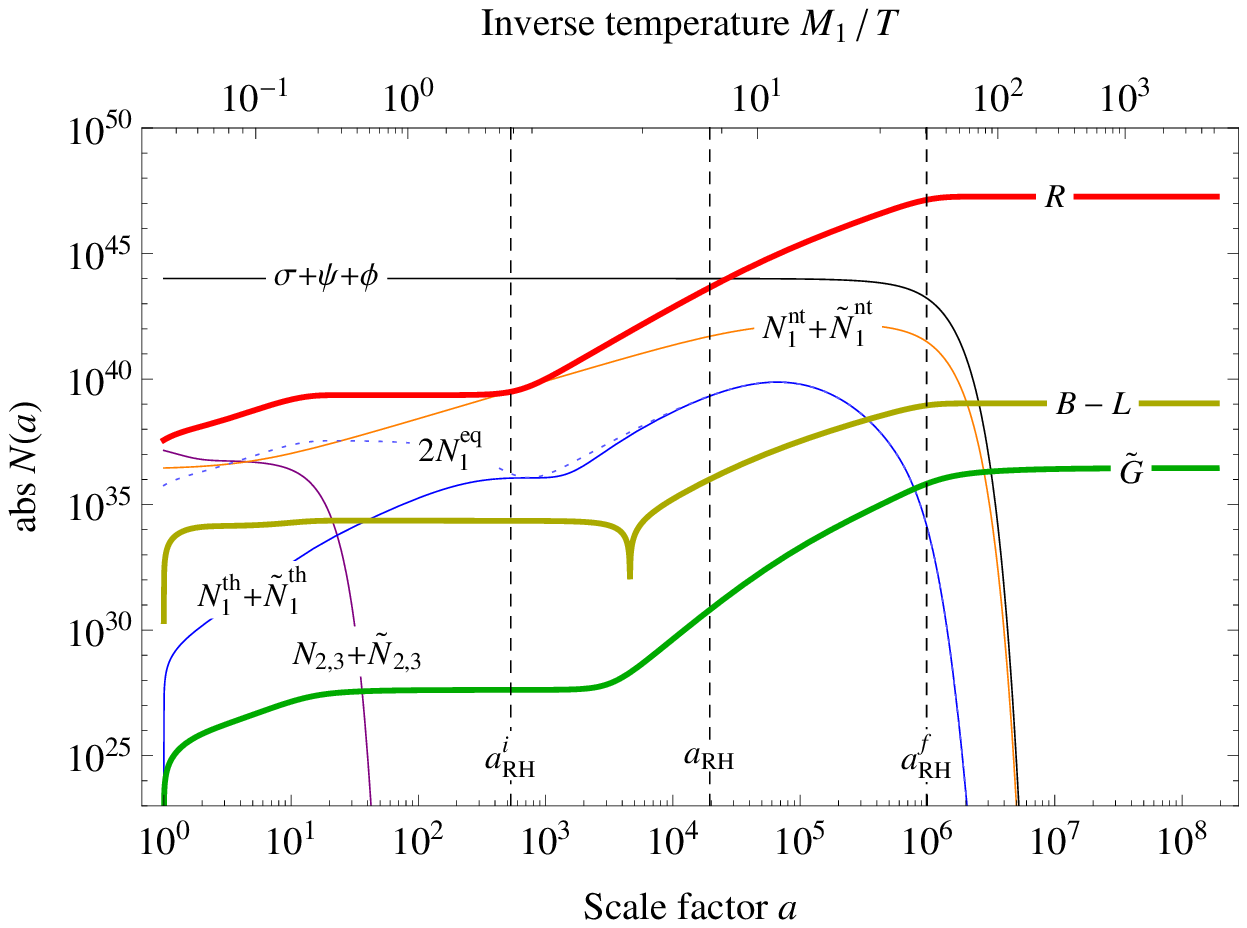}

\vspace{5mm}

\includegraphics[width=11.25cm]{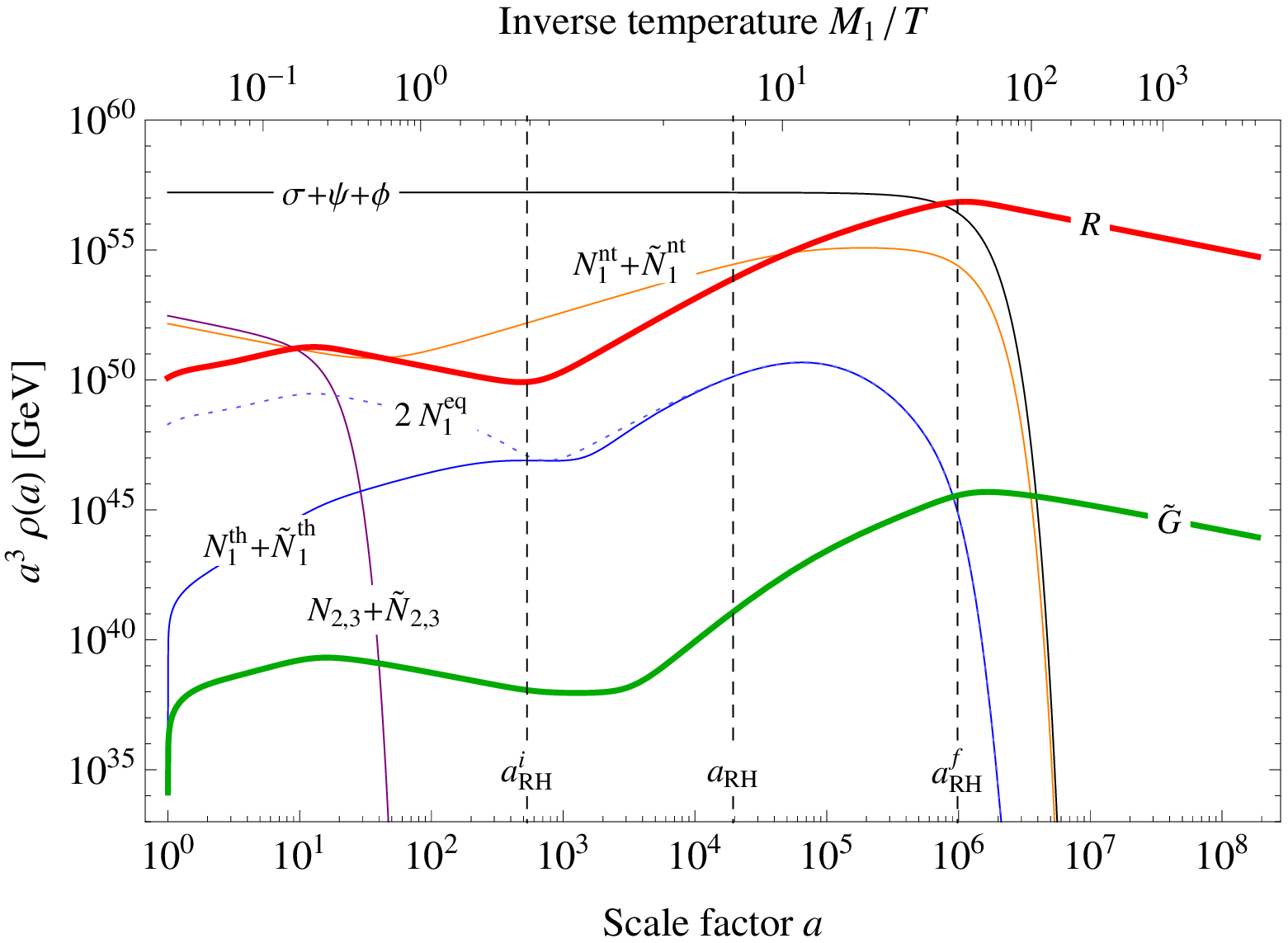}

\vspace{5mm}

\caption[Comoving number and energy densities]
{Comoving number densities \textbf{(upper panel)} and comoving
energy densities \textbf{(lower panel)} for particles from the SSB sector
(Higgs bosons $\sigma$ + higgsinos $\psi$ + inflatons $\phi$), (non)thermally
produced (s)neutrinos of the first generation
($N_{1}^{\textrm{th}} + \tilde{N}_{1}^{\textrm{th}}$, $N_{1}^{\textrm{nt}} + \tilde{N}_{1}^{\textrm{nt}}$),
(s)neutrinos of the first generation in thermal equilibrium
($2 N_1^\textrm{eq}$, for comparison),
(s)neutrinos of the second and third generation
($N_{2,3} + \tilde{N}_{2,3}$), the MSSM radiation ($R$),
the lepton asymmetry ($B$$-$$L$), and gravitinos ($\widetilde{G}$)
as functions of the scale factor $a$.
The vertical lines labeled $a_{\textrm{RH}}^i$, $a_{\textrm{RH}}$ and $a_{\textrm{RH}}^f$
mark the beginning, the middle and the end of the reheating process.
The corresponding values for the input parameters are
given in Eq.~\eqref{eq:exampleparameters}.
}
\label{fig:numengden}
\end{center}
\end{figure}

Fig.~\ref{fig:numengden} presents the comoving number and energy densities
of all relevant species as functions of the scale factor $a$.
In both panels of this figure, some of the displayed curves subsume a number of
closely related species.
These combined curves are broken down into their respective components in the
two panels of Fig.~\ref{fig:breakdown} and in the lower panel of Fig.~\ref{fig:tempasymm}.
The upper panel of Fig.~\ref{fig:tempasymm} presents the temperature of the
thermal bath as function of $a$.
In what follows, we will go through the various stages of the evolution
depicted in Figs.~\ref{fig:numengden}, \ref{fig:breakdown} and \ref{fig:tempasymm}
step by step.
Subsequent to that, we will, based on the plots in Fig.~\ref{fig:numdenwo},
discuss the impact of supersymmetry and the particles of the gauge sector on our results.


\subsection{Decay of the Massive Particles}
\label{subsec:decaymp}


\subsubsection{Initial conditions}


Tachyonic preheating transfers the bulk of the initial vacuum energy
into Higgs bosons, $\rho_\sigma\left(a_{\textrm{PH}}\right)/\rho_0 \simeq 1.0$,
and only small fractions of it into nonrelativistic higgsinos, inflatons,
gauge DOFs and (s)neutrinos $\big(N_i^{\text{PH}}, \tilde{N}_i^{\text{PH}}\big)$.
The particles in the gauge multiplet decay immediately afterwards
around $a = a_G$, giving rise to relativistic (s)neutrinos $\big(N_i^{G}, \tilde{N}_i^{G}\big)$
and an initial abundance of radiation, which thermalizes right away.
Initially, this thermal bath neither exhibits a lepton asymmetry, nor
are there any gravitinos present in it.
The cosmic expansion between preheating and the decay of the gauge DOFs is
practically negligible, $a_G \simeq a_{\textrm{PH}} \equiv 1$.
Note that technically all plots in
Figs.~\ref{fig:numengden}, \ref{fig:breakdown} and \ref{fig:tempasymm} start at $a = a_G$.


\subsubsection{Decay of the (S)neutrinos of the Second and Third Generation}


Among all particles present at $a = a_G$, the heavy (s)neutrinos of the second and
third generation have the shortest lifetimes (cf.\ Eq.~\eqref{eq:secondaryparameters}).
Due to time dilatation, the relativistic (s)neutrinos stemming from the decay of the 
gauge particles decay slower than the nonrelativistic (s)neutrinos produced during preheating.
The decay of the (s)neutrinos of the second and third generation is consequently responsible
for an increase in the radiation number and energy densities on two slightly distinct time scales.

\begin{figure}
\begin{center}
\includegraphics[width=11.25cm]{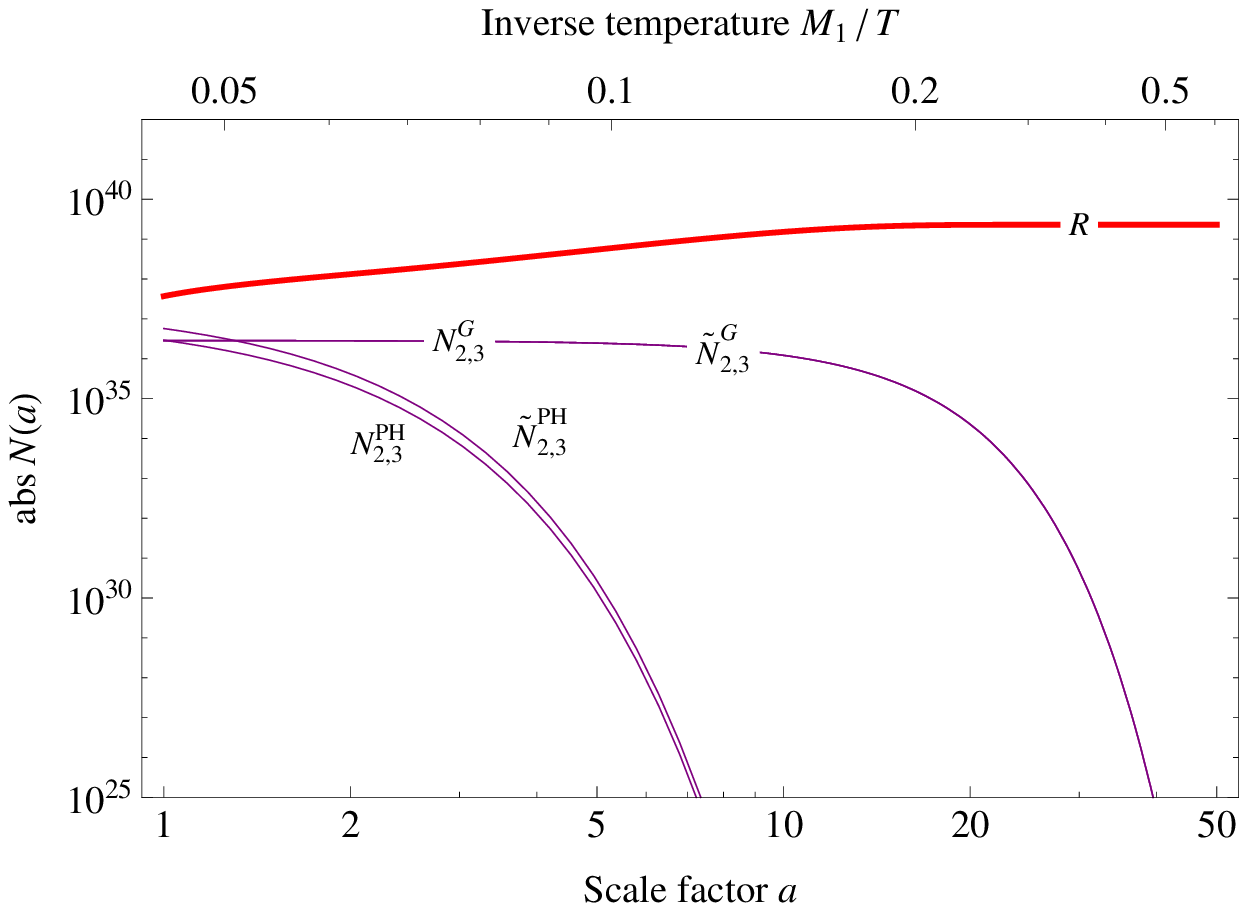}

\vspace{5mm}

\includegraphics[width=11.25cm]{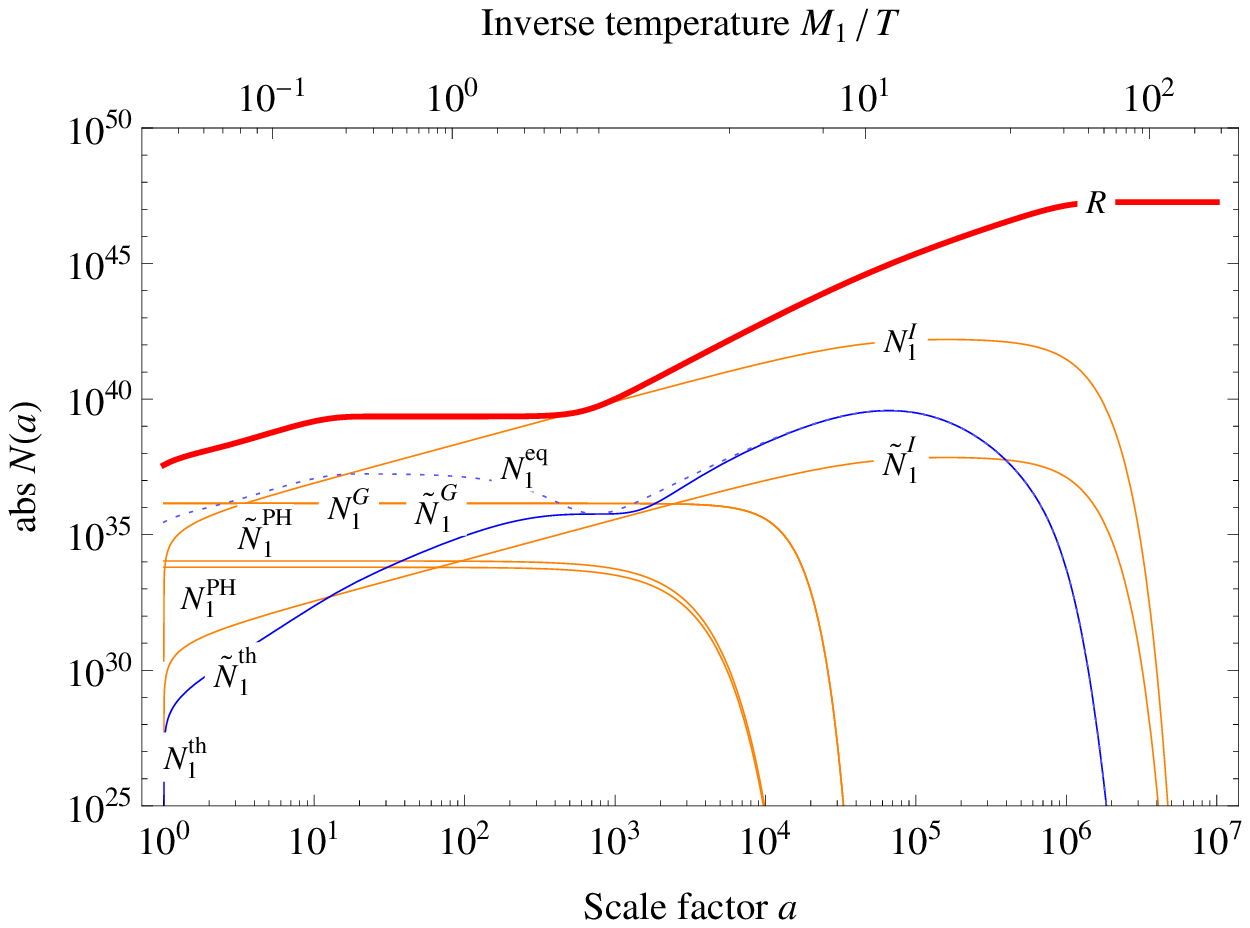}

\vspace{5mm}

\caption[Breakdown of the comoving number densities]
{Breakdown of the comoving number densities shown in the upper
panel of Fig.~\ref{fig:numengden}.
The (s)neutrinos of the second and third generation
($N_{2,3}$$+$$\tilde{N}_{2,3}$) \textbf{(upper panel)} split into
(s)neutrinos that are produced during preheating
($N_{2,3}^\textrm{PH}$, $\tilde{N}_{2,3}^\textrm{PH}$)
and in the decay of the gauge DOFs
($N_{2,3}^G$, $\tilde{N}_{2,3}^G$).
In all four cases, the sum of the contributions from both generations
is shown.
The (s)neutrinos of the first generation
($N_{1}^{\textrm{nt}}$$+$$\tilde{N}_{1}^{\textrm{nt}}$,
$N_{1}^{\textrm{th}}$$+$$\tilde{N}_{1}^{\textrm{th}}$)
\textbf{(lower panel)} split into
(s)neutrinos that are produced during preheating
($N_1^\textrm{PH}$, $\tilde{N}_1^\textrm{PH}$),
in the decay of the gauge DOFs
($N_1^G$, $\tilde{N}_1^G$),
in the decay of the particles from the SSB sector
($N_1^S$, $\tilde{N}_1^S$),
and from the thermal bath
($N_1^\textrm{th}$, $\tilde{N}_1^\textrm{th}$).
}
\label{fig:breakdown}
\end{center}
\end{figure}

The gauge particles decay in equal shares into
neutrinos and sneutrinos (cf.\ Sec.~\ref{subsec:BEneutrinos}).
Their number densities thus behave in exactly the same way, explaining
the overlapping curves in Fig.~\ref{fig:breakdown}.
The production of radiation through the decay of these $N_{2,3}^G$
neutrinos and $\tilde{N}_{2,3}^G$ sneutrinos is efficient, as long
as the radiation production rate $\hat \Gamma_R$ (cf.\ Eq.~\eqref{eq:GammaR})
exceeds the Hubble rate $H$.
At $a \simeq 11$, it drops below the Hubble rate, which roughly coincides with
the value of the scale factor at which the comoving energy density of radiation reaches its
first local maximum.
The period between preheating and this first maximum of the radiation energy density
can be regarded as the first stage of the reheating process.
In the following, we shall refer to it as the stage of $N_{2,3}$ reheating.


\subsubsection{Decay of the Particles of the Symmetry-Breaking Sector}


The production of higgsinos and inflatons during preheating is roughly equally efficient,
$N_\psi\left(a_\textrm{PH}\right) / N_\phi\left(a_\textrm{PH}\right) \simeq 1.0$.
Taking into account the kinematic constraints resulting from the mass
spectrum described in Sec.~\ref{subsec:flavour}, all particles from
the SSB sector exclusively decay into relativistic
(s)neutrinos of the first generation $\big(N_1^{S},\tilde{N}_1^{S}\big)$.


The majority of Higgs bosons, higgsinos and inflatons survives until
$t_S = t_{\text{PH}} + 1/\Gamma_S^0$ (cf.\ Eq.~\eqref{eq:GammaS0}),
which corresponds to a scale factor of $a_S \simeq 7.2 \times 10^5$.
Roughly up to this time, the main part of the total energy is stored in these particles.
At later times, i.e.\ for $a\gtrsim a_S$, the energy budget is dominated by the
energy in radiation.\footnote{Note that in general the value of the scale factor
at which the energy in radiation begins to dominate is determined by the lifetime
of the most long-lived particle.
In the case under study, the Higgs bosons have the longest lifetime.
But for other parameter choices, it may be instead the (s)neutrinos of the first generation.}
Higgs bosons which decay earlier than the average lifetime are responsible for
the generation of sizable abundances of $N_1^S$ neutrinos and $\tilde{N}_1^S$ sneutrinos.
The contributions from higgsino and inflaton
decays to this process are essentially negligible.


\subsubsection{Production and Decay of the Nonthermal (S)neutrinos of the First Generation}


The decay of the particles from the SSB sector is the most important
source for nonthermal (s)neutrinos.
According to our discussion in Sec.~\ref{subsec:BEneutrinos}, the ratio
between the number densities of $N_1^S$ neutrinos and $\tilde{N}_1^S$
sneutrinos is fixed to a constant value at all times (cf.\ Eq.~\eqref{eq:NtSNSratio}).
For our choice of parameters, we find $N_{\tilde N_1}^S/ N_{N_1}^S \simeq 4.4 \times 10^{-5}$.
Moreover, the large hierarchy between the two decay rates
$\Gamma_{N_1}^0$ and $\Gamma_S^0$ (cf.\ Eq.~\eqref{eq:secondaryparameters})
renders the $N_1^S$ and $\tilde{N}_1^S$ number densities unable
to exceed the number density of the Higgs bosons.
From the perspective of the rather long-lived Higgs bosons,
the (s)neutrinos essentially decay right after their production.
As long as they are efficiently fueled by Higgs decays,
the (s)neutrino number densities continue to rise.
But once the supply of Higgs bosons is on the decline, they die out as well.
The overall timescale of our scenario is hence controlled by the Higgs lifetime.
However, as we will see below, the characteristic temperature of the reheating process
is by contrast associated with the lifetime of the $N_1^S$ neutrinos.


Further contributions to the abundances of nonthermal (s)neutrinos
come from preheating as well as the decay of the gauge particles.
Just as in the case of the second and third generation,
the nonrelativistic (s)neutrinos produced during preheating decay at
the fastest rate and the number densities of $N_1^G$ neutrinos and
$\tilde{N}_1^G$~sneutrinos are always the same.


\subsection{Reheating and the Temperature of the Thermal Bath}
\label{subsec:RHandT}


\subsubsection{Reheating through the Decay of \boldmath{$N_1^S$} Neutrinos}


The energy transfer from the nonthermal (s)neutrinos of the first generation
to the thermal bath represents the actual reheating process.
It is primarily driven by the decay of the $N_1^S$ neutrinos, which soon exhibit the highest
abundance among all (s)neutrino species. 
In analogy to the notion of $N_{2,3}$ reheating, we may now speak
of $N_1$ reheating.
This stage of reheating lasts as long as $\hat \Gamma_R \geq H$ (cf.\ Eq.~\eqref{eq:GammaR}).
Let us denote the two bounding values of the scale factor at which $\hat \Gamma_R = H$ by
$a_{\textrm{RH}}^i$ and $a_{\textrm{RH}}^f$.
In the case of our parameter example, we find $a_{\textrm{RH}}^i \simeq 5.3 \times 10^2$ and
$a_{\textrm{RH}}^f \simeq 9.8 \times 10^5$.
Between these two values of the scale factor, the comoving number density
of radiation roughly grows like $N_R \propto a^3$.
Around $a = a_{\textrm{RH}}^i$, the comoving energy density of radiation reaches
a local minimum and around $a = a_{\textrm{RH}}^f$ a local maximum.
Similarly, we observe that the end of reheating nearly coincides with the time
at which the energy in radiation begins to dominate the total energy budget,
$a_{\textrm{RH}}^f \sim a_S$.


\subsubsection{Plateau in the Evolution of the Temperature}

\begin{figure}
\begin{center}
\includegraphics[width=11.25cm]{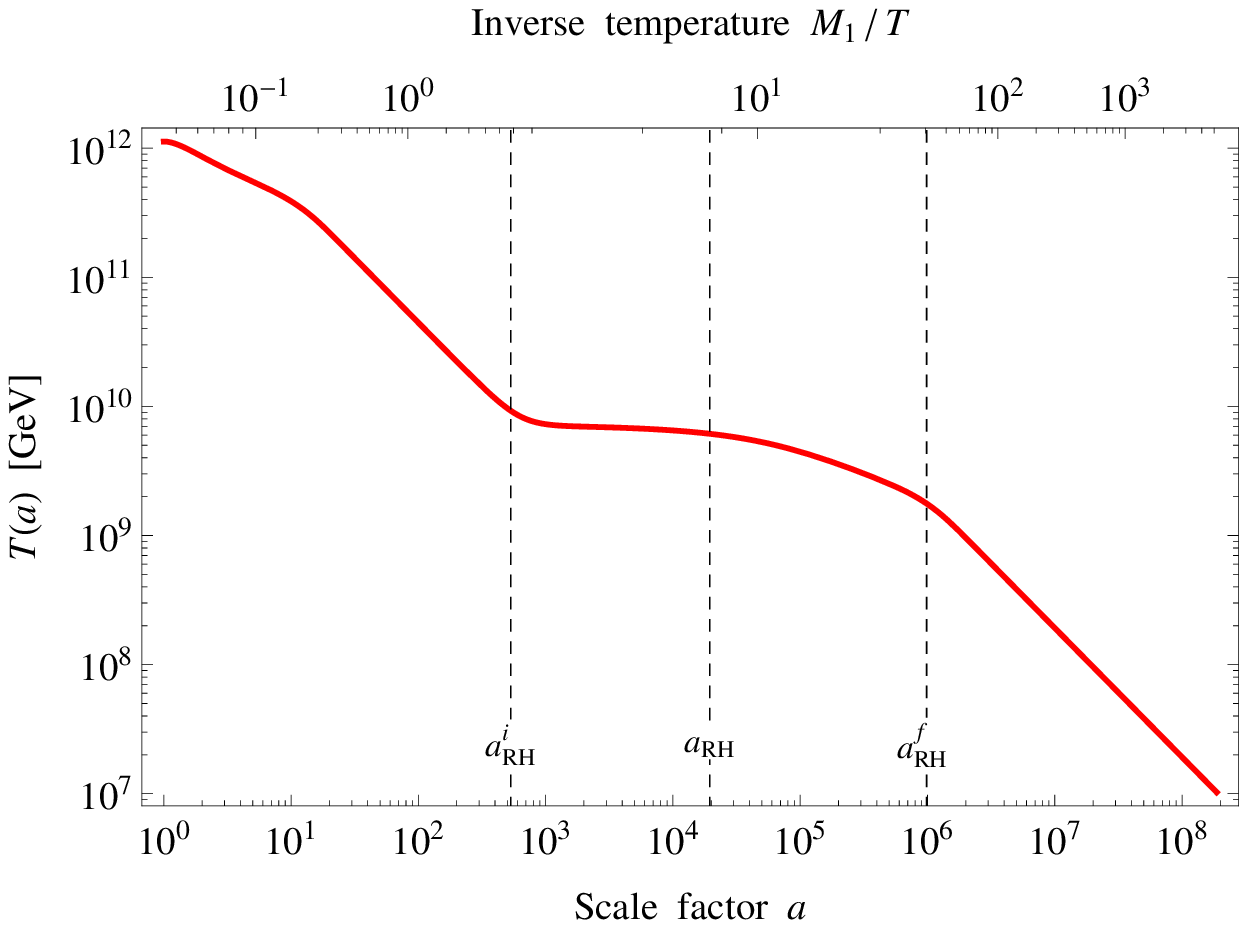}

\vspace{5mm}

\includegraphics[width=11.25cm]{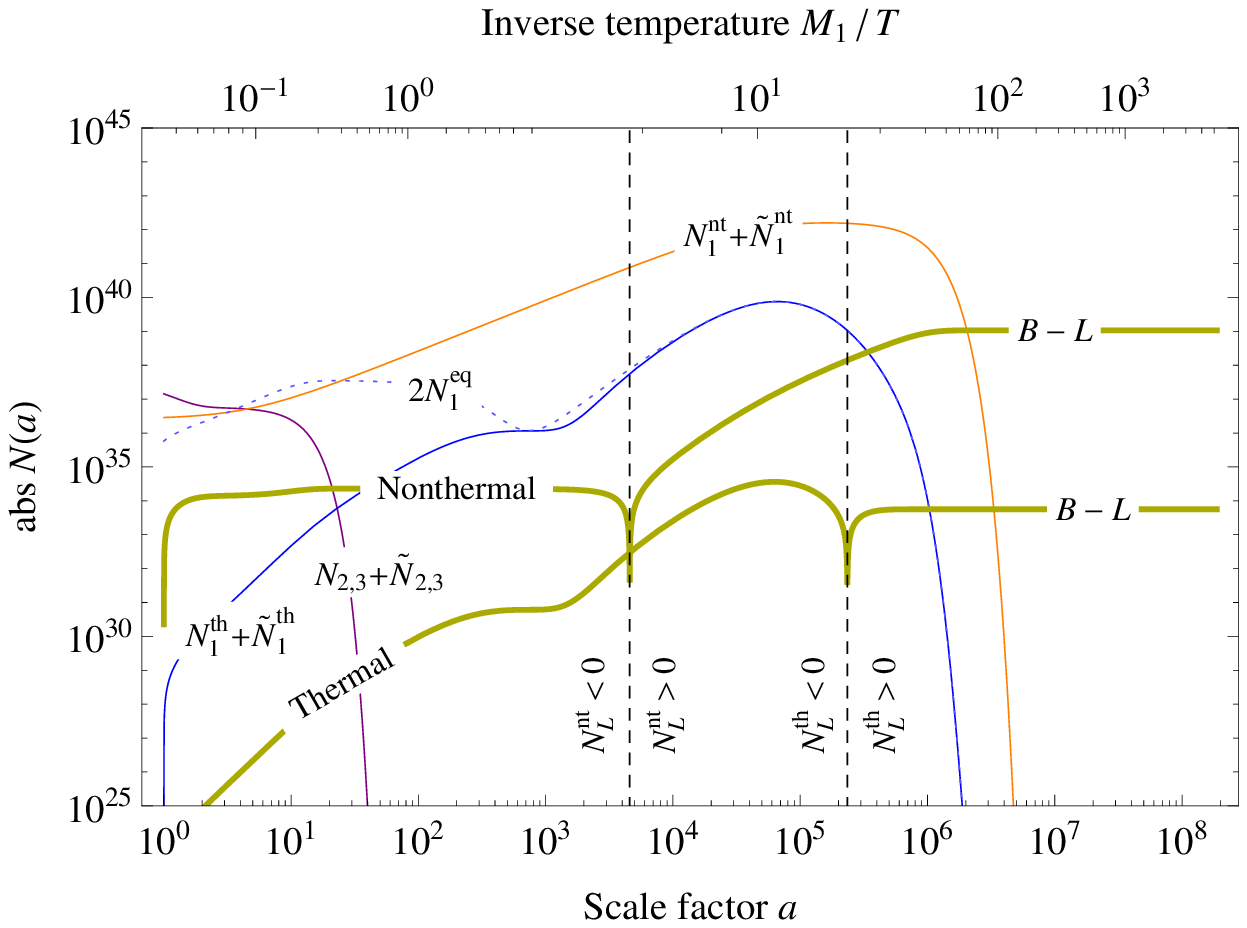}

\vspace{5mm}

\caption[Temperature of the thermal bath and lepton asymmetry]
{Temperature of the thermal bath $T$ \textbf{(upper panel)}
and comoving number densities for the nonthermal
($N_L^\textrm{nt}$) and thermal ($N_L^\textrm{th}$) contributions
to the total lepton asymmetry as well as all (s)neutrino species
($N_{1}^{\textrm{nt}} + \tilde{N}_{1}^{\textrm{nt}}$,
$N_{1}^{\textrm{th}} + \tilde{N}_{1}^{\textrm{th}}$,
$2 N_1^\textrm{eq}$ for comparison, and
$N_{2,3} + \tilde{N}_{2,3}$) \textbf{(lower panel)}
as functions of the scale factor $a$.
The vertical lines in the upper panel labeled $a_{\textrm{RH}}^i$,
$a_{\textrm{RH}}$ and $a_{\textrm{RH}}^f$ mark the beginning, the middle
and the end of the reheating process.
The vertical lines in the lower panel respectively mark the changes in
the signs of the two components of the lepton asymmetry.
}
\label{fig:tempasymm}
\end{center}
\end{figure}

The upper panel of Fig.~\ref{fig:tempasymm} displays the temperature of the thermal
bath $T$, calculated according to Eq.~\eqref{eq:tempNR}, as function of the scale factor $a$.
As a key result of our analysis, we find that during $N_1$ reheating
the temperature stays approximately constant.
For $a$ between $a_{\textrm{RH}}^i$ and $a_{\textrm{RH}}^f$,
it varies by less than an order of magnitude.
We thus conclude that in the first place $a_{\textrm{RH}}^i$ and $a_{\textrm{RH}}^f$
represent the limiting values for a plateau in the evolution of the radiation temperature.
The origin of this plateau is the continuous production of $N_1^S$ neutrinos
during reheating.
As long as these neutrinos are produced much faster than they decay,
their comoving number density grows linearly in time,
$N_{N_1}^S \propto \int_{t_{\textrm{PH}}}^t dt'$ (cf.\ Eq.~\eqref{eq:NXntsol}).
Taking into account that until $a \simeq a_S$ the expansion of the
universe is driven by the energy in the Higgs bosons, i.e.\ nonrelativistic matter,
this translates into $N_{N_1}^S \propto a^{3/2}$.
The $N_1^S$ number density in turn controls the scaling behaviour on the
right-hand side of the Boltzmann equation for radiation during $N_1$ reheating
(cf.\ Eq.~\eqref{eq:BERint}).
Using $H \propto a^{-3/2}$, we find
\begin{align}
a_{\textrm{RH}}^i \lesssim a \lesssim a_{\textrm{RH}}^f \,: \quad
a H \frac{d}{da} N_R  \propto N_{N_1}^S \propto a^{3/2} \,, \quad
N_R \propto a^3 \,, \quad
T \approx \textrm{const.}
\label{eq:plateau}
\end{align}


\subsubsection{Reheating Temperature}


The temperature at which the plateau in Fig.~\ref{fig:tempasymm} is located
sets the characteristic temperature scale of reheating.
In addition, it represents the highest temperature that is ever reached
in the thermal bath, as long as one restricts oneself to times at which
the bath contains a significant fraction of the total energy budget of the universe
(cf.\ lower panel of Fig.~\ref{fig:numengden}).
To turn this qualitative understanding of the
reheating temperature $T_{\textrm{RH}}$ into a precise prescription
for its calculation, we have to decide at which value
of the scale factor $a_{\textrm{RH}}$ we should read it
off from the curve in Fig.~\ref{fig:tempasymm}.
We choose the following intuitive definition:
$a_{\textrm{RH}}$ denotes the value of $a$ when the decay of the
$N_1^S$ neutrinos into radiation is about to become efficient,
which is the case once the Hubble rate $H$ has dropped to the
effective decay rate $\Gamma_{N_1}^S$,
\begin{align}
\Gamma_{N_1}^S \left(a_\textrm{RH}\right) = H \left(a_\textrm{RH}\right) \,,\quad
T_{\textrm{RH}} = T\left(a_\textrm{RH}\right) \,.
\label{eq:TRHdef}
\end{align}
This prescription yields a temperature which is representative for the
temperature plateau (cf.\ Fig.~\ref{fig:tempasymm}).
For the chosen set of parameters, Eq.~\eqref{eq:TRHdef} has the following solution,
\begin{align}
a_{\textrm{RH}} \simeq 1.9 \times 10^4 \,,\quad
H = \Gamma_{N_1}^S \simeq 3.5 \times 10^3 \,\textrm{GeV} \,,\quad
T_{\textrm{RH}} \simeq 6.1 \times 10^9 \,\textrm{GeV} \,.
\label{eq:TRHres}
\end{align}
In Figs.~\ref{fig:numengden} and \ref{fig:numdenwo} as well as in the upper panel of
Fig.~\ref{fig:tempasymm}, the three values of the scale factor marking the initial
($a_{\textrm{RH}}^i$), characteristic intermediate ($a_{\textrm{RH}}$) and final
($a_{\textrm{RH}}^f$) point of the reheating process are indicated by dashed vertical lines.


Apart from the definition of the reheating temperature in Eq.~\eqref{eq:TRHdef},
there are alternative ways to define the reheating temperature.
For instance, we could use the temperature at the beginning
($a = a_{\textrm{RH}}^i$) or the end of reheating ($a = a_{\textrm{RH}}^f$)
or the temperature when half of the total available energy has been transferred
to radiation ($a \simeq a_S$ for the parameter example discussed in this section).
In either case, although the respective value for $a_{\textrm{RH}}$ may
significantly vary, thanks to the temperature plateau during reheating,
the resulting reheating temperature would not change much,
\begin{align}
\frac{T\left(a_{\textrm{RH}}^i\right)}{T_{\textrm{RH}}} \simeq 1.5 \,,\quad
\frac{T\left(a_S\right)}{T_{\textrm{RH}}} \simeq \frac{1}{2.5} \,,\quad
\frac{T \big(a_{\textrm{RH}}^f\big)}  {T_{\textrm{RH}}} \simeq \frac{1}{3.0} \,.
\end{align}
Our definition of the reheating temperature may hence be regarded
as a compromise between several more extreme approaches.
But more important than that, it picks up on a physical feature
that other definitions would miss.
In Fig.~\ref{fig:tempasymm}, we observe that the temperature declines
less during the first part of reheating,
$a_{\textrm{RH}}^i \leq a \leq a_{\textrm{RH}}$, than during the second part,
$a_{\textrm{RH}}   \leq a \leq a_{\textrm{RH}}^f$.
The stage of $N_1$ reheating evidently splits up into two phases,
during the first of which the temperature is basically constant,
whereas during the second one the temperature slightly decreases.
The reason for this substructure in the temperature plateau is the following.
As soon as the $N_1^S$ neutrinos decay more efficiently, their comoving number
density starts to grow slower than $a^{3/2}$.
This diminishes the production rate of radiation.
According to Eq.~\eqref{eq:plateau}, a constant temperature
can then no longer be maintained.
The advantage of our definition for $T_{\textrm{RH}}$ now is that we read it off
from the curve in Fig.~\ref{fig:tempasymm} at exactly that value of the scale factor
at which the transition between these two phases of $N_1$ reheating takes place.
Our definition thus yields a temperature which is both representative, as it mediates
between several more extreme values, and especially singled out, as it is associated
with a prominent feature in the temperature curve.


For completeness, we should however mention that for other parameter choices
this picture may change.
If the Higgs decay rate $\Gamma_S^0$ is, for instance, larger than the neutrino decay
rate $\Gamma_{N_1}^S$, which can for example be achieved by going
to lower values of the \BmL scale, the scaling behaviour of the $N_1^S$ number density
changes when the neutrino production efficiency begins to cease and not when the decays
of the neutrinos themselves set in.
The slight kink in the temperature plateau is then located at $a \simeq a_S$, which is in
this case before the decay of the $N_1^S$ neutrinos has become fully efficient.
But the definition of the reheating temperature in Eq.~\eqref{eq:TRHdef} remains
reasonable nonetheless.
After all, if $\Gamma^0_S > \Gamma_{N_1}^S$, the bulk of the total energy is first
almost entirely accumulated in $N_1^S$ neutrinos, before it is passed on to radiation.
The energy in radiation thus receives its major contribution just when these neutrinos
decay with a sufficient efficiency.
The characteristic temperature at the time when this happens is then again obtained
from Eq.~\eqref{eq:TRHdef}.
Further details on the reheating temperature in regions in parameter space
in which $\Gamma^0_S > \Gamma_{N_1}^S$ can be found in Ref.~\cite{Buchmuller:2011mw}.


\subsubsection{Evolution of the Temperature off the Plateau}


During $N_{2,3}$ reheating, the temperature first increases up to a maximal value
and then decreases like $a^{-1/2}$.
The initial rise reflects the production of radiation through the decays
of the (s)neutrinos of the second and third generation as long as the expansion
of the universe is negligible.
The subsequent decrease then follows from the Boltzmann equation for
radiation (cf.\ Eq.~\eqref{eq:BERint}), using the fact that its right-hand side stays
almost constant up to the end of $N_{2,3}$ reheating,
\begin{align}
a_G \lesssim a \lesssim 11 \,: \quad
a H \frac{d}{da} N_R  \propto N_{N_{2,3}}^G \approx \textrm{const.} \,, \quad
N_R \propto a^{3/2} \,, \quad
T \propto a^{-1/2} \,.
\label{eq:TempN23Reheat}
\end{align}
Finally, we note that, between the two stages of reheating and after the end of reheating,
the temperature drops off like $a^{-1}$.
This is the usual adiabatic behaviour indicating that no radiation,
i.e.\ entropy, is being produced,
\begin{align}
11 \lesssim a \lesssim a_{\textrm{RH}}^i \:\:\textrm{and}\:\:
a_{\textrm{RH}}^f \lesssim a \,: \quad
a H \frac{d}{da} N_R  \approx 0 \,, \quad
N_R \approx \textrm{const.} \,, \quad
T \propto a^{-1} \,.
\label{eq:Tadiab}
\end{align}


\subsection{Small Departures from Thermal Equilibrium}


\subsubsection{Production and Decay of the Thermal (S)neutrinos of the First Generation}


Unlike the two heavier (s)neutrino flavours, the (s)neutrinos of the first
generation are also produced thermally $\big(N_i^{\text{th}}, \tilde{N}_i^{\text{th}}\big)$.
Thanks to supersymmetry, the evolution of the $N_1^{\textrm{th}}$ and $\tilde{N}_1^{\textrm{th}}$
number densities is governed by exactly the same Boltzmann equation
(cf.\ Eq.~\eqref{eq:BENth}), so that they are identical at all times.
As both species inherit their momentum distribution from the thermal bath,
they are always approximately in kinetic equilibrium.\footnote{For a more
detailed discussion, cf.\ Sec.~\ref{subsec:BEneutrinos}
and Appendix B of Ref.~\cite{Buchmuller:2011mw}.}
Simultaneously, the interplay between decays and inverse decays drives them
towards thermal equilibrium.
Initially, there are no thermal (s)neutrinos present in the thermal bath
and inverse decays result in a continuous rise of the thermal (s)neutrino
number densities until $a \sim a_{\textrm{RH}}^i$.
Around this time, the temperature drops significantly below the mass $M_1$
and the thermal (s)neutrinos become nonrelativistic.
The equilibrium number density $N_{N_1}^\textrm{eq}$ begins to decrease
due to Boltzmann suppression, until it almost reaches the actual number density
of the thermal (s)neutrinos.
The production of thermal (s)neutrinos can then no longer compete with the expansion of the
universe and their comoving number densities do not continue to grow.


This picture, however, soon changes because reheating sets in.
As the temperature remains almost perfectly constant until
$a \sim a_{\textrm{RH}}$, the equilibrium number density $N_{N_1}^\textrm{eq}$
is not diminished due to Boltzmann suppression any further up to this time.
Instead, it bends over and starts to increase like the volume,
$N_{N_1}^\textrm{eq} \propto a^3$.
The number densities of the thermal (s)neutrinos subsequently
follow this behaviour of the equilibrium number density.
During the second phase of $N_1$ reheating, the temperature slightly decreases
again, thereby reinforcing  the Boltzmann factor in $N_{N_1}^\textrm{eq}$.
Consequently, the equilibrium number density stops growing and
shortly afterwards starts to decline exponentially.
An instant after it has passed its global maximum, the number densities
of the thermal (s)neutrinos overshoot the equilibrium number density.
Due to their numerical proximity, the two values of the scale factor
at which $N_{N_1}^\textrm{eq}$ and $N_{N_1}^\textrm{th}$ respectively
reach their global maxima cannot be distinguished from
each other in Figs.~\ref{fig:numengden}.
Both events occur close to $a = 6.6 \times 10^4$.


\subsubsection{Generation of the Baryon Asymmetry}


The out-of-equilibrium decays of the heavy (s)neutrinos violate
$L$, $C$, and $CP$, thereby generating a lepton asymmetry in the thermal bath.
A first nonthermal asymmetry is introduced to the thermal
bath during $N_{2,3}$ reheating.
For $a_G \lesssim a \lesssim 2.2$, the decay of the (s)neutrinos
stemming from preheating leads to an increase of the absolute value
of the comoving number density $N_L^\textrm{nt}$.
In the interval $6.6 \lesssim a \lesssim 13$, the lepton asymmetry
is slightly augmented through the decay of the (s)neutrinos which
were produced in the decay of the gauge particles.
The main part of the nonthermal asymmetry is, however, generated
during $N_1$ reheating, while the scale factor takes values between
$a \simeq 2.0 \times 10^{3}$ and $a \simeq 1.3 \times 10^6$.
At all other times, the effective rate at which the nonthermal asymmetry
is produced is at least half an order of magnitude smaller than the Hubble rate.
Among all nonthermal (s)neutrinos of the first generation, only the
$N_1^S$ neutrinos contribute efficiently to the generation of the asymmetry.
Their decay results in a positive nonthermal asymmetry that gradually
overcompensates the negative asymmetry produced during $N_{2,3}$ reheating.
At $a \simeq 4.6 \times 10^3$, the entire initial asymmetry has been
erased and $N_L^\textrm{nt}$ changes its sign.


Washout processes almost do not have any impact on the evolution of the nonthermal asymmetry.
The rate $\hat \Gamma_W$, at which these processes occur (cf.\ Eq.~\eqref{eq:Lrates}),
is always smaller than the Hubble rate $H$ by a factor of at least $\mathcal{O}(10)$.
On top of that, at the time $\hat\Gamma_W$ is closest to $H$, which happens around
$a\simeq 4.0 \times 10^4$ when $\hat\Gamma_W / H \simeq 0.12$, the production rate
$\hat{\Gamma}_L^{\textrm{nt}}$ is constantly larger than $\hat \Gamma_W$ by a factor
of $\mathcal{O}(10)$, so that the effect of washout on
the nonthermal asymmetry is indeed always negligible.


The decays and inverse decays of thermal (s)neutrinos of the first generation
are responsible for the emergence of a thermal, initially negative asymmetry in the bath.
As long as the abundance of thermal (s)neutrinos is far away from the one
in thermal equilibrium, the absolute value of this asymmetry increases rapidly.
Around $a \sim a_{\textrm{RH}}^i$, this is not the case anymore, causing
the production of the thermal asymmetry to stall for a short moment.
At $a \simeq 6.3 \times 10^4$, the washout rate $\hat{\Gamma}_W$
overcomes the production rate $\hat{\Gamma}_L^{\textrm{th}}$ of the
thermal asymmetry and its absolute value begins to decline.
Note that, at this time, the rates $\hat{\Gamma}_L^{\textrm{th}}$
and $\hat{\Gamma}_W$ are smaller than $H$ by roughly a factor $9$.
Shortly afterwards, at $a \simeq 6.6 \times 10^4$, the number density
of thermal (s)neutrinos overshoots the equilibrium density, which results
in the asymmetry being driven even faster towards zero.
Already at $a \simeq 2.3 \times 10^5$, the initial thermal
asymmetry is completely erased.
Meanwhile, washout effects recede in importance.
From $a \simeq 6.9 \times 10^4$ onwards, $\hat \Gamma_L^{\textrm{th}}$ permanently
dominates over $\hat \Gamma_W$, which is why, once the thermal asymmetry has turned positive,
it does not decrease anymore.
Instead, it freezes out at its maximal value around $a \simeq 4.5 \times 10^5$,
which corresponds to the time when the ratio of $\Gamma_L^{\textrm{th}}$ and
the Hubble rate $H$ drops below $1/\sqrt{10}$.


The final values of $N_L^{\textrm{nt,th}}$
allow us to infer the present baryon asymmetry $\eta_B$ as well
as its composition in terms of a nonthermal ($\eta_B^{\textrm{nt}}$)
and a thermal ($\eta_B^{\textrm{th}}$) contribution (cf.\ Eq.~\eqref{eq:etaB0}),
\begin{align}
\eta_B = \frac{n_B^0}{n_\gamma^0} = \eta_B^{\textrm{nt}} + \eta_B^{\textrm{th}} \,,\quad
\eta_B^{\textrm{nt},\textrm{th}} =  C_{\textrm{sph}} \frac{g_{*,s}^0}{g_{*,s}}
\left.\frac{N_{L}^{\textrm{nt},\textrm{th}}}{N_\gamma}\right|_{a_f} \,.
\label{eq:etaBntth}
\end{align}
Here, $C_{\textrm{sph}} = 8/23$ denotes the sphaleron conversion factor
(cf.\ Eq.~\eqref{eq:CsphSMMSSM}), $g_{*,s} = 915/4$ and $g_{*,s}^0 = 43/11$
stand for the effective numbers of relativistic DOFs in the MSSM that enter
the entropy density $s_R$ of the thermal bath in the high- and low-temperature
regime, respectively (cf.\ Eqs.~\eqref{eq:sRgstars0} and \eqref{eq:sRgstars}),
and $N_\gamma = g_\gamma / g_{*,n} \, N_R$ is the comoving number density of photons.
As final value for the scale factor, we use  $a_f \simeq 1.9 \times 10^8$,
which is the maximal value depicted in the two plots of Fig.~\ref{fig:numengden}.
Since we are not able to predict the signs of the $CP$ violation parameters $\epsilon_i$
in any case, we do not bother about the relative sign between the lepton and the baryon
asymmetry and simply take $\eta_B$, $\eta_B^{\textrm{nt}}$ and $\eta_B^{\textrm{th}}$
to have the same signs as $N_L$, $N_L^{\textrm{nt}}$ and $N_L^{\textrm{th}}$.
In our parameter example, we then find
\begin{align}
\eta_B \simeq 3.7 \times 10^{-9} \,, \quad
\eta_B^{\textrm{nt}} \simeq 3.7 \times 10^{-9} \,, \quad
\eta_B^{\textrm{th}} \simeq 1.9 \times 10^{-14} \,.
\label{eq:etaBres}
\end{align}
Recall that in Sec.~\ref{subsec:flavour}, we set the $CP$ asymmetry
parameter $\epsilon_1$ to its maximal value (cf.\ Eq.~\eqref{eq:epsiloniest}).
In this sense, the resulting values for the baryon asymmetry must be
interpreted as upper bounds on the actually produced asymmetry and are
thus perfectly compatible with the observed value for the baryon
asymmetry, $\eta_B^{\textrm{obs}} \simeq 6.2 \times 10^{-10}$
(cf.\ Sec.~\ref{subsec:baryncmttr}).
We also recall that, in fact, the Froggatt-Nielsen model
typically predicts values for $\epsilon_1$ that are smaller than the maximal possible
value by roughly a factor of $\mathcal{O}(10)$ (cf.\ Eq.~\eqref{eq:m1tepsilon1}).
Using a generic value for $\epsilon_1$ according to the Froggatt-Nielsen model,
rather than estimating $\epsilon_1$ by means of its upper bound, would thus yield
an excellent agreement between prediction and observation in the context of our
parameter example, $\eta_B \simeq \eta_B^{\text{obs}}$.


Furthermore, we find that in the case under study, it is the nonthermal
contribution $\eta_B^{\textrm{nt}}$ that lifts the total baryon asymmetry
$\eta_B$ above the observational bound.
The thermal contribution $\eta_B^{\textrm{th}}$ is smaller than $\eta_B^{\textrm{nt}}$
by five orders of magnitude.
If we discarded the entire idea of nonthermally produced (s)neutrinos being the main
source of the lepton asymmetry and resorted to standard thermal leptogenesis, we would
struggle to reproduce the observed asymmetry.
For the chosen value of $\widetilde{m}_1$, standard leptogenesis would result in
$\eta_B^{\textrm{st}} \sim 10^{-10}$, which is almost an order of magnitude below the
observed value (cf.\ Ref.~\cite{Buchmuller:2004nz} for details).
By contrast, it is still much larger than our result for $\eta_B^{\textrm{th}}$.
\label{page:etabthdilu} This has mainly two reasons.
First, in our scenario, the decays of the nonthermal (s)neutrinos
continuously increase the entropy of the thermal bath (cf.\ Figs.~\ref{fig:numengden}
and \ref{fig:tempasymm}), which results in a nonstandard dilution of the thermal
asymmetry during and after its production.
Between, for instance, ${a \simeq 6.3 \times 10^4}$, which corresponds to the time when the
production of the negative asymmetry is reversed and the absolute value of the asymmetry starts
to decline, and $a = a_f$, the entropy of the thermal bath increases by a factor of
$\mathcal{O}(100)$.
Second, in consequence of the specific reheating mechanism at work, the generation of
the thermal asymmetry is delayed in time, so that it takes place at a lower
temperature than in the standard case.
This implies a correspondingly smaller abundance of thermal (s)neutrinos, rendering
our thermal mechanism for the generation of an asymmetry less efficient.
We will resume this comparison of the thermal asymmetry
$\eta_B^{\textrm{th}}$ with the expectation from standard leptogenesis $\eta_B^{\textrm{st}}$
in Sec.~\ref{subsec:baryonasym}, where we will discuss the respective
dependences on the neutrino mass parameters $\widetilde{m}_1$ and $M_1$.


\subsubsection{Production of Gravitino Dark Matter}


Inelastic $2$-to-$2$ scattering processes in the supersymmetric thermal plasma,
mediated predominantly via the strong interaction,
are responsible for the production of dark matter in the form of gravitinos.
As the right-hand side of the gravitino Boltzmann equation (cf.\ Eq.~\eqref{eq:BoltzGravi})
scales like $a^3 T^6$, the efficiency of gravitino production in the course of reheating
is directly controlled by the interplay between the expansion of the universe
and the evolution of the temperature of the thermal bath.


During $N_{2,3}$ reheating, the temperature roughly declines as
$T \propto a^{-1/2}$ (cf.\ Eq.~\eqref{eq:TempN23Reheat}), such
that in first approximation
\begin{align}
a H \frac{d}{da} N_{\widetilde{G}} =
\hat{\Gamma}_{\widetilde{G}} N_{\widetilde{G}} \propto a^3 T^6
\approx \textrm{const.} \,, \quad
\hat{\Gamma}_{\widetilde{G}} \propto H \propto a^{-3/2} \,, \quad
N_{\widetilde{G}} \propto a^{3/2} \,.
\end{align}
Once the decay of the (s)neutrinos of the second and third
generation has ceased, the temperature decreases adiabatically,
$T \propto a^{-1}$ or equivalently $a^3 T^6 \propto a^{-3}$
(cf.\ Eq.~\eqref{eq:Tadiab}).
The rate of gravitino production $\hat{\Gamma}_{\widetilde{G}}$ 
then begins to decrease much faster than the Hubble rate, in fact, initially
even slightly faster than $a^{-3}$, causing the comoving gravitino number density
$N_{\widetilde{G}}$ to approach a constant value.
The first stage of gravitino production is completed around $a \simeq 28$, which
corresponds to the time when $\hat{\Gamma}_{\widetilde{G}}$ is half an order of magnitude smaller
than $H$.
From this time onwards, $\hat{\Gamma}_{\widetilde{G}}$ scales like $a^{-3}$, the production
term in the Boltzmann equation is negligibly small and $N_{\widetilde{G}}$ is constant.


The decline of $\hat{\Gamma}_{\widetilde{G}}$ is reversed as soon as the
temperature plateau characteristic for the phase of $N_1$ reheating is reached,
such that approximately $a^3 T^6 \propto a^3$.
While ${\hat{\Gamma}_{\widetilde{G}}\ll H}$, the gravitino density $N_{\widetilde{G}}$ continues
to remain constant and $\hat{\Gamma}_{\widetilde{G}}$ increases almost as fast as $a^3$.
At $a \simeq 1.9 \times 10^3$, it has nearly caught up again with the Hubble rate,
i.e.\ the ratio $\hat{\Gamma}_{\widetilde{G}}/H$ reaches again a value of $1/\sqrt{10}$.
This time marks the beginning of the second stage of gravitino production.
The production term in the Boltzmann equation cannot be neglected any longer and,
assuming for a moment an exactly constant temperature during $N_1$ reheating, we have
\begin{align}
a H \frac{d}{da} N_{\widetilde{G}} =
\hat{\Gamma}_{\widetilde{G}} N_{\widetilde{G}} \propto a^3 T^6
\propto a^3 \,, \quad
\hat{\Gamma}_{\widetilde{G}} \propto H \propto a^{-3/2} \,, \quad
N_{\widetilde{G}} \propto a^{9/2} \,.
\end{align}
The gravitino density $N_{\widetilde{G}}$ hence begins to grow again, now even faster
than during $N_{2,3}$ reheating.
This terminates the rise of the rate $\hat{\Gamma}_{\widetilde{G}}$, turning it into
a decline proportional to $a^{-3/2}$.
We thus obtain the interesting result that, although the temperature evolves differently during
$N_{2,3}$ and $N_1$ reheating, the rate $\hat{\Gamma}_{\widetilde{G}}$ always runs parallel to the
Hubble rate during these two stages of the reheating process.


At the end of $N_1$ reheating, gravitino productions fades away in the same way as at
the end of $N_{2,3}$ reheating.
Around $a \simeq 3.5 \times 10^6$, when $\hat{\Gamma}_{\widetilde{G}}/H$ drops below
$1/\sqrt{10}$, the gravitino abundance freezes out.
The final value of $N_{\widetilde{G}}$ then allows us to calculate $\Omega_{\widetilde{G}} h^2$,
the present energy density of gravitinos $\rho_{\widetilde{G}}^0$  in units of $\rho_c/h^2$,
\begin{align}
\Omega_{\widetilde{G}} h^2 = \frac{\rho_{\widetilde{G}}^0}{\rho_c / h^2} =
\frac{m_{\widetilde{G}} n_\gamma^0}{\rho_c / h^{2}}
\frac{g_{*,s}^0}{g_{*,s}} \left.\frac{N_{\widetilde{G}}}{N_\gamma}\right|_{a_f} \,,
\label{eq:OmegaGth2}
\end{align}
where $\rho_c = 1.05 \times 10^{-5} \, h^2 \, \textrm{GeV}\, \textrm{cm}^{-3}$
denotes the critical energy density of the universe (cf.\ Eq.~\eqref{eq:rhoc0}),
$h = 0.70$ the Hubble rate $H$ in the units
$H = h \times 100\,\textrm{km} \, \textrm{s}^{-1} \, \textrm{Mpc}^{-1}$,
$n_\gamma^0 = 410\,\textrm{cm}^{-3}$ the number density of the CMB photons
(cf.\ Eq.~\eqref{eq:snrhogamm0}), and $g_{*,s}$, $g_{*,s}^0$, $N_\gamma$,
and $a_f$ are explained below Eq.~\eqref{eq:etaBntth}.
Recall that, after fixing $\widetilde{m}_1$, $m_{\widetilde{G}}$ and $m_{\tilde{g}}$, we adjusted
the heavy neutrino mass, $M_1 = 5.4 \times 10^{10}\,\textrm{GeV}$, such that we would
obtain the right abundance of gravitinos to account for the relic density
of dark matter, $\Omega_{\textrm{DM}}^{\textrm{obs}}h^2 \simeq 0.11$ (cf.\ Sec.~\ref{subsec:DM}).
By construction, we thus now find in our parameter example
\begin{align}
\Omega_{\widetilde{G}} h^2 \simeq 0.11\,.
\end{align}


In conclusion, we emphasize the intriguing simplicity of this mechanism
for the generation of dark matter.
Let us in particular focus on the physical picture behind the
second stage of gravitino production.
Initially, at the onset of $N_1$ reheating, the rate $\hat{\Gamma}_{\widetilde{G}}$
is still very small compared to the Hubble rate $H$.
But, given the constant spacetime density of gravitino production
$\gamma_{\widetilde{G}} = n_{\widetilde{G}}\,\hat{\Gamma}_{\widetilde{G}}\propto T^6$
during $N_1$ reheating and the rapid growth of the spatial volume due to the expansion,
$\hat{\Gamma}_{\widetilde{G}}$ rapidly grows sufficiently large to set the production of
gravitinos going.
During the remaining time of $N_1$ reheating, this production can then proceed without
further hindrance as the universe, although it is expanding, is filled by a thermal bath
at a constant temperature.
The continuous production of radiation nullifies the expansion and
gravitinos are produced as in a static universe.
In other words, one key feature of our scenario of reheating is that it turns
the universe into a chemistry laboratory, in which the temperature is fixed at a
certain value, so that dark matter can be cooked in it just to the right point.


\begin{figure}
\begin{center}
\includegraphics[width=10.75cm]{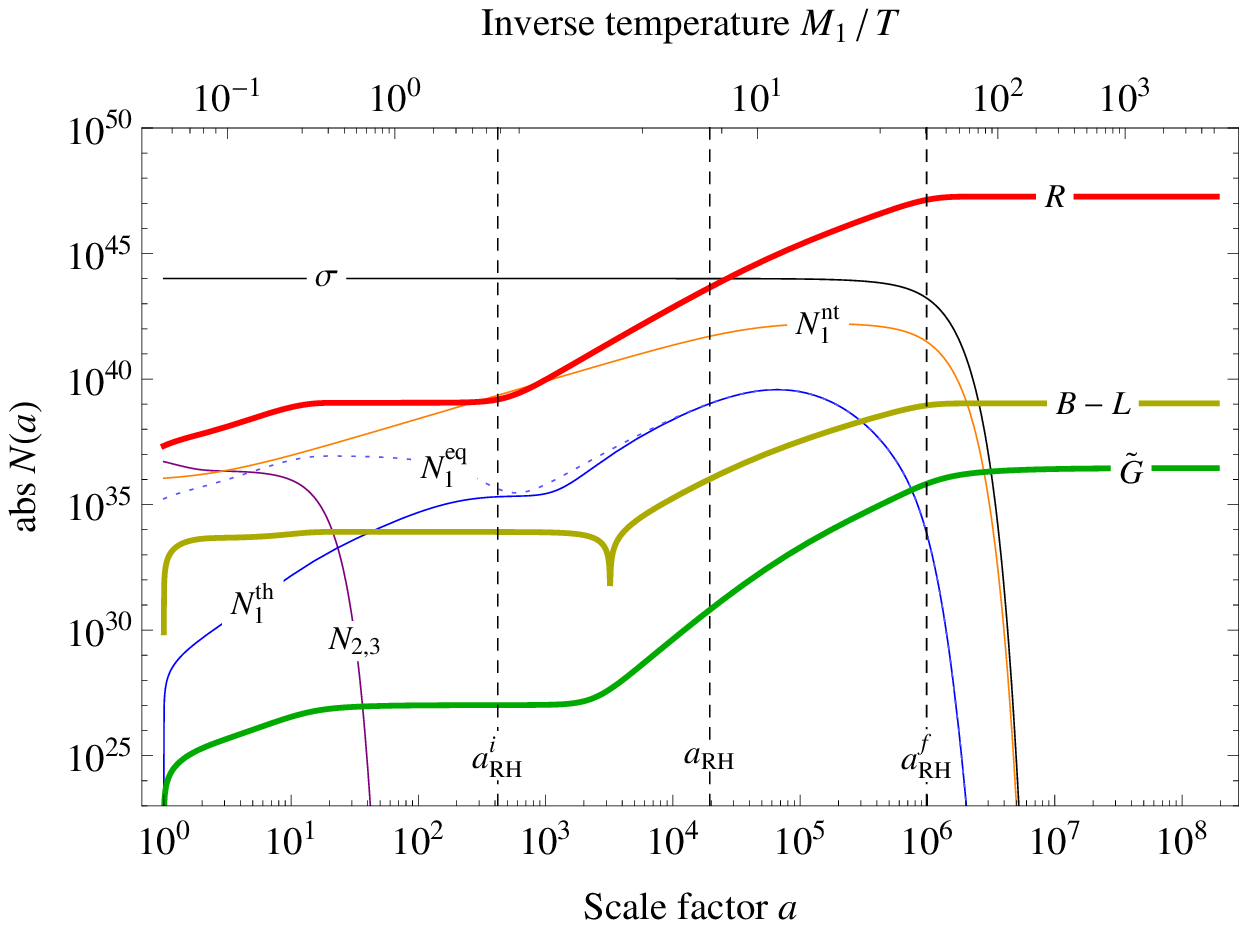}

\vspace{5mm}

\includegraphics[width=10.75cm]{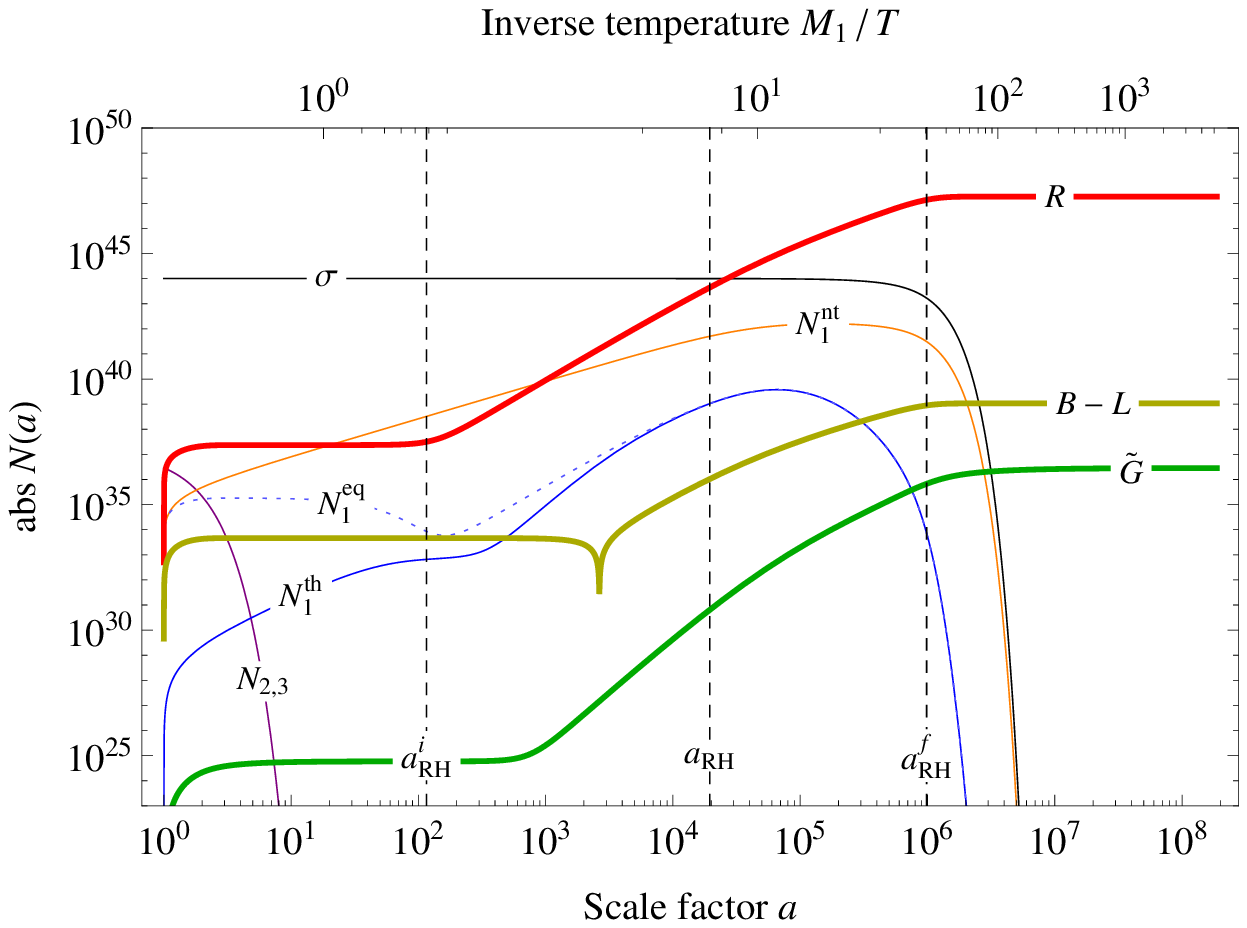}

\vspace{5mm}

\caption[Comoving number densities after slight theoretical modifications]
{Comoving number densities after omitting all massive superparticles
(\textbf{upper panel}) and in addition the \BmL vector boson (\textbf{lower panel}),
to be compared with the result of the full analysis in Fig.~\ref{fig:numengden}.
The individual curves show the comoving number densities
of the Higgs bosons~($\sigma$), nonthermally and thermally
produced neutrinos of the first generation ($N_1^{\text{nt}},\,N_1^{\text{th}}$),
neutrinos from the first generation in thermal equilibrium
($N_1^{\text{eq}}$, for comparison), neutrinos of the second
and third generation ($N_{2,3}$), the MSSM radiation ($R$),
the lepton asymmetry ($B$$-$$L$), and gravitinos ($\tilde G$)
as functions of the scale factor $a$.
The vertical lines labeled $a_{\textrm{RH}}^i$, $a_{\textrm{RH}}$
and $a_{\textrm{RH}}^f$ mark the beginning, the middle and the end
of the reheating process.
The corresponding values for the input parameters are
given in Eq.~\eqref{eq:exampleparameters}.
}
\label{fig:numdenwo}
\end{center}
\end{figure}


\subsection{Robustness against Theory Uncertainties}
\label{sec:robust}


In the previous part of this chapter, we discussed
in detail the emergence of the hot thermal universe after inflation.
The successful explanation of reheating as well the generation
of matter and dark matter by means of our scenario did, however, not rely on
any fortunate coincidence between certain particulars, but was a direct
consequence of the overall setup that we considered.
The essential steps in the evolution after symmetry breaking were the following.
Preheating results in an initial state the energy density of which
is dominated by nonrelativistic Higgs bosons.
These decay slowly into nonthermal neutrinos of the first generation,
which in turn decay into radiation, thereby reheating the universe,
generating a lepton asymmetry and setting the stage for the thermal
production of gravitinos.
At the same time, an additional contribution to the lepton asymmetry
is generated by thermally produced (s)neutrinos.
All further details which we took care of are, of course, important
for a complete understanding of the physical picture,
but merely have a small impact on the final outcome of our calculation.
In particular, as we will illustrate in this section, the
numerical results for the observables of interest, $T_{\textrm{RH}}$,
$\eta_B$, and $\Omega_{\widetilde{G}}h^2$, remain unaffected if one
neglects the superpartners of all massive particles or if one excludes the
gauge particles from the analysis (cf.\ Fig.~\ref{fig:numdenwo}, in which we
plot the corresponding comoving number densities of all remaining species
as functions of the scale factor).
This observation renders our scenario of reheating robust against uncertainties in
the underlying theoretical framework and opens up the possibility to connect it to
other models of inflation and preheating as long as these provide similar initial
conditions as spontaneous \BmL breaking after hybrid inflation.
In addition to that, the robustness of our scenario justifies to crudely simplify
its technical description.
If one is solely interested in the parameter dependence of the observables
and less in the exact evolution during reheating, one may simply omit effects due to
the gauge DOFs and supersymmetry, as we have done it in Refs.~\cite{Buchmuller:2010yy}
and \cite{Buchmuller:2011mw}.


\subsubsection{Nonsupersymmetric Analysis Including the Gauge Multiplet}


In a first step, in order to assess the impact
of supersymmetry on the reheating process,
we neglect the superpartners of all massive particles, i.e.\ the gauge scalar $C$, the gaugino
$\tilde{A}$, the higgsino $\psi$ as well as all heavy sneutrinos $\tilde{N}_i$.
Technically, this renders the inflaton $\phi$ stable, as it can only decay into a pair
of $\tilde{N_1}$ sneutrinos.
To avoid overclosure of the universe, we thus also omit the inflaton.
By contrast, we keep the full particle spectrum of the MSSM and the gravitino
because we still wish to account for dark matter by thermally produced gravitinos.
All in all, these simplifications imply drastically simpler Boltzmann equations
and induce small changes to the corresponding decay and production rates.

Again we solve the set of Boltzmann equations in combination with the initial
conditions set by preheating and the decay of the gauge DOFs.
For our key observables, we obtain
\begin{align}
T_{\textrm{RH}} & \: \simeq 6.1 \times 10^{9} \,\textrm{GeV} \,, \quad
\eta_B \simeq 3.7 \times 10^{-9} \,, \quad
\eta_B^{\textrm{nt}} \simeq 3.7 \times 10^{-9} \,, \label{eq:wosusyres}\\
\eta_B^{\textrm{th}} & \: \simeq 9.7 \times 10^{-15} \,, \quad
\Omega_{\widetilde{G}} h^2 \simeq 0.11 \,. \nonumber
\end{align}
With regard to their first two digits, these results for $T_{\textrm{RH}}$, $\eta_B$,
$\eta_B^{\textrm{nt}}$ and $\Omega_{\widetilde{G}}h^2$ are the same as in the
full analysis.
The result for $\eta_B^{\textrm{th}}$ is smaller by a factor $2$, reflecting
the missing contribution from the thermal sneutrinos of the first generation.
In the upper panel of Fig.~\ref{fig:numdenwo}, we present the corresponding
comoving number densities.
They behave very similarly to the original densities in the upper panel of
Fig.~\ref{fig:numengden}, the only minor differences being the following.
At early times, all densities but the one of the Higgs bosons are a bit smaller,
at most by a factor of $\mathcal{O}(10)$.
In turn, the density of the Higgs bosons is technically a bit larger.
But the relative change is of $\mathcal{O}\left(10^{-4}\right)$ and thus not
visible in Fig.~\ref{fig:numdenwo}.
The fact that initially more energy remains in the Higgs bosons has two reasons.
First, there are now simply less particle species present into which
the initial vacuum energy could be distributed.
Second, particles coupling to the gauge sector are produced
in smaller numbers after preheating due to the absence of
the superpartners of the \BmL vector boson.
A direct consequence of the densities being initially slightly smaller is
that they become sensitive to the decays of the nonthermal
$N_1^S$ neutrinos a bit earlier.
The onset of reheating and the inversion of the lepton asymmetry, for instance, take place
at $a_{\textrm{RH}}^i \simeq 4.2 \times 10^2$ and $a \simeq 3.2 \times 10^3$,
respectively, while these events occur later,
at $a_{\textrm{RH}}^i \simeq 5.3 \times 10^2$ and $a \simeq 4.6 \times 10^3$,
if supersymmetry is fully included.
However, as soon as the $R$ and \BmL abundances are dominated by the decay products of
the $N_1^S$ neutrinos, the differences between the two plots in the upper
panels of Figs.~\ref{fig:numengden} and \ref{fig:numdenwo} begin to vanish.
From $a \sim 10^4$ onwards, they are, apart from a factor 2 between the curves for the
thermal (s)neutrinos, at or below the percent level.


It is easy to understand why the omission of the heavy superparticles does not have
any effect on our final results.
According to Eq.~\eqref{eq:rhonBF}, the initial energy
densities of the gauge scalar $C$, the gaugino $\tilde{A}$,
the higgsino $\psi$, the inflaton $\phi$ as well as the heavy
sneutrinos $\tilde{N}_i$ are monotonic functions of the Higgs-inflaton coupling $\lambda$.
Setting $\lambda$ to its maximal value, $\lambda = 10^{-2}$,
we obtain upper bounds on these densities,
\begin{align}
\left.\frac{\rho_{\tilde{A}}}{\rho_0}\right|_{a_{\textrm{PH}}} \lesssim
\mathcal{O}\left(10^{-2}\right) \,,\quad
\left.\frac{\rho_{C,\psi,\phi,\tilde{N}_{2,3}}}{\rho_0}\right|_{a_{\textrm{PH}}} \lesssim
\mathcal{O}\left(10^{-3}\right) \,,\quad
\left.\frac{\rho_{\tilde{N}_1}}{\rho_0}\right|_{a_{\textrm{PH}}} \lesssim
\mathcal{O}\left(10^{-8}\right) \,.
\label{eq:boundrhorho}
\end{align}
We thus conclude that no matter how the dynamics of the above species look like in detail,
their influence on the reheating process will always be outweighed sooner or later by the
decay of the much more abundant Higgs bosons.
Hence, ignoring these particles does not affect the outcome of our calculation.
Similarly, we can show that only the fermionic decays of the Higgs bosons are relevant
for reheating.
The ratio of $\tilde{N}_1^S$ sneutrinos to $N_1^S$ neutrinos increases
monotonically with the mass $M_1$ (cf.\ Eq.~\eqref{eq:NtSNSratio}).
Our upper bound on this mass, $M_1 = 3 \times 10^{12}\,\textrm{GeV}$, then translates into
$N^S_{\tilde N_1}/N^S_{N_1} \lesssim \mathcal{O}\left(10^{-4}\right)$.
The nonthermal $\tilde{N}_1^S$ sneutrinos can therefore also be safely neglected.
In conclusion, our numerical results in Eqs.~\eqref{eq:boundrhorho} in
combination with the upper bound on $N^S_{\tilde N_1}/N^S_{N_1}$
substantiate our introductory comment at the beginning of this section.
The essential feature of our scenario of reheating is the Higgs boson decay chain,
$\sigma \rightarrow N_1^S \rightarrow R$.
From the point of view of the final results for the observables,
the inclusion of the full supersymmetric particle spectrum is rather a matter
of theoretical consistency than a numerical necessity.


\subsubsection{Nonsupersymmetric Analysis Neglecting the Gauge Multiplet}


Finally, we wish to demonstrate that one is also free to neglect the decay of
the gauge particles, if one is only interested in numerical results for the observables.
In addition to all massive superparticles, we now also exclude
the \BmL vector boson from our analysis.
Consequently, particle production in the decay of gauge particles does not take place any longer,
which simplifies our set of Boltzmann equations once more.
This time we find for our key observables
\begin{align}
T_{\textrm{RH}} & \: \simeq 6.1 \times 10^9 \,\textrm{GeV} \,, \quad
\eta_B \simeq 3.7 \times 10^{-9} \,, \quad
\eta_B^{\textrm{nt}} \simeq 3.7 \times 10^{-9} \,, \\
\eta_B^{\textrm{th}} & \: \simeq 9.7 \times 10^{-15} \,, \quad
\Omega_{\widetilde{G}} h^2 \simeq 0.11 \,. \nonumber
\end{align}
With regard to their first two digits, these results exactly match those in Eq.~\eqref{eq:wosusyres}.
The lower panel of Fig.~\ref{fig:numdenwo} displays the corresponding comoving number densities,
again to be compared with the original densities in the upper panel of
Fig.~\ref{fig:numengden}.
The absence of (s)neutrinos of the second and third generation produced through the decay
of gauge particles now results in a slightly smaller initial lepton asymmetry and, more
importantly, in drastically shorter $N_{2,3}$ reheating.
While this first stage of reheating still lasted until $a \simeq 11$ in our complete analysis
(cf.\ Sec.~\ref{subsec:decaymp}), it now comes to an end  already at $a \simeq 1.7$.
Before the onset of $N_1$ reheating, the abundances of radiation, thermal neutrinos and gravitinos
are hence significantly reduced.
For instance, at $a = 50$ the respective comoving number densities are suppressed
by factors of the following orders of magnitude,
\begin{align}
\textrm{$B$$-$$L$} \,:\: \mathcal{O}\left(10^{-1}\right) \,,\quad
R\,,\: N_1^{\textrm{th}}\,,\: N_1^{\textrm{eq}} \,:\: \mathcal{O}\left(10^{-2}\right) \,,\quad
\widetilde{G} \,:\: \mathcal{O}\left(10^{-3}\right) \,.\quad
\end{align}
As before, due to this initial suppression, these densities are earlier sensitive to the decay
of the $N_1^S$ neutrinos.
Now the onset of $N_1$ reheating and the inversion of the lepton asymmetry take place at
$a_{\textrm{RH}}^i \simeq 1.2 \times 10^2$ and $a \simeq 2.6 \times 10^3$, which is even earlier than
in our nonsupersymmetric analysis including the gauge multiplet.
However, during $N_1$ reheating the differences between the two plots in the upper panel
of Fig.~\ref{fig:numengden} and the lower panel of Fig.~\ref{fig:numdenwo} vanish again.
From $a \sim 10^4$ onwards, they are, apart from the factor $2$ between the curves
for the thermal (s)neutrinos, at or below the percent level.
In conclusion, we find that including the gauge DOFs has
a great impact on the dynamics at early times shortly after preheating,
but turns out be nonessential when calculating the final numerical results.


\section{Scan of the Parameter Space}
\label{sec:scan}


The value of the Boltzmann equations derived in Sec.~\ref{sec:boltzmann} is twofold.
On the one hand, as we have seen in the last section, they are the basis for a
detailed time-resolved description of the dynamics during reheating.
On the other hand, as we will demonstrate in this section,
solving them in the entire parameter space allows one to
study the quantitative dependence of our key quantities, $T_{\textrm{RH}}$, $\eta_B$, and
$\Omega_{\widetilde{G}}h^2$, on the parameters in the Lagrangian.


The relevant parameters of our model are the scale of \BmL breaking
$v_{B-L}$, the heavy neutrino mass $M_1$, the effective
neutrino mass $\widetilde{m}_1$, the gravitino mass $m_{\widetilde{G}}$, and the
gluino mass $m_{\tilde{g}}$.
Requiring consistency with hybrid inflation and the production of cosmic strings
fixes the \BmL breaking scale, $v_{B-L}  = 5 \times 10^{15}\,\textrm{GeV}$, and
limits the range of possible $M_1$ values (cf.\ Sec.~\ref{sec:strings}).
According to the Froggatt-Nielsen flavour model, $\widetilde{m}_1$ should be close
to $\overline{m}_{\nu} \simeq 3 \times 10^{-2} \,\textrm{eV}$.
However, in order to account for the uncertainties of the flavour model, we vary it
between $10^{-5}\,\textrm{eV}$ and $1\,\textrm{eV}$ (cf.\ Eq.~\eqref{eq:parameterspace}).
For the gravitino mass we consider typical values, as they arise in scenarios of gravity-
or gaugino-mediated supersymmetry breaking,
\begin{align}
30\,\textrm{MeV} \leq m_{\tilde{G}} \leq 700\,\textrm{GeV}\,.
\label{eq:Grmasses}
\end{align}
As for the gluino, we stick without loss of generality to the
mass that we used in the parameter example discussed in Sec.~\ref{sec:evolution},
$m_{\tilde{g}} = 1\,\textrm{TeV}$.
The generalization to different choices for $m_{\tilde{g}}$ is
straightforward (cf.\ App.~\ref{ch:gravitinos}) and simply amounts to
a rescaling of all values for the gravitino mass.
Gravitino masses as large as $700\,\textrm{GeV}$ are, in fact,
inconsistent with unified gaugino masses at the GUT scale.
If the gluino and the bino had the same mass at the GUT scale, the different running of
the respective renormalization group equations would then entail a mass ratio of roughly $6$
at low energies.
The gravitino, which we assume to be the LSP, would then have to
be lighter than the bino, resulting in an upper bound of
$m_{\widetilde G} \lesssim 170\,\textrm{GeV}$.
We however leave open the question whether gaugino mass unification takes
place at the GUT scale and work in the following with the full gravitino mass
range specified in Eq.~\eqref{eq:Grmasses}.


At each point of the parameter space defined by the above restrictions,
we solve the Boltzmann equations and record all important numerical results,
which we now discuss in turn.
In Sections~\ref{subsec:TRH} and \ref{subsec:baryonasym}, we study the parameter dependence
of the reheating temperature and the final baryon asymmetry, respectively.
In doing so, we devote particular attention to the composition of the asymmetry
in terms of a nonthermal and a thermal contribution.
By imposing the condition that the maximal possible baryon asymmetry be larger than
the observed one, we identify the region in parameter space that is consistent
with leptogenesis (cf.\ the comment below Eq.~\eqref{eq:etaBres}),
\begin{align}
\eta_B = \eta_B^{\textrm{nt}} + \eta_B^{\textrm{th}} \geq
\eta_B^{\textrm{obs}} \simeq 6.2 \times 10^{-10} \,.
\label{eq:condeta}
\end{align}
In Sec.~\ref{subsec:gravitinoDM}, we then turn to the generation of dark matter
in the form of gravitinos.
Requiring the final gravitino abundance to match the observed density
of dark matter,
\begin{align}
\Omega_{\widetilde{G}} h^2 = \Omega_{\textrm{DM}}^{\textrm{obs}} h^2 \simeq 0.11 \,,
\label{eq:condomega}
\end{align}
we are able to derive relations between the neutrino parameters $M_1$ and $\widetilde{m}_1$
and the superparticle masses $m_{\widetilde{G}}$ and $m_{\tilde{g}}$.
Combining the two conditions in Eqs.~\eqref{eq:condeta} and \eqref{eq:condomega},
we are eventually even able to set a lower bound on $m_{\widetilde{G}}$ in terms
of $\widetilde{m}_1$.


Note that in all plots in this section
(cf.\ Figs.~\ref{fig:tempRH}, \ref{fig:asym} and \ref{fig:mGbounds})
the position of the parameter point which we investigated in
Sec.~\ref{sec:evolution} is marked by a small white circle.


\subsection{Reheating Temperature}
\label{subsec:TRH}


\begin{figure}[t]
\begin{center}
\includegraphics[width=8.75cm]{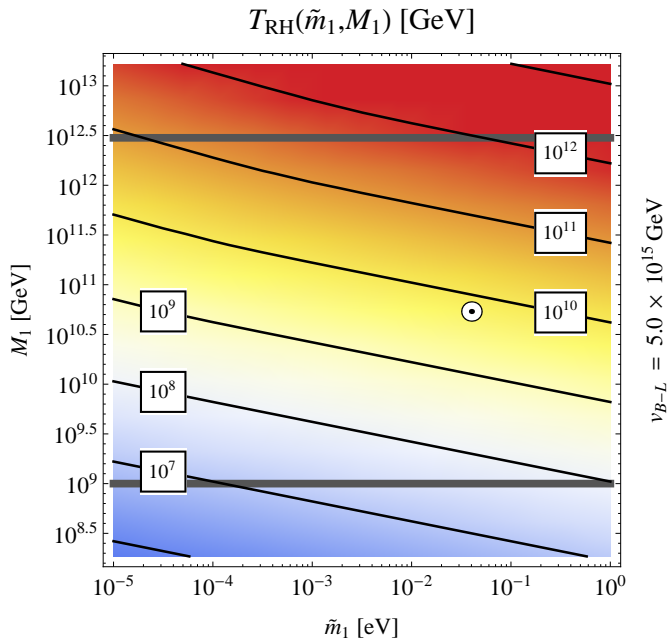}

\vspace{5mm}

\caption[Contour plot of the reheating temperature]
{Contour plot of the reheating temperature $T_{\textrm{RH}}$ as a function
of the effective neutrino mass~$\widetilde{m}_1$ and the heavy neutrino mass $M_1$.
The reheating temperature is calculated according to Eq.~\eqref{eq:TRHdef}
after solving the Boltzmann equations,
cf.\ Sec.~\ref{subsec:RHandT} for a comparison of our definition of the reheating
temperature with other common approaches.
The thick horizontal gray lines represent the lower and the upper bound on $M_1$, respectively,
which arise from requiring consistency with hybrid inflation and the production of cosmic strings
during the \BmL phase transition (cf.\ Eq.~\eqref{eq:parameterspace}).
The small white circle marks the position of the parameter point discussed in
Sec.~\ref{sec:evolution}.
\label{fig:tempRH}
}
\end{center}
\end{figure}


The process of reheating after the \BmL phase transition is
accompanied by an intermediate plateau in the decline of the temperature,
which determines the characteristic temperature scale of reheating.
In Sec.~\ref{subsec:RHandT}, we concretized this intuitive notion
and defined the reheating temperature $T_{\textrm{RH}}$ as the temperature
of the thermal bath at the moment when the decay of the $N_1^S$ neutrinos
into radiation is about to become efficient (cf.\ Eq.~\eqref{eq:TRHdef}),
\begin{align}
\Gamma_{N_1}^S \left(a_\textrm{RH}\right) = H \left(a_\textrm{RH}\right) \,,\quad
T_{\textrm{RH}} = T\left(a_\textrm{RH}\right) \,.\nonumber
\end{align}
We also argued in Sec.~\ref{subsec:RHandT} that this definition is particularly
convenient compared to alternative approaches, because it is not only representative
for the temperature plateau during reheating, but also associated with a physical feature
in the temperature curve.


Having the solutions of the Boltzmann equations for all allowed values of
$\widetilde{m}_1$ and $M_1$ at hand, Eq.~\eqref{eq:TRHdef} enables us to determine the
reheating temperature as a function of these two parameters,
$T_{\textrm{RH}} = T_{\textrm{RH}}\left(\widetilde{m}_1,M_1\right)$.
As the reheating process is solely controlled by Higgs and neutrino decays,
$T_{\textrm{RH}}$ obviously does not depend on the gravitino or gluino mass.
In Fig.~\ref{fig:tempRH}, we present the result of our analysis.
We find that, within the considered range of neutrino parameters,
the reheating temperature varies by almost five orders of magnitude.
For $\widetilde{m}_1 = 10^{-4}\,\textrm{eV}$ and $M_1 = 10^9 \,\textrm{GeV}$,
we have, for instance, ${T_{\textrm{RH}} \sim 10^7 \,\textrm{GeV}}$, while for
$\widetilde{m}_1 = 10^{-1}\,\textrm{eV}$ and $M_1 = 10^{12} \,\textrm{GeV}$
we obtain $T_{\textrm{RH}} \sim 3 \times 10^{11}\,\textrm{GeV}$.
Remarkably, the reheating temperature never exceeds the neutrino mass~$M_1$.
Instead, it is typically smaller than $M_1$ by one or even two orders of magnitude.
As the ratio $M_1 / T_{\textrm{RH}}$ controls the strength of washout
processes during reheating, we conclude that the effect of washout on the
generation of the lepton asymmetry is in most cases negligible (cf.
Sec.~\ref{subsec:baryonasym}, where we will come back to this observation).


The reheating temperature increases monotonically with both neutrino parameters,
$\widetilde{m}_1$ and $M_1$, with the dependence on $M_1$ being much more pronounced
than the dependence on $\widetilde{m}_1$.
In the following, we will derive a simple semianalytical approximation
for $T_{\textrm{RH}}$, by means of which this behaviour can be easily understood.
A more detailed discussion can be found in Appendix~C of Ref.~\cite{Buchmuller:2011mw}.
By definition, $T_{\textrm{RH}}$ corresponds to the decay temperature of $N_1$ neutrinos
decaying with the effective rate $\Gamma_{N_1}^S$.
To first approximation, we may thus write
\begin{align}
T_{\textrm{RH}} \approx \left(\frac{90}{8 \pi^3 g_{*,\rho}}\right)^{1/4} \sqrt{\Gamma_{N_1}^S M_P}
\,=\, \gamma^{-1/2}\left(\frac{90}{8 \pi^3 g_{*,\rho}}\right)^{1/4} \sqrt{\Gamma_{N_1}^0 M_P}\,,
\label{eq:TRH1}
\end{align}
where $\gamma = \gamma\left(\widetilde{m}_1,M_1\right)$ denotes the
Lorentz factor relating $\Gamma_{N_1}^S$ to the zero-temperature
decay rate $\Gamma_{N_1}^0$ evaluated at $a=a_{\textrm{RH}}$.
This first estimate of the reheating temperature fails to accurately reproduce our
numerical results because of two imprecisions.
First, Eq.~\eqref{eq:TRH1} is based on the assumption that, at $a = a_{\textrm{RH}}$, the dominant
contribution to the total energy is contained in radiation.
This is, however, never the case.
At $a = a_{\textrm{RH}}$, the decays of the $N_1^S$ neutrinos have just set in,
so that at this time a significant fraction of the total energy is hence always
still stored in these neutrinos.
On top of that, for $\Gamma_S^0 \ll \Gamma_{N_1}^S$, which is the case in almost the
entire parameter space, the Higgs bosons have not decayed yet at $a = a_{\textrm{RH}}$,
so that in the end they dominate the total energy density at the time of reheating.
To remedy this first imprecision, we have to multiply Eq.~\eqref{eq:TRH1} by $\alpha^{-1/4}$,
where $\alpha = \alpha \left(\widetilde{m}_1,M_1\right) =
\rho_{\textrm{tot}}\left(a_{\textrm{RH}}\right)/ \rho_R\left(a_{\textrm{RH}}\right)$.
The second imprecision is related to the fact that we do not explicitly solve
the Friedmann equation to determine the Hubble parameter, but rather calculate it
as $\dot{a}/a$ with the scale factor $a$ being constructed as described in
Sec.~\ref{sec:boltzmann} (cf.\ Eq.~\eqref{eq:scalefac}).
As a consequence of this procedure, $H$ does not always exactly fulfill the Friedmann equation.
We account for this technical imprecision by multiplying Eq.~\eqref{eq:TRH1} by $\beta^{-1/2}$,
where $\beta = \beta\left(\widetilde{m}_1,M_1\right)$ relates $\dot{a}/a$ to the exact solution
of the Friedmann equation at $a = a_{\textrm{RH}}$.
For appropriate functions $\alpha$, $\beta$ and $\gamma$,  we can then write $T_{\textrm{RH}}$ as
\begin{align}
T_{\textrm{RH}} = & \: \alpha^{-1/4} \beta^{-1/2} \gamma^{-1/2}
\left(\frac{90}{8 \pi^3 g_{*,\rho}}\right)^{1/4} \sqrt{\Gamma_{N_1}^0 M_P}\\
= & \: 7.1 \times 10^{11}\,\textrm{GeV} \times \alpha^{-1/4} \beta^{-1/2} \gamma^{-1/2}
\left(\frac{\widetilde{m}_1}{0.04\,\textrm{eV}}\right)^{1/2}
\left(\frac{M_1}{10^{11}\,\textrm{GeV}}\right) \,.\nonumber
\end{align}
The dependence of $\alpha$, $\beta$ and $\gamma$ on $\widetilde{m}_1$ and $M_1$
follows from the solutions of the Boltzmann equations.
Restricting ourselves to the region in parameter space in which
${\Gamma_{N_1}^0/\Gamma_S^0 \gtrsim \mathcal{O}(100)}$, we find
that $\beta$ and $\gamma$ are basically constant.
We obtain $\beta \simeq 0.99$ and $\gamma \simeq 85$ with deviations around
these values of a few percent.
The dependence of the correction factor $\alpha$ on $\widetilde{m}_1$ and $M_1$
is well described by
\begin{align}
\alpha \simeq 1.2 \times 10^3 \times
\left(\frac{\widetilde{m}_1}{0.04\,\textrm{eV}}\right)
\left(\frac{10^{11}\,\textrm{GeV}}{M_1}\right) \,.
\label{eq:alphapara}
\end{align}
Such a behaviour directly follows from the interplay of the decay rates
$\Gamma_{N_1}^0$ and $\Gamma_S^0$.
For large $\Gamma_{N_1}^0$ and small $\Gamma_S^0$, reheating takes place
quite early, at a time when most Higgs bosons have not decayed yet.
For small $\Gamma_{N_1}^0$ and large $\Gamma_S^0$, reheating takes place
later and not as many Higgs bosons are present anymore at $a = a_{\textrm{RH}}$.
The magnitude of $\alpha$ is hence controlled by the ratio $\Gamma_{N_1}^0/
\Gamma_S^0$ which scales like $\widetilde{m}_1/M_1$.
This explains the parameter dependence in Eq.~\eqref{eq:alphapara}.
Putting all these results together yields a fitting formula for $T_{\textrm{RH}}$
that reproduces our numerical results with an error of less than a percent
in almost the entire parameter space,
\begin{align}
T_{\textrm{RH}} \simeq 1.3 \times 10^{10}\,\textrm{GeV}
\left(\frac{\widetilde{m}_1}{0.04\,\textrm{eV}}\right)^{1/4}
\left(\frac{M_1}{10^{11}\,\textrm{GeV}}\right)^{5/4} \,.
\label{eq:TRHfit}
\end{align}


\subsection{Baryon Asymmetry}
\label{subsec:baryonasym}


\begin{figure}[t]
\begin{center}
\includegraphics[width=11.8cm]{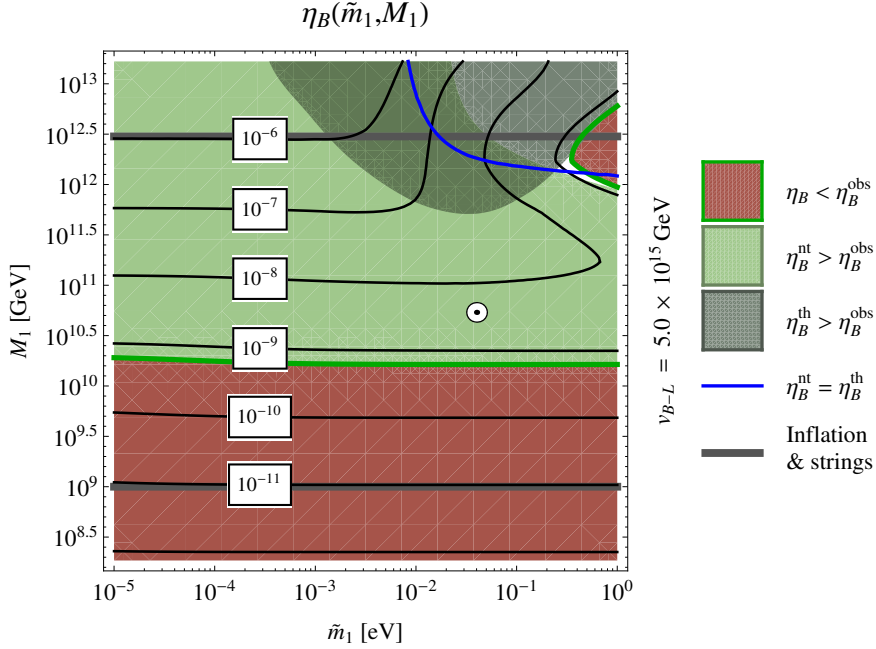}

\vspace{5mm}

\caption[Contour plot of the baryon asymmetry]
{Contour plot of the baryon asymmetry $\eta_B$ as a function
of the effective neutrino mass~$\widetilde{m}_1$ and the heavy neutrino mass $M_1$.
The baryon asymmetry is calculated according to Eq.~\eqref{eq:etaBntth}
after solving the Boltzmann equations.
In the bright green (gray green) region the nonthermal (thermal) asymmetry is consistent with
the observed asymmetry. In the red region the total asymmetry falls short of the observational
bound. Below (above) the thin blue line the nonthermal (thermal) asymmetry dominates over the
thermal (nonthermal) asymmetry.
The thick horizontal gray lines represent the lower and the upper bound on $M_1$, respectively,
which arise from requiring consistency with hybrid inflation and the production of cosmic strings
during the \BmL phase transition (cf.\ Eq.~\eqref{eq:parameterspace}).
The small white circle marks the position of the parameter point discussed in
Sec.~\ref{sec:evolution}.
\label{fig:asym}
}
\end{center}
\end{figure}


Based on the solutions of the Boltzmann equations, we calculate the
nonthermal and thermal contributions to the final baryon asymmetry
(cf.\ Eq.~\eqref{eq:etaBntth}) for all values of the neutrino parameters
$\widetilde{m}_1$ and $M_1$.
We present the result of this analysis in Fig.~\ref{fig:asym}.
The parameter regions in Fig.~\ref{fig:asym} where the nonthermal and thermal baryon
asymmetries $\eta_B^{\textrm{nt}}$ and $\eta_B^{\textrm{th}}$ are consistent with
the observational bound $\eta_B^{\textrm{obs}}$ are shaded in bright green and gray green,
respectively.
The overlap of these two regions is coloured in dark green.
In the white patch around $\widetilde{m}_1 \sim 0.3\,\textrm{eV}$ and
$M_1 \sim 10^{12}\,\textrm{GeV}$, the total asymmetry $\eta_B = \eta_B^{\textrm{nt}} +
\eta_B^{\textrm{th}}$ is larger than $\eta_B^{\textrm{obs}}$, but neither of its
two contributions is.
Below the solid blue line in Fig.~\ref{fig:asym}, the nonthermal
asymmetry dominates over the thermal one.
Above the solid blue line, it is the other way around.
We conclude that in the part of parameter space that we are
interested in the thermal asymmetry is almost always outweighed by its nonthermal counterpart.
Especially in the region in which leptogenesis is consistent with gravitino
dark matter, where $M_1$ is typically of $\mathcal{O}\left(10^{11}\right)\,\textrm{GeV}$
(cf.\ Sec.~\ref{subsec:gravitinoDM}), the thermal asymmetry is negligibly small.


In most of the parameter space the nonthermal asymmetry is
insensitive to $\widetilde{m}_1$ and thus solely controlled by $M_1$.
Only for large values of $\widetilde{m}_1$ and $M_1$, it depends on both neutrino
mass parameters.
This behaviour is directly related to the efficiency of the washout processes
in the respective parameter regions.
Let us suppose for a moment that washout does not take place.
The final nonthermal asymmetry then only depends on the total number of
(s)neutrinos produced during reheating and the amount of $CP$ violation per
(s)neutrino decay.
Neither of these two quantities is, however, affected by changes in $\widetilde{m}_1$,
so that the asymmetry, indeed, ends up being a function of $M_1$ only.
From this perspective, the insensitivity of $\eta_B^{\textrm{nt}}$ to $\widetilde{m}_1$
signals that the effect of washout on the generation of the asymmetry is negligible
for most values of the neutrino parameters.
This result is consistent with our findings for the reheating temperature and
in particular the ratio $M_1 / T_{\textrm{RH}}$ as a function of $\widetilde{m}_1$ and $M_1$
(cf.\ Sec.~\ref{subsec:TRH}).
To see this, note that, for temperatures $T \lesssim M_1$, the effective washout rate
$\hat{\Gamma}_W$ decreases exponentially when raising the ratio $M_1/T$,
\begin{align}
T \lesssim M_1 \,:\quad \hat\Gamma_W =
\frac{N_{N_1}^{\textrm{eq}}}{2 N_{\ell}^{\textrm{eq}}} \Gamma_{N_1}^{\textrm{th}}
\propto \left(\frac{M_1}{T}\right)^{3/2} e^{-M_1/T} \,\Gamma_{N_1}^0 \,,
\label{eq:GammaWprop}
\end{align}
which readily follows from Eqs.~\eqref{eq:GammathN1} and \eqref{eq:nrhoXeq}.
The fact that $M_1/T_{\textrm{RH}}$ is of $\mathcal{O}(10)$ or even larger for most
parameter values then explains why the impact of washout is typically vanishingly small.
In turn, Eq.~\eqref{eq:GammaWprop} also illustrates the importance of washout at
very large values of $\widetilde{m}_1$ and $M_1$, for which the ratio $M_1/T_{\textrm{RH}}$
approaches values of $\mathcal{O}(1)$.
Comparing our results for the reheating temperature and the baryon asymmetry
in Figs.~\ref{fig:tempRH} and \ref{fig:asym}, respectively, we find that washout
only plays a significant role if $M_1/T_{\textrm{RH}} \lesssim 10$ and
$M_1 \gtrsim 10^{11}\,\textrm{GeV}$.
Interestingly, the parameter region defined by these two conditions
covers the entire range of parameters, in which the thermal asymmetry exceeds
the observed asymmetry.


If washout is negligible, the nonthermal asymmetry can be reproduced to good
approximation by assuming that all $N_1^S$ neutrinos decay instantaneously
at time $t_1 = t_S + 1/\Gamma_{N_1}^0$ into radiation.
The resultant baryon asymmetry is then given by
\begin{align}
\eta_B^{\mathrm{nt}} \approx \frac{3\pi^4 g_{*,s}^0}{90 \zeta(3) g_\gamma} C_{\mathrm{sph}}
\,\epsilon_1 \left.\frac{T}{\varepsilon_{N_1}^S}\right|_{t=t_1}\,,
\end{align}
where $\varepsilon_{N_1}^S$ denotes the average energy per $N_1^S$ neutrino.
The ratio $T / \varepsilon_{N_1}^S$ is proportional to $N_{N_1}^S/N_R$,
the number density of $N_1^S$ neutrinos at the time when these decay normalized
to the radiation number density.
It directly follows from the solutions of the Boltzmann equations and is well described by
\begin{align}
\label{eq:Tovereps}
\left.\frac{T}{\varepsilon_{N_1}^S}\right|_{t=t_1} \simeq 3.7 \times 10^{-4}
\left(\frac{M_1}{10^{11}\,\textrm{GeV}}\right)^{1/2}\,.
\end{align}
Together with the expression for $\epsilon_1$ in Eq.~\eqref{eq:epsiloniest}, this
yields the following fitting formula for the nonthermal asymmetry in the case of
weak washout,
\begin{align}
\eta_B^{\mathrm{nt}} \simeq 6.7 \times 10^{-9}
\left(\frac{M_1}{10^{11}\,\textrm{GeV}}\right)^{3/2} \,.
\label{eq:etantFit}
\end{align}
It reproduces our numerical results for $\eta_B^{\mathrm{nt}}$ within a factor of 2 for
most values of $M_1$.


The requirement that the maximal possible asymmetry be larger than the observed one
constrains the allowed range of $M_1$ values.
Fig.~\ref{fig:asym} implies the following lower bound,
\begin{align}
\eta_B \geq \eta_B^{\textrm{obs}} \simeq 6.2 \times 10^{-10}
\quad\longrightarrow\quad
M_1 \geq M_1^{\textrm{min}} \simeq 1.7 \times 10^{10}\,\textrm{GeV} \,,
\label{eq:M1bound}
\end{align}
where we have averaged out the slight dependence on $\widetilde{m}_1$.
If $M_1$ is chosen below this minimal value, the asymmetry falls below the
observational bound for two reasons.
On the one hand, small $M_1$ implies a small $CP$ parameter $\epsilon_1$
(cf.\ Eq.\eqref{eq:epsiloniest}).
On the other hand, according to Eq.~\eqref{eq:Tovereps}, a small $M_1$ value also
entails a small ratio $T / \varepsilon_{N_1}^S$, i.e.\ a small abundance of (s)neutrinos
at the time the asymmetry is generated.
The combination of both effects then renders the successful generation of the lepton asymmetry
impossible.


The thermal asymmetry has, to first approximation, the same parameter dependence
as the asymmetry generated in standard leptogenesis.
It increases monotonically with $M_1$.
If $M_1$ is kept fixed at some value $M_1 \gtrsim 10^{12}\,\textrm{GeV}$,
it is largest for $\widetilde{m}_1$ values
of $\mathcal{O}\left(10^{-2}\right)\,\textrm{eV}$.
The monotonic behaviour in $M_1$ is a direct consequence of the fact that
the $CP$ parameter $\epsilon_1$ scales linearly with $M_1$.
The preference for intermediate values of $\widetilde{m}_1$ has the same
reason as in the standard case.
Large $\widetilde{m}_1$ corresponds to strong washout, at least for the
high values of $M_1$ at which the thermal generation of the asymmetry carries weight.
Small $\widetilde{m}_1$ results in a low temperature and a
small neutrino decay rate $\Gamma_{N_1}^0$, such that the
thermal production of (s)neutrinos is suppressed.
Especially in the parameter region in which the thermal
asymmetry dominates over the nonthermal asymmetry, the expectation
from standard leptogenesis $\eta_B^{\mathrm{st}}$ approximates our numerical results
reasonably well,
\begin{align}
\eta_B^{\mathrm{th}} \approx \eta_B^{\mathrm{st}} =
\frac{3}{4} \frac{g_{*,s}^0}{g_{*,s}} C_{\mathrm{sph}} \epsilon_1
\kappa_f (\widetilde{m}_1)\,.
\label{eq:etathermal}
\end{align}
Here, $\kappa_f = \kappa_f \left(\widetilde{m}_1\right)$ denotes the final efficiency factor.
In the strong washout regime, $\widetilde m_1 \gg 10^{-3}\,\textrm{eV}$, it is
inversely proportional to $\widetilde{m}_1$ and independent of the initial conditions
at high temperatures \cite{Buchmuller:2004nz},
\begin{align}
\kappa_f (\widetilde{m}_1) \simeq 2\times 10^{-2}\left(
 \frac{10^{-2}\,\textrm{eV}}{\widetilde m_1}\right)^{1.1}\,.
\label{eq:kappaf}
\end{align}
Combining Eqs.~\eqref{eq:etathermal} and \eqref{eq:kappaf} with the expression
for $\epsilon_1$ in \eqref{eq:epsilon1}, we obtain
\begin{align}
\eta_B^{\mathrm{th}} \simeq 7.0 \times 10^{-10}
\left(\frac{0.1\,\textrm{eV}}{\widetilde{m}_1}\right)^{1.1}
\left(\frac{M_1}{10^{12}\,\textrm{GeV}}\right) \,.
\end{align}
In the region in parameter space where $\eta_B^{\textrm{th}} > \eta_B^{\textrm{nt}}$,
this fitting formula reproduces our numerical results within a factor of $2$.


Despite these similarities it is, however, important to note that our thermal
mechanism for the generation of the lepton asymmetry differs from
the standard scenario in two important aspects.
First, our variant of thermal leptogenesis is accompanied by continuous entropy
production, while one assumes an adiabatically expanding thermal bath in the case of
standard leptogenesis.
Consequently, our thermal asymmetry experiences an additional dilution during and after its
generation (cf.\ the comment on page~\pageref{page:etabthdilu}).
Second, our scenario of reheating implies a particular relation between the temperature
at which leptogenesis takes place, which is basically $T_{\textrm{RH}}$ in our case, and the neutrino
mass parameters (cf.\ Sec.~\ref{subsec:TRH}) that differs drastically from the corresponding relation implied by standard
leptogenesis.
This translates into a different parameter dependence of the ratio $M_1/T$ as a function of $\widetilde{m}_1$ and $M_1$, which in turn alters the efficiency of washout process and the
production of thermal (s)neutrinos from the bath in the respective regions of parameter space.
In the end, our thermal asymmetry therefore rather corresponds to a distorted version
of the asymmetry generated by standard leptogenesis.
As we have remarked above, in the parameter region where the thermal asymmetry is larger
than the nonthermal asymmetry, $\eta_B^{\textrm{th}}$ hardly deviates from $\eta_B^{\textrm{st}}$.
But as soon as we go to smaller values of $\widetilde{m}_1$ and $M_1$,
the difference between the two asymmetries grows.
The minimal value of $M_1$ for which the thermal asymmetry is still able to exceed the
observational bound, for instance, turns out to be much larger in our scenario than
in standard leptogenesis.
We find an absolute lower bound on $M_1$ of roughly $5.1 \times 10^{11}\,\textrm{GeV}$
at an effective neutrino mass $\widetilde{m}_1 \simeq 3.3 \times 10^{-2}\,\textrm{eV}$,
while standard leptogenesis only constrains $M_1$ to values larger than
$M_1 \sim 10^9\,\textrm{GeV}$.
Lowering $M_1$ below $5.1 \times 10^{11}\,\textrm{GeV}$
either implies a larger ratio $M_1/T_{\textrm{RH}}$ or
a larger effective neutrino mass $\widetilde{m}_1$ (cf.\ Fig.~\ref{fig:tempRH}).
In either case the thermal asymmetry is reduced, so that it drops below the observed value.


In conclusion, we emphasize that the generation of the lepton asymmetry is
typically dominated by the decay of the nonthermal (s)neutrinos.
Only in the parameter region of strong washout, which is characterized by
a small ratio $M_1/T_{\textrm{RH}}$, the nonthermal asymmetry is suppressed
and the thermal asymmetry has the chance to dominate.
Related to that, we find that the viable region in parameter space
governed by the nonthermal mechanism is significantly larger than
the corresponding region for the thermal mechanism.
Independently of $\widetilde{m}_1$, the neutrino mass $M_1$ can be as small as
$M_1^{\textrm{min}} \simeq 1.7 \times 10^{10}\,\textrm{GeV}$, which is an order
of magnitude below the bound of ${5.1 \times 10^{11}\,\textrm{GeV}}$ which one
obtains in the purely thermal case.


\subsection{Gravitino Dark Matter}
\label{subsec:gravitinoDM}


\begin{figure}
\begin{center}
\includegraphics[width=11.5cm]{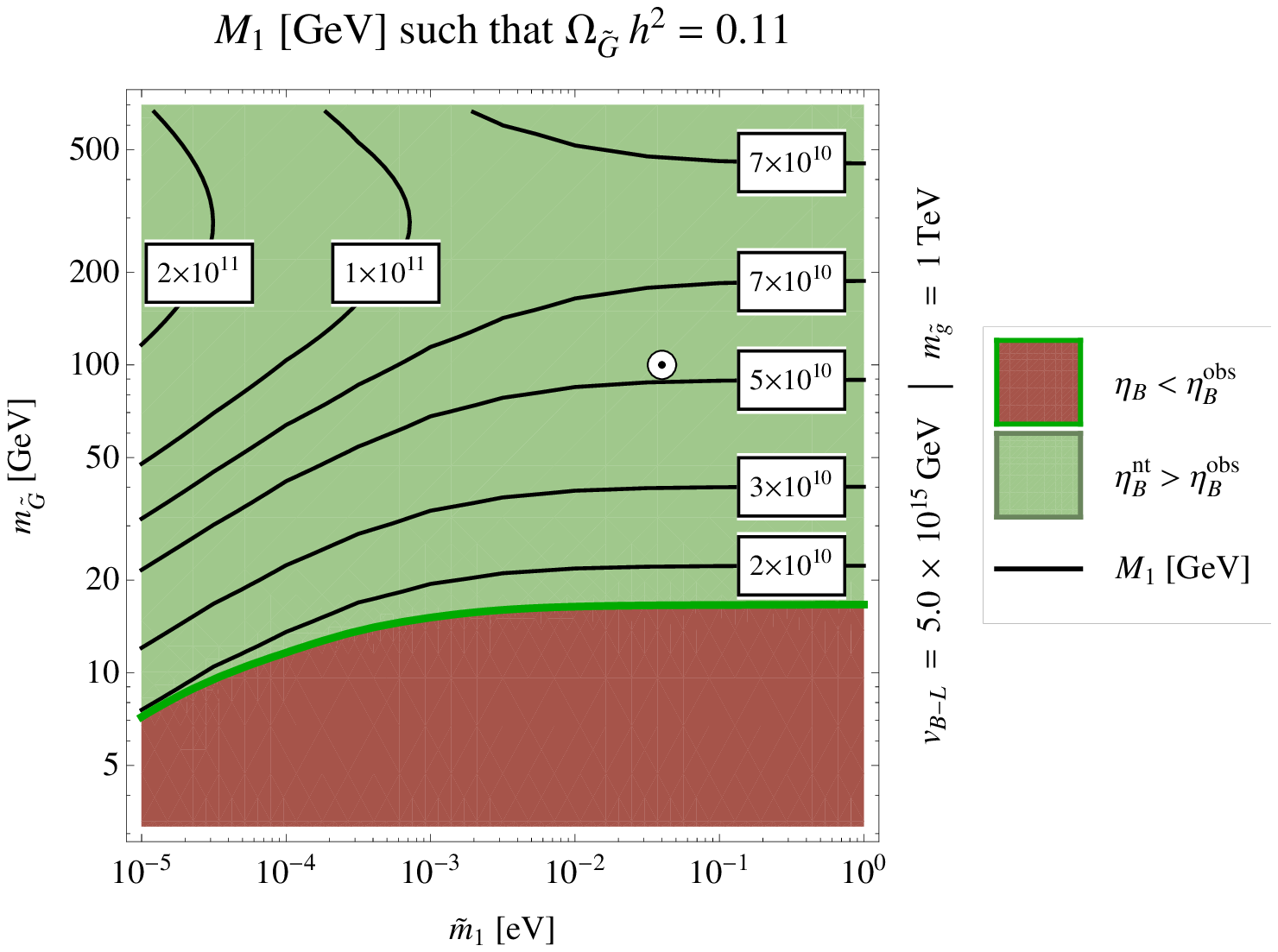}

\vspace{5mm}

\includegraphics[width=11.5cm]{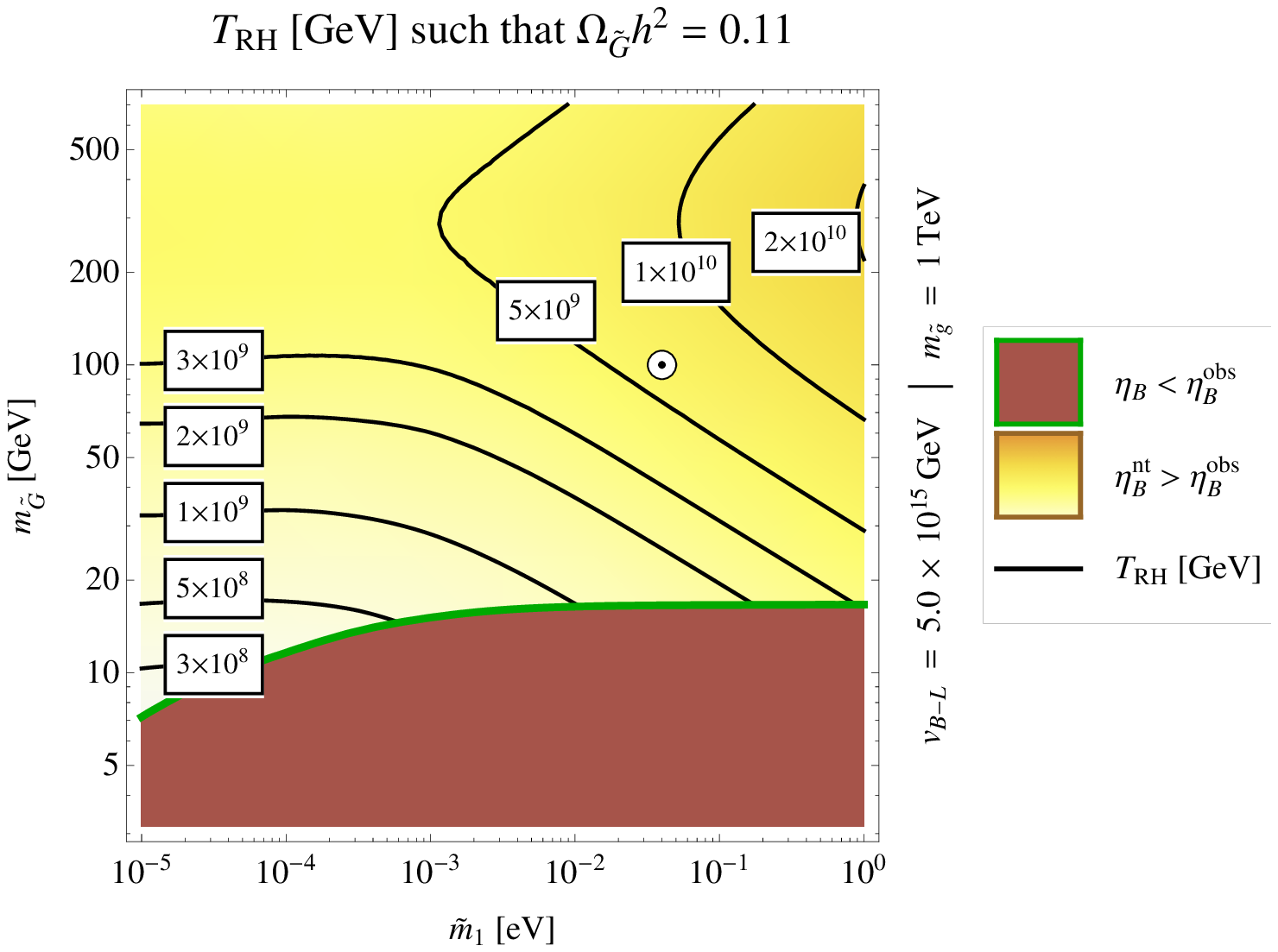}

\vspace{5mm}

\caption[Relations between neutrino parameters and the gravitino mass]
{Contour plots of the heavy neutrino mass $M_1$ \textbf{(upper panel)}
and the reheating temperature $T_{\textrm{RH}}$ \textbf{(lower panel)} as functions
of the effective neutrino mass~$\widetilde{m}_1$ and the gravitino mass $m_{\widetilde{G}}$,
such that the relic density of dark matter is accounted for by gravitinos
(cf.\ Eqs.~\eqref{eq:solveM1} and \eqref{eq:TRHfunc}).
In the red region, the lepton asymmetry generated by leptogenesis is smaller than the
observed one, providing us with a lower bound on the gravitino mass in dependence
on $\widetilde{m}_1$.
The colour code is the same as in Figs.~\ref{fig:tempRH} and \ref{fig:asym}.
The small white circles mark the position of the parameter point discussed in
Sec.~\ref{sec:evolution}.
\label{fig:mGbounds}
}
\end{center}
\end{figure}


The final abundance of gravitinos $\Omega_{\widetilde G} h^2$
depends on three parameters:
the reheating temperature $T_{\textrm{RH}}$ as well as the two superparticle
masses $m_{\widetilde{G}}$ and $m_{\tilde{g}}$.
A key result of our reheating scenario is that $T_{\textrm{RH}}$ is determined
by the neutrino mass parameters $\widetilde{m}_1$ and $M_1$.
As we keep the gluino mass fixed at $1\,\textrm{TeV}$, the gravitino
abundance thus ends up being a function of $\widetilde{m}_1$, $M_1$ and $m_{\widetilde{G}}$.
Based on the solutions of the Boltzmann equations, we calculate
$\Omega_{\widetilde G} h^2$ according to Eq.~\eqref{eq:OmegaGth2} for all values of
these three masses.
By imposing the condition that gravitinos be the constituents
of dark matter, we can then eliminate one of the free mass parameters,
for instance the neutrino mass $M_1$,
\begin{align}
\Omega_{\widetilde{G}} h^2 \left(\widetilde{m}_1,M_1,m_{\widetilde{G}}\right)
=  \Omega_{\textrm{DM}}^{\textrm{obs}} h^2 \quad\longrightarrow\quad
M_1 = M_1 \left(\widetilde{m}_1,m_{\widetilde{G}}\right)\,.
\label{eq:solveM1}
\end{align}
The physical picture behind this step is the following.
For given $m_{\widetilde{G}}$, the reheating
temperature has to have one specific value, so that the abundance of gravitinos
comes out right.
Each choice for $\widetilde{m}_1$ then implies one particular
value of $M_1$ for which this desired reheating temperature is obtained.
Solving Eq.~\eqref{eq:solveM1} for $M_1$ yields this value
as a function of $\widetilde{m}_1$ and $m_{\widetilde{G}}$.
The corresponding reheating temperature follows immediately,
\begin{align}
T_{\textrm{RH}} = T_{\textrm{RH}}
\left(\widetilde{m}_1,M_1 \left(\widetilde{m}_1,m_{\widetilde{G}}\right)\right)
\quad\longrightarrow\quad
T_{\textrm{RH}} = T_{\textrm{RH}}\left(\widetilde{m}_1,m_{\widetilde{G}}\right)\,.
\label{eq:TRHfunc}
\end{align}
In summary, combining the requirement that gravitinos make up the dark matter
with the fact that the reheating temperature is determined by neutrino parameters
allows us to infer relations between these neutrino parameters and superparticle masses.
The lower bound on $M_1$ induced by leptogenesis (cf.\ Eq.~\eqref{eq:M1bound}) can then be
translated into a constraint
on the mass parameters $\widetilde{m}_1$ and $m_{\widetilde{G}}$.
\begin{align}
\eta_B = \eta_B
\left(\widetilde{m}_1,M_1 \left(\widetilde{m}_1,m_{\widetilde{G}}\right)\right)
\geq \eta_B^{\textrm{obs}}
\quad\longrightarrow\quad
\eta_B = \eta_B\left(\widetilde{m}_1,m_{\widetilde{G}}\right) \geq \eta_B^{\textrm{obs}}\,.
\label{eq:condeta2}
\end{align}
We present our results for the functions $M_1 \left(\widetilde{m}_1,m_{\widetilde{G}}\right)$
and $T_{\textrm{RH}} \left(\widetilde{m}_1,m_{\widetilde{G}}\right)$ in the two panels of
Fig.~\ref{fig:mGbounds}, respectively.
Furthermore, we indicate in both plots the constraint arising from the requirement
of successful leptogenesis.


We observe the following trends in the two plots of Fig.~\ref{fig:mGbounds}.
Both quantities, $M_1$ and $T_{\textrm{RH}}$, show a stronger dependence on
the gravitino mass than on the effective neutrino mass.
For $\widetilde{m}_1 \lesssim 10^{-3}\,\textrm{eV}$, the reheating temperature is
almost completely insensitive to $\widetilde{m}_1$.
The neutrino mass $M_1$ slightly increases, when lowering the value of
$\widetilde{m}_1$.
For large values of the effective neutrino mass, $\widetilde{m}_1 \gtrsim 10^{-3}\,\textrm{eV}$,
the exact opposite is the case.
$M_1$ does not depend on $\widetilde{m}_1$ anymore and $T_{\textrm{RH}}$
slightly rises when increasing $\widetilde{m}_1$.
In the following, we will construct semianalytical approximations for $M_1$ and
$T_{\textrm{RH}}$, which will allow us to get some intuition for this behaviour.
The final gravitinos abundance $\Omega_{\widetilde{G}} h^2$ can be parametrized
as (cf.\ App.~\ref{ch:gravitinos})
\begin{align}
\Omega_{\widetilde{G}} h^2 = \varepsilon \, C_1
\left(\frac{T_{\textrm{RH}}}{10^{10}\,\textrm{GeV}}\right) \left[
C_2\left(\frac{m_{\widetilde{G}}}{100\,\textrm{GeV}}\right)
+ \Big(\frac{100\,\textrm{GeV}}{m_{\widetilde{G}}}\Big)
\left(\frac{m_{\tilde{g}}}{1\,\textrm{TeV}}\right)^2\right] \,.
\label{eq:OmegaAnaly}
\end{align}
Here, the two coefficient functions $C_{1,2} = C_{1,2}\left(T_{\textrm{RH}}\right)$
subsume all factors contributing to $\Omega_{\widetilde{G}} h^2$ which can be taken
care of analytically,
\begin{align}
C_1 = & \: 10^{14}\,\textrm{GeV}^2 \,\frac{n_\gamma^0}{\rho_c/h^2}
\frac{g_{*,s}^0}{g_{*,s}} \left(\frac{90}{8\pi^3 g_{*,\rho}}\right)^{1/2}
\frac{18 g_s^6\left(T_{\textrm{RH}}\right)}{g_\gamma g_s^4\left(\mu_0\right) M_P} \left[
\log\left(\frac{T_{\textrm{RH}}^2}{m_g^2\left(T_{\textrm{RH}}\right)}\right) + 0.8846\right] \,,
\nonumber \\
C_2 = & \: \frac{3 g_s^4\left(\mu_0\right)}{100 g_s^4\left(T_{\textrm{RH}}\right)}\,,
\end{align}
They both depend only very weakly on the reheating temperature, so that
for our purposes it will suffice to treat them as constants, $C_1 \simeq 0.26$ and
$C_2 \simeq 0.13$.
The factor $\varepsilon$ parametrizes all effects that cannot be accounted for analytically in
the derivation of Eq.~\eqref{eq:OmegaAnaly}, i.e.\ the amount of energy in radiation at
$a = a_{\textrm{RH}}$, the ratio $\hat{\Gamma}_{\widetilde{G}}/H$ at $a = a_{\textrm{RH}}$ as well as
the increase in the comoving number densities of gravitinos and radiation after $a = a_{\textrm{RH}}$.
In principle, it depends on all mass parameters.
In practice, after solving the Boltzmann equations,
we find that it is mainly controlled by $\widetilde{m}_1$,
\begin{align}
\varepsilon \left(\widetilde{m}_1\right) \simeq 1.2 \left(\frac{10^{-3}\,\textrm{eV}}{\widetilde{m}_1}\right)^c \,,
\label{eq:epsilonfit}
\end{align}
where the exponent $c$ is $c \simeq 0.21$ for $\widetilde{m}_1 \gtrsim 10^{-3}\,\textrm{eV}$
and $c \simeq -0.01$ for $\widetilde{m}_1 \lesssim 10^{-3}\,\textrm{eV}$.
We insert our results for $C_{1,2}$ and $\varepsilon$ into Eq.~\eqref{eq:OmegaAnaly},
set $\Omega_{\widetilde{G}}h^2$ to $\Omega_{\textrm{DM}}^\textrm{obs}h^2$ and solve for
$T_{\textrm{RH}}$,
\begin{align}
T_{\textrm{RH}}\simeq
3.5 \times 10^{9}\,\textrm{GeV}
\left(\frac{\widetilde{m}_1}{10^{-3}\,\textrm{eV}}\right)^c
\left[0.13\left(\frac{m_{\widetilde{G}}}{100\,\textrm{GeV}}\right)
+ \Big(\frac{100\,\textrm{GeV}}{m_{\widetilde{G}}}\Big)\right]^{-1} \,.
\label{eq:TRHm1tmGfit}
\end{align}
The corresponding expression for $M_1$ can then be obtained by exploiting Eq.~\eqref{eq:TRHfit},
\begin{align}
M_1 \simeq 7.2 \times 10^{10}\,\textrm{GeV}
\left(\frac{\widetilde{m}_1}{10^{-3}\,\textrm{eV}}\right)^d
\left[0.13\left(\frac{m_{\widetilde{G}}}{100\,\textrm{GeV}}\right)
+ \Big(\frac{100\,\textrm{GeV}}{m_{\widetilde{G}}}\Big)\right]^{-4/5} \,,
\label{eq:M1m1tmGfit}
\end{align}
where the exponent $d$ is given as $4c/5$$-$$1/5$,
so that $d \simeq -0.03$ for $\widetilde{m}_1 \gtrsim 10^{-3}\,\textrm{eV}$
and $d \simeq -0.20$ for $\widetilde{m}_1 \lesssim 10^{-3}\,\textrm{eV}$.
These two fitting formulae reproduce our numerical results with deviations of $\mathcal{O}(10\%)$
and nicely illustrate the different dependence of $T_{\textrm{RH}}$ and $M_1$ on $\widetilde{m}_1$
for small and large values of $\widetilde{m}_1$, respectively.
As expected, they show that the dependence on $\widetilde{m}_1$ is always very mild and
solely stems from the factor~$\varepsilon$, i.e.\ corrections beyond the purely analytical
result for $\Omega_{\widetilde{G}}h^2$.
If we were to omit these corrections and set $\varepsilon$ to $1$, the reheating temperature
required for gravitino dark matter would be a function of $m_{\widetilde{G}}$
only, $T_{\textrm{RH}} = T_{\textrm{RH}}\left(\widetilde{m}_G\right)$,
in accordance with the fact that the only parameters entering the gravitino production
rate $\hat\Gamma_{\widetilde{G}}$ are the masses of the gravitino and the gluino.


Another interesting feature of the two plots in Fig.~\ref{fig:mGbounds} is
that for fixed $\widetilde{m}_1$ the neutrino mass as well as the reheating
temperature always reach their respective maximal values around gravitino masses
of $280\,\textrm{GeV}$.
The extremal values of $M_1$ and $T_{\textrm{RH}}$ depend on the corresponding
choice for $\widetilde{m}_1$, but are typically of order $10^{11}\,\textrm{GeV}$ and
$10^{10}\,\textrm{GeV}$, respectively.
The fact that both $M_1$ and $T_{\textrm{RH}}$ increase with $m_{\widetilde{G}}$
at small gravitino mass and decrease with $m_{\widetilde{G}}$ at large gravitino mass
is also reflected in our two fitting functions in Eqs.~\eqref{eq:TRHm1tmGfit} and
\eqref{eq:M1m1tmGfit}.
The expressions in these two equations have their respective maxima at
$m_{\widetilde{G}} = 0.13^{-1/2} \times 100\,\textrm{GeV} \simeq 280\,\textrm{GeV}$.
The physical origin of this behaviour and also the reason why $\Omega_{\widetilde{G}}h^2$
in Eq.~\eqref{eq:OmegaAnaly} receives two contributions, one of which is proportional to
$m_{\widetilde{G}}$ and the
other of which is inversely proportional to $m_{\widetilde{G}}$, is the composition of
the gravitino in terms of two transverse DOFs, corresponding to
helicity~$\pm\frac{3}{2}$ states, and two goldstino DOFs, corresponding to
helicity~$\pm\frac{1}{2}$ states.
Recall that the rate $\hat{\Gamma}_{\widetilde{G}}$ contains the following
factor (cf.\ Eq.~\eqref{eq:GammaG}),
\begin{align}
\hat{\Gamma}_{\widetilde{G}} = \hat{\Gamma}_{\widetilde{G}} \left(T,m_{\widetilde{G}},m_{\tilde{g}}\right)
\propto \left(1 + \frac{m_{\tilde{g}}^2(T)}{3 m_{\widetilde{G}}^2}\right)\,.
\label{eq:GammaGprop}
\end{align}
The first term in this factor accounts for the production of the transverse
gravitino components, the second for the production of the goldstino components.
For $m_{\widetilde{G}} \lesssim m_{\tilde{g}}\left(T\right)/\sqrt{3}$,
goldstino production dominates.
An increase in $m_{\widetilde{G}}$ then has to be compensated by an appropriately
larger temperature, so that the final gravitino abundance remains constant.
For $m_{\widetilde{G}} \gtrsim m_{\tilde{g}}\left(T\right)/\sqrt{3}$,
mainly states with helicity $\pm\frac{3}{2}$ are populated, turning the
rate $\hat{\Gamma}_{\widetilde{G}}$ into a function of the temperature only.
In such a case the final gravitino abundance $\Omega_{\widetilde{G}}h^2$ simply
scales linearly with $m_{\widetilde{G}}$ (cf.\ Eq.~\eqref{eq:OmegaGth2}) and
larger gravitino masses have to be balanced by smaller temperatures.
The particular gravitino mass which separates the two regimes of gravitino production
directly follows from the gluino mass at $T \sim 10^{10}\,\textrm{GeV}$ (cf.\ Eq.~\eqref{eq:mgtmg}),
\begin{align}
m_{\widetilde{G}} \simeq \frac{m_{\tilde{g}}\left(T\right)}{\sqrt{3}} =
\frac{g_s^2\left(T\right)}{g_s^2\left(\mu_0\right)} \frac{m_{\tilde{g}}}{\sqrt{3}} \simeq
480/\sqrt{3}\,\textrm{GeV} \simeq 280 \,\textrm{GeV}\,.
\label{eq:mGt280}
\end{align}


The relation between the gravitino mass and the
neutrino parameters $\widetilde{m}_1$ and $M_1$ translates the lower
bound on $M_1$ imposed by the requirement of successful leptogenesis (cf.\ Eq.~\eqref{eq:M1bound})
into a lower bound on $m_{\widetilde{G}}$.
As we can read off from Fig.~\ref{fig:mGbounds}, $m_{\widetilde{G}}$
must be at least of $\mathcal{O}(10)\,\textrm{GeV}$ to obtain consistency
between leptogenesis and gravitino dark matter.
In fact, the bound on $m_{\widetilde{G}}$ slightly varies with $\widetilde{m}_1$.
For $\widetilde{m}_1$ values between $10^{-5}\,\textrm{eV}$ and $10^{-2}\,\textrm{eV}$,
it monotonically increases from roughly $7\,\textrm{GeV}$ to $17\,\textrm{GeV}$,
from $\widetilde{m}_1 \sim 10^{-2}\,\textrm{eV}$ onwards it remains at
$m_{\widetilde{G}} \simeq 17\,\textrm{GeV}$.
For such low gravitino masses, the first term in the brackets on the right-hand side of
Eq.~\eqref{eq:M1m1tmGfit} is negligibly small,\footnote{In physical terms this means that for
small gravitino masses mainly the goldstino DOFs of the
gravitino rather than its transverse DOFs are excited.}
so that the fitting formula for $M_1$ can be easily solved for $m_{\widetilde{G}}$,
\begin{align}
m_{\widetilde{G}} \simeq 8\,\textrm{GeV}
\left(\frac{M_1}{10^{10}\,\textrm{GeV}}\right)^{5/4}
\left(\frac{\widetilde{m}_1}{10^{-3}\,\textrm{eV}}\right)^{1/4-c}\,.
\end{align}
Imposing the condition that $M_1$ be larger than
$M_1^{\textrm{min}} \simeq 1.7 \times 10^{10} \,\textrm{GeV}$ (cf.\ Eq.~\eqref{eq:M1bound})
provides us with an analytical expression for the lower bound on $m_{\widetilde{G}}$,
\begin{align}
m_{\widetilde{G}} \geq m_{\widetilde{G}}^{\textrm{min}} \simeq 16\,\textrm{GeV}
\left(\frac{\widetilde{m}_1}{10^{-3}\,\textrm{eV}}\right)^{1/4-c}\,.
\end{align}
This estimate reproduces our numerical results with a precision at the level of
$\mathcal{O}\left(10\%\right)$.
Physically, the connection between the bounds on $m_{\widetilde{G}}$
and $M_1$ is the following.
For gravitino masses below $\mathcal{O}\left(10\right)\,\textrm{GeV}$, a
reheating temperature $T_{\textrm{RH}} \lesssim \mathcal{O}\left(10^{8..9}\right)\,\textrm{GeV}$
is required to avoid overproduction of gravitinos.
According to our reheating mechanism, such low reheating temperatures are associated
with comparatively small values of the neutrino mass, $M_1 \lesssim \mathcal{O}\left(10^{10}\right)\,\textrm{GeV}$.
The low temperature and low mass then entail a small abundance of (s)neutrinos
at the time the asymmetry is generated as well as a small $CP$ parameter $\epsilon_1$
(cf.\ Eqs.~\eqref{eq:Tovereps} and \eqref{eq:epsilon1}, respectively).
Both effects combine and result in an insufficient lepton asymmetry, rendering
dark matter made of gravitinos with a mass below $\mathcal{O}\left(10\right)\,\textrm{GeV}$ inconsistent with leptogenesis.


In conclusion, we find that our scenario of reheating can be easily realized
in a large fraction of parameter space.
The two conditions of successful leptogenesis and gravitino dark matter,
in combination with constraints from hybrid inflation and the production of
cosmic strings, allow us to interconnect parameters of the neutrino and supergravity sector.
In particular, we are able to determine the neutrino mass $M_1$ and the
reheating temperature $T_{\textrm{RH}}$ as functions of the the effective neutrino mass
$\widetilde{m}_1$ and the gravitino mass $m_{\widetilde{G}}$.
Furthermore, the consistency between all ingredients of our scenario
indicates preferences for $M_1$ and $T_{\textrm{RH}}$,
namely $M_1$ values close to $10^{11}\,\textrm{GeV}$
and $T_{\textrm{RH}}$ values close to $3 \times 10^{9}\,\textrm{GeV}$.
Finally, we obtain a lower bound on the gravitino mass of roughly $10\,\textrm{GeV}$.

\cleardoublepage


\chapter[WIMP Dark Matter from Heavy Gravitino Decays]
{WIMP Dark Matter\newline from Heavy Gravitino Decays}
\label{ch:wimp}


In Ch.~\ref{ch:reheating}, we have demonstrated that the \BmL phase transition,
in combination with the subsequent reheating process, is indeed capable of
generating the initial conditions for the hot early universe.
In the context of the supersymmetric Abelian Higgs model, which we discussed
in Ch.~\ref{ch:model}, and with \BmL breaking taking place at the GUT scale,
an initial phase of unbroken \BmL yields hybrid inflation, ending in tachyonic
preheating in the course of which \BmL is spontaneously broken.
If the gravitino is the LSP, the entropy of the thermal bath, the baryon asymmetry
as well as gravitino dark matter are successfully produced during reheating.


In this chapter we now point out that the spontaneous breaking of
\BmL can also ignite the thermal phase of the universe, if the gravitino
is the \textit{heaviest} superparticle.
This possibility is realized in anomaly mediation \cite{Giudice:1998xp,Randall:1998uk}
and has recently been reconsidered in the case of wino \cite{Ibe:2011aa},
higgsino \cite{Jeong:2011sg} and bino \cite{Krippendorf:2012ir} LSP,
motivated by hints of the LHC experiments ATLAS and CMS that the Higgs
boson may have a mass of about $125\,\textrm{GeV}$
\cite{ATLAS:2012ae,Chatrchyan:2012tx}.
It is known that a gravitino heavier than about $10\,\textrm{TeV}$ can be
consistent with primordial nucleosynthesis and leptogenesis \cite{Weinberg:1982zq,Gherghetta:1999sw,Ibe:2004tg}
(cf.\ Sec.~\ref{subsec:candidates}).
In the following we shall investigate the restrictions on the mass of
a WIMP as LSP, which are imposed by the consistency of hybrid inflation,
leptogenesis, BBN and the dark matter density.
As a preparation for our analysis, we first discuss the mechanisms
contributing to the relic LSP abundance (cf.\ Sec.~\ref{sec:production}).
Then, we present our results and illustrate how the bounds on the various
parameters under study are related to each other (cf.\ Sec.~\ref{sec:relations}).
Lastly, we comment very briefly on the prospects for the experimental
confirmation of our scenario (cf.\ Sec.~\ref{sec:prospects}).


The results presented in this chapter were first published in Ref.~\cite{Buchmuller:2012bt}.


\newpage


\section{Thermal and Nonthermal Neutralino Production}
\label{sec:production}


During the radiation dominated era, WIMPs are produced through inelastic scatterings
in the thermal bath as well as in gravitino decays.
Let us now discuss in turn the thermal and nonthermal contributions to the
final WIMP abundance.


\subsubsection{Thermal Freeze-Out}


The WIMP abundance from thermal freeze-out strongly depends on
the nature of the LSP.
Motivated by anomaly mediation and the present hints for the Higgs boson mass from LHC,
we shall assume in this chapter that the superparticle mass spectrum exhibits
the following characteristic hierarchy \cite{Ibe:2011aa,Jeong:2011sg,Krippendorf:2012ir},
\begin{align}
m_{\textrm{LSP}}  \ll m_{\textrm{squark},\textrm{slepton}} \ll m_{\widetilde{G}} \,.
\label{eq:masshierarchy}
\end{align}
Due to this hierarchy, the LSP is typically a \textit{pure} gaugino or higgsino.
It is well known that in this situation the thermal abundance of a bino LSP
is generically too large, which is therefore disfavoured.
Hence, the case  of a light wino \cite{Ibe:2011aa} or higgsino
\cite{Jeong:2011sg} is preferred.\footnote{Note that a \textit{pure} higgsino
also occurs as next-to-lightest superparticle along with multi-TeV coloured
particles in hybrid  gauge-gravity mediation, however with the gravitino
as LSP \cite{Brummer:2011yd}.\smallskip}
A pure neutral wino $(\widetilde{w})$ or higgsino $(\widetilde{h})$ is almost
mass degenerate with a chargino belonging to the same $SU(2)$ multiplet.
The current lower bound on chargino masses \cite{Nakamura:2010zzi}
thus also applies to the LSP.
The thermal abundance of a pure wino\footnote{Compared to
Ref.~\cite{ArkaniHamed:2006mb}, we have reduced the thermal wino abundance
$\Omega^{\textrm{th}}_{\widetilde{w}} h^2$ by $30\,\%$ to account
for the Sommerfeld enhancement effect \cite{Hisano:2006nn,Cirelli:2007xd}.}
or higgsino LSP becomes only significant for masses above $1~\textrm{TeV}$,
where it is well approximated by \cite{ArkaniHamed:2006mb}
\begin{align}
\Omega_{\widetilde{w},\widetilde{h}}^{\textrm{th}} h^2
= c_{\widetilde{w},\widetilde{h}}
\left(\frac{m_{\widetilde{w},\widetilde{h}}}{1\,\textrm{TeV}}\right)^2 \,,
\quad c_{\widetilde{w}} = 0.014 \,, \quad c_{\widetilde{h}} = 0.10  \,.
\label{eq:omegaDMth}
\end{align}


\subsubsection{Heavy Gravitino Decay}


In this chapter, we shall consider gravitino masses in the range from
$10\,\textrm{TeV}$ to $10^3\,\textrm{TeV}$, as they are suggested by anomaly mediation.
The gravitino lifetime is then given by (cf.\ Eq.~\eqref{eq:Glife280d})
\begin{align}
\tau_{\widetilde{G}} =
\left[\frac{1}{4}\left(n_v + \frac{n_m}{12}\right) \frac{m_{\widetilde{G}}^3}{M_P^2}\right]^{-1}
\simeq 24 \left(\frac{10\,\textrm{TeV}}{m_{\widetilde{G}}}\right)^3 \,\textrm{sec} \,,
\label{eq:Glife24sec}
\end{align}
which corresponds to a gravitino decay temperature $T_{\widetilde{G}}$ of
(cf.\ Eq.~\eqref{eq:Ttrelation})
\begin{align}
T_{\widetilde{G}} = \left(\frac{90 \, M_P^2}
{32\pi^3 \, g_{*,\rho}(T_{\widetilde{G}})\,\tau_{\widetilde{G}}^2}\right)^{1/4}
\simeq 0.24 \left(\frac{43/4}{g_{*,\rho}(T_{\widetilde{G}})}\right)^{1/4}
\bigg(\frac{m_{\widetilde{G}}}{10\,\textrm{TeV}}\bigg)^{3/2} \,\textrm{MeV}\,.
\end{align}
For gravitino masses between $10\,\textrm{TeV}$ to $10^3\,\textrm{TeV}$,
the temperature $T_{\widetilde{G}}$ varies between $0.2\,\textrm{MeV}$
and $200\,\textrm{MeV}$, i.e.\ roughly between the temperatures of BBN
(cf.\ Sec.~\ref{subsec:BBN}) and the QCD phase transition (cf.\ Sec.~\ref{subsec:transitions}).
In this temperature range, the entropy increase due to gravitino decays
and hence the corresponding dilution of the baryon asymmetry are negligible.


For such heavy gravitinos as we consider them in this chapter,
the gravitino production rate $\hat{\Gamma}_{\widetilde{G}}$ becomes
independent of the gravitino mass (cf.\ Eq.~\eqref{eq:GammaG}).
The present-day gravitino number density
$n_{\widetilde{G}}^0 \propto \Omega_{\widetilde{G}}^0 h^2 / m_{\widetilde{G}}$
is hence solely determined by the reheating temperature $T_{\textrm{RH}}$.
Solving the Boltzmann equations governing the reheating process
(cf.\ Sec.~\ref{sec:boltzmann}) for a heavy gravitino,
$m_{\widetilde{G}} \gg 1\,\textrm{TeV}$, and neutrino mass parameters
$\widetilde{m}_1$ and $M_1$ which result in reheating temperatures
$T_{\textrm{RH}} = 10^8 .. 10^{11}\,\textrm{GeV}$ we find
\begin{align}
\left(\frac{100\,\textrm{GeV}}{m_{\widetilde{G}}}\right)\Omega_{\widetilde{G}}^0 h^2
\simeq 2.7 \times 10^{-2} \,\left(\frac{T_{\textrm{RH}}\left(\widetilde{m}_1,M_1\right)}
{10^{10}\,\textrm{GeV}}\right) \,.
\label{eq:OmegaGTRH}
\end{align}
Note that the rate $\hat{\Gamma}_{\widetilde{G}}$ of thermal gravitino production,
and thus also the numerical prefactor on the right-hand of this relation, has a
theoretical uncertainty of at least a factor of $2$ (cf.\ App.~\ref{ch:gravitinos}
for an analytical reconstruction of Eq.~\eqref{eq:OmegaGTRH}).
The decay of a heavy gravitino, $m_{\widetilde{G}} \gg m_{\textrm{LSP}}$,
produces approximately one LSP.
This yields the nonthermal contribution to the WIMP abundance
$\Omega_{\textrm{LSP}}^{\widetilde{G}} h^2$.
Assuming that the gravitinos are thermally produced during reheating,
i.e.\ employing the relation in Eq.~\eqref{eq:OmegaGTRH}, we obtain
\begin{align}
\Omega_{\textrm{LSP}}^{\widetilde{G}} h^2 =
\frac{m_{\textrm{LSP}}}{m_{\widetilde{G}}} \,\Omega_{\widetilde{G}} h^2
\simeq 2.7\times 10^{-2} \,\bigg(\frac{m_{\textrm{LSP}}}{100\,\textrm{GeV}}\bigg)
\left(\frac{T_{\textrm{RH}}\left(\widetilde{m}_1,M_1\right)}
{10^{10}\,\textrm{GeV}}\right) \,.
\label{eq:omegaDMG}
\end{align}
For LSP masses below $1\,\textrm{TeV}$, which are most interesting for
the LHC as well as for direct searches, the total LSP abundance,
\begin{align}
\Omega_{\widetilde{w},\widetilde{h}} h^2 =
\Omega_{\widetilde{w},\widetilde{h}}^{\widetilde{G}} h^2 +
\Omega_{\widetilde{w},\widetilde{h}}^{\textrm{th}} h^2 \,,
\label{eq:omegaDMtot}
\end{align}
is thus dominated by the contribution from gravitino decay.


\begin{figure}
\centering
%
\includegraphics[width=10cm]{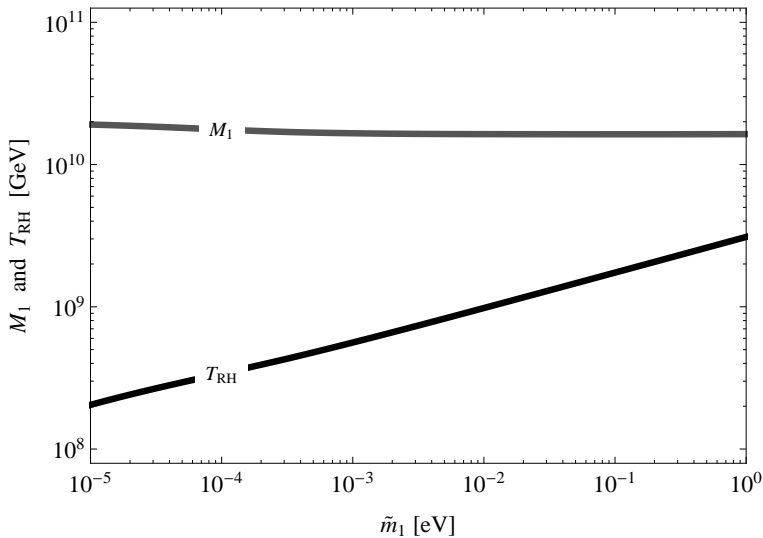}

\vspace{5mm}

\caption[Lower bounds on $M_1$ and $T_{\textrm{RH}}$ as functions of $\widetilde{m}_1$]
{Lower bounds on the heavy (s)neutrino mass $M_1$ and the reheating
temperature $T_{\textrm{RH}}$ as functions of the effective neutrino
mass $\widetilde{m}_1$ from successful leptogenesis.
\label{fig:lbound}
}
%
\end{figure}


Finally, we point out that due to the large mass hierarchy,
$m_{\widetilde{G}} \gg m_{\textrm{LSP}}$, the LSPs are produced relativistically.
They thus form \textit{warm} dark matter, which can affect structure formation
on small scales.
A straightforward calculation yields the free-streaming length $\lambda_{\textrm{FS}}$,
\begin{align}
\lambda_{\textrm{FS}} = \int_{\tau_{\widetilde{G}}}^{t_0} dt \,\frac{v_{\textrm{LSP}}}{a}
\simeq \left(\frac{3}{4}\right)^{2/3} \frac{m_{\widetilde{G}}}{2\, m_{\textrm{LSP}}}
\left(\tau_{\widetilde{G}} \, t_{\textrm{eq}}\right)^{1/2}
\left(\frac{t_0}{t_{\textrm{eq}}}\right)^{2/3}\left[
\ln \frac{16 \,t_{\textrm{eq}} \,  m_{\textrm{LSP}}^2}{\tau_{\widetilde{G}} \, m_{\widetilde{G}}^2}
+ 4 \right] \,,
\end{align}
where $v_{\textrm{LSP}}$ denotes the time-dependent absolute value of the
three-velocity of an LSP which is produced in the decay of a heavy gravitino
at time $\tau_{\widetilde{G}}$.
Furthermore, $t_{\textrm{eq}}$ and $t_0$ are the time of radiation-matter equality
and the age of the universe, respectively (cf.\ Eq.~\eqref{eq:tLambdateq}).
For the gravitino and LSP masses that we consider in this chapter,
one finds $\lambda_{\textrm{FS}} \lesssim 0.1\,\textrm{Mpc}$, which is below
the scales relevant for structure formation \cite{Borzumati:2008zz}.


\section[Relations between Neutralino, Gravitino and Neutrino Masses]
{Relations between Neutralino, Gravitino\newline and Neutrino Masses}
\label{sec:relations}

Requiring consistency between all ingredients of our
scenario, we are able to deduce
(i) constraints on the reheating temperature
as well as bounds on the neutralino and the gravitino mass,
which (ii) mutually depend on each other
and (iii) also vary as functions of the effective neutrino mass $\widetilde{m}_1$.
We shall now discuss these results one after another.


\subsubsection{Bounds on the Reheating Temperature}


\begin{figure}[t]
\centering
%
\includegraphics[width=12cm]{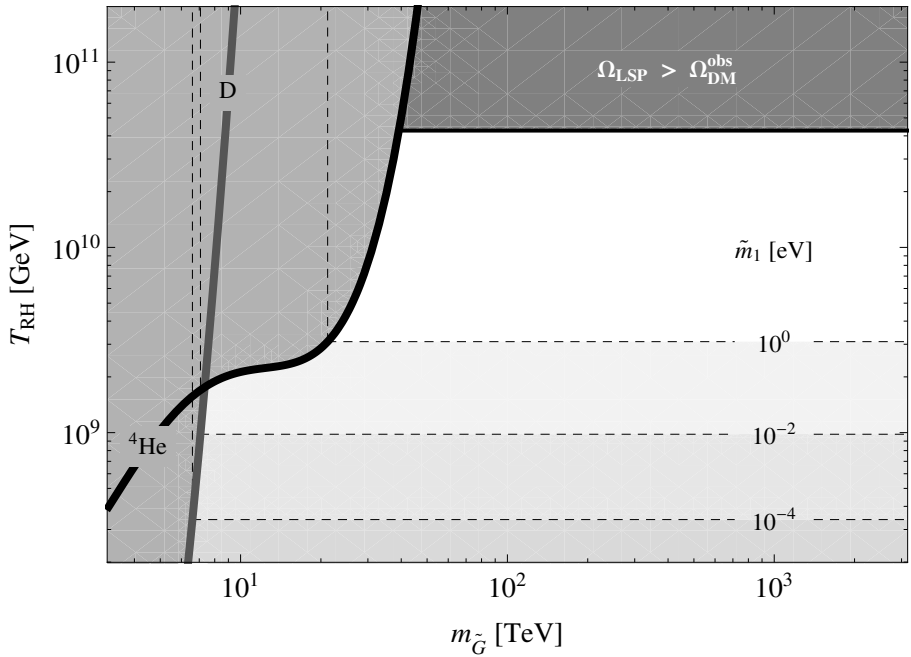}

\vspace{5mm}

\caption[Upper and lower bounds on $T_{\textrm{RH}}$ as functions of $\widetilde{m}_1$ and
$m_{\widetilde{G}}$]
{Upper and lower bounds on the reheating
temperature as functions of the gravitino mass.
The horizontal dashed lines denote lower bounds imposed by
successful leptogenesis for different values of the effective
neutrino mass $\widetilde{m}_1$, cf.~Fig.~\ref{fig:lbound}.
The curves labeled $^4\textrm{He}$ and $\textrm{D}$ denote upper bounds
originating from the primordial helium-4 and deuterium abundances
created during BBN, which are taken from \cite{Kawasaki:2008qe}
(case~2, which gives the most conservative bounds).
The vertical dashed lines represent the absolute lower bounds on
the gravitino mass for fixed effective neutrino mass $\widetilde{m}_1$
and minimal reheating temperature.
The shaded region marked $\Omega_{\textrm{LSP}} > \Omega_{\text{DM}}^{\textrm{obs}}$
is excluded, as it corresponds to overproduction of dark matter, taking into account
that the LSP mass is bounded from below, $m_{\textrm{LSP}} \geq 94\,\textrm{GeV}$.
\label{fig:Thwbounds1}
}
%
\end{figure}


Successful leptogenesis implies a lower bound on the heavy (s)neutrino
mass $M_1$, which slightly depends on $\widetilde{m}_1$ (cf.\ Fig.~\ref{fig:asym}).
After averaging out the dependence on $\widetilde{m}_1$, we already stated
the rough magnitude of this bound in Eq.~\eqref{eq:M1bound}.
Now we fully present its behaviour as a function of $\widetilde{m}_1$ in Fig.~\ref{fig:lbound}.
As each pair of $\widetilde{m}_1$ and $M_1$ values corresponds to a specific value of the
reheating temperature $T_{\textrm{RH}}$ (cf.\ Fig.~\ref{fig:tempRH}),
the lower bound on $M_1$ is readily translated into a lower bound on
$T_{\textrm{RH}}$, which is also displayed in Fig.~\ref{fig:lbound}.


The LSP has to be heavier than $94~\textrm{GeV}$, the current lower bound
on chargino masses \cite{Nakamura:2010zzi}. From the requirement of LSP
dark matter, i.e.\
$\Omega_{\textrm{LSP}} h^2 = \Omega_{\textrm{DM}} h^2 \simeq 0.11$, one then
obtains an upper bound on the reheating 
temperature, $T_{\textrm{RH}} < 4.2\times 10^{10}~\textrm{GeV}$.
For gravitino masses below $40~\textrm{TeV}$, primordial nucleosynthesis
provides a more stringent upper bound on the reheating temperature 
\cite{Kawasaki:2008qe}.
In Fig.~\ref{fig:Thwbounds1}, we compare  upper and lower
bounds on the reheating temperature from dark matter density, nucleosynthesis
and leptogenesis, respectively, as functions of the gravitino mass. It is
remarkable that for the entire mass range, 
$10~\textrm{TeV} \lesssim m_{\widetilde{G}} \lesssim 10^3~\textrm{TeV}$,
nucleosynthesis, dark matter and leptogenesis can be consistent.


\subsubsection{Relation between the Neutralino and the Gravitino Mass}


\begin{figure}
\centering
%
\includegraphics[width=9.75cm]{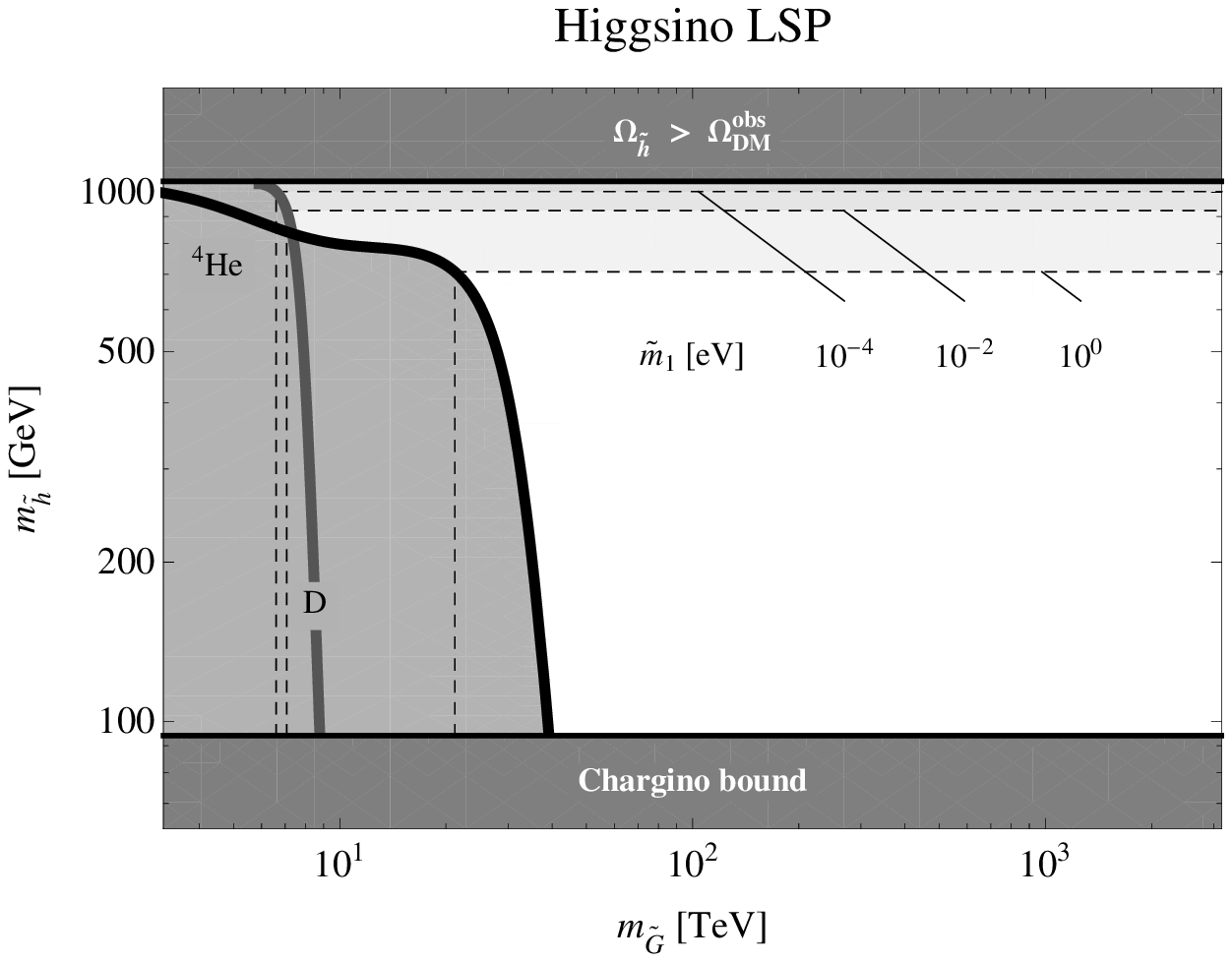}

\vspace{5mm}

\includegraphics[width=9.75cm]{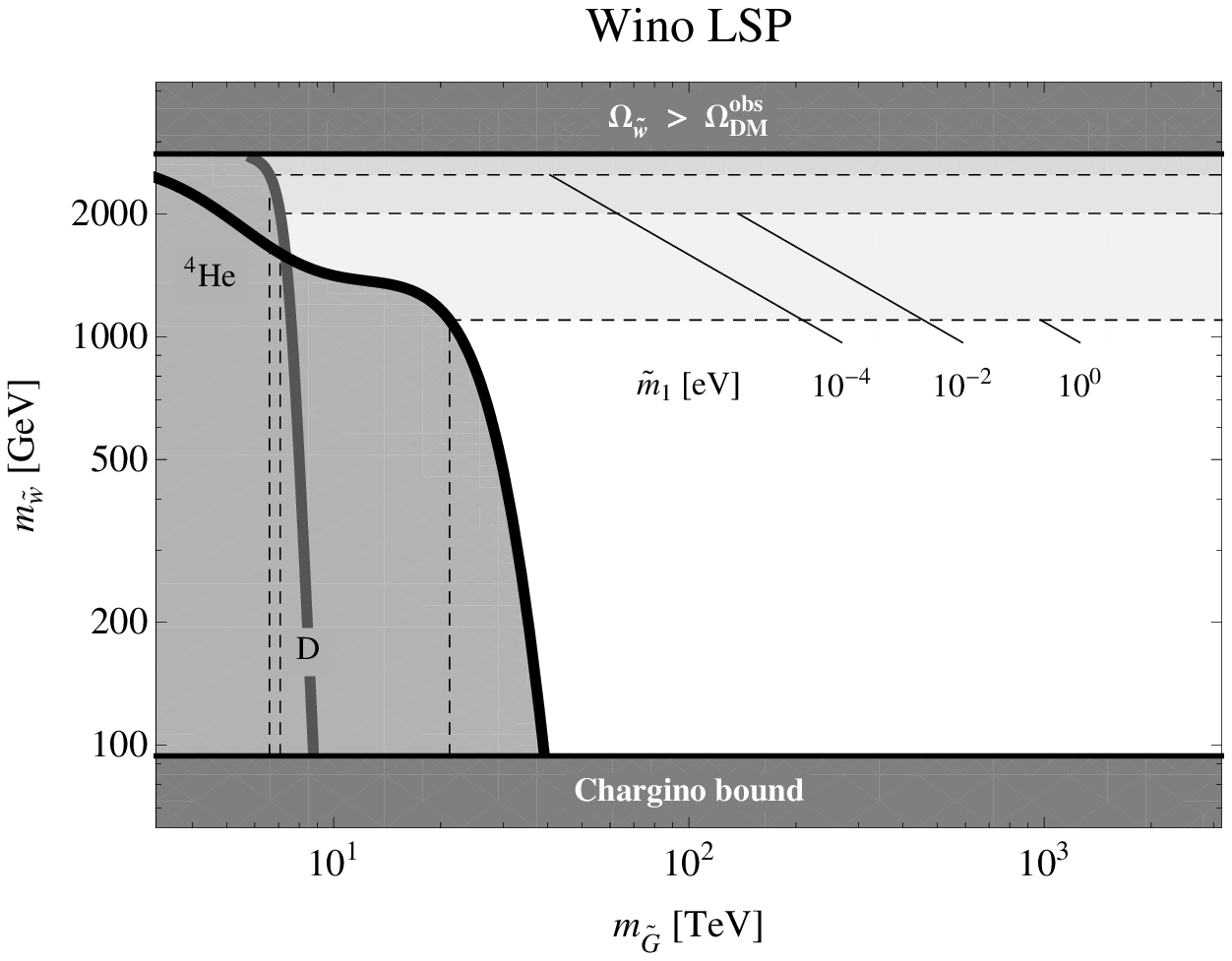}

\vspace{5mm}

\caption[Upper and lower bounds on $m_{\textrm{LSP}}$ and lower bounds on $m_{\widetilde{G}}$.]
{Upper and lower bounds on the LSP mass in the higgsino and wino
case, respectively, and lower bounds on the gravitino mass.
These bounds are in one-to-one correspondence with the bounds on the
reheating temperature and the gravitino mass in Fig.~\ref{fig:Thwbounds1}.
The horizontal dashed lines denote the upper bounds on the LSP mass
imposed by successful leptogenesis for different values of the effective neutrino mass $\widetilde{m}_1$.
The curves labeled $^4\text{He}$ and $\textrm{D}$ denote
lower bounds on the LSP as well as on the gravitino mass originating
from the primordial helium-4 and deuterium abundances created during BBN.
The vertical dashed lines represent the absolute lower bounds on
the gravitino mass for fixed effective neutrino mass $\widetilde{m}_1$ and maximal LSP mass.
The dark shaded regions on the upper edge of the plots
correspond to thermal overproduction of dark matter and are hence excluded.
We do not consider LSP masses below $94\,\textrm{GeV}$ due to the present lower bound on the
chargino mass.
\label{fig:Thwbounds2}
}
%
\end{figure}


The dark matter constraint $\Omega_{\textrm{LSP}} h^2 = \Omega_{\textrm{DM}} h^2 \simeq 0.11$,
with $\Omega_{\textrm{LSP}} h^2$ calculated according
to Eqs.~(\ref{eq:omegaDMth}), (\ref{eq:omegaDMG}) and Eq.~(\ref{eq:omegaDMtot}),
establishes a one-to-one connection between LSP masses and values
of the reheating temperature.
This relation maps the viable region in the
$\left(m_{\widetilde{G}},T_{\textrm{RH}}\right)$-plane for a given effective
neutrino mass $\widetilde{m}_1$ into the corresponding viable region in the
$\left(m_{\widetilde{G}},m_{\textrm{LSP}}\right)$-plane.
We present our results for higgsino and wino LSP in the two panels of
Fig.~\ref{fig:Thwbounds2}, respectively.
The upper bound on the LSP mass is a consequence of the lower bound on the reheating
temperature from leptogenesis, which is why it depends on the effective neutrino mass $\widetilde{m}_1$.
The lower bound on the LSP mass corresponds to the upper bound on the reheating temperature
from BBN and hence depends on the gravitino mass $m_{\widetilde{G}}$.
This latter relation between $m_{\textrm{LSP}}$ and $m_{\widetilde{G}}$ can also be
interpreted the other way around.
As each LSP mass is associated with a certain reheating temperature, we find for
each value of $m_{\textrm{LSP}}$ a lower bound on the gravitino mass.
For given $\widetilde{m}_1$, we then obtain an absolute lower bound on the gravitino mass
by raising the LSP mass to its maximal possible value.


\subsubsection{Dependence on the Effective Neutrino Mass}


\begin{figure}[t]
\centering
%
\includegraphics[width=12cm]{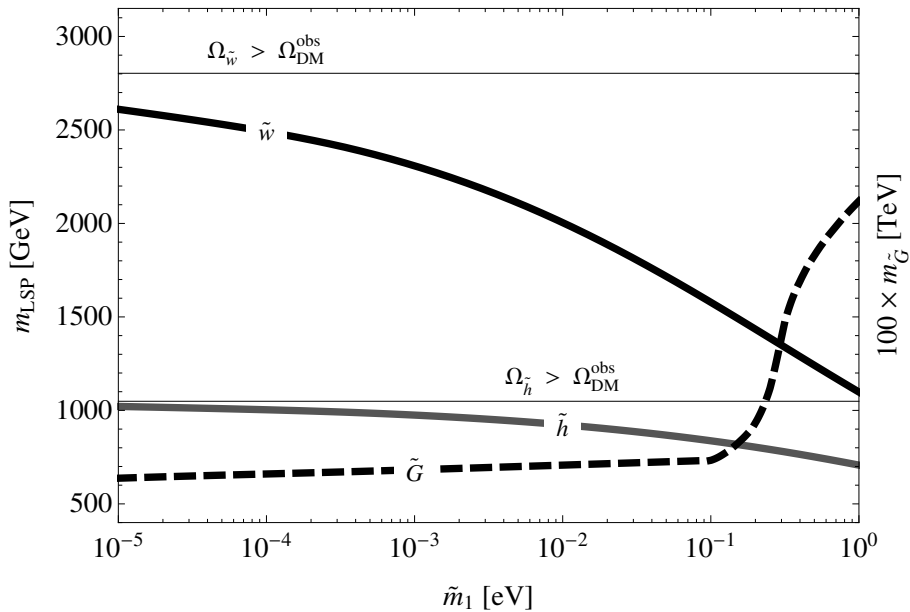}

\vspace{5mm}

\caption[Upper bounds on $m_{\textrm{LSP}}$ and lower bound on $m_{\widetilde{G}}$
as functions of $\widetilde{m}_1$]
{Upper bounds on wino ($\widetilde{w}$) and higgsino
($\widetilde{h}$) LSP masses imposed by successful leptogenesis as well as absolute
lower bound on the gravitino mass according to BBN
as functions of the effective neutrino mass $\widetilde{m}_1$.
Note that in Fig.~\ref{fig:Thwbounds2} these bounds are indicated by horizontal
and vertical dashed lines, respectively, for different value for $\widetilde{m}_1$.
Wino masses larger than $2.8\,\textrm{TeV}$ and higgsino masses larger than
$1.0\,\textrm{TeV}$ result in thermal overproduction.
\label{fig:lbound2}
}
%
\end{figure}


The upper bound on the LSP mass as well as the absolute lower bound on the gravitino
mass both depend on the effective neutrino mass $\widetilde{m}_1$.
In Fig.~\ref{fig:lbound2}, we now finally show the explicit dependence of
these bounds on $\widetilde{m}_1$.
The upper bound on the LSP mass imposed by successful leptogenesis increases when lowering
$\widetilde{m}_1$, i.e.\ when extending the range of allowed reheating temperatures to lower
values.
For very small $\widetilde{m}_1$, it approaches the upper bound on the LSP mass above which
thermal freeze-out leads to an overabundance of LSPs.
At large values of $\widetilde{m}_1$, the bound on the LSP mass from leptogenesis
becomes stronger.
Furthermore, we find that the absolute lower bound on the gravitino mass is
rather insensitive to the effective neutrino mass for $\widetilde m_1 \lesssim 10^{-1}$~eV,
but rapidly increases as a function of $\widetilde{m}_1$ for larger values of $\widetilde m_1$.
This reflects the fact that small values of $\widetilde m_1$ correspond to low
reheating temperatures, for which the allowed range of gravitino masses,
being determined by the BBB abundance of deuterium, hardly changes when
varying the temperature.
It turn, when the allowed range of gravitino masses is determined by the BBN abundance
of helium-4, which is the case for very large $\widetilde{m}_1$, the absolute lower bound on
$m_{\widetilde{G}}$ increases with $\widetilde{m}_1$.


\section{Prospects for Direct Detection and Collider Experiments}
\label{sec:prospects}


We conclude with a few remarks on the prospects for the
confirmation of our scenario in direct detection and/or collider experiments.


\subsubsection{Direct Detection Experiments}


For pure wino and higgsino LSPs, the exchange of the
lightest Higgs boson yields at tree level for the spin-independent
elastic scattering cross section \cite{Hisano:2004pv}
\begin{align}
\sigma^{\widetilde{w}}_{\textrm{SI}} \sim & \: 2\times 10^{-43}\,\textrm{cm}^2
\left(\frac{125\,\textrm{GeV}}{m_{h^0}}\right)^4
\left(\frac{100\,\textrm{GeV}}{m_{\widetilde{h}}}\right)^2
\left(\sin{2\beta}+\frac{m_{\widetilde{w}}}{m_{\widetilde{h}}}\right)^2 \,,\\
\sigma^{\widetilde{h}}_{\textrm{SI}} \sim & \: 7\times 10^{-44}\,\textrm{cm}^2
\left(\frac{125\,\textrm{GeV}}{m_{h^0}}\right)^4
\left(\frac{100\,\textrm{GeV}}{m_{\widetilde{w}}}\right)^2 \,, \nonumber
\end{align}
where $m_{h^0}$ is the mass of the lightest Higgs boson.
For the hierarchical mass spectrum of Eq.~\eqref{eq:masshierarchy}, one has
$r_{\widetilde{w}} \equiv m_{\widetilde{w}}/m_{\widetilde{h}} \ll 1$ for wino LSP and
$r_{\widetilde{h}} \equiv m_{\widetilde{h}}/m_{\widetilde{w}} \ll 1$ for higgsino LSP,
respectively.
Hence, the spin-independent scattering cross sections are significantly below
the present experimental sensitivity for LSP masses below $1\,\textrm{TeV}$.


\subsubsection{Collider Signatures}


For the considered hierarchy of superparticle masses, gluinos and squarks
are heavy.
Hence, the characteristic missing energy signature of events with LSPs in
the final state may be absent and the discovery of winos or higgsinos therefore
very challenging.\footnote{For recent discussions, cf.\ for instance
Refs.~\cite{Baer:2011ec,Bobrovskyi:2011jj,Moroi:2011ab}.}
In both cases the neutral LSP is almost mass degenerate with a chargino,
which increases the discovery potential.
One may hope for macroscopic charged tracks of the produced charginos.
A generic prediction is also the occurrence of monojets caused by
the Drell-Yan production of higgsino/wino pairs
in association with initial state gluon radiation.



\chapter{Conclusions and Outlook}
\label{ch:conclusions}


Cosmological phase transitions might be common events in the history
of the hot early universe.
Extrapolating the evolution of the universe back in time beyond
primordial nucleosynthesis, we expect for instance the QCD as well as the
electroweak phase transition to take place at temperatures around the QCD
and the electroweak scale, respectively.
In this thesis, we have now proposed the idea that also the very origin
of the hot early universe as well as the generation of its initial
conditions are connected to a phase transition, \textit{viz.}\ the \BmL phase transition,
which represents the cosmological realization of spontaneous \BmL breaking.


The false vacuum phase of unbroken \BmL symmetry drives a stage of hybrid inflation,
which ends in a waterfall transition that is accompanied by tachyonic preheating and
the production of topological defects in the form of cosmic strings.
Successful inflation and the nonobservation of cosmic strings require the
\BmL phase transition to occur at the GUT scale, which strengthens the supposition
that the breaking of \BmL at the end of inflation might be embedded
in the breaking scheme of some more comprehensive theory of grand unification.
Tachyonic preheating, the decay of the \BmL gauge DOFs,
the decay of the \BmL Higgs bosons and its superpartners as well as thermal processes
produce an abundance of heavy (s)neutrinos.
These (s)neutrinos decay into the lepton-Higgs pairs of the supersymmetric
standard model, which reheats the universe and generates a primordial lepton asymmetry.
At the same time, inelastic scattering processes in the thermal bath unavoidably
lead to a thermal abundance of gravitinos.
The initial conditions of the hot thermal phase of the early universe
hence end up being completely determined by
the parameters of the fundamental Lagrangian which governs the dynamics of the \BmL
phase transition as well as the interactions of the heavy (s)neutrinos.
This is to say that the temperature scale of reheating, the primordial baryon asymmetry as well
as the thermal gravitino abundance are no longer unknown cosmological parameters.
Instead, they are related to the masses and couplings of elementary particles, which
can in principle be measured in particle physics experiments and astrophysical observations.


We have studied the \BmL phase transition in the full supersymmetric Abelian
Higgs model and given a detailed time-resolved description of the reheating
process based on the complete set of Boltzmann equation.
A notable result of our analysis is that the competition of cosmic expansion
and entropy production leads to an intermediate plateau of constant temperature,
during which baryon asymmetry and gravitinos are produced.
Remarkably, the final asymmetry as well as gravitino abundance are rather insensitive
to many of the theoretical uncertainties associated with the \BmL phase transition and
the subsequent reheating process.
We have explicitly checked that the final outcomes of our calculations
are robust against modifications of the theoretical framework pertaining to
(i) the production and relaxation of cosmic strings, (ii) the massive
superparticles, and (iii) the \BmL gauge DOFs.
For instance, even if $50\,\%$ of the false vacuum energy density is initially
stored in strings, they quickly loose most of their energy and the effect on
the final baryon asymmetry and gravitino abundance is negligible.
This robustness is due to the fact that after all most of the energy of the false vacuum
is transferred to the \BmL Higgs bosons, whose slow decay, via heavy (s)neutrinos,
dominates the reheating process.


In order to circumvent the cosmological gravitino problems, we have considered two,
in a sense quite opposite superparticle mass spectra.
In our first scenario, we took the gravitino to be the LSP.
In this case, the requirement of consistency between hybrid inflation,
leptogenesis and gravitino dark matter provided us with relations between
neutrino and superparticle masses.
For a gluino with a mass of $1\,\textrm{TeV}$, we a find a lower bound on the gravitino
mass of about $10\,\textrm{GeV}$.
The mass of the lightest of the heavy neutrinos $M_1$ typically has to have a value
of order $10^{11} \,\textrm{GeV}$.
For a wide range of light neutrino masses, this results in a reheating temperature
of order $10^9..10^{10}\,\textrm{GeV}$.
Our second scenario was motivated by hints for a $125\,\textrm{GeV}$ Higgs boson at the LHC
and featured the gravitino as the heaviest superparticle along with WIMP dark matter
in the form of pure wino or higgsino LSPs.
In this scenario, heavy gravitinos, which are thermally produced during reheating,
decay at some time between the QCD phase transition and BBN
into LSPs, thereby nonthermally generating the dark matter.
This time, the requirement of consistency between hybrid inflation, leptogenesis,
WIMP dark matter and BBN allowed us to derive upper and lower bounds on
the LSP mass as well as lower bounds on the gravitino mass, all of which depend on
the lightest neutrino mass.


On the way towards these phenomenological results, we completed a number of
technical tasks, some of which deserve particular attention.
First of all, we derived the Lagrangian of a general supersymmetric Abelian gauge
theory in arbitrary gauge and concretized it for the Abelian Higgs model
of the \BmL phase transition in unitary gauge.
Furthermore, we gave a detailed discussion of the nonperturbative dynamics during the
\BmL phase transition and generalized Linde's waterfall conditions
for hybrid inflation to the supersymmetric case.
Next, we devoted ourselves to the Boltzmann equations governing the
reheating process.
Partly, we were able to solve these equations analytically.
Apart from that, we developed techniques for the treatment of (i) the various contributions
to the heavy (s)neutrino abundances, (ii) the evolution of the gravitational background,
and (iii) the evolution of the temperature of the thermal bath.
Finally, we addressed several important technical issues in the appendices
such as the $CP$ violation in $2$-to-$2$ scatterings with intermediate (s)neutrino states
or the generalization of our analysis to other gluino masses.


Besides its cosmological consequences, we have also studied the implications
of the \BmL phase transition for the standard model neutrino sector.
The breaking of \BmL during the decay of the false vacuum sets
the stage for the seesaw mechanism.
Upon the seesaw model we imposed a flavour structure of the Froggatt-Nielsen type,
which naturally accounts for the measured quark and lepton mass hierarchies and the large
neutrino mixing angles.
Combining this flavour structure with the present knowledge on neutrino parameters,
we were able to derive precise predictions for yet unknown observables, in particular
the smallest mixing angle $\theta_{13}$, the smallest neutrino mass $m_1$, and the
Majorana phase $\alpha_{21}$.
This statement is based on a Monte-Carlo study:
treating unspecified $\mathcal{O}(1)$ parameters of the considered Froggatt-Nielsen
model as random variables, we found that the observables of interest are sharply peaked
around certain central values.


In this thesis, we have made use of the fact that, in the context of supersymmetric
hybrid inflation, the amplitude of the scalar power spectrum requires the \BmL breaking
scale to be of order the GUT scale (cf.\ Eqs.~\eqref{eq:vBLAs} and \eqref{eq:parameterspace}).
However, one may also ignore this requirement and merely assume that the inflationary
dynamics being responsible for the primordial scalar perturbations are in fact more complicated
than in the simple hybrid inflation model considered in the present case.
Under this assumption, the \BmL phase transition may equally take place at a scale
much below the GUT scale.
In Ref.~\cite{Buchmuller:2011mw}, we investigate this possibility in more detail
and arrive at the result that lower values of the \BmL breaking scale entail
weaker bounds on the gravitino mass.
If the \BmL phase transition occurs for instance at a scale $v_{B-L}\sim 10^{12}\,\textrm{GeV}$,
the gravitino could have a mass of $\mathcal{O}(100)\,\textrm{MeV}$.
Similarly, for a lower \BmL breaking scale, reheating would occur at a higher temperature
because of faster Higgs decays.
This would result in a stronger washout of the lepton asymmetry generated in (s)neutrino
decays.
Small $v_{B-L}$ hence implies an upper bound on the effective neutrino
mass $\widetilde{m}_1$ of about $0.1\,\textrm{eV}$.
In this thesis, we have by contrast demonstrated that, if the \BmL phase transition
takes place at the GUT scale, this restriction does no longer apply, rendering the
proposed reheating mechanism viable for all reasonable masses of the light neutrinos.
On the other hand, reducing the gravitino mass from $\mathcal{O}(10)\,\textrm{GeV}$
to $\mathcal{O}(100)\,\textrm{MeV}$ significantly shortens the lifetime of the NLSP,
which may soften the bound on the NLSP mass imposed by the requirement that the
late-time decays of the NLSP must not spoil the success of BBN
(cf.\ Sec.~\ref{subsec:candidates}).
A lower \BmL breaking scale thus possibly entails a simple solution to the NLSP decay
problem, albeit at the cost of a more complicated inflationary sector.

\newpage


An interesting alternative to supersymmetric $F$-term hybrid inflation is
supersymmetric $D$-term hybrid inflation.
In this inflationary model, inflation ends in the same manner
as in the $F$-term model discussed in this thesis.
The \BmL phase transition is in particular still required to occur at the GUT scale.
But, allowing for noncanonical terms in the K\"ahler potential, the $D$-term variant
of hybrid inflation may possibly improve upon its $F$-term sibling in terms of the
predicted value for the scalar spectral index (cf.\ Sec.~\ref{subsec:infltn}).
In an ongoing research project, we currently consider a model
in which the structure of the
nonminimal K\"ahler potential is determined by the requirement
that the superconformal invariance inherent in the superpotential
of $D$-term hybrid inflation be only slightly broken during inflation.%
\footnote{As of July 2013, the work on this project has
been completed; its results have been published in Ref.~\cite{Buchmuller:2012ex}.\smallskip}


Further important questions which have remained unanswered in this thesis
concern the role of the inflaton field during tachyonic preheating, if one goes beyond
the quench approximation.
In particular, it is necessary to investigate how the false vacuum energy is
distributed among the waterfall and the inflaton field when oscillations
in field configuration space in the direction of the inflaton field are also taken into account.
Moreover, one may introduce a further constant term $W_0$ in the superpotential,
which might unavoidably arise in the course of spontaneous supersymmetry breaking
\cite{Nakayama:2010xf},
\begin{align}
W_0 = m_{\widetilde{G}} \, M_P^2 \,.
\end{align}
This superpotential induces a mass mixing term for the waterfall and the inflaton field,
which may partly affect the details of the reheating process.
However, as the waterfall and the inflaton field both decay into the same chiral
multiplet, \textit{viz.} the heavy (s)neutrinos of the first generation,
we do not expect this mass mixing and in fact the entire superpotential $W_0$
to lead to any qualitative changes of the overall picture presented in this thesis.
Related to the occurrence of the term $W_0$ in the superpotential,
one may also ask whether there potentially is a connection between the spontaneous
breaking of \BmL at the end of hybrid inflation and the spontaneous breaking
of supersymmetry.
Assuming that the NLSP decay problem is avoided due to a small amount of $R$-parity
breaking, this question equally applies to the mechanism for the breaking of
$R$ parity.


One may also address the production of gravitational waves during the
\BmL phase transition.
In an ongoing research project, we currently attempt to
calculate the spectrum of gravitational waves that is generated during
the decay of the false vacuum.%
\footnote{As of July 2013, this project is almost complete;
a preprint summarizing our results is available online~\cite{Buchmuller:2013lra}.}
Finally, we point out that the warm WIMP dark matter scenario
which we discussed in Ch.~\ref{ch:wimp} might after all have interesting consequences
for the formation of matter structures on small scales.
It seems worthwhile to go further into this question as well.


In summary, we conclude that the decay of a false vacuum of unbroken \BmL symmetry
represents indeed an intriguing possibility to implement the transition between inflation
and the hot thermal phase of the early universe.
Tachyonic preheating after hybrid inflation and the dynamics of the \BmL gauge DOFs set
the stage for a matter dominated phase, whose evolution towards a hot thermal
state is described by means of Boltzmann equations.
We have carefully studied this process, putting particular emphasis on the various
nonthermal and thermal contributions to the abundances of the heavy (s)neutrinos,
and eventually arrived at a consistent picture of the early universe, whose properties
are largely determined by the parameters of the neutrino sector.
Measurements of the absolute neutrino mass scale and superparticle masses
consistent with our predictions would hence provide important indirect evidence for
the \BmL phase transition as the origin of the hot early universe.

\cleardoublepage



\chapter*{Acknowledgements}
\phantomsection
\addcontentsline{toc}{chapter}{Acknowledgements}


In the first place, I wish to express my gratitude to Wilfried Buchm\"uller,
who opened the door for me into particle cosmology.
Having always been enthusiastic about cosmology and particle physics,
I consider myself fortunate to have been given the opportunity
to work with him on the physics of the early universe.
He has been an excellent advisor, teacher and mentor.


I am also especially grateful to Valerie Domcke and Gilles Vertongen
for their collaboration at different stages of the work presented in this thesis.
Furthermore, I would like to thank Guido Altarelli, Torsten Bringmann, Felix Br\"ummer,
Laura Covi, Kohei Kamada, Graham Ross, Fuminobu Takahashi, David Wark, Walter Winter,
and Tsutomu Yanagida for helpful discussions and comments.
My sincere gratitude goes to Jan Louis and G\"unter Sigl for agreeing
to act as referees of my disputation and my dissertation, respectively.
Moreover, I acknowledge support by the German Science Foundation (DFG) within
the Collaborative Research Center (SFB) 676 ``Particles, Strings and the Early Universe''.


I highly appreciate the fact that DESY is a unique research facility offering doctoral
candidates in particle physics a stimulating and inspiring environment.
I benefited countless times from the high level of expertise which can be found everywhere
at DESY and I am thankful to all people at DESY who helped and supported me
in the course of the last three years.
Many thanks go out to all members of the DESY Theory Group for the pleasant time
and in particular to my office mates Jasper Hasenkamp, Michael Grefe, Vladimir Mitev,
Sebastian Schmidt, and V\'aclav Tlap\'ak for the enjoyable atmosphere in our office.
My time at DESY would not have been the same without Kranzzeit and our common lunch seminars.


Finally, I wish to thank my friends and fellows Sergei Bobrovskyi, Martin Krasny,
Falk Lindner, and Markus Rummel, who stood by me through all ups and downs.
It is thanks to them that my time in Hamburg became a once-in-a-lifetime experience
and I am excited to see how our story continues.
The same holds true for my friends and my family in and from Berlin.
Throughout the years, I could always count on my fellow students
Christopher Bronner, Frank Essenberger, Max Hoffmann, Kai Kornhuber, and Andreas Linscheid,
on~my friends Monia Gl\"aske and Wolfgang Hasse as well as on my brother, my sisters and
my parents.

\pagestyle{plain}
\cleardoublepage
\pagestyle{fancy}


\begin{appendix}
%

\chapter{Statistical Thermodynamics}
\label{ch:conventions}

In Ch.~\ref{ch:reheating}, we derive and solve the coupled system of semiclassical
Boltzmann equations describing the cosmic evolution after the
\BmL phase transition.
As a supplement to this analysis, we summarize the underlying formalism of Boltzmann
equations in this appendix (cf.\ Sec.~\ref{sec:kinetic}).
In addition to that, we also discuss the properties of particle species in kinetic
or thermal equilibrium (cf.\ Sec.~\ref{sec:equilibrium}).


\section{Kinetic Theory in the Expanding Universe}
\label{sec:kinetic}


The Boltzmann equation for a particle species $X$ describes the
time evolution of its phase space distribution function
$f_X$ in the one-particle phase space $\Phi_X$~\cite{Kolb:1990vq}.
The distribution function $f_X$ is defined such that $f_X d\Phi_X$
gives the average number of $X$ particles in the phase space volume
$d\Phi_X$ at time $t$.
In general, $f_X$ is a function of time $t$ as well as of all phase space
coordinates.
But imposing homogeneity and isotropy of spacetime, $f_X$ ends up solely
depending on time $t$ and the absolute value $p$ of the physical
three-momentum $\mathbf{p}$.
In the Friedmann-Lema\^itre framework, the Boltzmann equation for $f_X$ reads
\begin{align}
\hat{\mathcal{L}} f_X(t,p) =
E_X \left(\frac{\partial}{\partial t} - H p \frac{\partial}{\partial p}\right)
f_X(t,p) =  \mathcal{C}_X\,, \quad E_X = \sqrt{p^2 +m_X^2} \,,
\label{eq:BEX}
\end{align}
where $\hat{\mathcal{L}}$ denotes the Liouville operator,
$\mathcal{C}_X$ stands for the total collision operator,
$H$ is the Hubble rate and $m_X$ the mass of an $X$ particle.
$\mathcal{C}_X$ keeps track of changes in $f_X$ due to elastic and inelastic
interactions and may be decomposed into individual collision operators $C_X$,
respectively accounting for all the different processes through which an
$X$ particle may interact with other particles $a,b,..$ and $i,j,..$,
\begin{align}
\mathcal{C}_X =
\sum_{ab..}\sum_{ij..} C_X(X ab.. \leftrightarrow ij..)
= \sum_{ij..} C_X(X \leftrightarrow ij..) +
\sum_{a}\sum_{ij..} C_X(X a \leftrightarrow ij..) + .. \,.
\end{align}


The operators $C_X$ are obtained from integrating quantum mechanical transition
probabilities $\left(2\pi\right)^4\delta^{(4)}\left|\mathcal{M}\right|^2$
over the multi-particle phase space,
\begin{align}
C_X(X ab.. \leftrightarrow ij..) = & \:
\frac{1}{2 g_X} \int d\Pi\left(X|a,b,..;i,j,..\right) \left(2\pi\right)^4
\delta^{(4)}\left(\textstyle \sum p_{\textrm{out}} - \textstyle \sum p_{\textrm{in}}\right) \label{eq:CX}\\
&\times  \: \big[f_i f_j..\left(1\pm f_X\right)\left(1\pm f_a\right)\left(1\pm f_b\right) ..
\left|\mathcal{M}\left(i j .. \rightarrow X a b ..\right)\right|^2 \nonumber\\
& \: - f_X f_a f_b ..\left(1\pm f_i\right)\left(1\pm f_j\right) ..
\left|\mathcal{M}\left(X a b .. \rightarrow i j ..\right)\right|^2\big]\,, \nonumber
\end{align}
where $g_X$ is the number of internal DOFs of $X$ and
$d\Pi$ subsumes all Lorentz-invariant momentum space elements
$d\tilde{p} = \left(2\pi\right)^{-3} d^3 p/2E$ along with a statistical factor $S$,
preventing us from double counting in the case of identical particles,
\begin{align}
d\Pi(X|a,b,..;i,j,..) = S(X,a,b,..;i,j,..)
d\tilde{p}_a d\tilde{p}_b ..  d\tilde{p}_i  d\tilde{p}_j ..\,.
\end{align}
The amplitudes squared $\left|\mathcal{M}\right|^2$ are
understood to be summed over all internal DOFs of the particles in
the initial and in the final state.
$(1+f)$ and $(1-f)$ are quantum statistical factors, respectively
implementing the effects of the Bose enhancement and Pauli
blocking.\footnote{The Boltzmann equation is actually
a \textit{classical} evolution equation for $f_X$.
However, using quantum mechanical $S$ matrix elements as well as including
the quantum statistical factors $(1\pm f)$ in the calculation of
the actually \textit{classical} collision operators $C_X$
renders it \textit{semiclassical}.}
These factors are expected to have only a minor impact
on the evolution of the distribution function~\cite{HahnWoernle:2009qn}.
In particular, their influence may partly be canceled by other
quantum corrections like off-shell effects \cite{Anisimov:2010dk}.
We shall thus neglect the quantum statistical factors in this thesis.
Finally, we note that the operators $C_X$ may be split into two parts,
respectively accounting for the two directions in which the corresponding
processes may proceed,
\begin{align}
C_X(X a b.. \leftrightarrow ij..) =
C_X(ij.. \rightarrow X a b..) - C_X(X a b.. \rightarrow i j ..)\,.
\end{align}


The number density $n_X$ of the particle species $X$ follows from
integrating its distribution function $f_X$ over the momentum space
element $g_X d^3p_X/\left(2\pi\right)^3$,
\begin{align}
n_X(t) = & \: \frac{g_X}{\left(2\pi\right)^3} \int d^3 p \; f_X(t,p) \,. \label{eq:numdendef}
\end{align}
Similarly, the corresponding energy density $\rho_X$ is given as the integral over $E_X f_X$,
\begin{align}
\rho_X(t) = & \: \frac{g_X}{\left(2\pi\right)^3} \int d^3 p \; E_X f_X(t,p) \,.
\label{eq:engdendef}
\end{align}
Dividing the Boltzmann equation in Eq.~\eqref{eq:BEX} by $E_X$ on both sides
and again integrating over $g_X d^3p_X/\left(2\pi\right)^3$ yields the
first moment of the Boltzmann equation or simply the
\textit{integrated Boltzmann equation} for the number density $n_X$,
\begin{align}
\dot{n}_X + 3 H n_X = \hat{\gamma}_X =
\sum_{ab..}\sum_{ij..} \gamma(X a b.. \leftrightarrow ij..) \,,
\label{eq:BEnX}
\end{align}
with $\hat{\gamma}$ being the total spacetime density of inelastic
interaction events involving particles of species $X$.
The individual interaction densities $\gamma$ are defined as
\begin{align}
\gamma(X ab.. \leftrightarrow ij..) =
\frac{g_X}{(2\pi)^3} \int \frac{d^3 p}{E_X}\, C_X(Xab.. \leftrightarrow ij..)\,.
\label{eq:gammadef}
\end{align}
The Boltzmann equation in Eq.~\eqref{eq:BEnX} can alternatively
be written as an equation for the comoving number
density $N_X = a^3 n_X$ as a function of the scale factor $a$,
\begin{align}
a H \frac{d}{da} N_X =  a^3 \hat{\gamma}_X =
a^3 \sum_{ab..}\sum_{ij..} \gamma(X a b.. \leftrightarrow ij..) = \hat{\Gamma}_X N_X \,,
\end{align}
where we have introduced $\hat{\Gamma}_X$ as the total effective production rate
of $X$ particles,
\begin{align}
\hat{\Gamma}_X = \frac{\hat{\gamma}_X}{n_X} = \frac{1}{N_X} \,a^3 \hat{\gamma}_X =
\frac{1}{N_X} \, a^3\sum_{ab..}\sum_{ij..} \gamma(X a b.. \leftrightarrow ij..) \,.
\end{align}


\section{Kinetic and Thermal Equilibrium}
\label{sec:equilibrium}


The phase distribution function of a bosonic particle species $X$
in kinetic equilibrium is given by the Bose-Einstein distribution  $(-1)$,
whereas the phase distribution function of a fermionic particle species $X$
in kinetic equilibrium corresponds to the Fermi-Dirac distribution $(+1)$,
\begin{align}
f_X(t,p) = \frac{1}{e^{\left(E_X-\mu_X\right)/T_X} \pm 1} \,.
\end{align}
Here, $\mu_X$ denotes the chemical potential of the particle species $X$ and $T_X$
is its equilibrium temperature.
If the interactions between $X$ particles and photons are fast enough,
the temperature $T_X$ coincides with the photon temperature $T$, i.e.\ the temperature
of the thermal bath.
The chemical potentials of the MSSM particles are an important ingredient to
the calculation of the sphaleron conversion factor $C_{\textrm{sph}}$
(cf.\ Eq.~\ref{eq:BLeqCsph}).
If inelastic processes between different particle species are sufficiently
fast, their respective chemical potentials are related to each other
and the involved species are said to be in chemical equilibrium.
For each independently conserved particle number, there exists in particular
one independent chemical potential.
Conversely, if there are no constraints on the number of $X$ particles
enforced by conservation laws, the chemical potential of species $X$ vanishes.
In this thesis, we refer to a collection of particle species with negligibly
small chemical potentials that is in chemical equilibrium as being in thermal equilibrium.
The phase space distribution function of a bosonic or fermionic species $X$
in thermal equilibrium is therefore given as
\begin{align}
f_X^{\textrm{eq}}(t,p) \approx \frac{1}{e^{E_X/T_X} \pm 1} \,.
\label{eq:fBEFD}
\end{align}
Furthermore, for all particles acquiring a mass in the course of \BmL breaking,
we approximate the respective distributions in kinetic and thermal
equilibrium, $f_X$ and $f_X^{\textrm{eq}}$, by classical
Maxwell-Boltzmann distributions,
\begin{align}
f_X(t,p) \approx e^{-\left(E_X-\mu_X\right)/T_X} \,,\quad f_X^{\textrm{eq}}(t,p) \approx e^{-E_X/T_X} \,.
\label{eq:fMB}
\end{align}


The number and energy densities of \textit{massless} bosons and fermions
in thermal equilibrium are readily obtained from Eqs.~\eqref{eq:numdendef}, \eqref{eq:engdendef}
and \eqref{eq:fBEFD},
\begin{align}
\textrm{Bosons: } \qquad & \:
n_X^{\textrm{eq}} = \phantom{\frac{3}{4}}  \, g_X \frac{\zeta(3)}{\pi^2} T_X^3 \,,\quad
\rho_X^{\textrm{eq}} = \phantom{\frac{7}{8}} \, g_X \frac{\pi^2}{30} T_X^4 \,,
\label{eq:nrhoBFeq}\\
\textrm{Fermions: } \qquad & \:
n_X^{\textrm{eq}} = \frac{3}{4}  \, g_X \frac{\zeta(3)}{\pi^2} T_X^3 \,,\quad
\rho_X^{\textrm{eq}} = \frac{7}{8} \, g_X \frac{\pi^2}{30} T_X^4 \,, \nonumber
\end{align}
where $\zeta$ is the Riemann zeta function.
With the aid of these expressions, we are able to write down $n_R$ and
$\rho_R$, the number and the energy density of MSSM radiation quanta in the
thermal bath,
\begin{align}
n_R = g_{*,n} \frac{\zeta(3)}{\pi^2} T^3 \,,\quad
\rho_R = g_{*,\rho} \frac{\pi^2}{30} T^4 \,, \label{eq:nRrhoR}
\end{align}
with $g_{*,n}$ and $g_{*,\rho}$ being the corresponding effective sums of
relativistic DOFs,
\begin{align}
g_{*,n} = & \: \sum_{\textrm{bosons}} g_X\left(\frac{T_X}{T}\right)^3 +
\frac{3}{4} \sum_{\textrm{fermions}} g_X\left(\frac{T_X}{T}\right)^3 \,, \\
g_{*,\rho} = & \: \sum_{\textrm{bosons}} g_X\left(\frac{T_X}{T}\right)^4 +
\frac{7}{8} \sum_{\textrm{fermions}} g_X\left(\frac{T_X}{T}\right)^4 \,. \nonumber
\end{align}
Setting all equilibrium temperatures $T_X$ to the photon temperature $T$,
the MSSM values of $g_{*,n}$ and $g_{*,\rho}$ turn out to be $427/2$ and $915/4$,
respectively.
From the result for $\rho_R$ in Eq.~\eqref{eq:nRrhoR}, we can
easily deduce an expression for the radiation entropy density $s_R$.
Making use of the equation of state of radiation, $\omega_R = p_R/\rho_R = 1/3$,
where $p_R$ is the radiation pressure, we find
\begin{align}
s_R = & \: \frac{\rho_R + p_R}{T} = \frac{4}{3}\frac{\rho_R}{T} =
g_{*,s}\frac{2\pi^2}{45}T^4 \,, \label{eq:sRgstars}\\
g_{*,s} = & \: \sum_{\textrm{bosons}} g_X\left(\frac{T_X}{T}\right)^3 +
\frac{7}{8} \sum_{\textrm{fermions}} g_X\left(\frac{T_X}{T}\right)^3 \,. \nonumber
\end{align}
With all equilibrium temperatures $T_X$ coinciding with the photon temperature $T$,
$g_{*,s}$ equals $g_{*,\rho}$.
In the MSSM, we then have $g_{*,s} = 915/4$.
For \textit{massive} particles in thermal equilibrium, now employing
the approximation in Eq.~\eqref{eq:fMB}, we obtain,
\begin{align}
n_X^{\textrm{eq}} = g_X \frac{m_X^3}{2\pi^2 z_X} K_2(z_X) \,, \quad
\rho_X^{\textrm{eq}} = g_X \frac{m_X^4}{2\pi^2}\left[\frac{1}{z_X}K_1(z_X) +
\frac{3}{z_X^2}K_2(z_X)\right] \,. \label{eq:nrhoXeq}
\end{align}
Here, $z_X = m_X/T_X$ represents the inverse temperature in units of $m_X^{-1}$,
while $K_n$ denotes the modified Bessel function of the second kind of order $n$.


Finally, we note that, approximating the distribution functions $f_X$ and $f_X^{\textrm{eq}}$
by Maxwell-Boltzmann distributions as in Eq.~\eqref{eq:fMB},
implies that $f_X$ is proportional to $f_X^{\textrm{eq}}$,
\begin{align}
f_X(t,p) = R_X(t) f_X^{\textrm{eq}}(t,p) \,,\quad R_X(t) = e^{\mu_X/T_X} \,, \label{eq:kineq}
\end{align}
According to Eq.~\eqref{eq:numdendef}, $R_X$ is nothing but the ratio
of $n_X$ to the number density in thermal equilibrium, $R_X = n_X/ n_X^{\textrm{eq}}$.
Here, the number density $n_X$ may have any value.
Furthermore, one can easily show that $R_X$ also corresponds to the ratio of
the energy density $\rho_X$ and the corresponding energy density in thermal
equilibrium $\rho_X^{\textrm{eq}}$,
\begin{align}
\rho_X = R_X(t) \rho_X^{\textrm{eq}} = 
\frac{n_X}{n_X^{\textrm{eq}}} \rho_X^{\textrm{eq}} \,.
\end{align}
This relation shows in particular that, within the Maxwell-Boltzmann approximation,
the average energy per particle $\varepsilon_X = \rho_X/n_X$ in kinetic equilibrium  is
the same as in thermal equilibrium.

\cleardoublepage


\chapter[\textit{CP} Violation in 2-to-2 Scattering Processes]
{\textit{CP} Violation\newline in 2-to-2 Scattering Processes}
\label{ch:scatterings}


The Boltzmann equation for the lepton asymmetry $L$ corresponds to the difference
of the respective equations for the lepton multiplet $\ell$ and the antilepton multiplet
$\bar{\ell}$ (cf.\ Eq.~\eqref{eq:BEL}).
The ordinary collision operators in the Boltzmann equations for $\ell$ and $\bar{\ell}$
account for the decays and inverse decays of heavy (s)neutrinos \textit{on the mass shell}.
In the Boltzmann equation for the lepton asymmetry, these collision operators induce
terms of $\mathcal{O}\left(\epsilon_i\right)$.
In addition to that, it turns out that $\Delta L = 2$ scatterings with MSSM
lepton-Higgs pairs in the external states and \textit{off-shell} (s)neutrinos in the
intermediate state yield contributions of the same order in the $CP$ violation
parameters $\epsilon_i$ to the Boltzmann equation for the lepton asymmetry.
To consistently calculate the lepton asymmetry up to first order
in the parameters $\epsilon_i$, one therefore has to add the reduced collision operators
$\mathcal{C}_\ell^{\textrm{red}}$ and $\mathcal{C}_{\bar{\ell}}^{\textrm{red}}$ to
the equations for $\ell$ and $\bar{\ell}$ (cf.\ Eq.~\eqref{eq:Clred}).
These operators incorporate the off-shell contributions to all relevant
$\Delta L = 2$ scatterings and are hence related to the full $2$-to-$2$ scattering
operators in the following way,
\begin{align}
\mathcal{C}_X = \mathcal{C}_X^{\textrm{on}} + \mathcal{C}_X^{\textrm{red}}
\,,\qquad X = \ell, \bar{\ell} \,,
\label{eq:CConCred}
\end{align}
where $\mathcal{C}_X^{\textrm{on}}$ are the on-shell collision operators accounting
for scattering processes with real intermediate states.
In the derivation of the Boltzmann equation for the lepton asymmetry
in Sec.~\ref{subsec:BELR}, we neglect the $CP$-preserving parts of all operators
in Eq.~\eqref{eq:CConCred} (cf.\ Eq.~\eqref{eq:CXCPpres}) and make use of the fact
that the $CP$-violating parts of $\mathcal{C}_X$ vanish up to corrections of
$\mathcal{O}\left((h^\nu)^4\right)$ (cf.\ Eq.~\eqref{eq:CXCPviol}).
The purpose of this appendix is to show that this is indeed the case,
\begin{align}
\mathcal{C}_{X,\cancel{CP}} = 0 + \mathcal{O}\left((h^\nu)^4\right) \,.
\label{eq:CXCPprove}
\end{align}
For the nonsupersymmetric case, a prove of this statement is given in
Refs.~\cite{Buchmuller:1997yu} and \cite{Roulet:1997xa}.
In this appendix we shall now extent it to the supersymmetric case.


The full $2$-to-$2$ scattering operator $\mathcal{C}_X$ is composed of
individual collision operators $C_X$, which are all of the following form,
\begin{align}
C_X(I \leftrightarrow \bar{F}) = & \: C_X(\bar{F} \rightarrow I) - C_X(I \rightarrow \bar{F})
\phantom{\int}\\
C_X(I \rightarrow \bar{F}) = & \:
\frac{1}{2g_X} \int d\Pi\left(X|a;ij\right) f_X f_a
\left(2\pi\right)^4 \delta^{(4)}\left|M\left(Xa\rightarrow \bar{i}\bar{j}\right)\right|^2
\,, \nonumber \\
C_X(\bar{F} \rightarrow I) = & \:
\frac{1}{2g_X} \int d\Pi\left(X|a;ij\right) f_{\bar{i}} f_{\bar{j}}
\left(2\pi\right)^4 \delta^{(4)}\left|M\left(\bar{i}\bar{j}\rightarrow Xa\right)\right|^2
\,. \nonumber
\end{align}
Here, $I$ denotes the pair of initial state particles,
$I = Xa = \ell H_u,\tilde{\ell}\tilde{H}_u,\tilde{\ell}H_u,\ell\tilde{H}_u$,
while $F$ stands for the pair of
antiparticles corresponding to the pair of final state particles $\bar{F}$.
For $I = \ell H_u,\tilde{\ell}\tilde{H}_u$, we have $F = I,\tilde{I}$, where
$\tilde{I}$ is the pair of superparticles corresponding to the pair of particles $I$,
while for $I = \tilde{\ell} H_u,\ell\tilde{H}_u$, the only possibility is $F = \tilde{I}$.
Note that each pair of particles $I$, $F$ implicitly carries a weak isospin as well as a
flavour index.
The matrix element squared $\left|M\right|^2$ is understood to be summed over all
internal DOFs of the particles in the external states.
It is directly related to the corresponding $S$ matrix element squared,
\begin{align}
\left|S\left(Xa\rightarrow ij\right)\right|^2 =
\left(2\pi\right)^4 \delta^{(4)}\left|M\left(Xa\rightarrow ij\right)\right|^2 \,,
\end{align}
where $\left|S\right|^2$ represents the probability per spacetime unit volume
for the occurrence of the process $Xa\rightarrow ij$.
In the case of distinct particles in the initial and final state,
the matrix element squared $\left|M\right|^2$ reduce to the ordinary
transition amplitude squared $\left|\mathcal{M}\right|^2$ (cf.\ Sec.~\ref{sec:kinetic}).


To prove Eq.~\eqref{eq:CXCPprove}, it is sufficient to show that
the $CP$-violating contribution to the sum over all operators $C_X(I \rightarrow \bar{F})$
vanishes up to corrections of $\mathcal{O}\left((h^\nu)^4\right)$.
In order to do so, we demonstrate that difference between the sum over all
operators $C_X(I \rightarrow \bar{F})$ and the $CP$ conjugate of this sum
vanishes up to corrections of $\mathcal{O}\left((h^\nu)^4\right)$,
\begin{align}
\sum_{I,F} \left[C_X(I \rightarrow \bar{F}) - C_X(\bar{I} \rightarrow F)\right]
= 0 + \mathcal{O}\left((h^\nu)^4\right) \,. \label{eq:sumC0}
\end{align}
If we manage to confirm this statement, we have simultaneously shown
that the $CP$-violating contribution to the sum over all operators
$C_X(\bar{F} \rightarrow I)$ vanishes up to $\mathcal{O}\left((h^\nu)^4\right)$, since
\begin{align}
\sum_{F,I} \left[C_X(\bar{F} \rightarrow I) - C_X(F \rightarrow \bar{I})\right] = -
\sum_{I,F} \left[C_X(I \rightarrow \bar{F}) - C_X(\bar{I} \rightarrow F)\right] \,.
\label{eq:sumCbar0}
\end{align}
The combination of Eqs.~\eqref{eq:sumC0} and \eqref{eq:sumCbar0} then entails that
$\mathcal{C}_{X,\cancel{CP}}$ is of the same order of magnitude as the sum over all
operators $C_X(I \rightarrow \bar{F})$, i.e.\ zero up to corrections of
$\mathcal{O}\left((h^\nu)^4\right)$.


To show that Eq.~\eqref{eq:sumC0} holds, we first rewrite the operators
$C_X$ in Eq.~\eqref{eq:sumC0} as integrals over $S$ matrix elements squared.
In a self-explanatory shorthand notation, we may write
\begin{align}
\sum_{I,F} \left[C_X(I \rightarrow \bar{F}) - C_X(\bar{I} \rightarrow F)\right]
= \sum_{I,F}\int \frac{d\Pi\left(X|a;ij\right)}{2g_X}
\left[\left(ff\right)_I \left|S_{I\bar{F}}\right|^2 -
\left(ff\right)_{\bar{I}}  \left|S_{\bar{I}F}\right|^2 \right] \,.
\label{eq:sumCS2}
\end{align}
The $CPT$ invariance of the $S$ matrix implies that
$\left|S_{IF}\right|^2 = \left|S_{\bar{F}\bar{I}}\right|^2$.
Similarly, the fact that, up to corrections of $\mathcal{O}\left(\epsilon_i\right)$,
all particles in the MSSM lepton and Higgs multiplets are in
thermal equilibrium provides us with $\left(ff\right)_I \approx \left(ff\right)_{\bar{F}}$.
In fact, as we are eventually interested in the $CP$-violating parts of the $S$ matrix
elements squared, which are of $\mathcal{O}\left(\epsilon_i\right)$ themselves,
it is for our purposes sufficient to work with $\left(ff\right)_I = \left(ff\right)_{\bar{F}}$
in the following.
We then find\footnote{The scattering processes with heavy \textit{neutrinos}
in the intermediate state feature different initial and final states $I$ and $F$
than the processes with heavy \textit{sneutrinos} or heavy \textit{antisneutrinos}
in the intermediate state.
The entire discussion in this appendix would hence remain valid, if we were to restrict
ourselves solely to scattering processes with either only \textit{neutrinos},
\textit{sneutrinos} or \textit{antisneutrinos} in the intermediate state.}
\begin{align}
\sum_{I,F} C_X(I \rightarrow F) = \sum_{I,F} C_X(\bar{F} \rightarrow \bar{I}) =
\sum_{I,F} C_X(\bar{I} \rightarrow \bar{F}) \,,
\end{align}
which allows us to include the collision operators into the sum in Eq.~\eqref{eq:sumCS2}
that account for the lepton number-conserving processes $I \rightarrow F$
and $\bar{I}\rightarrow\bar{F}$,
\begin{align}
& \: \sum_{I,F} \left[C_X(I \rightarrow \bar{F}) - C_X(\bar{I} \rightarrow F)\right]
= \frac{1}{2g_X}  \sum_{I,F}\int d\Pi\left(X|a;ij\right) \label{eq:CsumLcons}\\
\times & \: \left[
\left(ff\right)_I \left|S_{I\bar{F}}\right|^2
+ \left(ff\right)_I \left|S_{IF}\right|^2
- \left(ff\right)_{\bar{I}}  \left|S_{\bar{I}F}\right|^2
- \left(ff\right)_{\bar{I}} \left|S_{\bar{I}\bar{F}}\right|^2
\right] \,, \nonumber
\end{align}
Owing to the unitary of the $S$ matrix, the integration of
the $S$ matrix elements squared for a fixed initial state $I$
over all possible final-state configurations $F$ and $\bar{F}$ yields unity,
\begin{align}
\sum_{F}\int d\Pi\left(ij\right)
\left[\left|S_{I\bar{F}}\right|^2 + \left|S_{IF}\right|^2\right] =
\sum_{F}\int d\Pi\left(ij\right)
\left[\left|S_{\bar{I}F}\right|^2 + \left|S_{\bar{I}\bar{F}}\right|^2\right] =
1 + \mathcal{O}\left((h^\nu)^4\right) \,. \label{eq:Sunitarity}
\end{align}
Since we only integrate over all possible \textit{two-particle} final states and hence
omit all possible \textit{multi-particle} final states, we obtain corrections to
the exact result of $\mathcal{O}\left((h^\nu)^4\right)$.
The leading corrections are due to four-particle final states,
which are of $\mathcal{O}\left((h^{\nu})^8\right)$, if the heavy (s)neutrino
in the intermediate state is off-shell~\cite{Roulet:1997xa}.
Close to the resonance pole, the corrections due to processes with four particles in
the final state are enhanced, so that they reach a magnitude
of $\mathcal{O}\left((h^{\nu})^4\right)$~\cite{Covi:1996wh,Buchmuller:2004nz}.
This observation concludes our argument.
Substituting Eq.~\eqref{eq:Sunitarity} back into Eq.~\eqref{eq:CsumLcons},
we end up with the statement in Eq.~\eqref{eq:sumC0}, which we intended to prove.


In the literature, the fact that the $CP$-violating contributions to the collision operator
$\mathcal{C}_X$ vanish up to corrections of $\mathcal{O}\left((h^{\nu})^4\right)$ is
often formulated in terms of other quantities, which are closely related to $\mathcal{C}_X$.
To facilitate the comparison of our analysis with other works, we point out
that the collision operators $C_X$ contained in $\mathcal{C}_X$ can also be written
as momentum space integrals over reduced cross sections $\hat{\sigma}$,
\begin{align}
C_X(I \rightarrow \bar{F}) = & \:
\frac{1}{2g_X} \int d\Pi\left(X|a\right) f_X f_a
\,\hat{\sigma}\left(Xa\rightarrow \bar{i}\bar{j}\right) \,, \\
\hat{\sigma}\left(Xa\rightarrow \bar{i}\bar{j}\right) = & \:
\int d\Pi\left(ij\right)
\left(2\pi\right)^4 \delta^{(4)}\left|M\left(Xa\rightarrow \bar{i}\bar{j}\right)\right|^2 =
\int d\Pi\left(ij\right) \left|S\left(Xa\rightarrow \bar{i}\bar{j}\right)\right|^2
\,. \nonumber
\end{align}
These reduced cross sections $\hat{\sigma}$ are related to the ordinary
cross sections $\sigma$ as follows,
\begin{align}
\hat{\sigma}\left(Xa\rightarrow \bar{i}\bar{j}\right) = 2 s \,
\lambda^{1/2}\left(1,m_X^2/2,m_a^2/s\right)
g_X g_a \,\sigma\left(Xa\rightarrow \bar{i}\bar{j}\right) \,,
\end{align}
where $s$ is the Mandelstam variable corresponding to the square of
the center-of-mass energy, while $\lambda$ is given as
$\lambda(a,b,c) = a^2+b^2+c^2-2ab-2ac-2bc$.
The collision operators $C_X$ may hence also be written as
\begin{align}
C_X(I \rightarrow \bar{F}) =
g_a \int d\Pi\left(X|a\right) f_X f_a
\,s \, \lambda^{1/2}\left(1,m_X^2/2,m_a^2/s\right)
\,\sigma\left(Xa\rightarrow \bar{i}\bar{j}\right) \,.
\end{align}
Finally, integrating over $g_X/E_X \,d^3p/\left(2\pi\right)^3$
yields the interaction densities $\gamma$ (cf.\ Eq.~\eqref{eq:gammadef}),
\begin{align}
\gamma(I \rightarrow \bar{F}) = & \:
\frac{g_X}{(2\pi)^3} \int \frac{d^3 p}{E_X} \,C_X(I \rightarrow \bar{F}) \,, \\
= & \: \int d\Pi\left(Xa;ij\right) f_X f_a
\left(2\pi\right)^4 \delta^{(4)}\left|M\left(Xa\rightarrow \bar{i}\bar{j}\right)\right|^2
\,, \nonumber \\
= & \: \int d\Pi\left(Xa;ij\right) f_X f_a
\left|S\left(Xa\rightarrow \bar{i}\bar{j}\right)\right|^2
\,, \nonumber \\
= & \: \int d\Pi\left(Xa\right) f_X f_a \,\hat{\sigma}\left(Xa\rightarrow \bar{i}\bar{j}\right)
\,, \nonumber \\
= & \: 2 g_X g_a \int d\Pi\left(Xa\right) f_X f_a
\, s \,\lambda^{1/2}\left(1,m_X^2/2,m_a^2/s\right)
\,\sigma\left(Xa\rightarrow \bar{i}\bar{j}\right) \,. \nonumber
\end{align}


In conclusion, we summarize the central implication of our result in Eq.~\eqref{eq:CXCPprove},
which we require in the derivation of the Boltzmann equation for the lepton asymmetry.
From the combination of Eqs.~\eqref{eq:CConCred} and \eqref{eq:CXCPprove},
we directly infer that (cf.\ Eq.~\eqref{eq:CXCPviol})
\begin{align}
\mathcal{C}_{X,\cancel{CP}}^{\textrm{red}} =
- \mathcal{C}_{X,\cancel{CP}}^{\textrm{on}} + \mathcal{O}\left((h^\nu)^4\right) \,, \nonumber
\end{align}
which implies that
\begin{align}
& \: \sum_{I,F} C_{X,\cancel{CP}}^{\textrm{red}}(I\rightarrow\bar{F}) =
\sum_{I,F} \sum_i \textrm{Br}(V_i \rightarrow \bar{F})\,
C_{X,\cancel{CP}}^{\textrm{red}}(I\rightarrow V_i) = \\
- & \: \sum_{I,F} C_{X,\cancel{CP}}^{\textrm{on}}(I\rightarrow\bar{F}) =
\sum_{I,F} \sum_i \textrm{Br}(R_i \rightarrow \bar{F})\,
C_{X,\cancel{CP}}^{\textrm{on}}(I\rightarrow R_i) + \mathcal{O}\left((h^\nu)^4\right)
\,, \nonumber
\end{align}
with $R_i$ and $V_i$ denoting real and virtual heavy (s)neutrinos in the intermediate
state.
For $I =\ell H_u,\tilde{\ell}\tilde{H}_u$, we have $R_i,V_i = N_i$, while
for $I =\tilde{\ell} H_u$ and $I =\ell\tilde{H}_u$, the intermediate
states are given by $R_i,V_i = \tilde{N}_i$ and by $R_i,V_i = \tilde{N}_i^*$, respectively.
The branching ratios summed over all final states $\bar{F}$
cancel on both sides, so that we end up with
\begin{align}
\sum_{I} \sum_i C_{X,\cancel{CP}}^{\textrm{red}}(I\rightarrow V_i) =
- \sum_{I} \sum_i C_{X,\cancel{CP}}^{\textrm{on}}(I\rightarrow R_i)
+ \mathcal{O}\left((h^\nu)^4\right) \,.
\end{align}
Hence, thanks to the fact the on- and off-shell contributions to
$\mathcal{C}_{X,\cancel{CP}}$ cancel each other up to $\mathcal{O}\left((h^\nu)^4\right)$,
\textit{adding} the \textit{off-shell} operators $C_{X,\cancel{CP}}^{\textrm{red}}$
to the Boltzmann equations for $\ell$ and $\bar{\ell}$ is equivalent to
\textit{subtracting} the \textit{on-shell} operators $C_{X,\cancel{CP}}^{\textrm{on}}$.



\chapter{Thermal Gravitino Production}
\label{ch:gravitinos}


In Sec.~\ref{subsec:gravitinoDM}, in the derivation of approximate formulae
for the reheating temperature $T_{\textrm{RH}}$ and the heavy (s)neutrino mass $M_1$
as functions of the effective neutrino mass $\widetilde{m}_1$ and the gravitino mass
$m_{\widetilde{G}}$, we require an analytical expression for the thermal gravitino
abundance $\Omega_{\widetilde{G}}^0 h^2$ generated in the course of reheating
(cf.\ Eq.~\eqref{eq:OmegaAnaly}).
Similarly, in Sec.~\ref{sec:production}, we make use of a fit formula for
$\Omega_{\widetilde{G}}^0h^2$ as a function of $T_{\textrm{RH}}$, when computing
the nonthermal LSP abundance produced in gravitino decays
(cf.\ Eq.~\eqref{eq:OmegaGTRH}).
In this appendix, we now explicitly derive the expression for $\Omega_{\widetilde{G}}^0h^2$
in Eq.~\eqref{eq:OmegaAnaly}, which automatically provides us with an analytical approximation
for the numerical relation in Eq.~\eqref{eq:OmegaGTRH}.
Our quantitative analysis in Secs.~\ref{sec:evolution} and
\ref{sec:scan} is based on the assumption of a gluino mass of $1\,\textrm{TeV}$.
In this appendix, we therefore also illustrate how our results are easily generalized
to other values of the gluino mass.

\begin{figure}
\begin{center}
\includegraphics[width=11cm]{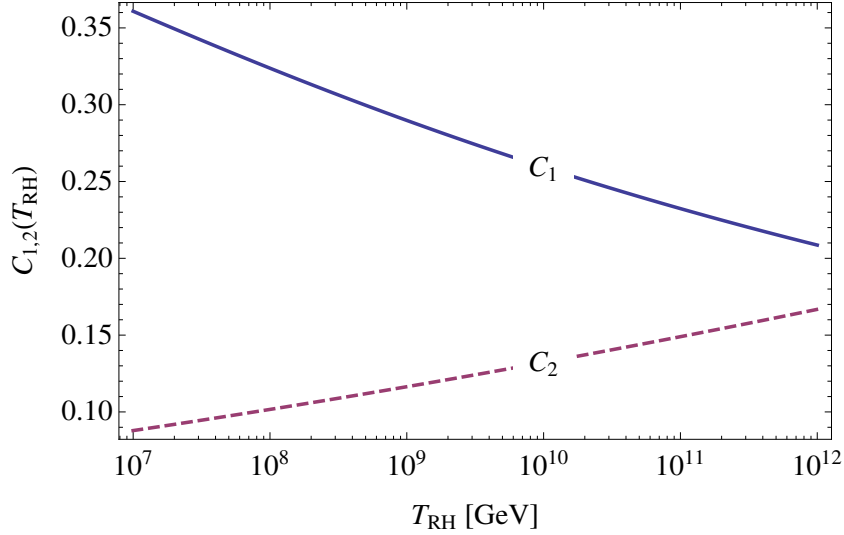}

\vspace{5mm}

\caption[$C_1$ and $C_2$ as functions of $T_{\textrm{RH}}$]
{The coefficients $C_1$ and $C_2$ as functions of the reheating
temperature $T_{\textrm{RH}}$.
\label{fig:C1C2coefficients}}
\end{center}
\end{figure}

In the current epoch, gravitinos are nonrelativistic.
Their present contribution to the energy density of the universe is hence given by
\begin{align}
\Omega_{\widetilde{G}}^0 h^2 =
\Omega_{\widetilde{G}}^0 h^2
(\widetilde{m}_1,M_1,m_{\widetilde{G}},m_{\tilde{g}})
= m_{\widetilde{G}} \,\eta_{\widetilde{G}}^0 \,n_\gamma^0 \,h^2 / \rho_c^0 \,,
\quad
\eta_{\widetilde{G}}^0 = n_{\widetilde{G}}^0 / n_\gamma^0\,.
\label{eq:OmegaG}
\end{align}
In order to relate the gravitino-to-photon ratio $\eta_{\widetilde{G}}^0$
to the corresponding number densities during reheating, we make two simplifying
assumptions.
First, we assume that after $a = a_{\textrm{RH}}$ the entropy of the thermal bath is not
increased much further, which leads us to
\begin{align}
\quad n_{\gamma}^0 = \delta_1  \left(\frac{a_{\textrm{RH}}}{a_0}\right)^3
\frac{g_{*,s}}{g_{*,s}^0}\, n_\gamma(a_{\textrm{RH}})\,.
\label{eq:ngamma0}
\end{align}
Second, we assume that at $a = a_{\textrm{RH}}$ the gravitino production
becomes inefficient such that at later times not many
further gravitinos are produced,
\begin{align}
n_{\widetilde{G}}^0 = \delta_2  \left(\frac{a_{\textrm{RH}}}{a_0}\right)^3
n_{\widetilde{G}}(a_{\textrm{RH}})\,.
\label{eq:ngravi0}
\end{align}
This second assumption also implies that at $a = a_{\textrm{RH}}$
the gravitino production rate $\hat{\Gamma}_{\widetilde{G}}$
is of the same order of magnitude as the Hubble rate $H$,
\begin{align}
\hat{\Gamma}_{\widetilde{G}}(a_{\textrm{RH}}) =
\frac{\gamma_{\widetilde{G}}(a_{\textrm{RH}})}{n_{\widetilde{G}}(a_{\textrm{RH}})}
= \delta_3^{-1} H(a_{\textrm{RH}}) \,,\quad
n_{\widetilde{G}}(a_{\textrm{RH}}) = \delta_3 \,
\frac{\gamma_{\widetilde{G}}(a_{\textrm{RH}})}{H(a_{\textrm{RH}})}\,.
\label{eq:ngraviRH}
\end{align}
The three correction factors $\delta_1 \gtrsim 1$,
$\delta_2 \gtrsim 1$ and $\delta_3 \sim \mathcal{O}(1)$,
introduced in Eqs.~\eqref{eq:ngamma0}, \eqref{eq:ngravi0}
and \eqref{eq:ngraviRH}, respectively, quantify the deviations
of the actual values of $n_\gamma^0$, $n_{\widetilde{G}}^0$ and
$n_{\widetilde{G}}(a_{\textrm{RH}})$ from our approximations.
Combining them into one factor $\delta = \delta_2 \delta_3 / \delta_1$,
we may write for $\eta_{\widetilde{G}}^0$
\begin{align}
\eta_{\widetilde{G}}^0 = \delta \, \frac{g_{*,s}^0}{g_{*,s}}
\frac{\gamma_{\widetilde{G}}(a_{\textrm{RH}})}
{n_\gamma(a_{\textrm{RH}})H(a_{\textrm{RH}})}\,,
\label{eq:etaG}
\end{align}
where $n_\gamma(a_{\textrm{RH}})$, $\gamma_{\widetilde{G}}(a_{\textrm{RH}})$
and $H(a_{\textrm{RH}})$ directly follow from
Eqs.~\eqref{eq:nrhoBFeq}, \eqref{eq:GammaG} and the Friedmann equation
(cf.\ Sec.~\ref{subsec:TRH}).
Inserting Eq.~\eqref{eq:etaG} back into Eq.~\eqref{eq:OmegaG},
we find for $\Omega_{\widetilde{G}}^0h^2$
\begin{align}
\Omega_{\widetilde{G}}^0h^2 = \varepsilon 
f_{\widetilde{G}}(T_{\textrm{RH}}) \left(m_{\widetilde{G}} +
\frac{m_{\tilde{g}}^2 (T_{\textrm{RH}})}{3 m_{\widetilde{G}}}\right) T_{\textrm{RH}}\,,
\quad \varepsilon = \alpha^{-1/2} \beta^{-1} \delta\,,
\label{eq:OmegaGRes}
\end{align}
where $f_{\widetilde{G}}(T_{\textrm{RH}})$ stands for
\begin{align}
f_{\widetilde{G}}(T_{\textrm{RH}}) = \frac{n_\gamma^0}{\rho_c/h^2}
\frac{g_{*,s}^0}{g_{*,s}} \left(\frac{90}{8 \pi^3 g_{*,\rho}}\right)^{1/2}
\frac{54 \,g_s^2(T_{\textrm{RH}})}{g_\gamma M_p} \left[\ln\left(\frac{T_{\textrm{RH}}^2}
{m_g^2(T_{\textrm{RH}})}\right) + 0.8846 \right]\,.
\end{align}
Eq.~\eqref{eq:OmegaGRes} may conveniently be rewritten as
\begin{align}
\Omega_{\widetilde{G}} h^2 = \varepsilon \, C_1
\left(\frac{T_{\textrm{RH}}}{10^{10}\,\textrm{GeV}}\right) \left[
C_2\left(\frac{m_{\widetilde{G}}}{100\,\textrm{GeV}}\right)
+ \Big(\frac{100\,\textrm{GeV}}{m_{\widetilde{G}}}\Big)
\left(\frac{m_{\tilde{g}}}{1\,\textrm{TeV}}\right)^2\right] \,.
\label{eq:OmegaGRes2}
\end{align}
with $C_1$ and $C_2$ being defined as
\begin{align}
C_1 = & \: 10^{14}\,\textrm{GeV}^2 \,\frac{n_\gamma^0}{\rho_c/h^2}
\frac{g_{*,s}^0}{g_{*,s}} \left(\frac{90}{8\pi^3 g_{*,\rho}}\right)^{1/2}
\frac{18 g_s^6\left(T_{\textrm{RH}}\right)}{g_\gamma g_s^4\left(\mu_0\right) M_P} \left[
\log\left(\frac{T_{\textrm{RH}}^2}{m_g^2\left(T_{\textrm{RH}}\right)}\right) + 0.8846\right] \,,
\nonumber \\ \label{eq:C2def}
C_2 = & \: \frac{3 g_s^4\left(\mu_0\right)}{100 g_s^4\left(T_{\textrm{RH}}\right)}\,.
\end{align}
The expressions for $\Omega_{\widetilde{G}} h^2$, $C_1$ and $C_2$ in Eqs.~\eqref{eq:OmegaGRes2}
and \eqref{eq:C2def} are exactly those which we employ in our analysis in
Sec.~\ref{subsec:gravitinoDM}.
A fit formula for the correction factor $\varepsilon$, which we are not able to
determine analytically, is provided in Eq.~\eqref{eq:epsilonfit}.
The dependence of $C_1$ and $C_2$ on the reheating
temperature is presented in Fig.~\ref{fig:C1C2coefficients}.
We find that $C_2 \ll 1$, which means that for $m_{\widetilde{G}} \ll m_{\tilde{g}}$
the term linear in $m_{\widetilde{G}}$ in Eq.~\eqref{eq:OmegaGRes2} can usually
be neglected.
Notice that doing so and setting $\varepsilon = 1$
turns Eq.~\eqref{eq:OmegaGRes2} into Eq.~\eqref{eq:OmegaGth2Approx}.


Finally, our results may be easily generalized to gluino masses other than
$1\,\textrm{TeV}$.
In fact, for given values of $\widetilde{m}_1$, $M_1$
and $m_{\widetilde{G}}$, it is possible to keep
$\eta_B^0$ and $\Omega_{\widetilde{G}}^0 h^2$ constant when changing $m_{\tilde{g}}$
by simply rescaling the gravitino mass,
\begin{align}
m_{\widetilde{G}}^0 \rightarrow m_{\widetilde{G}} =
m_{\widetilde{G}}\left(m_{\tilde{g}},m_{\widetilde{G}}^0\right)\,, \qquad
m_{\widetilde{G}}\left(1\,\textrm{TeV},m_{\widetilde{G}}^0\right) = m_{\widetilde{G}}^0\,.
\end{align}
As for the baryon asymmetry, this is a trivial consequence
of the fact that $\eta_B^0$ is a function of $\widetilde{m}_1$
and $M_1$ only.
In the case of the gravitino abundance, we observe that for
fixed reheating temperature,
$T_{\textrm{RH}} = T_{\textrm{RH}}\left(\widetilde{m}_1,M_1\right)$,
$\Omega_{\widetilde{G}}^0h^2$ remains constant
as long as $m_{\widetilde{G}}\big(m_{\tilde{g}},m_{\widetilde{G}}^0\big)$
is chosen such that the term in square brackets in
Eq.~\eqref{eq:OmegaGRes2} does not change,
\begin{align}
\left[C_2\left(\frac{m_{\widetilde{G}}^0}{100\,\textrm{GeV}}\right)
+ \left(\frac{100\,\textrm{GeV}}{m_{\widetilde{G}}^0}\right)\right] =
\left[C_2\bigg(\frac{m_{\widetilde{G}}}{100\,\textrm{GeV}}\bigg)
+ \left(\frac{100\,\textrm{GeV}}{m_{\widetilde{G}}}\right)
\left(\frac{m_{\tilde{g}}}{1\,\textrm{TeV}_{}}\right)^2\right]\,.
\label{eq:mgscaled}
\end{align}
From this condition, we can determine
the rescaled gravitino mass $m_{\widetilde{G}}$
as a function of the rescaled gluino mass $m_{\tilde{g}}$ and
the original gravitino mass $m_{\widetilde{G}}^0$.
As Eq.~\eqref{eq:mgscaled} is a quadratic equation in $m_{\widetilde{G}}$,
it generically has two solutions $m_{\widetilde{G}}^\pm$,
one of which is typically closer to the original gravitino mass than the other.
$m_{\widetilde{G}}^0$ lies right in between $m_{\widetilde{G}}^-$
and $m_{\widetilde{G}}^+$ once the two terms in square
brackets in Eq.~\eqref{eq:OmegaGRes2} are of equal size, i.e.\ when
gravitinos in helicity $\pm\frac{1}{2}$ states contribute exactly as much to
the total abundance as gravitinos in helicity $\pm\frac{3}{2}$ states.
One easily sees that this is the case when $m_{\widetilde{G}}^0 \simeq 280\,\textrm{GeV}$
(cf. Eq.~\eqref{eq:mGt280}).
When going to values of $m_{\tilde{g}}$ larger than $1\,\textrm{TeV}$, we have
$m_{\widetilde{G}}^0 \gtrsim m_{\widetilde{G}}^+ \gg m_{\widetilde{G}}^-$
above $280\,\textrm{GeV}$ and $m_{\widetilde{G}}^0 \lesssim m_{\widetilde{G}}^- \ll m_{\widetilde{G}}^+$
below $280\,\textrm{GeV}$.
At $m_{\tilde{g}}$ smaller than $1\,\textrm{TeV}$, we always
find $m_{\widetilde{G}}^- < m_{\widetilde{G}}^0 < m_{\widetilde{G}}^+$.


\begin{figure}
\begin{center}
\includegraphics[width=8.25cm]{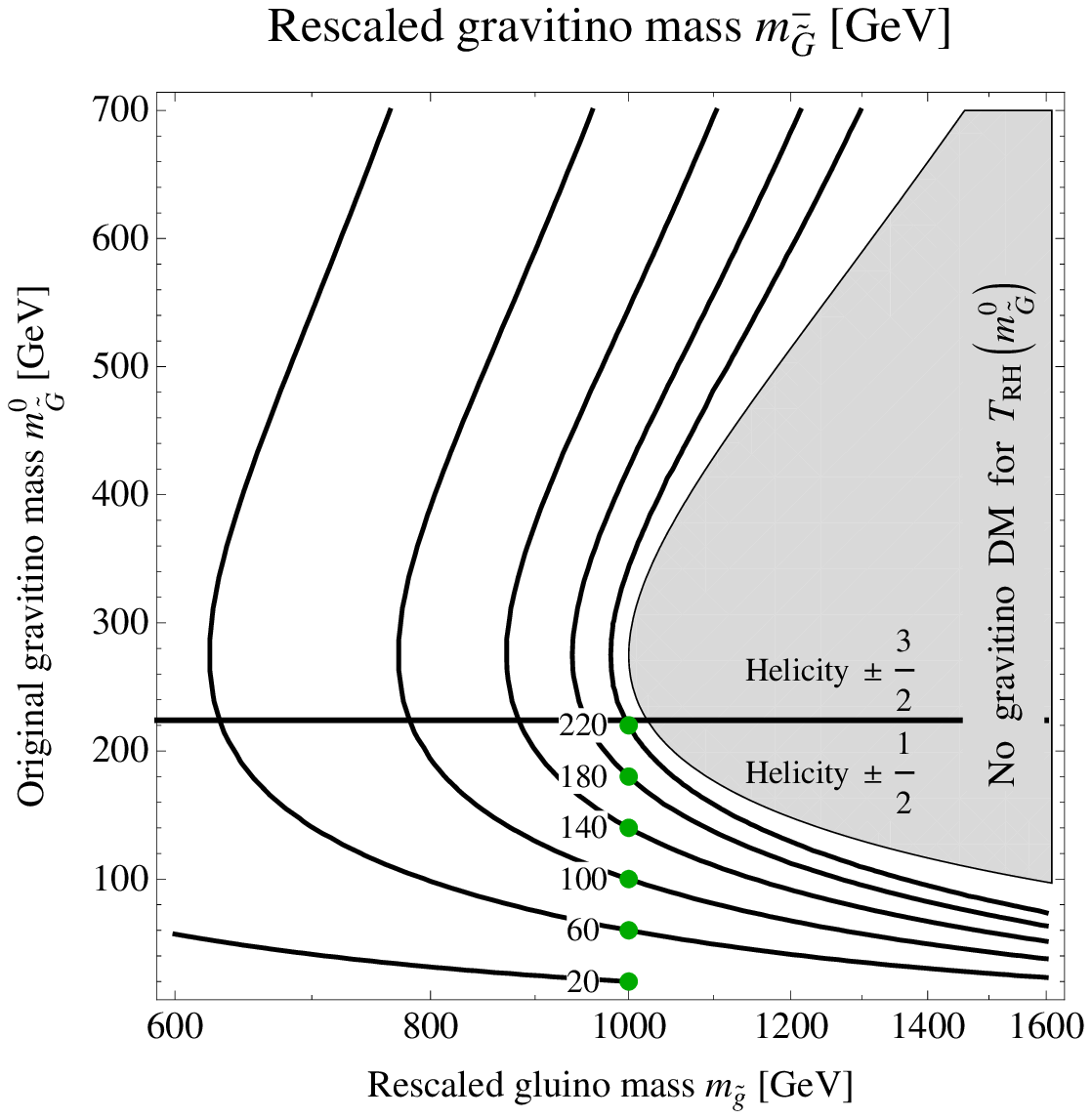}

\vspace{5mm}

\includegraphics[width=8.25cm]{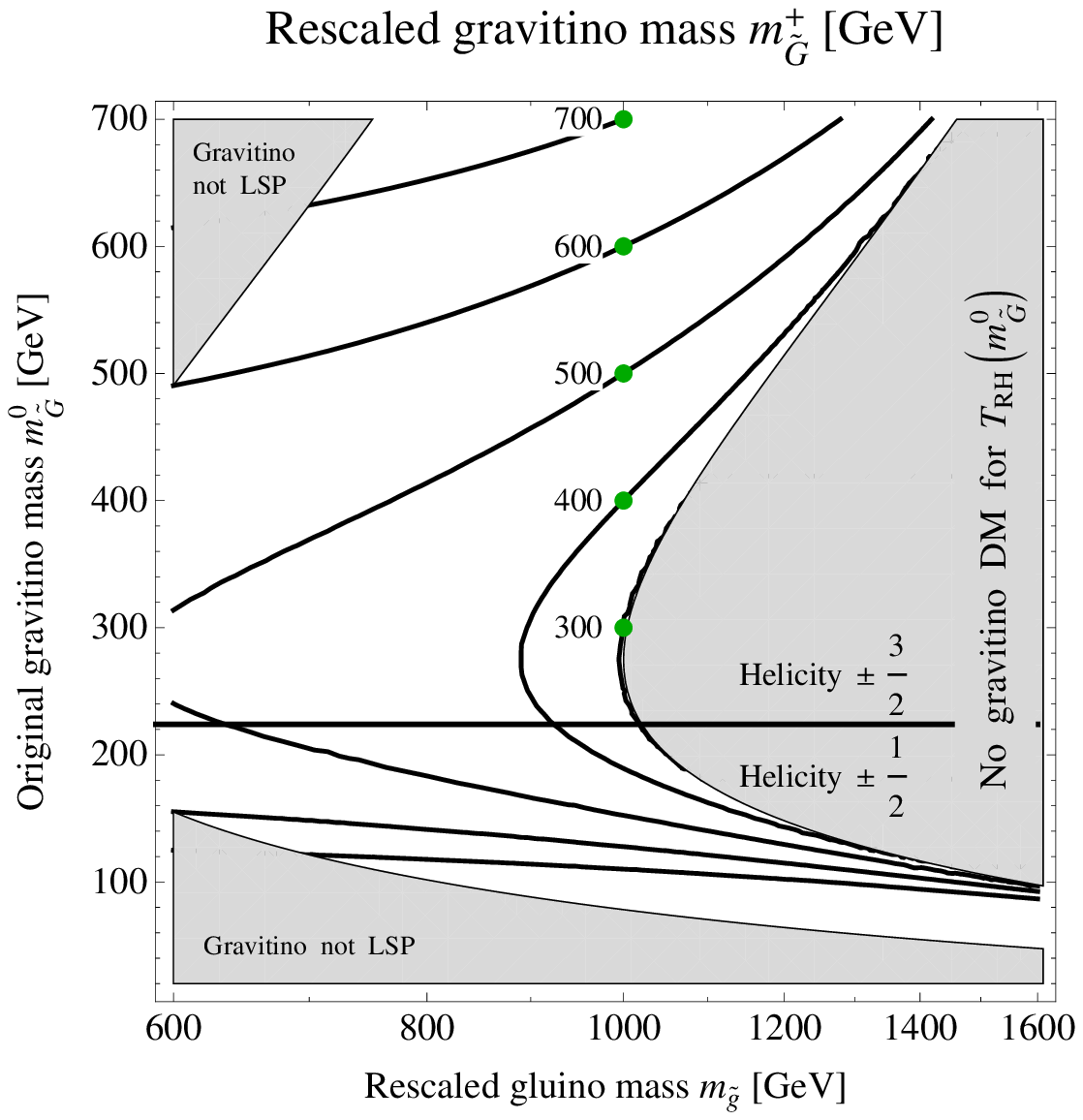}

\vspace{5mm}

\caption[Rescaling prescription for the gravitino mass]
{Contour plots of the two solutions $\big(m_{\widetilde{G}}^\pm\big)$
of Eq.~\eqref{eq:mgscaled}
for the rescaled gravitino mass $m_{\widetilde{G}}$ as a function of the
rescaled gluino mass $m_{\tilde{g}}$ and the 
original gravitino mass $m_{\widetilde{G}}^0$.
The black solid contours correspond to constant values of $m_{\widetilde{G}}$
(given next to the green dots).
They serve as level curves that allow a determination
of $m_{\widetilde{G}}$ for arbitrary points in the $\big(m_{\tilde{g}},m_{\widetilde{G}}^0\big)$-plane.
They can also be regarded as function graphs of $m_{\widetilde{G}}^0$
as a function of $m_{\tilde{g}}$ for constant $m_{\widetilde{G}}$.
We restrict ourselves to the interval
$20\,\textrm{GeV} \leq m_{\widetilde{G}}^0 \leq 700\,\textrm{GeV}$
in this figure.
Below $20\,\textrm{GeV}$, Eq.~\eqref{eq:naivesc} provides
an excellent approximation.
\label{fig:mGscaled}}
\end{center}
\end{figure}


If the gravitino mass is much smaller than the gluino mass,
almost only the goldstino part of the gravitino is produced
and the term linear in $m_{\widetilde{G}}$ in Eq.~\eqref{eq:OmegaGRes2}
can be neglected.
The scaling behaviour of the gravitino mass then becomes trivial,
\begin{align}
m_{\widetilde{G}}^0 \ll m_{\tilde{g}}\:: \qquad
m_{\widetilde{G}} =
m_{\widetilde{G}}^0 \left(\frac{m_{\tilde{g}}}{1\,\textrm{TeV}}\right)^2\,.
\label{eq:naivesc}
\end{align}

\newpage


Actually, the rescaled gravitino mass $m_{\widetilde{G}}$ also is a function
of $T_{\textrm{RH}}$, as it depends on the coefficient $C_2(T_{\textrm{RH}})$.
As apparent from Fig.~\ref{fig:mGbounds} and Eq.~\eqref{eq:TRHm1tmGfit}, this
dependence on $T_{\textrm{RH}}$ directly translates into a dependence on the
effective neutrino mass $\widetilde{m}_1$.
In order to solve Eq.~\eqref{eq:mgscaled}, we set $\widetilde{m}_1$
to $0.04\,\textrm{eV}$ and compute
$T_{\textrm{RH}}$ according to Eq.~\eqref{eq:TRHm1tmGfit} as a function of the input gravitino
mass, $T_{\textrm{RH}} = T_{\textrm{RH}}\big(m_{\widetilde{G}}^0\big)$.
Our solutions $m_{\widetilde{G}}^\pm$ for the rescaled gravitino mass are
presented in the two panels of Fig.~\ref{fig:mGscaled}, respectively.
In the gray shaded regions, there are either no real solutions of
Eq.~\eqref{eq:mgscaled} or the rescaled gravitino mass is larger
than the corresponding gluino mass, $m_{\widetilde{G}} > m_{\tilde{g}}$.
The former case implies that it is impossible to keep the gravitino
abundance constant, when going to larger $m_{\tilde{g}}$
while sticking to the reheating temperature $T_{\textrm{RH}}\big(m_{\widetilde{G}}^0\big)$.
In the latter case, the gravitino would not be the LSP any longer.

\cleardoublepage
 \end{appendix}
 %

\phantomsection
\addcontentsline{toc}{chapter}{Bibliography}
\bibliography{bib}
\cleardoublepage

\end{document}